# On the Backreaction of Scalar and Spinor Quantum Fields in Curved Spacetimes

vorgelegt von
Thomas-Paul Hack
aus Timisoara

Hamburg
2010

# Zusammenfassung

In der vorliegenden Arbeit werden zunächst einige Konstruktionen und Resultate in Quantenfeldtheorie auf gekrümmten Raumzeiten, die bisher nur für das Klein-Gordon Feld behandelt und erlangt worden sind, für Dirac Felder verallgemeinert. Es wird im Rahmen des algebraischen Zugangs die erweiterte Algebra der Observablen konstruiert, die insbesondere normalgeordnete Wickpolynome des Diracfeldes enthält. Anschließend wird ein ausgezeichnetes Element dieser erweiterten Algebra, der Energie-Impuls Tensor, analysiert. Unter Zuhilfenahme ausführlicher Berechnungen der Hadamardkoeffizienten des Diracfeldes wird gezeigt, dass eine lokale, kovariante und kovariant erhaltene Konstruktion des Energie-Impuls Tensors möglich ist. Anschließend wird das Verhältnis der mathematisch fundierten Hadamardregularisierung des Energie-Impuls Tensors mit der mathematisch weniger rigorosen DeWitt-Schwinger Regularisierung verglichen. Man findet, dass die beiden Regularisierungen im wesentlichen äquivalent sind, insbesondere lässt sich die DeWitt-Schwinger Regularisierung mathematisch exakt formulieren. Während die bisher angeführten Resultate auf allgemeinen gekrümmten Raumzeiten gültig sind, werden zusätzliche Untersuchungen auf einer Klasse von flachen Robertson-Walker Raumzeiten, die eine lichtartige Urknallhyperfläche besitzen, angestellt. Mit hilfe holographischer Methoden werden auf solchen Raumzeiten Hadamardzustände für das Klein-Gordon- und Diracfeld konstruiert, die dadurch ausgezeichnet sind, dass sie im Sinne eines Limes zur Urknallhyperfläche asymptotische Gleichgewichtszustände darstellen. Abschließend werden Lösungen der semiklassischen Einsteingleichungen für Quantenfelder mit beliebigem Spin im flachen Robertson-Walker Fall untersucht. Es stellt sich heraus, dass die gefundene Lösungen Supernova Typ Ia Messdaten ebenso gut wie das $\Lambda$CDM Modell erklären. Damit ist eine natürliche Erklärung für dunkle Energie und ein einfaches Modell für kosmologische dunkle Materie gefunden.

# Abstract

In the first instance, the present work is concerned with generalising constructions and results in quantum field theory on curved spacetimes from the well-known case of the Klein-Gordon field to Dirac fields. To this end, the enlarged algebra of observables of the Dirac field is constructed in the algebraic framework. This algebra contains normal-ordered Wick polynomials in particular, and an extended analysis of one of its elements, the stress-energy tensor, is performed. Based on detailed calculations of the Hadamard coefficients of the Dirac field, it is found that a local, covariant, and covariantly conserved construction of the stress-energy tensor is possible. Additionally, the mathematically sound Hadamard regularisation prescription of the stress-energy tensor is compared to the mathematically less rigorous DeWitt-Schwinger regularisation. It is found that both prescriptions are essentially equivalent, particularly, it turns out to be possible to formulate the DeWitt-Schwinger prescription in a well-defined way. While the aforementioned results hold in generic curved spacetimes, particular attention is also devoted to a specific class of Robertson-Walker spacetimes with a lightlike Big Bang hypersurface. Employing holographic methods, Hadamard states for the Klein-Gordon and the Dirac field are constructed. These states are preferred in the sense that they constitute asymptotic equilibrium states in the limit to the Big Bang hypersurface. Finally, solutions of the semiclassical Einstein equation for quantum fields of arbitrary spin are analysed in the flat Robertson-Walker case. One finds that these solutions explain the measured supernova Ia data as good as the $\Lambda$CDM model. Hence, one arrives at a natural explanation of dark energy and a simple quantum model of cosmological dark matter.

*It seems plain and self-evident, yet it needs to be said: the isolated knowledge obtained by a group of specialists in a narrow field has in itself no value whatsoever, but only in its synthesis with all the rest of knowledge and only inasmuch as it really contributes in this synthesis toward answering the demand τίνες δὲ ἡμεῖζ; ('who are we?')*

Erwin Schrödinger, [Sch96, p. 109]

# Contents

# Introduction

Quantum field theory in curved spacetimes is a framework in which the spacetime is treated as a curved Lorentzian manifold in accord with General Relativity, whereas matter is treated as a quantum field. This framework is expected to provide a semiclassical approximation to a full quantum theory of both gravity and matter, and, hence, finds its applications in cases where the spacetime curvature is low enough for quantum gravity effects to be negligible, but too large for Minkowskian quantum field theory to make sense.

Well, these are the words one often finds as introductory phrases in works on quantum field theory in curved spacetimes. Unfortunately, they somehow suggest that quantum field theory on curved spacetimes ultimately needs to be legitimated by a full-fledged quantum gravity theory. Although this is certainly as correct as it is already for flat spacetime quantum field theory, this point of view seems to disregard a maybe obvious fact. Namely, we inevitably live in a curved spacetime, although our local curvature is quite low. Hence, if one wants to treat matter as quantum fields in our whole universe at all, then one is forced to consider quantum field theory in curved spacetime. Therefore, quantum field theory on curved spacetimes is more fundamental than quantum field theory on Minkowski spacetime, the latter being undoubtedly a good local approximation in cases where the relevant energy scales are much larger than the curvature scales. Thus, whichever new lessons quantum field theory on curved space teaches us, we should take them serious and should not raise an eyebrow if they seem to contradict our Minkowskian intuition.

Two situations in which the curvature of the spacetime can certainly not be neglected are the early universe and the vicinity of black holes. In fact, quantum field theory in curved spacetimes is certainly best known for Hawking's famous discovery of black hole radiation [Haw75] (see also [Wal95], and [Wal06] for a historical account) and this phenomenon is thought to teach us a lot about a possible theory of quantum gravity. Hence, one has tried to reproduce it in two candidate theories of quantum gravity, namely, loop quantum gravity and string theory. Regarding the early universe, the current paradigm in inflationary cosmology is that the quantum fluctuations of a classical field are responsible for the structure formation in our universe (see, *e.g.* [Dod03]), and so it seems that quantum field theory in curved spacetimes has found its applications in both scenarios where one considers it to have non-trivial effects at first glance. However, dwelling further on the cosmological applications, one should not only be concerned with the *local* effects of quantum fields (which are thought to be 'stretched' to global classical effects in the inflationary scenario), but also their inherent *global* effects. Namely, the globally distributed energy density of a quantum field in our universe could have effects both in the low and in the high curvature regimes. In this respect, let us consider our more recent cosmological history. It seems that it is in principle widely accepted that the currently observed cosmological constant is possibly the 'dark' vacuum energy of a quantum field, however, we also have the impression that only very few serious attempts to investigate this question have been taken, see [PaRa99, PaRa01] and references therein. Indeed, one often reads statements such as 'quantum field theory predicts a Planck scale, and, hence, way too large cosmological constant'. We strongly disagree. The above statement is a result of arguing with Minkowskian quantum field theory and not taking the more fundamental quantum field theory in curved spacetimes serious. Namely, one of the lessons quantum field theory in curved spacetimes teaches us, is that renormalisation ambiguities occur already on the level of free fields [Wal95, HoWa01]. Hence, a free quantum field theory in curved spacetimes has free parameters in addition to the parameters of the theory entering the Lagrangean. As a result, *quantum field theory does not predict a large cosmological constant whatsoever.* In fact, it does not predict any cosmological constant, but one needs to fix the renormalisation freedom of the quantum energy density and other observables by comparison with experiment. Note that this is by no means related to any anthropic principle. In contrast, it is conceptually the same as measuring any parameter of the standard model of particle physics.

Regarding potential global effects of quantum fields in the early universe, it seems that these are not considered at all. In inflationary cosmology, the cosmological evolution is thought to be dominated by one (or more) classical fields, and quantum field theoretical effects play only a secondary role. This is to some extent astonishing, because, in the context of flat spacetimes, there is undoubtedly a large agreement that quantum fields are more fundamental than classical fields. Moreover, the first inflationary model ever proposed was the purely quantum field theoretical model of Starobinsky [Sta80]. Notwithstanding, it seems that the model has not been very convincing, and also recent papers appearing on the subject, see for instance [Kok09], draw the conclusion that (even a generalisation of) Starobinsky's model does not have much to say about the cosmological reality. We hope to make it clear here and in the following that the somehow ancillary role of quantum field theory on curved spacetimes in the present understanding of our universe is not due to the weakness and triviality of the theory, but due to the fact that its full implications have never been properly investigated. In the currently preferred scenario of inflationary cosmology, the direct role of quantum fields on the global evolution of the universe is apparently discarded at all, whereas in Starobinsky's original model and most of its further analyses, one has only considered the effect of the so-called trace anomaly. As we will discuss in more detail in chapter IV of our thesis, this is an undoubtedly important component of the vacuum energy density of a quantum field, but by far not the only one. In fact, we shall see that, once the full possibilities and properties of quantum field theory on curved spacetime are taken into account, one finds that *both the rapid expansion of the early universe and its late time acceleration can naturally be explained from first principles within quantum field theory in curved spacetimes.* We will even find that the thermal energy density of curved spacetime quantum fields can model the cosmological effects of (dark) classical matter. These statements will both be derived in general and successfully compared with experimental data in particular.

After these rather general and foundational paragraphs, let us go into details. In Minkowski spacetime, quantum field theoretical treatments are mostly based on the Fock space, or equivalently, particle picture, related to the unique, Poincaré-invariant Minkowski vacuum. However, a general curved spacetime does not have any symmetries at all, and, hence, no unique vacuum state exists. Note that this holds even in the highly symmetric case of Robertson-Walker spacetimes. The lack of time translation symmetry due to the time variations of a generic scale factor $a(t)$ permit the existence of any preferred global state, though, locally, one might consider the Minkowskian situation, as we implicitly do by probing nature with particle accelerators. Since we have no preferred state and, hence, no unique notion of 'particles' at hand, it seems sensible to try to formulate quantum field theory in terms of fields and without any initial reference to a state. In fact, one does not only have to dispose of a preferred state, but also of a Hilbert space in general. This stems from the fact that a Hilbert space constitutes a representation space of a quantum field algebra. But, as a quantum field has infinitely many degrees of freedom, different Hilbert space representations are in general not related by unitary equivalence [Haa92, Wal95]; a Hilbert space is, hence, a secondary ingredient of any quantum field theory. The lack of a preferred state and particle picture suggests that a quantum field theory should be formulated in terms of fields and without having recourse to a vacuum state. Hence, we shall work within the *algebraic approach* to quantum field theory [Haa92, Wal95], and we claim that this way to formulate quantum field theory is the only one that can grasp all features of quantum field theory in curved spacetimes.

The algebraic construction of quantum field theory on curved spacetimes is by now completely understood in conceptual terms, at least as far perturbative constructions of interacting quantum field theories are concerned [BFK95, BrFr00, HoWa01, HoWa02, HoWa03, HoWa05, HoRu01, Hol07, ChFr08], see also the recent introductory review in [BrFr09]. To understand the essential concepts, it is advisable to investigate the easiest field model, and, hence, most conceptual works have discussed the Hermitean scalar field. Let us briefly review how a Klein-Gordon field $\phi$ propagating on a curved spacetime $M$ is treated in the algebraic framework. The starting point is the basic free field algebra $\mathscr{A}(M)$, which consists of sums of products of elements like $\phi(f)$. Here, $f$ is a compactly supported, complex-valued smooth function, and it is interpreted as localising the observable $\phi(x)$. It is required that $\phi(x)$ satisfies the Klein-Gordon equation $P\phi(x) = (-\Box + m^2 + \xi R)\phi(x)$ in the sense $\phi(Pf) = 0$, and that it fulfils canonical commutation relations $[\phi(x), \phi(y)] = i\Delta(x, y)$, where $\Delta(x, y)$ is the unique *causal propagator* or *commutator function* of the Klein-Gordon operator $P$. To have a notion of 'taking the adjoint', one introduces a $*$-operation given as $(\phi(f)\phi(g))^* = \phi(g)^*\phi(f)^* = \phi(\overline{g})\phi(\overline{f})$. A state $\omega$ on $\mathscr{A}(M)$ is a linear functional $\omega : \mathscr{A}(M) \to \mathbb{C}$, which is positive and normalised, namely, $\omega(A^*A) \geq 0$ for all $A \in \mathscr{A}(M)$, and $\omega(\mathbb{I}) = 1$, where $\mathbb{I}$ is the unit element in $\mathscr{A}(M)$. In this context, $\omega(A)$ for $A \in \mathscr{A}(M)$ has the physical interpretation of being the expectation value of $A$. Given an algebraic state $\omega$, one obtains a canonical representation on a Hilbert space via the *GNS construction* [Haa92]. Moreover, given a Hilbert space with the Klein-Gordon field realised as an operator (valued distribution) on it, the algebra constituted by these operators together with a normalised Hilbert space state are naturally a field algebra and a state in the abstract sense. Finally, an algebraic state $\omega$ on $\mathscr{A}(M)$ is uniquely determined, once all its $n$-point functions $\omega_n(x_1, \cdots, x_n) = \omega(\phi(x_1) \cdots \phi(x_n))$ are known.

The algebra $\mathscr{A}(M)$ contains only basic fields and no Wick polynomials. To obtain the latter, it is necessary to have a notion of normal ordering. However, as we would like to construct everything independent of a state, we also need to consider a way to formulate normal ordering in a way which is state-independent. Moreover, in Minkowski spacetime, normal ordering removes the basic UV divergences of a quantum field theory in a consistent way. Namely, normal ordered Wick polynomials such as $:\phi^2(x):$ can be multiplied in a meaningful way, and the result is given by the famous Wick theorem. Hence, Wick polynomials in Minkowski spacetime constitute an algebra themselves. To treat interacting field theories on curved spacetimes, we therefore need to provide an algebra $\mathscr{W}(M)$ of Wick polynomials also in the curved context. This has been achieved only at a relatively recent date, and the main pillar of the construction of $\mathscr{W}(M)$ are *Hadamard states*. These are states that possess the same UV-singularity structure like the Minkowski vacuum state, and they can be characterised in two ways. On the one hand, they have a very specific functional behaviour. Namely, their two-point function $\omega_2(x, y)$ is, for $x$ and $y$ close to each other, given as

$$\omega_2(x, y) = \frac{1}{8\pi^2} \left( \frac{u(x, y)}{\sigma(x, y)} + v \log(\sigma(x, y)) + w(x, y) \right) = \frac{1}{8\pi^2} \left( h(x, y) + w(x, y) \right),$$

where $\sigma(x, y)$ is one half the squared geodesic distance, and $u$, $v$, $w$ are smooth functions. In this functional form, the UV divergences of $\omega_2(x, y)$ are clearly visible, and they are completely contained in $h(x, y)$. One can compare the above form with the massless Minkowskian vacuum state, whose two-point function is given as

$$\omega_2^{\text{Mink}}(x, y) = \frac{1}{8\pi^2} \frac{1}{(x - y)^2}.$$

It turns out that $u$, $v$, and, hence, $h(x, y)$ are completely specified by the parameters in the Klein-Gordon operator $P$ and the local curvature in the neighbourhood of the points $x$ and $y$. Hence, $h$ is completely state-independent, and two Hadamard states differ only by the regular part $w$. The reason why normal ordering works in Minkowski spacetime is the UV singularity structure of the Minkowskian vacuum state. Hence, it seems that, taking into account that $h$ has the same singularity structure as the Minkowski vacuum and is in addition state-independent, it should be a good candidate for the definition of normal ordering. However, to fully understand this, one needs the mentioned second characterisation of Hadamard states, and this is given in terms of a local Fourier spectrum condition, the so-called *wave front set* $WF(h)$ of a distribution $h$. Essentially, this object stores both the singular points in position space, and the 'singular directions' in momentum space of $h$. In fact, as shown by Radzikowski in his seminal paper [Rad96a], Hadamard states do not only look like the Minkowski vacuum state in their functional form, but also have the same wave front set, and, hence, really the same singularity structure. Obviously, the same holds also for $h(x, y)$, as it is $\omega_2$ minus a regular term. Based on the work of Radzikowski, the Wick polynomial algebra has been constructed in [BFK95, BrFr00, HoWa01, HoWa05]. It turns out that this is most efficiently done by directly encoding the Wick theorem as an algebra product. This *deformation quantization* approach has been introduced in [DüFr01a] and fully developed in [BDF09]. As is well-known, the Wick theorem says that a product of normal-ordered quantities like, *e.g.*

$$:\phi^2(x)::\phi^2(y):$$

is given in terms of other Wick polynomials and contractions, which in the Minkowskian case are contractions with respect to the vacuum state two-point function $\omega_2^{\text{Mink}}$. However, to obtain a normal ordering which is state independent, one defines a *$\star$-product* $\star_h$ which encodes the Wick theorem contractions with $h$ on an abstract algebraic level. As a result, the quantities in an algebra endowed with the product $\star_h$ have the interpretation to be regularised by means of $h$ in a way implicitly given by Wick's theorem. Moreover, the power of the knowledge of the wave front set and the related mathematical branch of *microlocal analysis* allow one to actually compute on an abstract level that all contractions are in fact well-defined in a mathematical sense. In addition, microlocal analysis does not only reveal that the contractions of basic Wick polynomials with $h$ are well-defined, but also tells us how 'large' the Wick polynomial algebra $\mathscr{W}(M)$ can be, in the sense that one knows which field quantities in addition to simple Wick polynomials can be contracted meaningfully with $h$. In this respect, one finds that $\mathscr{W}(M)$ also contains time-ordered products of Wick polynomials and, hence, already allows one to construct perturbative interacting quantum field theories on curved spacetimes [BrFr00, HoWa01]. Moreover, it turns out that quantum field theories which are perturbatively renormalisable in Minkowski spacetime retain this property in curved spacetimes [BrFr00, HoWa01, Hol07]. However, as already anticipated, a fundamentally new phenomenon arises. Namely, it turns out that perturbative interacting quantum field theories have, in addition to their renormalisation degrees of freedom known from Minkowski spacetime, new degrees of freedom in terms of curvature tensors. While this may not be so surprising for the quantities of the interacting theory like time-ordered products, it turns out that this freedom is already present on the level of quantities of the free theory, namely, for Wick polynomials such as $:\phi^2(x):$. In fact, if one takes a fundamental point of view and defines objects like $:\phi^2(x):$ by physically sensible requirements, like locality, scaling behaviour, commutation relations with the single free field $\phi$, covariance, and others, then it turns out that an object like $:\phi^2(x):$ is not defined uniquely, but two possible incarnations of $:\phi^2(x):$ differ as

$$\widetilde{:\phi^2(x):} := :\phi^2(x): + \left(\alpha R + \beta m^2\right)\mathbb{I},$$

where $\alpha$ and $\beta$ are dimensionless constants [HoWa01]. Here, locality and covariance are the most important requirements. The former demands that a Wick polynomial like $:\phi^2(x):$ is only constructed by means of quantities which are completely determined by the knowledge of the spacetime in a small neighbourhood of $x$; this is particularly fulfilled if we define $:\phi^2(x):$ by a subtraction of the Hadamard singularity $h$. Moreover, covariance essentially asks for $:\phi^2(x):$ to transform like a scalar in the sense of General Relativity. The above discussion may seem to suggest that one can dispose completely of states, but this is of course not the case. If one wants to make contact with observations, one needs to consider expectation values of elements of the Wick polynomial algebra $\mathscr{W}(M)$, and then Hadamard states are essential because only states which have the same singularity structure like $h$ can give regular expectation values of elements in $\mathscr{W}(M)$. As an example, the expectation value $\omega(:\phi^2(x):)$ of $:\phi^2(x):$ in a Hadamard state $\omega$ reads

$$\omega(:\phi^2(x):) = \lim_{x \to y}(\omega_2(x,y) - h(x,y)) + \alpha R + \beta m^2.$$

The brief account of algebraic quantum field theory given above calls for two lines of further research. One the one hand, the full explicit perturbative construction of an interacting quantum field theory had only been given for the Hermitean scalar field to date, and, although it has been in principle clear how one should proceed, the extended Wick polynomial algebra of, say, the Dirac field had not been constructed yet explicitly. On the other hand, it is clear that a sensible quantum field theory model on curved spacetimes requires the knowledge of explicit examples of Hadamard states. In this thesis, we seek to add to both aforementioned lines of research. First, we construct the extended algebra of observables of the Dirac field $\mathscr{W}(DM)$ via a deformation quantization approach. To achieve such a task, one starts from the algebra of classical observables of the Dirac field, which is constituted by antisymmetric compactly supported and smooth test sections of the Dirac bundles and is endowed with a classical anticommuting product given in terms of an antisymmetrised tensor product $\cdot_a$. The basic algebra of quantum observables $\mathscr{A}(DM)$ is obtained from the algebra of classical observables by replacing the anticommuting classical product $\cdot_a$ with a quantum product $\star_S$ constituted by contractions with the unique *causal propagator* or *anticommutator function* $S$ of the Dirac operator $D$. Based on the equivalence of the functional form of Hadamard states on $\mathscr{A}(DM)$ and their wave front set, which has been known due to the works of [Köh95, Kra00, Hol99, SaVe01, San08, San09a], we provide a construction of the extended algebra of quantum observables $\mathscr{W}(DM)$ by first supplementing the smooth test sections constituting $\mathscr{A}(DM)$ with compactly supported test distributions with a specific wave front set and then deforming the product $\star_S$ to a product $\star_H$ provided by a bidistribution $H$ having the Hadamard singularity. We argue that the resulting algebra product $\star_H$ has good properties, namely, the algebra $\mathscr{W}(DM)$ is closed with respect to $\star_H$, $\star_H$ enjoys specific continuity properties, and it encodes the Wick theorem known from Fermionic flat spacetime quantum field theory in a state-independent way. Moreover, extending the results of [HoRu01], it is argued that Hadamard states on $\mathscr{A}(DM)$ can be naturally and meaningfully extended to the larger algebra $\mathscr{W}(DM)$. Finally, by employing only the singular part $H$ of any Hadamard state on $\mathscr{A}(DM)$ in the construction of $\mathscr{W}(DM)$, one obtains Wick polynomials of the Dirac field which are local and covariant objects.

Subsequently, the briefly anticipated construction of explicit examples of Hadamard states will be performed for a special class of flat Robertson-Walker spacetimes, whose scale factor $a(\tau)$ given in terms of the *conformal time* $\tau$ is such that it is exponentially vanishing towards $\tau \to -\infty$. In terms of the *cosmological time* $t$, this implies that $a(t)$ shows a *power-law inflationary behaviour* at early times, and possesses a *Big Bang singularity* at a finite time $t_0$ in particular. As is well-known [LuMa85], power-law inflation models solve the horizon problem of the *standard model of cosmology* by having an infinite *particle horizon*, which, in more geometrical terms, entails that the Big Bang equal-time hypersurface is a *lightlike* or *null* surface. These features lead us to term these specific flat Robertson-Walker spacetimes *null Big Bang spacetimes* (NBB). The relevance of NBB spacetimes in the context of quantum field theory on curved spacetime is provided by the observation [DFP08, Pin10] that these spacetimes arise naturally as solutions of the *semiclassical Einstein equation*

$$G_{\mu\nu}(x) = 8\pi G \omega(:T_{\mu\nu}(x):),$$

which describes the effect of quantum matter-energy on the curvature of spacetime. On conceptual grounds, the geometric features of null Big Bang spacetimes make it possible to construct preferred states by employing holographic methods. Namely, as realised in [Pin10], the Big Bang hypersurface of an NBB spacetime has the same geometric properties as the *conformal boundary* of an asymptotically flat spacetime. Hence, the bulk-to-boundary methods successfully developed and applied in [DMP06, DMP09a, DMP09b, DMP09c, Mor06, Mor08] can be naturally extended to the case of null Big Bang spacetimes, and in [Pin10] this has been done by constructing a preferred Hadamard state for the quantized Klein-Gordon field. The underlying observation in this construction is the following. Being a flat Robertson-Walker spacetime, an NBB spacetime is not time-translation-invariant, but it possesses a conformal Killing isometry provided by translations in conformal time $\tau$. As is well-known [BiDa82], this conformal isometry can be implemented in massless quantum field theoretical models to obtain conformally invariant theories and, hence, preferred conformal ground and thermal equilibrium states, where the latter are just appropriate conformal rescalings of ground and thermal equilibrium states in Minkowski. However, this is not possible in the case of massless fields, because the mass breaks the conformal invariance. Hence, neither preferred ground states, nor preferred equilibrium states have been known in the massive case. In fact, there has been no sensible and general notion of thermal equilibrium states at all for massive quantum fields. The crucial observation in quantum field theories on NBB spacetimes is that, in the limit towards the Big Bang, the conformal non-covariance of massive quantum fields disappears on account of the properties of $a(\tau)$, hence, massive quantum field theories are conformally invariant in an asymptotic sense. As a matter of fact, it is exactly this asymptotic conformal invariance which makes it possible to construct preferred states also in the case of massive fields. Based on this insight, the mentioned preferred Hadamard state for the massive Klein-Gordon field on NBB spacetimes has been constructed in [Pin10], and the natural physical interpretation of this state is that it is an *asymptotic conformal ground state*. In the present work, we extend this result by constructing preferred *asymptotic conformal equilibrium states* for the massive Klein-Gordon field on NBB spacetimes, and constructing both preferred asymptotic conformal ground states and preferred asymptotic equilibrium states for the massive Dirac field on such spacetimes. On the technical side, this construction requires to map the quantum field theory in an NBB spacetime to its Big Bang hypersurface, which is strictly speaking a conformal boundary and not a part of the spacetime itself. This conformal boundary is highly symmetric, particularly, it possesses a form of time-translation invariance which can be understood as a limiting case of the conformal translation invariance of the bulk NBB spacetime. Exploiting this time-translation invariance, one constructs unique ground states and equilibrium states for the boundary quantum field theory, and obtains the mentioned preferred states for the bulk quantum field theory by a suitable pull-back. Additionally, using the methods introduced in [Mor08] and already applied in [Pin10], we are able to prove that the two-point functions of the found preferred states possess a specific wave front set and are, hence, indeed Hadamard states.

We have already mentioned that the *backreaction*, *i.e.* the influence of the stress-energy of a quantum field on the curvature of the underlying background spacetime is described by the semiclassical Einstein equation. As already remarked in a conceptually clear way in [Wal77], the very form of the semiclassical Einstein equation imposes restrictions on its right hand side. Namely, in the equation under consideration, one equates a classical quantity – the Einstein tensor $G_{\mu\nu}(x)$ – with a probabilistic quantity, *i.e.* the expectation value of the regularised stress-energy tensor of a quantum field $\omega(:T_{\mu\nu}(x):)$. Hence, one has to assure that the right hand side has at least finite fluctuations, and, as realised in [Wal77], this is situation is given once the expectation value $\omega(:T_{\mu\nu}(x):)$ is taken with respect to a Hadamard state $\omega$. Further requirements on $\omega(:T_{\mu\nu}(x):)$ are the demand of a local and covariant dependence on the local curvature in the neighbourhood of $x$ and the covariant conservation of $\omega(:T_{\mu\nu}(x):)$, namely, $\nabla^{\mu}\omega(:T_{\mu\nu}(x):) = 0$. While the former requirement is motivated by the fact that one wants to describe the influence of $\omega(:T_{\mu\nu}(x):)$ on the local background geometry without knowing the global background geometry beforehand, the latter is a simple consistency condition, as $\nabla^{\mu}G_{\mu\nu}(x) = 0$ generally holds on account of the *Bianchi identities*. We have already remarked that normal ordering, *i.e.*, the definition of a Wick polynomials such as $:T_{\mu\nu}(x):$, is ambiguous in curved spacetimes, and one might hope that this ambiguity is large enough to fulfil the consistency conditions necessary for the expectation value $\omega(:T_{\mu\nu}(x):)$ appearing in the semiclassical Einstein equation. Indeed, as shown in [Wal77, Wal78a] this is the case for Klein-Gordon fields, and a conceptually improved version of Wald's construction has been given in [Mor03]. Again, these results have only been known for the case of scalar fields, and we thus extend them in the present thesis to the case of Dirac fields. The hereby obtained local, covariant, and covariantly conserved construction of a stress-energy tensor of the Dirac field requires on the technical side the detailed computation of the Dirac Hadamard coefficients. Finally, the just described procedure of the mathematically well-defined *Hadamard regularisation prescription* of $\omega(:T_{\mu\nu}(x):)$ is compared with the non-rigorous *DeWitt-Schwinger regularisation prescription* of the same object. It has been well-known that these two regularisation methods must be related in some way, but its seems that an explicit comparison has not been given in past works. We close this gap by proving that the DeWitt-Schwinger prescription can be formulated in a mathematically sound way by using the same concepts which assure the mathematical rigour of the Hadamard prescription.

The above mentioned construction of a stress-energy tensor of quantum fields is valid in generic curved spacetimes, but, to consider the application of quantum field theory in curved spacetime to cosmology, we close our thesis by specialising to the case of flat Robertson-Walker spacetimes. Generalising the results in [DFP08], we find that some solutions of the semiclassical Einstein equation in this scenario, where $\omega(:T_{\mu\nu}(x):)$ is considered as the expectation value of the stress-tensor of an arbitrary number of free quantum fields of arbitrary spin, generically display the already anticipated power-law inflationary behaviour at early times. Moreover, at late times, all solutions display the behaviour of an effective cosmological constant. We stress that these results are a fundamental generalisation of Starobinsky's original model [Sta80] and its extensions, *e.g.* [Kok09], in that one takes the *full renormalisation freedom* of $\omega(:T_{\mu\nu}(x):)$ and its *full state dependence* into account, and not only its trace anomaly part, or a partial state dependence, *cf.* [PaRa99, PaRa01]. As a result, we are able to fit supernova Ia measurements *as good as the cosmological concordance model*. Hence, we find that quantum field theory on curved spacetimes offers *a natural and generic explanation of the origin of Dark Energy*. Additionally, we observe that temperature-dependent part of $\omega(:T_{\mu\nu}(x):)$ evaluated in an asymptotic conformal equilibrium state *mimics the behaviour of (dark) matter on cosmological scales*. This opens the possibility that no additional matter field beyond the ones present in the standard model of particle physics is necessary to explain the effects of dark matter.

A comment on the organisation of our thesis is in order. In chapter I, we describe the geometrical preliminaries necessary for a formulation of both classical and quantum field theories on curved spacetimes. Building on this, we lay the foundation to the formulation of quantum theories in curved backgrounds by first describing how classical field theories on curved spacetimes can be formulated in chapter II. In chapter III, we initially review the algebraic formulation of the quantized Klein-Gordon and its Wick polynomials in curved spacetimes and then construct both the Wick polynomials of the Dirac field in general spacetimes, and preferred Hadamard states for Klein-Gordon fields and Dirac fields in NBB spacetimes. Finally, in chapter IV, the general construction of the stress-energy tensor of the Klein-Gordon field is reviewed and extended to the Dirac case, and a rigorous version of the DeWitt-Schwinger renormalisation prescription is presented. The thesis closes with the analysis of the solutions of the semiclassical Einstein equation in the flat Robertson-Walker scenario and their comparison with experimental data.

# I

# Spacetimes and Spacetime Structures

As the topic of the current thesis are quantized fields on classical curved spacetimes, we are bound to start by explaining which class of spacetimes we will be considering. Furthermore, we have to introduce and explain the related notions and structures which are necessary both to define physical theories on curved manifolds and to investigate their properties. The incipient chapter is, hence, devoted to these subjects.

## I.1 Spacetimes

In this thesis, a *spacetime* $(M, g)$ is meant to be a Hausdorff, connected, smooth manifold $M$, endowed with a Lorentzian metric $g$, the invariant volume measure of which shall be denoted by $d_g x \doteq \sqrt{|\det g|}\, dx$. We will only consider four-dimensional spacetimes, since this seems to be the situation favoured by experimental data. However, most notions and results can be formulated and obtained for Lorentzian spacetimes with a dimension $d$ differing from four and we will try to point out how the spacetime dimension affects them whenever it seems interesting and possible. We will follow the highly recommendable monograph by Wald [Wal84] regarding most conventions and notations and, hence, work with the metric signature $(-, +, +, +)$. It is often required that a spacetime be second countable, or, equivalently, paracompact, *i.e.* that its topology has a countable basis. Though, as proven by Geroch in [Ger68], paracompactness already follows from the properties of $(M, g)$ listed above. In addition to the attributes already required, we demand that the spacetime under consideration is orientable and time-orientable and that an orientation has been chosen in both respects. We will often omit the spacetime metric $g$ and denote a spacetime by $M$ in brief.

For a point $x \in M$, $T_x M$ denotes the tangent space of $M$ at $x$ and $T_x^* M$ denotes the respective cotangent space; the tangent and cotangent bundles of $M$ shall be denoted by $TM$ and $T^* M$, respectively. If $\chi : M_1 \to M_2$ is a diffeomorphism, we denote by $\chi^*$ the *pull-back* of $\chi$ and by $\chi_*$ the *push-forward* of $\chi$. $\chi^*$ and $\chi_*$ map tensors on $M_2$ to tensors on $M_1$ and tensors on $M_1$ to tensors on $M_2$, respectively; they furthermore satisfy $\chi_* = (\chi^{-1})^*$ [Wal84, app. C]. In case $g_1$ and $g_2$ are the chosen Lorentzian metrics on $M_1$ and $M_2$ and $\chi_* g_1 = g_2$, we call $\chi$ an *isometry*; if $\chi_* g_1 = \Omega^2 g_2$ with a strictly positive smooth function $\Omega$, $\chi$ shall be called a *conformal isometry* and $\Omega^2 g$ a *conformal transformation* of $g$. Note that this definition differs from the one often used in the case of highly symmetric or flat spacetimes since one does not rescale coordinates, but the metric. A conformal transformation according to our definition is sometimes called *Weyl transformation* in the literature. If $\chi$ is an embedding $\chi : M_1 \hookrightarrow M_2$, *i.e.* $\chi(M_1)$ is a submanifold of $M_2$ and $\chi$ a diffeomorphism between $M_1$ and $\chi(M_1)$, it is understood that a push-forward $\chi_*$ of $\chi$ is only defined on $\chi(M_1) \subset M_2$. In case an embedding $\chi : M_1 \hookrightarrow M_2$ between the manifolds of two spacetimes $(M_1, g_1)$ and $(M_2, g_2)$ is an isometry between $(M_1, g_1)$ and $(\chi(M_1), g_2 \restriction_{\chi(M_1)})$, we call $\chi$ an *isometric embedding*, whereas an embedding which is a conformal isometry between $(M_1, g_1)$ and $(\chi(M_1), g_2 \restriction_{\chi(M_1)})$ shall be called a *conformal embedding*.

Some works make extensive use of the *abstract index notation*, *i.e.* they use Latin indices to denote tensorial identities which hold in any basis to distinguish them from identities which hold only in specific bases. As this distinction will not be necessary in the present work, we will not use abstract index notation, but reserve Latin indices to denote tensorial components in a *frame* or *tetrad* basis, where a frame $\{e_a\}_{a=0,1,2,3}$ is a collection of four, in general only locally defined, vector fields satisfying $g(e_a, e_b) \equiv \eta_{ab}$ and $\eta_{ab}$ is the Minkowski metric $\mathrm{diag}(-1, 1, 1, 1)$. Tensorial components in a coordinate basis $\{\partial_\mu\}_{\mu=0,1,2,3}$ will be denoted be Greek indices and the coordinate basis components of a frame $e_a^\mu$, defined by $e_a \doteq e_a^\mu \partial_\mu$, constitute the transformation matrix relating the two aforementioned bases; here, we have employed the Einstein summation convention as will be the case throughout this work. We shall lower Greek indices by means of $g_{\mu\nu} \doteq g(\partial_\mu, \partial_\nu)$ and raise them by $g^{\mu\nu} \doteq (g^{-1})_{\mu\nu}$, while the same actions are performed on frame indices employing $\eta_{ab} = \eta^{ab}$.

Every smooth Lorentzian manifold has a unique metric-compatible and torsion-free linear connection, the *Levi-Civita connection*, and we shall denote the associated *covariant derivative* along a vector field $v$ by $\nabla_v$. We will abbreviate $\nabla_{e_a}$ by $\nabla_a$ and $\nabla_{\partial_\mu}$ by $\nabla_\mu$ and furthermore use the shorthand notation $T_{;\mu_1\cdots\mu_n} \doteq \nabla_{\mu_1}\cdots\nabla_{\mu_n} T$ for covariant derivatives of a tensor field $T$. Our definitions for the *Riemann tensor* $R_{\alpha\beta\gamma\delta}$, the *Ricci tensor* $R_{\alpha\beta}$, and the *Ricci scalar* $R$ are

$$v_{\alpha;\beta\gamma} - v_{\alpha;\gamma\beta} \doteq R_\alpha{}^\lambda{}_{\beta\gamma} v_\lambda, \qquad R_{\alpha\beta} \doteq R_\alpha{}^\lambda{}_{\beta\lambda}, \qquad R \doteq R^\alpha{}_\alpha, \tag{I.1}$$

where $v_\alpha$ are the components of an arbitrary covector. The Riemann tensor possesses the symmetries

$$R_{\alpha\beta\gamma\delta} = -R_{\beta\alpha\gamma\delta} = R_{\gamma\delta\alpha\beta}, \qquad R_{\alpha\beta\gamma\delta} + R_{\alpha\delta\beta\gamma} + R_{\alpha\gamma\delta\beta} = 0 \tag{I.2}$$

and fulfils the *Bianchi identity*

$$R_{\alpha\beta\gamma\delta;\varepsilon} + R_{\alpha\beta\varepsilon\gamma;\delta} + R_{\alpha\beta\delta\varepsilon;\gamma} = 0. \tag{I.3}$$

Moreover, its trace-free part, the *Weyl tensor*, is defined as

$$C_{\alpha\beta\gamma\delta} = R_{\alpha\beta\gamma\delta} - \frac{1}{6}\left(g_{\alpha\delta}g_{\beta\gamma} - g_{\alpha\gamma}g_{\beta\delta}\right)R - \frac{1}{2}\left(g_{\beta\delta}R_{\alpha\gamma} - g_{\beta\gamma}R_{\alpha\delta} - g_{\alpha\delta}R_{\beta\gamma} + g_{\alpha\gamma}R_{\beta\delta}\right),$$

where the appearing coefficients differ in spacetimes with $d \neq 4$. In addition to the covariant derivative, we can define the notion of a *Lie derivative* along a vector field $v$: the integral curves $c(s)$ of $v$ with respect to a curve parameter $s$ define, in general only for small $s$ and on an open neighbourhood of $c(0)$, a one-parameter group of diffeomorphisms $\chi_s^v$ [Wal84, chap. 2.2]. Given a tensor field $T$ of arbitrary rank, we can, hence, define the Lie derivative of $T$ along $v$ as

$$\pounds_v T \doteq \lim_{s\to 0}\left(\frac{(\chi_{-s}^v)^* T - T}{s}\right).$$

If $\chi_s^v$ is a one-parameter group of isometries, we call $v$ a *Killing vector field*, while in case of $\chi_s^v$ being a one-parameter group of conformal isometries, we shall call $v$ a *conformal Killing vector field*. It follows that a Killing vector field $v$ fulfils $\pounds_v g = 0$, while a conformal Killing vector field $v$ fulfils $\pounds_v g = f g$ with some smooth function $f$ [Wal84, app. C.3].

Up to now, we have tacitly disregarded an essential property of $(M, g)$ which is a prerequisite for the formulation of physical theories incorporating causality, namely, *global hyperbolicity*. To define this concept, we need a few additional standard notions related to Lorentzian spacetimes. To wit, following our sign convention, we call a vector $v_x \in T_x M$ *timelike* if $g(v_x, v_x) < 0$, *spacelike* if $g(v_x, v_x) > 0$, *lightlike* or *null* if $g(v_x, v_x) = 0$, and *causal* if it is either timelike or null. Extending this, we call a vector field $v : M \to TM$ spacelike, timelike, lightlike, or causal if it possesses this property at every point. Finally, we call a curve $c : \mathbb{R} \supset I \to M$, with $I$ an interval, spacelike, timelike, lightlike, or causal if its tangent vector field bears this property. Note that, according to our definition, a trivial curve $c \equiv x$ is lightlike. As $(M, g)$ is time orientable, we can split the lightcones in $TM$ at all points in $M$ into 'future' and 'past' in a consistent way and say that a causal curve is *future directed* if its tangent vector field at a point is always in the future lightcone at this point; *past directed* causal curves are defined analogously.

For the definition of global hyperbolicity, we need the notion of inextendible causal curves; these are curves that 'run off to infinity' or 'run into a singular point'. Hence, given a future directed curve $c$ parametrised by $s$, we call $x$ a *future endpoint* of $c$ if, for every neighbourhood $\mathcal{O}$ of $x$, there is an $s_0$ such that $c(s) \in \mathcal{O}$ for all $s > s_0$. With this in mind, we say that a future directed causal curve is *future inextendible* if, for all possible parametrisations, it has *no* future endpoint and we define *past inextendible* past directed causal curves similarly. A related notion is the one of a *complete geodesic*. A geodesic $c$ is called complete if, in its *affine parametrisation* defined by $\nabla_{dc/ds} \frac{dc}{ds} = 0$, the affine parameter $s$ ranges over all $\mathbb{R}$. A manifold $M$ is called *geodesically complete* if all geodesics on $M$ are complete.

In the following, we are going to define the generalisations of flat spacetime lightcones in curved spacetimes. By $I^+(x, M)$ we denote the *chronological future* of a point $x$ relative to $M$, *i.e.* all points in $M$ which can be reached by a future directed timelike curve starting from $x$, while $J^+(x, M)$ denotes the *causal future* of a point $x$, *viz.* all points in $M$ which can be reached by future directed causal curve starting from $x$. Notice that, generally, $x \in J^+(x, M)$ and $I^+(x, M)$ is an open subset of $M$ while the situations $x \notin I^+(x, M)$ and $J^+(x, M)$ being a closed subset of $M$ are not generic, but for instance present in globally hyperbolic spacetimes [Wal84]. In analogy to the preceding definitions, we define the *chronological past* $I^-(x, M)$ and *causal past* $J^-(x, M)$ of a point $x$ by employing past directed timelike and causal curves, respectively. We extend this definition to a general subset $\mathcal{O} \subset M$ by setting

$$I^{\pm}(\mathcal{O}, M) \doteq \bigcup_{x \in \mathcal{O}} I^{\pm}(x, M) \qquad J^{\pm}(\mathcal{O}, M) \doteq \bigcup_{x \in \mathcal{O}} J^{\pm}(x, M);$$

additionally, we define $I(\mathcal{O}, M) \doteq I^+(\mathcal{O}, M) \cup I^-(\mathcal{O}, M)$ and $J(\mathcal{O}, M) \doteq J^+(\mathcal{O}, M) \cup J^-(\mathcal{O}, M)$. As the penultimate prerequisite for the definition of global hyperbolicity, we say that a subset $\mathcal{O}$ of $M$ is *achronal* if $I^+(\mathcal{O}, M) \cap \mathcal{O}$ is empty, *i.e.* an achronal set is such that every timelike curve meets it at most once. Given a closed achronal set $\mathcal{O}$, we define its *future domain of dependence* $D^+(\mathcal{O}, M)$ as the set containing all points $x \in M$ such that every past inextendible causal curve through $x$ intersects $\mathcal{O}$. By our definitions, $D^+(\mathcal{O}, M) \subset J^+(\mathcal{O}, M)$, but note that $J^+(\mathcal{O}, M)$ is in general considerably larger than $D^+(\mathcal{O}, M)$. We define $D^-(\mathcal{O}, M)$ analogously and set $D(\mathcal{O}, M) \doteq D^+(\mathcal{O}, M) \cup D^-(\mathcal{O}, M)$. $D(\mathcal{O}, M)$ is sometimes also called the *Cauchy development* of $\mathcal{O}$. With this, we are finally in the position to state the definition of global hyperbolicity.

**Definition I.1.1** *A **Cauchy surface** is a closed achronal set $\Sigma \subset M$ with $D(\Sigma, M) = M$. A spacetime $(M, g)$ is called **globally hyperbolic** if it contains a Cauchy surface.*

Since the preceding discussion of the causal structure of a Lorentzian spacetime has been completely independent of its dimension, the same holds for the above definition.

Although the geometric intuition sourced by our knowledge of Minkowski spacetime can fail us in general Lorentzian spacetimes, it is essentially satisfactory in globally hyperbolic spacetimes. According to definition I.1.1, a Cauchy surface is a 'non-timelike' set on which every 'physical signal' or 'worldline' must register exactly once. This is reminiscent of a constant time surface in flat spacetime and one can indeed show that this is correct. In fact, Geroch has proved in [Ger70] that globally hyperbolic spacetimes are topologically $\mathbb{R} \times \Sigma$ and Bernal and Sanchez [BeSa03, BeSa05, BeSa06] have been able to improve on this and to show that every globally hyperbolic spacetime has a *smooth* Cauchy surface $\Sigma$ and is, hence, even diffeomorphic to $\mathbb{R} \times \Sigma$. This implies in particular the existence of a (non-unique) smooth global *time function* $t : M \to \mathbb{R}$, *i.e.* $t$ is a smooth function with a timelike and future directed gradient field $\nabla t$; $t$ is, hence, strictly increasing along any future directed timelike curve. In the following, we shall always consider smooth Cauchy surfaces, even in the cases where we do not mention it explicitly.

In the remainder of this thesis, we will gradually see that globally hyperbolic curved spacetimes have many more nice properties well-known from flat spacetime and, hence, seem to constitute the perfect compromise between a spacetime which is generically curved and one which is physically sensible. Particularly, it will turn out that second order, linear, hyperbolic partial differential equations have well-defined global solutions on a globally hyperbolic spacetime. Hence, whenever we speak of a spacetime in the following and do not explicitly demand it to be globally hyperbolic, this property shall be understood to be present implicitly. A prominent example for a non-globally hyperbolic spacetime is Anti de Sitter space, see, *e.g.* [BGP07, chap. 3.5].

On globally hyperbolic spacetimes, there can be no closed timelike curves, otherwise we would have a contradiction to the existence of a smooth and strictly increasing time function. There is a causality condition related to this which can be shown to be weaker than global hyperbolicity, namely, *strong causality*. A spacetime is called strongly causal if it can not contain almost closed timelike curves, *i.e.* for every $x \in M$ and every neighbourhood $\mathscr{O}_1 \ni x$, there is a neighbourhood $\mathscr{O}_2 \subset \mathscr{O}_1$ of $x$ such that no causal curve intersects $\mathscr{O}_2$ more than once. One might wonder if this weaker condition can be filled up to obtain full global hyperbolicity and indeed some references, *e.g.* [BGP07, HaEl73], define a spacetime $(M, g)$ to be globally hyperbolic if it is strongly causal and $J^+(x) \cap J^-(y)$ is compact for all $x, y \in M$. One can show that the latter definition is equivalent to definition I.1.1 [BGP07, Wal84] which is, notwithstanding, the more intuitive one in our opinion.

We close this subsection by introducing a few additional sets with special causal properties. To this avail, we denote by $\exp_x$ the exponential map at $x \in M$. A set $\mathscr{O} \subset M$ is called *geodesically starshaped* with respect to $x \in \mathscr{O}$ if there is an open subset $\mathscr{O}'$ of $T_x M$ which is starshaped with respect to $0 \in T_x M$ such that $\exp_x : \mathscr{O}' \to \mathscr{O}$ is a diffeomorphism. We call a subset $\mathscr{O} \subset M$ *geodesically convex* if it is geodesically starshaped with respect to all its points. This entails in particular that each points $x, y$ in $\mathscr{O}$ are connected by a unique geodesic which is completely contained in $\mathscr{O}$. A related notion are *causal domains*, these are subsets of geodesically convex sets which are in addition globally hyperbolic. Finally, we would like to introduce *causally convex regions*, a generalisation of geodesically convex sets. They are open, non-empty subsets $\mathscr{O} \subset M$ with the property that, for all $x, y \in \mathscr{O}$, all causal curves connecting $x$ and $y$ are entirely contained in $\mathcal{O}$. One can prove that every point in a spacetime lies in a geodesically convex neighbourhood and in a causal domain [Fri75] and one might wonder if the case of a globally hyperbolic spacetime which is geodesically convex is not quite generic. However, even de Sitter spacetime, which is both globally hyperbolic and maximally symmetric and could, hence, be expected to share many properties of Minkowski spacetime, is not geodesically convex.

## I.2 Spinors on Curved Spacetimes

The previous section I.1 contains essentially everything we need to know about curved spacetimes in order to construct classical and quantized field theories of integer spin on them. However, for Fermionic theories, we need additional input. In this section, we will, hence, discuss how to define a spinor field and its covariant derivative on curved spacetimes.

### I.2.1 Spin Structures

Let us recall that, in Minkowski spacetime, a spinor field transforms under the double covering group of the *proper, orthochronous Poincaré group*. While global Poincaré symmetry is not realised in generic curved spacetimes, the *proper, orthochronous Lorentz group* $SO_0(3,1)$ is still a meaningful *local* symmetry group by the *Einstein equivalence principle*. It therefore seems sensible to define a spinor field locally by its transformation properties under the double covering group of $SO_0(3,1)$, namely, the *identity component of the spin group* $Spin_0(3,1)$ which is isomorphic to $SL(2,\mathbb{C})$. To have a sensible *global* notion of a spinor field, it is then necessary to make sure that the local double covering of $SO_0(3,1)$ can be consistently performed on the full spacetime $M$. If this is possible, we say that $M$ has a *spin structure*. In the following, we shall discuss these issues in more detail.

This subsection will make extensive use of the language of *fibre bundles*, see [Hus96, KoNo63] for a stringent mathematical treatment and [CoWi06, Nak03] for an introduction well-motivated from physics. A reference for spin geometry is [LaMi89], but we also recommend the exposition in the PhD thesis of Sanders [San08, chap. 4].

To encode the local Lorentz symmetry of a spacetime in a meaningful mathematical object, let us recall that we can define Lorentz frames $e_a$ with $g(e_a, e_b) = \eta_{ab} = \text{diag}(-1, 1, 1, 1)$ at every point of a Lorentzian spacetime, though in general not globally. Since we only consider oriented and time-oriented spacetimes, we can restrict our attention to oriented frames with a future pointing $e_0$. Of course we can have many different such frames at every point, and they are related by a proper, orthochronous Lorentz transformation. In fact, $SO_0(3,1)$ acts on the frames at one point from the *right*, *i.e.* given a frame $e_a$ and $\Lambda^a{}_b \in SO_0(3,1)$, we obtain a new oriented and time oriented frame by $e_a \Lambda^a{}_b$. Moreover, this *right action* is *free* as only the identity in $SO_0(3,1)$ leaves a frame unchanged and *transitive* since each pair of frames is related by an element of $SO_0(3,1)$. The collection of all oriented and time-oriented Lorentz frames, hence, constitutes a *principle fibre bundle* over $M$ with *typical fibre* $SO_0(3,1)$.

**Definition I.2.1.1** *The **Lorentz frame bundle** of an oriented and time-oriented spacetime $M$ is the principle fibre bundle $LM \doteq LM[SO_0(3,1), R_L, \pi_L, M]$ with typical fibre $SO_0(3,1)$, right action $R_L : SO_0(3,1) \times LM \to LM$, base space $M$, and base projection $\pi_L : LM \to M$.*

We are now in the position to formulate the globally consistent local double covering of the proper, orthochronous Lorentz group in precise terms. To achieve this, let us denote the covering Lie group homomorphism $Spin_0(3,1) \to SO_0(3,1)$ by $\Pi$. Initially, $\Pi$ is defined to be a double covering homomorphism, but, in four spacetime dimensions, $\Pi$ constitutes the universal covering of $SO_0(3,1)$ as $Spin_0(3,1) = SL(2,\mathbb{C})$ is already simply connected. A standard way to realise $\Pi$ is to map a Minkowski vector $x$ to a $2 \times 2$ matrix as $\widetilde{x} \doteq x^0\mathbb{I}_2 + \sum_{i=1}^{3} x^i\sigma_i$, where $\mathbb{I}_n$ denotes the $n$-dimensional identity matrix and $\sigma_i$ are the *Pauli matrices*

$$\sigma_1 \doteq \begin{pmatrix} 0 & 1 \\ 1 & 0 \end{pmatrix}, \quad \sigma_2 \doteq \begin{pmatrix} 0 & -i \\ i & 0 \end{pmatrix}, \quad \sigma_3 \doteq \begin{pmatrix} 1 & 0 \\ 0 & -1 \end{pmatrix}. \tag{I.4}$$

Then, one can show that the adjoint action of $SL(2,\mathbb{C}) \ni \widetilde{\Lambda}$ on $\widetilde{x}$, i.e. $\widetilde{\Lambda}\widetilde{x}\widetilde{\Lambda}^{-1}$, defines a covering homomorphism $SL(2\mathbb{C}) \to SO_0(3,1)$ with kernel $\pm\mathbb{I}_2$. After discussing the spin structure, we will also regard a different, albeit equivalent, way to define $\Pi$.

**Definition I.2.1.2** *Given an oriented and time-oriented spacetime $M$, a **spin structure** is a pair $(SM, \rho)$, where $SM \doteq SM[Spin_0(3,1), R_S, \pi_S, M]$ is a principle fibre bundle over $M$ with typical fibre $Spin_0(3,1)$, right action $R_S : Spin_0(3,1) \times SM \to SM$, and base projection $\pi_S : SM \to M$, and $\rho : SM \to LM$, is a smooth bundle morphism that fulfils the following two conditions:*

*a) $\rho$ is base point preserving, i.e. $\pi_L \circ \rho = \pi_S$,*

*b) $\rho$ intertwines $R_S$ and $R_L$ by implementing the covering homomorphism $\Pi$ as*

$$\rho \circ R_S(\widetilde{\Lambda}) = R_L(\Lambda) \circ \rho,$$

*where $\widetilde{\Lambda} \in Spin_0(3,1)$ and $\Lambda \in SO_0(3,1)$ are related by $\Pi(\widetilde{\Lambda}) = \Lambda$.*

We call $SM$ a **spin frame bundle** and say that two spin structures $(SM_1, \rho_1)$ and $(SM_2, \rho_2)$ are **equivalent** if there is a base point preserving bundle isomorphism $\widetilde{\rho} : SM_1 \to SM_2$ fulfilling $\rho_2 \circ \widetilde{\rho} = \rho_1$.

A spacetime $(M, g)$ endowed with a spin structure $(SM, \rho)$ shall sometimes be denoted by $(M, g, SM, \rho)$, but we will often denote such spacetime by $M$ in brief.

Now we know precisely what kind of an additional structure we need in order to define spinor fields on our spacetime $M$. This of course raises the ardent question whether such a structure really exists on $M$. The most general answer to this question is due to Borel and Hirzebruch [BoHi58] and given in terms of *Stiefel-Whitney classes*, see, i.e. [Hus96, chap. 17]. These are initially defined to be characteristic classes of $M$ with values in the *Čech cohomology groups* $\check{H}^n(M, \mathbb{Z}_2)$, but one can show that the latter coincide with the *de Rham cohomology groups* $H^n(M, \mathbb{Z}_2)$ in the case of a smooth manifold $M$. Since the situation is much simpler in the present case of a four-dimensional, globally hyperbolic spacetime, we will not go into details here, but only state the general result for the sake of completeness. In addition to the monographs already cited, we recommend the exposition in chapter 2 and appendix A of [Köh95] to the reader interested in further details.

***Theorem* I.2.1.3** *An oriented and time oriented smooth spacetime admits a spin structure if and only if its second Stiefel-Whitney class* $w_2(M) \in H^2(M, \mathbb{Z}_2)$ *is trivial.*

Let us understand why the question of existence of a spin structure can be answered in easier terms in our case. If $M$ were parallelisable, *i.e.* if $M$ would admit a *global* Lorentz frame $e_a$, then the existence of a spin structure would follow immediately, since one could just cover the Lorentz group at one point and extend this to the full spacetime by means of $e_a$. In fact, one can even prove that the existence of a spin structure implies the parallelisability of $M$, *cf.* [Ger68, Hus96]. The last ingredient we, hence, need is the insight that an oriented and time-oriented, four-dimensional, globally hyperbolic spacetime $M$ is always parallelisable. This follows from the fact that such $M$ is diffeomorphic to $\mathbb{R} \times \Sigma$, with a three-dimensional, oriented Cauchy surface $\Sigma$, and the standard result of differential geometry that all oriented threefolds are parallelisable. Altogether, we arrive at the following result, which strengthens the already obtained insight that globally hyperbolic spacetimes are physically meaningful.

***Lemma* I.2.1.4** *Every four-dimensional, oriented and time-oriented, globally hyperbolic spacetime admits a spin structure.*

One might wonder whether a spin structure is unique (up to equivalence); it can be shown that this is the case if and only if $M$ is simply connected. We refer to [Ger68, Ger70] for this and other interesting facts about spin structures. For examples of spacetimes *not* admitting spin structures, see [Ger70, Ohl92].

Bosonic fields are tensor fields and therefore sections in some tensor powers of $TM$ and $T^*M$, which are in turn vector bundles *associated* to $LM$, see for example [Nak03, p. 370]. The natural bundles for spinor fields to be sections of should thus be tensor powers of vector bundles associated to $SM$; the definition of the latter is thus in order. To this avail, let us recall that, in flat spacetime, Dirac spinors transform under the four-dimensional reducible representation of $SL(2, \mathbb{C})$ given by the direct sum $\pi_\oplus \doteq D^{(\frac{1}{2}, 0)} \oplus D^{(0, \frac{1}{2})}$ of the fundamental representation of $SU(2)$ and its complex conjugate. Consequently, it seems advisable to construct the spinor bundles employing $\pi_\oplus$ also in curved spacetimes.

***Definition* I.2.1.5** *The* **Dirac spinor bundle** *of a four-dimensional spacetime $M$ is the associated vector bundle $DM \doteq SM \times_{\pi_\oplus} \mathbb{C}^4$. $DM$ is thus the set of equivalence classes $[(p, z)]$ where $p \in SM$, $z \in \mathbb{C}^4$, and $(p_1, z_1)$, $(p_2, z_2)$ are defined to be equivalent if there exists a $\widetilde{\Lambda} \in SL(2, \mathbb{C})$ such that $R_S(\widetilde{\Lambda})p_1 = p_2$ and $\pi_\oplus(\widetilde{\Lambda}^{-1})z_1 = z_2$. This entails that $DM$ is a fibre bundle over $M$ with typical fibre $\mathbb{C}^4$, structure group $\pi_\oplus(SL(2, \mathbb{C}))$, and the base point projection $\pi_D$ defined as $\pi_D([(p, z)]) \doteq \pi_S(p)$. A* **Dirac spinor field** *is a section of $DM$.*

*In analogy, denoting by* $*$ *the adjoint with respect to the inner product on* $\mathbb{C}^4$, *we call* **Dirac cospinor bundle** *of M the* $\mathbb{C}^{4*}$-*bundle associated to SM by identifying* $(p, z^*)$ *with* $(R_S(\widetilde{\Lambda})p, z^*\pi_{\oplus}(\widetilde{\Lambda}))$; *a section of* $D^*M$ *shall be called* **Dirac cospinor field**.

*We define a* **fibrewise dual pairing** *of DM and* $D^*M$ *by extending the dual pairing of* $\mathbb{C}^4$ *and* $\mathbb{C}^{4*}$: *for* $Z_1 \doteq [(p, z_1)]$, $Z_2^* \doteq [(p, z_2^*)]$, *we set* $Z_2^*(Z_1) \doteq z_2^*(z_1)$.

In Minkowski spacetime $\mathbb{M}$, a Dirac spinor field is just a map $\mathbb{M} \to \mathbb{C}^4$. The above definition implies that, locally, we have the same situation in a globally hyperbolic, four-dimensional spacetime. But, in fact, on simply connected spacetimes, this holds even *globally*. We have learned that a four-dimensional, globally hyperbolic spacetime is parallelisable and $LM$ is therefore a trivial bundle. Consequently, $SM$ is a trivial bundle as well if $M$ is simply connected. Altogether, on simply connected, four-dimensional spacetimes, we are in the nice situation that all vector bundles we have introduced up to now, *i.e.* $TM$, $T^*M$, $DM$, and $D^*M$, are trivial and we can define global frames for all of them. If $M$ is not simply connected, only $TM$ and $T^*M$ are trivial bundles, and $DM$ and $D^*M$ are in general not trivial. Moreover, to be in accord with the spin structure, we have to require that the spin and Lorentz frames are related. Let us therefore choose a consistent set of frames.

**Definition** I.2.1.6

a) *Let* $E : M \to SM$ *be an arbitrary but fixed smooth section of SM. We define a* **spin frame** $\{E_A\}_{A=1,2,3,4}$ *as a set of four sections of DM by* $E_A(x) \doteq [(E(x), z_A)]$, *where* $z_A$ *is the standard basis of* $\mathbb{C}^4$. *If M is simply connected, we can choose E and, hence,* $E_A$ *to be global sections.*

b) *We can then construct a* **dual spin frame** $\{E^B\}_{B=1,2,3,4}$ *as a set of four sections of* $D^*M$ *by requiring that* $E^B(x)(E_A(x)) = \delta_A^B$ *for all* $x \in M$. *Again, we can choose* $E^B$ *to be global if M is simply connected.*

c) *Employing the spin structure morphism* $\rho$, *we obtain a smooth section* $e : M \to LM$ *by setting* $e \doteq \rho \circ E$. *We can choose e to be a global section. If E is already global, this is straightforward. If we have only local spin frames, we can choose them such that their (sign) ambiguity does not matter while projecting to e. Out of e we define, in analogy to a), a* **Lorentz frame** $\{e_a\}_{a=0,1,2,3}$ *as a set of four global sections of TM making use of the fact that TM is an* $\mathbb{R}^4$-*bundle associated to LM.*

d) *Similar to b), we define a* **dual Lorentz frame** $\{e^b\}_{b=0,1,2,3}$ *as a set of four global sections of* $T^*M$ *by demanding that* $e^b(x)(e_a(x)) = \delta_a^b$ *holds at all points of M.*

Henceforth, we shall denote spinor indices by capital Roman letters and we shall raise and lower them by $(\mathbb{I}_4)_{AB} = (\mathbb{I}_4)^{AB} = \delta_A^B$.

Using the just defined global frames, we can decompose every *spinor-tensor*, *i.e.* every section

$$f : M \to \underbrace{TM \otimes \cdots \otimes TM}_{i} \otimes \underbrace{T^*M \otimes \cdots \otimes T^*M}_{j} \otimes \underbrace{DM \otimes \cdots \otimes DM}_{k} \otimes \underbrace{D^*M \otimes \cdots \otimes D^*M}_{l}$$

as

$$f = f^{a_1 \cdots a_i A_1 \cdots A_k}_{b_1 \cdots b_j B_1 \cdots B_l} \, e_{a_1} \otimes \cdots \otimes e_{a_i} \otimes e^{b_1} \otimes \cdots \otimes e^{b_j} \otimes E_{A_1} \otimes \cdots \otimes E_{A_k} \otimes E^{B_1} \otimes \cdots \otimes E^{B_l},$$

where the coefficients $f'^{a'_1,\ldots,a'_i,A'_1,\ldots,A'_k}_{b'_1,\ldots,b'_j,B'_1,\ldots,B'_l}$ are functions on $M$. One could in principle certainly choose a different sections $E'$ of $SM$ and thus obtain different spin and Lorentz frames which are related to the previous ones by local spin and Lorentz transformations. On the level of coefficients, such a change of frames results in

$$f'^{a'_1,\ldots,a'_i,A'_1,\ldots,A'_k}_{b'_1,\ldots,b'_j,B'_1,\ldots,B'_l} = \left(\Lambda^{-1}\right)^{a'_1}_{a_1} \cdots \left(\Lambda^{-1}\right)^{a'_i}_{a_i} \left(\widetilde{\Lambda}^{-1}\right)^{A'_1}_{A_1} \cdots \left(\widetilde{\Lambda}^{-1}\right)^{A'_k}_{A_k} \Lambda^{b_1}_{b'_1} \cdots \Lambda^{b_j}_{b'_j} \widetilde{\Lambda}^{B_1}_{B'_1} \cdots \widetilde{\Lambda}^{B_k}_{B'_l} f^{a_1 \cdots a_i A_1 \cdots A_k}_{b_1 \cdots b_j B_1 \cdots B_l}, \quad \text{(I.5)}$$

where $\widetilde{\Lambda}$ is a smooth map $\widetilde{\Lambda} \colon M \to Spin_0(3,1)$, $\Lambda = \Pi \circ \widetilde{\Lambda}$ is a smooth map $\Lambda \colon M \to SO_0(3,1)$, and we have omitted the representation $\pi_\oplus$ for the sake of notational simplicity.

### I.2.2 The Spin Connection

We have seen how to define a Dirac spinor field on globally hyperbolic spacetimes in a sensible way, using as much as we could save in the passage from flat spacetime to a curved one. In the long run, this is of course not sufficient, since we will be interested in defining some dynamics for the Dirac field. In Minkowski spacetime, Poincaré symmetry and the requirement that the dynamically allowed Dirac field constitutes an *irreducible* representation of the double covering group of the Poincaré group uniquely select the Dirac equation as the equation of motion [Fol63]. In curved spacetimes, such strong requirements are of course not at our disposal and the best we can do is to take the covariant generalisation of the Minkowskian Dirac equation as the equation of motion. To this avail, we shall need both *γ-matrices* and a covariant derivative for spinors, *i.e.* a *spin connection*. We will start by discussing the former.

**Definition I.2.2.1** *The **Dirac algebra** $Cl(3,1)$ is the **Clifford algebra** of Minkowski space, i.e. it is the real, associative algebra generated by the identity $\mathbb{I}$ and a set of elements $\{\gamma_a\}_{a=0,1,2,3}$ satisfying*

$$\{\gamma_a, \gamma_b\} \doteq \gamma_a \gamma_b + \gamma_b \gamma_a = 2\eta_{ab} \mathbb{I}.$$

As is well known due to Pauli [Pau36], see also [LaMi89, chap. 4 & 5] and [San08, chap. 4.1], we have the following important result.

**Theorem I.2.2.2** *There is, up to equivalence, only one irreducible, real-linear, representation of $Cl(3,1)$ as complex $4 \times 4$ matrices. It can be specified as*

$$\gamma_0 = i \begin{pmatrix} \mathbb{I}_2 & 0 \\ 0 & -\mathbb{I}_2 \end{pmatrix}, \quad \gamma_i = i \begin{pmatrix} 0 & \sigma_i \\ -\sigma_i & 0 \end{pmatrix}.$$

*We shall henceforth call the matrices $\{\gamma_a\}_{a=0,1,2,3}$ **γ-matrices**.*

The following discussion of Dirac fields will of course depend on the choice of representation of $Cl(3,1)$. However, one can show that the resulting field theories are equivalent up to gauge transformations [San08]. Moreover, if one restricts attention to observables of the Dirac field, the choice of representation is even irrelevant, because observables are always traces over spinor indices, and these are independent of the representation on account of the cyclicity of the trace. We shall therefore use the representation mentioned in theorem I.2.2.2 throughout this work without loss of generality. Note that our $\gamma$-matrices differ from the usual ones by an imaginary unit $i$, this is due to our sign convention $(-,+,+,+)$, which requires in particular that $\gamma_0^2 = -\mathbb{I}_4$.

A quick look at the anticommutation relations of the Dirac algebra in definition I.2.2.1 tells us that both the complex conjugate $\overline{\gamma_a}$ of a representation of $Cl(3,1)$ and its Hermitian adjoint $\gamma_a^*$ are valid representations as well. These representations must therefore be related to $\gamma_a$ by an invertible 4x4 matrix.

**Definition I.2.2.3** *We define the **Dirac conjugation matrix** $\beta \in SL(4,\mathbb{C})$ and the **charge conjugation matrix** $\mathscr{C} \in SL(4,\mathbb{C})$ by*

$$\beta^* = \beta, \quad \gamma_a^* = -\beta \gamma_a \beta^{-1}, \quad -i\beta N^a \gamma_a \text{ is a positive matrix,}$$

$$\overline{\mathscr{C}}\mathscr{C} = \mathbb{I}_4, \quad \overline{\gamma_a} = \mathscr{C}\gamma_a\mathscr{C}^{-1},$$

*where $N = N^a e_a$ is a future pointing timelike vector field.*

These matrices of course depend on the choice of representation of the Dirac algebra. One can show that [San08, thm. 4.1.6], in our representation,

$$\beta = -i\gamma_0 = \beta^{-1}, \quad \mathscr{C} = -i\gamma_2 = -\mathscr{C}^{-1}, \quad \mathscr{C} = -\beta\mathscr{C}\beta. \tag{I.6}$$

To define the already anticipated spin connection, we have to discuss the relation between the Dirac algebra $Cl(3,1)$ and the identity component of the spin group $Spin_0(3,1)$. To this avail, let us first note that the anticommutation relations of $Cl(3,1)$ imply that the $\gamma$-matrices $\{\gamma_a\}_{a=0,1,2,3}$ are linearly independent. Hence, we can identify their linear span with Minkowski spacetime $\mathbb{M}$ and regard the latter as a subset of $Cl(3,1)$. Furthermore, let us recall that the same anticommutation relations relate *even* elements of the Dirac algebra, *i.e.* elements of $Cl(3,1)$ which are sums of a products of an even number of $\gamma_a$. These considerations lead to the following definitions.

**Definition I.2.2.4**

a) *By $Cl_0(3,1)$ we denote the **even subalgebra** of $Cl(3,1)$.*

b) *The **pin group** is the set*

$$Pin(3,1) \doteq \{\tilde{\Lambda} \in Cl(3,1) \mid \tilde{\Lambda} = c_1 \cdots c_k, \; c_i \in \mathbb{M}, \; c_i^2 = \pm\mathbb{I}\}.$$

*c) The **spin group** can then be defined as $Spin(3,1) \doteq Pin(3,1) \cap Cl_0(3,1)$.*

To relate the above defined group $Spin(3,1)$ to our previous discussion, we have to prove that its identity component is the double covering group of $SO_0(3,1)$. This can be proven in two steps, and we refer to [San08, chap. 4.1] for the details.

**Proposition** I.2.2.5

*a) $Pin(3,1) = \{ \widetilde{\Lambda} \in Cl(3,1) \mid det \widetilde{\Lambda} = 1, \, \forall v \in \mathbb{M} \; \widetilde{\Lambda} v \widetilde{\Lambda}^{-1} \in \mathbb{M} \}$.*

*b) The map $\Pi : Pin(3,1) \to O(3,1)$ defined as*

$$\widetilde{\Lambda} \gamma_a \widetilde{\Lambda}^{-1} \doteq \gamma_b \left[ \Pi(\widetilde{\Lambda}) \right]^b_{\ a}$$

*is a double covering Lie group homomorphism which restricts to a double covering Lie group homomorphism $\Pi : Spin_0(3,1) \to SO_0(3,1)$.*

The outcome of the above proposition is an expression for the covering homomorphism $\Pi$ which is better suited to our purposes than the one briefly mentioned at the beginning of the present section. In standard treatments of Dirac fields on Minkowski spacetime, the expression derived in proposition I.2.2.5 if often obtained by requiring that the Dirac equation transforms covariantly under Lorentz transformations.

Having a clear-cut relation between the Dirac algebra and the spin group at hand, one might wonder how the representation of $Cl(3,1)$ in terms of $\gamma$-matrices is related to the representation $\pi_\oplus = D^{(\frac{1}{2},0)} \oplus D^{(0,\frac{1}{2})}$ of $Spin_0(3,1) = SL(2,\mathbb{C})$ chosen in the definition of the Dirac spinor bundles. In this respect, one can show that the restriction of a Clifford algebra respresentation on a vector space to the related spin group is always the sum of two inequivalent representations of such group, see [LaMi89, prop 5.15]; in the case of $Cl(3,1)$, this induced representation of $Spin_0(3,1)$ coincides, up to equivalence, with $\pi_\oplus$.

The last step we need to take in order to obtain a covariant derivative for spinors is to compute the derivative of $\Pi$ at the identity. We shall compute this explicitly, since the sign of the outcome depends on the sign convention of the metric and, as also remarked and corrected by [San08], some previous works on Dirac fields in curved spacetimes, *i.e.* [Dim82], have been disregarding this fact. To this end, let us recall that the Lie algebras of $Pin(3,1)$ and $O(3,1)$, $\mathfrak{pin}(3,1)$ and $\mathfrak{o}(3,1)$ respectively, equal the Lie algebras of their respective identity components.

**Lemma** I.2.2.6 *The derivative $d\Pi : \mathfrak{pin}(3,1) \to \mathfrak{o}(3,1)$ of the covering morphism $\Pi : Spin_0(3,1) \to SO_0(3,1)$ at the identity fulfils*

$$d\Pi^{-1}(\lambda^a_{\ b}) = \frac{1}{4} \lambda^{ab} \gamma_a \gamma_b$$

*for all $\lambda^a_{\ b} \in \mathfrak{o}(3,1)$.*

*Proof.* Let us take an arbitrary differentiable curve $\widetilde{\Lambda} : [0,1] \rightarrow Spin_0(3,1)$, $\widetilde{\Lambda}(0) = \mathbb{I}_4$ whose projection on $SO_0(3,1)$, $\Lambda(s) \doteq \Pi(\widetilde{\Lambda}(s))$, is differentiable as well; the following identity holds

$$\widetilde{\Lambda}(s) \gamma_a \widetilde{\Lambda}(s)^{-1} = \gamma_b \Lambda(s)^b{}_a \,.$$

Deriving with respect to $s$, we obtain

$$\frac{d\widetilde{\Lambda}(s)}{ds} \gamma_a \widetilde{\Lambda}(s)^{-1} + \widetilde{\Lambda}(s) \gamma_a \frac{d\widetilde{\Lambda}(s)^{-1}}{dt} = \gamma_b \left( \frac{d\Lambda(s)}{ds} \right)^b{}_a$$

$$\Leftrightarrow \quad \frac{d\widetilde{\Lambda}(s)}{ds} \gamma_a \widetilde{\Lambda}(s)^{-1} - \widetilde{\Lambda}(s) \gamma_a \widetilde{\Lambda}(s)^{-1} \frac{d\widetilde{\Lambda}(s)}{ds} \widetilde{\Lambda}(s)^{-1} = \gamma_b \left( \frac{d\Lambda(s)}{ds} \right)^b{}_a , \quad (I.7)$$

where, in the second step, we have exploited the fact that the derivation of $\widetilde{\Lambda}(s)\widetilde{\Lambda}(s)^{-1} = \mathbb{I}_4$ yields

$$\frac{d\widetilde{\Lambda}(s)}{ds} \widetilde{\Lambda}(s)^{-1} = -\widetilde{\Lambda}(s) \frac{d\widetilde{\Lambda}(s)^{-1}}{ds} \,.$$

Setting $s = 0$, $\widetilde{\lambda} \doteq \frac{d\widetilde{\Lambda}(s)}{ds}|_{s=0}$, and $\lambda = \frac{d\Lambda(s)}{ds}|_{s=0}$, it holds $\widetilde{\lambda} \gamma_a - \gamma_a \widetilde{\lambda} = \gamma_b \lambda^b{}_a$. Right-multiplication with $\gamma^a$ and $\gamma^a \gamma_a = \eta^{ab} \gamma_a \gamma_b = \eta^{ab} \eta_{ab} \mathbb{I}_4 = 4\mathbb{I}_4$ results in

$$4\widetilde{\lambda} - \gamma_a \widetilde{\lambda} \gamma^a = \lambda_{ab} \gamma^a \gamma^b . \quad (I.8)$$

Taking into account the antisymmetry of $\lambda \in \mathfrak{o}(3,1)$ and the identity

$$\gamma^a [\gamma^b, \gamma^c] \gamma_a \doteq \gamma^a (\gamma^b \gamma^c - \gamma^c \gamma^b) \gamma_a = 0 \,,$$

a possible solution of (I.8) is

$$\widetilde{\lambda} = \frac{1}{4} \lambda_{ab} \gamma^a \gamma^b \,. \quad (I.9)$$

As we are solving a linear, inhomogeneous equation, the above solution is only a particular one, and we can in principle add arbitrary solutions of the homogeneous equation

$$4\overline{\lambda} - \gamma_a \overline{\lambda} \gamma^a = 0 \,. \quad (I.10)$$

To assure uniqueness of (I.9), we have thus to show that the homogeneous equation has only trivial solutions. Let us therefore take an arbitrary solution $\overline{\lambda}$ of (I.10). It fulfils $\gamma^a [\gamma_a, \overline{\lambda}] = 0$, but, since the $\gamma$-matrices are linearly independent, the last equality is equivalent to $[\gamma_a, \overline{\lambda}] = 0$. Schur's lemma [Hal03, thm. 4.26] together with the fact that $\gamma_{a_1} \cdots \gamma_{a_n}$ with $a_1 < \cdots < a_n$ and $n \leq 4$ are a basis of $M(4, \mathbb{C})$ then entails that $\overline{\lambda} = k\mathbb{I}_4$ with some $k \in \mathbb{C}$. The value of $k$ can be unambiguously determined if we recall that $\overline{\lambda}$ as an element in the Lie algebra of $Spin_0(3,1) = SL(2, \mathbb{C})$ has vanishing trace. This implies $k = 0$ and closes the proof. $\quad \square$

To define a spin connection and a related covariant derivative, it seems natural to just take the unique Levi-Civita connection already at our disposal on $TM$ or, equivalently, $LM$ and try to lift it to $SM$ to obtain a covariant derivative on sections of $DM$. Moreover, this option is in fact the only one we have if we want the resulting covariant derivative on mixed spinor-tensors to be compatible with contractions of spinor indices.

**Definition I.2.2.7** *Let $\Gamma_{ac}^{\ \ b}$ denote the* **connection coefficients of the Levi-Civita connection** *and let the associated covariant derivative be specified as $\nabla e_c \doteq \Gamma_{ac}^{\ \ b} e_b \otimes e^a$. Moreover, let $\omega : LM \to T^*LM \otimes \mathfrak{o}(3,1)$ denote the connection 1-form on $LM$ induced by $\Gamma_{ac}^{\ \ b}$ as*

$$\Gamma_{ac}^{\ \ b} \doteq e^b \left( (e^*\omega)(e_a) e_c \right),$$

*with $e^* : T^*LM \to T^*M$ denoting the pull-back of $e$ in the sense of cotangent vectors.*

*We define the connection 1-form $\Omega$ of the* **spin connection** *by the pull-back*

$$\Omega \doteq (d\Pi)^{-1} \circ \rho^* \circ \omega \,.$$

*Let $E^* : T^*SM \to T^*M$ denote the pull-back of $E$ in the sense of cotangent vectors. Via the* **spin connection coefficients**

$$\sigma_{aA}^{\ \ B} \doteq E^B \left( (E^*\Omega)(e_a) E_A \right),$$

*we can define a covariant derivative associated to the spin connection as*

$$\nabla E_A = \sigma_{aA}^{\ \ B} e^a \otimes E_B \,.$$

*Let $V = V^A E_A$ be an arbitrary differentiable section of $DM$. We define the* **spin curvature tensor** $\mathfrak{R}_{\ \ Bab}^{A}$ *via*

$$V_{\ ;ab}^{A} - V_{\ ;ba}^{A} \doteq \mathfrak{R}_{\ \ Bab}^{A} V^B \,.$$

We can straightforwardly extend the covariant derivatives in definition I.2.2.7 to a covariant derivative $\nabla$ on arbitrary spinor-tensors, *i.e.* sections of arbitrary tensor products of $TM$, $T^*M$, $DM$, and $D^*M$. As an example, the action of $\nabla$ on

$$f \doteq f_{bB}^{aA} e_a \otimes e^b \otimes E_A \otimes E^B$$

reads

$$\nabla f = e^c \nabla_c \left( f_{bB}^{aA} e_a \otimes e^b \otimes E_A \otimes E^B \right) =$$
$$\left[ \partial_c f_{bB}^{aA} - \Gamma_{cb}^{\ \ d} f_{dB}^{aA} + \Gamma_{cd}^{\ \ a} f_{bB}^{dA} + \sigma_{cC}^{\ \ A} f_{bB}^{aC} - \sigma_{cB}^{\ \ C} f_{bC}^{aA} \right] e^c \otimes e_a \otimes e^b \otimes E_A \otimes E^B \,.$$

For actual computations, it will be helpful to have an explicit expression for the spin connection coefficients and the spin curvature tensor. To this end, a comment on the nature of the connection coefficients $\Gamma_{ac}^{\ \ b}$ is in order. $\Gamma_{ac}^{\ \ b}$ are *not* the well-known *Christoffel symbols, i.e.*

the connection coefficients of the Levi-Civita connection in the coordinate basis. Particularly, the $\Gamma^{\;b}_{ac}$ are *antisymmetric* in the upper and lower right index, as these indices correspond to indices of a matrix in $\mathfrak{o}(3,1)$, while the Christoffel symbols would be symmetric in the lower two indices. The distinction between the two connection coefficients implies in particular that the expression for $\sigma^A_{aB}$ stated in the forthcoming lemma I.2.2.8 is only valid in a frame basis. This is related to that fact that both the Christoffel symbols and $\Gamma^{\;b}_{ac}$ are *not* the coefficients of a tensor. However, the identity for the spin curvature tensor $\mathfrak{R}^A_{\;\;Bab}$ we will present is valid in all bases, as it involves only tensorial quantities, see also the discussion following lemma I.2.2.8. Due to the non-tensorial nature of $\Gamma^{\;b}_{ac}$, we can, given a point $x \in M$, always find a frame such that $\Gamma^{\;b}_{ac}(x) = 0$. This fact and a direct application of lemma I.2.2.6 yield the following identities [Lic64, sec. 14].

**Lemma** I.2.2.8  *The spin connection coefficients $\sigma^B_{aA}$ and the spin curvature tensor $\mathfrak{R}^A_{\;\;Bab}$ fulfil:*

*a)* $\sigma^A_{aB} = \frac{1}{4}\Gamma^{\;b}_{ad}(\gamma_b\gamma^d)^A_{\;\;B}$,

*b)* $\mathfrak{R}^A_{\;\;Bab} = \frac{1}{4}R_{abcd}(\gamma^c\gamma^d)^A_{\;\;B}$,

*c)* $\mathfrak{R}_{ABab} = -\mathfrak{R}_{BAab} = -\mathfrak{R}_{ABba}$.

Calculations with Dirac fields usually require some serious $\gamma$-matrix jugglery, and the present work makes no exception. We shall now state a few identities used in the remainder of this thesis, which can all be derived in a straightforward manner employing lemma I.2.2.8, the anticommutation relations of the $\gamma$-matrices, the cyclicity of the trace, the symmetries and the Bianchi identity of the Riemann tensor, and finally the fact that a single $\gamma$-matrix has vanishing trace. Before stating the promised identities, let us remark that we will now and often omit spinor indices to simplify notation and let us compute an example to give a brief glimpse into the necessary calculations, namely, the identity $\mathfrak{R}_{ab}\gamma^b = \frac{1}{2}R_{ab}\gamma^b$. Due to lemma I.2.2.8, we know that the starting point of the computation is $\mathfrak{R}_{ab}\gamma^b = \frac{1}{4}R_{abcd}\gamma^b\gamma^c\gamma^d$. By (I.2) and a change of summation indices, we get

$$\mathfrak{R}_{ab}\gamma^b = \frac{1}{4}R_{abcd}\gamma^c\gamma^d\gamma^b = \frac{1}{4}\left(R_{acbd} + R_{adcb}\right)\gamma^c\gamma^d\gamma^b = \frac{1}{4}R_{abcd}\left(\gamma^b\gamma^d\gamma^c + \gamma^c\gamma^b\gamma^d\right).$$

We now use the Clifford relations and the antisymmetry of the Riemann tensor to commute the $\gamma$-matrices in the two summands of the last term until we reach the initial order.

$$\frac{1}{4}R_{abcd}\left(\gamma^b\gamma^d\gamma^c + \gamma^c\gamma^b\gamma^d\right) = \frac{1}{4}R_{abcd}\left(\gamma^c\gamma^d\gamma^b + 2\eta^{bd}\gamma^c - 2\eta^{bc}\gamma^d + \gamma^c\gamma^d\gamma^b + 2\eta^{bd}\gamma^c\right).$$

The last identity and the definition of the Ricci tensor (I.1) entail

$$\mathfrak{R}_{ab}\gamma^b = \frac{3}{2}R_{ab}\gamma^b - 2\mathfrak{R}_{ab}\gamma^b \qquad \Leftrightarrow \qquad \mathfrak{R}_{ab}\gamma^b = \frac{1}{2}R_{ab}\gamma^b.$$

Related considerations lead to the following identities.

*Lemma* **I.2.2.9**

a) *The trace of any product of an odd number of $\gamma$-matrices vanishes. Moreover, if we denote by Tr the trace over spinor indices, and by $_{[\cdot,\cdot]}$ the idempotent antisymmetrisation of indices, then*

$$Tr\,\gamma_a\gamma_b = 4\eta_{ab}\,, \qquad Tr\,\gamma_a\gamma_b\gamma_c\gamma_d = 4\left(\eta_{ab}\eta_{cd} - \eta_{ac}\eta_{bd} + \eta_{ad}\eta_{bc}\right),$$

$$Tr\,\gamma_{[a}\gamma_{b]}\gamma_{[c}\gamma_{d]}\gamma_e\gamma_f = 4\left(\eta_{f[a}\eta_{b][c}\eta_{d]e} + \eta_{e[a}\eta_{b][d}\eta_{c]f} + \eta_{d[a}\eta_{b]c}\eta_{ef}\right).$$

b) *Products of $\gamma$-matrices with two of the appearing indices contracted fulfil*

$$\gamma^a\gamma_a = 4\mathbb{1}_4\,, \qquad \gamma^a\gamma^b\gamma_a = -2\gamma^b\,, \qquad \gamma^a\gamma^b\gamma^c\gamma_a = 4\eta^{bc}\mathbb{1}_4\,,$$

$$\gamma^a\gamma^b\gamma^c\gamma^d\gamma_a = -2\gamma^d\gamma^c\gamma^b\,, \qquad \gamma^a\gamma^b\gamma^c\gamma^d\gamma^e\gamma_a = 2\left(\gamma^e\gamma^b\gamma^c\gamma^d + \gamma^d\gamma^c\gamma^b\gamma^e\right).$$

c) *Let us denote by $[\cdot\cdot]$ the commutator of two matrices. The spin curvature tensor fulfils the following identities*

$$-\gamma^b\mathfrak{R}_{ab} = \mathfrak{R}_{ab}\gamma^b = \frac{1}{2}R_{ab}\gamma^b\,, \qquad \left[\mathfrak{R}_{ab},\gamma_c\right] = R_{abdc}\gamma^d\,,$$

$$Tr\,\mathfrak{R}_{ab}\gamma_c\gamma_d = -2R_{abcd}\,, \quad \mathfrak{R}_{ab;}{}^{ab} = 0\,, \qquad Tr\,\mathfrak{R}_{ab}\mathfrak{R}^{ab} = -\frac{1}{2}R_{abcd}R^{abcd}\,,$$

$$Tr\,\mathfrak{R}_{ab}\mathfrak{R}^{ab}\gamma_c\gamma_d = Tr\,\mathfrak{R}_{ab}\mathfrak{R}^{ab}\eta_{cd}\,.$$

To close this section, let us note that we can promote the initially constant matrices $\gamma_a$ to a smooth section $\gamma : M \to T^*M \otimes DM \otimes D^*M$ by setting

$$\gamma \doteq \gamma_{aB}^A\, e^a \otimes E_A \otimes E^B, \tag{I.11}$$

where $\gamma_{aB}^A$ are the matrix elements of $\gamma_a$. Note that this definition of $\gamma$ depends manifestly on the chosen spin frame $E_A$, but this shall not trouble us since all sensible observables of a spinor field should be frame-independent. With the above definition at hand, we can contract $\gamma$ with a vector field $v = v^a e_a$ to obtain a section of $DM \otimes D^*M$. As it is customary, we will denote this by *Feynman slash notation, i.e.*

$$\slashed{v} \doteq \gamma(v) = v^a\gamma_{aB}^A\, E_A \otimes E^B\,.$$

Moreover, we can obtain coordinate basis $\gamma$-matrices as $\gamma_\mu \doteq e^a_\mu \gamma_a$. These fulfil anticommutation relations with respect to $g_{\mu\nu}$, *viz.*

$$\{\gamma_\mu,\gamma_\nu\} = 2g_{\mu\nu}.$$

Finally, we would like to state an important fact, which is sometimes used to *define* the covariant derivative on spinor-tensors in more pragmatic treatments of Dirac fields on curved spacetimes. To wit, employing lemma I.2.2.8 and the antisymmetry of $\Gamma_{ac}^b$, one can straightforwardly compute [Lic64, sec. 13]:

*Lemma* **I.2.2.10** *The section of $\gamma$-matrices is covariantly constant, i.e. $\nabla\gamma = 0$.*

## I.3 Null Big Bang Spacetimes

Up to now we have discussed generic four-dimensional, globally hyperbolic spacetimes and these will indeed be the backgrounds for the general constructions of scalar and spinor field theories we are going to describe in the following three chapters of this thesis. However, we will also be interested in cosmological applications of the aforementioned general constructions. Particularly, we are aiming for concrete calculations in *Friedmann-Lemaître-Robertson-Walker spacetimes*, henceforth abbreviated by FLRW. It is thus essential to see how the general setup described in the previous two sections manifests and simplifies in such spacetimes. Moreover, as already anticipated by the title of the present section, we shall focus on a restricted class of FLRW spacetimes, as we will explain and motivate in the following.

### I.3.1 Cosmological Spacetimes and Power-Law Inflation

According to the well-known *cosmological principle*, our universe is *homogeneous* and *isotropic*. This postulate implies that, on large scales, the cosmos looks 'the same' everywhere and in all directions, see [Wal84, chap. 5] for a precise definition and a discussion of these issues. A remarkable confirmation of the isotropy of our universe is the fact that the temperature of the *Cosmic Microwave Background* (CMB) is isotropic up to relative fluctuations of the order $10^{-5}$ [Jar10]. Based on the cosmological principle[1], we shall regard Friedmann-Lemaître-Robertson-Walker spacetimes as the curved manifolds describing our universe on large scales. The underlying manifold of such spacetimes is $I_t \times \Sigma_\varkappa$, where $I_t$ denotes an open interval in $\mathbb{R}$ and $\Sigma_\varkappa$ is a three-dimensional manifold of constant curvature $\varkappa$, and their metric is given by the line element

$$ds^2 = -dt^2 + a^2(t)\left(\frac{dr^2}{1-\varkappa r^2} + r^2 d\mathbb{S}^2(\theta, \varphi)\right).$$

Here, $a(t)$ is a strictly positive smooth function called the *scale factor*, $t \in I_t$ denotes *cosmological time*, and $\theta, \varphi$ are coordinates on the 2-sphere $\mathbb{S}^2$, the canonical line element of which is denoted by $d\mathbb{S}^2$. If we recall the discussion in section I.1, we immediately realise that $\Sigma_\varkappa$ is a Cauchy surface, and hence all FLRW spacetimes are globally hyperbolic. In the following, we shall restrict attention to $\varkappa = 0$ as this is the situation strongly favoured by experimental data, the CMB measurements in particular [Kom10, Amm10]. In this case, one speaks of *flat* FLRW spacetimes and $\Sigma_\varkappa = \mathbb{R}^3$, while $r$ denotes the Euclidean distance in $\mathbb{R}^3$, which in our scenario is called the *comoving distance*. Obviously, a possible time function on a FLRW spacetime is given by the cosmological time itself, but there is another possibility called the *conformal time* $\tau$. It is defined as

$$\tau(t) \doteq \int_{t_0}^{t} \frac{1}{a(t')} dt',$$

---

[1] Note that the mentioned postulate is not beyond any doubt, see for instance [CaSt08, Cla07, HBG04].

which allows us to view flat FLRW spacetimes as a manifold $I_\tau \times \mathbb{R}^3$ with metric line element

$$ds^2 = a^2(\tau)\left(-d\tau^2 + dr^2 + r^2 d\mathbb{S}^2(\theta, \varphi)\right). \tag{I.12}$$

Here, $I_\tau \ni \tau$ denotes the open interval in $\mathbb{R}$ obtained from $I_t$ by the transition from cosmological time to conformal time.

The functional behaviour of the scale factor $a$ describes the 'history' of our universe, which, according to General Relativity, is completely determined by the specification of the matter-energy content of our universe in terms of the *stress-energy tensor* $T_{\mu\nu}$ and its coupling to gravity via the *Einstein equation*

$$G_{\mu\nu} \doteq R_{\mu\nu} - \frac{1}{2}R g_{\mu\nu} = 8\pi G T_{\mu\nu}. \tag{I.13}$$

Here, $G_{\mu\nu}$ denotes the *Einstein tensor*, $G$ is *Newton's constant*, and $T_{\mu\nu}$ is, in the present case of a FLRW spacetime, given by the stress-tensor of a *perfect fluid*, viz.

$$T^\mu{}_\nu = \begin{pmatrix} -\varrho & \\ & p\mathbb{I}_3 \end{pmatrix}. \tag{I.14}$$

In (I.14), the *energy density* $\varrho$ and the *pressure* $p$ of matter-energy are related by the *equation of state*

$$p = w\varrho \tag{I.15}$$

and we have written $T_{\mu\nu}$ with one upper index to display it in its most simple form. The most prominent types of matter-energy are *dust* (i.e. non-relativistic, classical matter) with $w \equiv 0$ and *radiation* with $w \equiv \frac{1}{3}$. Sometimes, a *cosmological constant* $\Lambda$ is added to the left hand side of the Einstein equation via $\Lambda g_{\mu\nu}$, but, in view of chapter IV, we consider it as a part of the matter-energy content with $w \equiv -1$.

Observations indicate that the current matter-energy content of the universe is comprised of roughly 27% dust and 73% *dark energy*, while the relative contribution of radiation is only of order $10^{-3}$ [Kom10, Amm10]. The measured $w$ of dark energy is in good agreement with a constant $w = -1$ [Kom10, Amm10] and can therefore be interpreted as a cosmological constant, though its origin remains unclear. As we shall discuss in great detail in chapter IV, quantum matter-energy has a time-varying equation of state that approaches $w = -1$ at late times; it is thus a possible explanation for dark energy. Moreover, we shall see that, in a suitable thermal state, the temperature-dependent part of quantum matter-energy mimics the behaviour of dust.

To solve the Einstein equation in flat FLRW spacetimes, one first computes [Wal84, chap. 5.2]

$$R^\mu{}_\nu = \begin{pmatrix} 3\frac{\ddot{a}}{a} & \\ & \left(2\frac{\dot{a}^2}{a^2} + \frac{\ddot{a}}{a}\right)\mathbb{I}_3 \end{pmatrix}, \quad R = 6\left(\frac{\dot{a}^2}{a^2} + \frac{\ddot{a}}{a}\right), \quad G^\mu{}_\nu = \begin{pmatrix} -3\frac{\dot{a}^2}{a^2} & \\ & \left(-\frac{\dot{a}^2}{a^2} - 2\frac{\ddot{a}}{a}\right)\mathbb{I}_3 \end{pmatrix}, \tag{I.16}$$

where a dot indicates taking the derivative with respect to $t$. The above system of ordinary differential equations reduces to the *Friedmann equations*

$$3\frac{\dot{a}^2}{a^2} = 8\pi G\varrho, \qquad 3\frac{\ddot{a}}{a} = -4\pi G(\varrho + 3p). \tag{I.17}$$

As the Bianchi identity (I.3) implies $\nabla^\mu G_{\mu\nu} = 0$, the Einstein equation only makes sense if the stress-energy tensor $T_{\mu\nu}$ is covariantly conserved as well, *i.e.* $\nabla^\mu T_{\mu\nu} = 0$. In (not necessarily flat) FLRW spacetimes, the covariant conservation of the stress-energy tensor implies

$$\dot{\varrho} + 3\frac{\dot{a}}{a}(\varrho + p) = 0, \tag{I.18}$$

and this equation can be obtained directly from the Friedmann equations since $\nabla^\mu T_{\mu\nu} = 0$ has been implicitly assumed in their derivation. Under the assumption that each matter-energy component is conserved on its own, (I.18) implicates that

$$\varrho_{\mathrm{rad.}} a^4 = \text{constant}, \qquad \varrho_{\mathrm{dust}} a^3 = \text{constant}, \qquad \varrho_{\mathrm{c.c.}} = \text{constant}. \tag{I.19}$$

We see that dust, radiation, and the cosmological constant have very different scaling behaviours with respect to $a$. The first Friedmann equation implies that, if $\varrho > 0$ for all times and $\dot{a} > 0$ at one instant of time, then $a$ will be strictly increasing for all times. Consequently, if we consider the present matter-energy content described above and assume that dark energy is a cosmological constant, then our universe must have had two phases of evolution preceding the present era dominated by dark energy: a phase where radiation has determined the behaviour of $a$ followed by an era dominated by dust. This motivates examining the solutions of the Friedmann equations separately for each matter-energy component and one finds [Wal84, chap 5.2]

$$a_{\mathrm{rad.}} \propto (t - t_0)^{\frac{1}{2}}, \qquad a_{\mathrm{dust}} \propto (t - t_0)^{\frac{2}{3}}, \qquad a_{\mathrm{c.c.}} \propto e^{\sqrt{\frac{\Lambda}{3}}\, t}. \tag{I.20}$$

The outcome of the preceding discussion is that, under the mentioned assumptions, our universe must have inevitably faced a *Big Bang* at some point of time in the past, *i.e.* there has been a $t_0 > -\infty$ with $a(t_0) = 0$ and we shall set $t_0 = 0$ in the following. Note that the occurrence of a Big Bang follows already from the second Friedmann equation and the assumptions $\dot{a} > 0$, $\varrho > 0$, since then $\varrho + 3p > 0$ and therefore $\ddot{a} < 0$ if we take into account the sum of all three matter-energy constituents in the radiation-dominated era at early times.

The scenario described above is part of the present *standard model of cosmology*, see for instance [Dod03] for a detailed discussion, and it is known that it has (at least) one hitch, usually termed *horizon problem*. We refer to [Dod03, chap. 6.2] for a quantitative discussion of this issue and only consider its qualitative aspects here. To wit, the isotropy of the temperature proper to the CMB radiation entails that the so-called *last scattering surface*, *i.e.* the region from where the CMB photons we see today have been emitted, must have lied in the forward lightcone of some event responsible for the thermal equilibrium of such region. The size of the last scattering surface is the radius $r_{\mathrm{em}}$ of our past lightcone at the time $t_{\mathrm{em}}$ of CMB photon emission, namely, the speed of light times the conformal time difference $\tau(t_{\mathrm{now}}) - \tau(t_{\mathrm{em}})$. The isotropy of the CMB therefore entails that the following inequality must hold

$$\tau(t_{\mathrm{em}}) - \tau(0) = \int_0^{t_{\mathrm{em}}} \frac{dt'}{a(t')} \geq r_{\mathrm{em}} \tag{I.21}$$

and one can compute that this is not the case in the standard model of cosmology; this is the horizon problem. A prominent possibility to solve the horizon problem is *inflation*, see for instance [Dod03, chap. 6.2]. In this scenario, one usually assumes that, in the very early universe, there has been an additional matter-energy component mimicking a large cosmological constant and thus leading to phase of exponential expansion. Inserting this assumption into (I.21) leads to a large negative $\tau(0)$ and therefore allows for (I.21) to be fulfilled. However, one does not have to assume an exponential expansion of the universe at early times to solve the horizon problem. Let us note that, in the standard model of cosmology, $\tau(0)$ in (I.21) is small and in particular finite because $a(t) \propto \sqrt{t}$ in the radiation dominated phase close to the Big Bang and the integral in (I.21) thus converges at the lower limit. If we assume that the very early universe was dominated by a matter-energy component leading to a scale factor fulfilling $a(t) \leq Ct$ for some $C > 0$, then $\tau(0) = -\infty$ and we have found another solution of the horizon problem. This scenario is called *power-law inflation* [LuMa85] and, as shown for the first time in [DFP08, Pin10] and analysed further in chapter IV, quantum fields on curved spacetimes constitute a matter-energy component leading to power-law inflation. This fact leads us to pay particular attention to quantum field theories on the class of Big Bang spacetimes incorporating power-law inflation in the remainder of this thesis.

In addition to the aforementioned reasons, there is another strong motivation to consider quantum field theories on power-law inflationary spacetimes. To understand it, let us redefine the conformal time as

$$\tau(t) \doteq -\int\limits_{t}^{t_1} \frac{1}{a(t')} dt',  \tag{I.22}$$

with some (possibly infinite) $t_1 > 0$ chosen such that the integral converges at the upper limit. Having done this, we are in the following situation: from (I.12), we can infer that, by means of a conformal embedding, the Big Bang spacetimes $M$ under consideration correspond to a portion of Minkowski spacetime $\mathbb{M}$. If $\tau(0)$ is finite, the 'surface' $t = 0$ of $M$ corresponds to a *spacelike* constant-time surface of $\mathbb{M}$, while in the case of an infinite $\tau(0)$, the time-zero surface of $M$ corresponds to the $\tau = -\infty$ 'surface' of $\mathbb{M}$. Technically, the $t = 0$ and $\tau = -\infty$ surfaces of $M$ and $\mathbb{M}$ respectively are certainly not a part of the particular spacetimes. We will discuss this issues in more precise terms in the next subsection, where we will also see that the $\tau = -\infty$ 'surface' of $\mathbb{M}$ is a *null surface*, which is indeed the anticipated second motivation to analyse quantum field theories on power-law inflationary spacetimes. Namely, as shown in a series of papers [DMP06, DMP09a, DMP09b, DMP09c, DPP10, Mor06, Mor08], it is possible to map quantum field theories in particular (not necessarily FLRW) spacetimes to distinguished codimension one null surfaces in a very controllable way. The advantage of this is that, while the spacetimes $M$ under consideration may not contain isometries, or the field theories on $M$ may not be invariant under the isometries of $M$, the mentioned null surfaces are always highly symmetric and the field theories mapped to them turn out to be invariant under these symmetries quite generically. This exploitation of the *holographic principle* introduced by 't Hooft in [tHo93] allows one to extend constructions well-known from Minkowski spacetime, like the selection of distinguished ground states or thermal states, to a large class of non-trivial curved spacetimes, as displayed in [DMP06, DMP09a, DMP09b, DMP09c, DPP10, Mor06, Mor08]. One part of this thesis is concerned with adding to these results by constructing ground states and thermal states for quantum field theories in Big Bang spacetimes of the kind described above. Before we lay the foundation to this by describing the geometric setup in detail in the next subsection, we close the present subsection by merging the above considerations into a definition. The additional technical restrictions we shall impose on $a(t)$ will ensure nice properties of $a(\tau)$, as we shall see in the following, and will furthermore be proper to a scale factor driven by quantum matter-energy, see [Pin10] and the last chapter of this thesis.

**Definition I.3.1.1** *A **null Big Bang spacetime (NBB)** is a flat FLRW spacetime with a scale factor $a(t)$ fulfilling:*

$$\text{there exist constants} \quad C_0 \in (0, \infty), \quad t_0 > 0 \quad \text{such that} \quad a(t) \le C_0 t \quad \forall \, t \le t_0.$$

*This implies that there is a (possibly infinite) $\tau_{max} > -\infty$ such that the conformal time $\tau$ defined by (I.22) ranges over $I_\tau = (-\infty, \tau_{max})$.*

*Moreover, we demand that, for all $\varepsilon \in (0, 1)$, there exist constants $C_\varepsilon \in (0, \infty)$, $t_\varepsilon > 0$ fulfilling*

$$C_\varepsilon \, t^{1+\varepsilon} \le a(t) \quad \forall \, t \le t_\varepsilon$$

*and that all derivatives of $a$ with respect to $t$ are bounded at the origin.*

### I.3.2 The Conformal Boundary of Null Big Bang Spacetimes

As already remarked in the preceding subsection, we will be concerned with mapping field theories on an NBB spacetime, henceforth denoted by $M_B$, to the $t = 0$, or equivalently, $\tau = -\infty$ boundary of $M_B$. The first task is thus to have a clear understanding of the structure and properties of this boundary. As we already know that an NBB spacetime is conformally embeddable in Minkowski spacetime $\mathbb{M}$, we can expect that the past conformal boundary $\partial^- M_B$ of $M_B$ is a part of the full conformal boundary $\partial \mathbb{M}$ of $\mathbb{M}$; we shall therefore proceed by discussing the latter.

Upon choosing polar coordinates, the Minkowski metric $\eta$ is specified by the line element

$$ds^2 = -d\tau^2 + dr^2 + r^2 d\mathbb{S}^2(\theta, \varphi)$$

where the notation is the same as in (I.12). To analyse, or, rather, to *define* $\partial \mathbb{M}$, we have to compactify it in order to be able to understand it as an open subset of a larger manifold $\widetilde{M}$; $\partial \mathbb{M}$ will then be defined as the topological boundary of $\mathbb{M}$ in $\widetilde{M}$. To achieve this, let us introduce the *retarded* and *advanced light coordinates* $u$ and $v$ respectively by setting

$$u \doteq \tau - r, \qquad v \doteq \tau + r.$$

In these coordinates, the Minkowskian line element reads

$$ds^2 = -du\,dv + \left(\frac{v-u}{2}\right)^2 d\mathbb{S}^2(\theta, \varphi).$$

In order to obtain a compactification of $\mathbb{M}$ which preserves the causal structure of $\mathbb{M}$, we define the compactified light coordinates $U$ and $V$ as

$$U = \tan^{-1}(u), \qquad V = \tan^{-1}(v)$$

and therefore transform the line element into

$$ds^2 = -\frac{dU\,dV}{\cos^2(U)\cos^2(V)} + \frac{\sin^2(V-U)}{4\cos^2(U)\cos^2(V)} d\mathbb{S}^2(\theta, \varphi).$$

If we finally define time and radial coordinates in the compactified space by

$$T \doteq V + U, \qquad R \doteq V - U,$$

then we obtain the line element

$$ds^2 = \frac{-dT^2 + dR^2 + \sin^2(R)d\mathbb{S}^2(\theta, \varphi)}{4\cos^2\left(\frac{T-R}{2}\right)\cos^2\left(\frac{T+R}{2}\right)}.$$

The enumerator of the above line element is nothing but the line element of the metric $g_E$ proper to the *Einstein static universe* $(M_E, g_E)$, sometimes also called *Einstein cylinder*. This spacetime is the result of Einstein's famous introduction of the cosmological constant into General Relativity and its manifold $M_E$ is the Cartesian product of $\mathbb{R}$ and the three sphere $\mathbb{S}^3$, see, *e.g.* [HaEl73] for further details. Note that the singularities of $g_E$ at $R \in \{0, \pi\}$ are nothing but the usual 'pole singularities' of polar coordinates. Altogether we see that, via a judicious choice of coordinates, we have been able to map Minkowski spacetime $\mathbb{M}$ into an open subset of $M_E$ and $\eta$ to a conformal transformation of $g_E$. If we call the corresponding conformal embedding $\chi^{\mathbb{M}} : (\mathbb{M}, \eta) \hookrightarrow (M_E, g_E)$, trace back our steps, and recall $r \geq 0$, we have that $(\chi_*^{\mathbb{M}}\eta)\!\restriction_{\chi^{\mathbb{M}}(\mathbb{M})} = (\Omega_{\mathbb{M}}^{-2} g_E)\!\restriction_{\chi^{\mathbb{M}}(\mathbb{M})}$, with the conformal factor specified as

$$\Omega_{\mathbb{M}}^2 = 4\cos^2\left(\frac{T-R}{2}\right)\cos^2\left(\frac{T+R}{2}\right) = \frac{4}{(1+v^2)(1+u^2)}, \tag{I.23}$$

and we can see that $\chi^{\mathbb{M}}(\mathbb{M})$ is the open subset of $M_E$ specified by the inequalities

$$-\pi < T + R < \pi, \qquad -\pi < T - R < \pi, \qquad R \geq 0.$$

We have depicted the situation in figure I.1, a so-called *Carter-Penrose diagram*. Note that the usual Carter-Penrose diagram of Minkowski spacetime encountered in the literature is the 'doubling' of figure I.1 and, hence, a 'diamond'. However, we have chosen to not draw it in this way, since $R$ corresponds to a 'generalised azimuthal angle' of $\mathbb{S}^3$ ranging only from $0$ to $\pi$. We consider a diamond picture misleading because of this fact and also in view of the forthcoming discussion of $\partial\mathbb{M}$. Minkowski spacetime, *i.e.* $\chi^{\mathbb{M}}(\mathbb{M})$, corresponds to the interior and left edge of the triangle displayed in figure I.1, while its conformal boundary corresponds to the topological boundary of $\chi^{\mathbb{M}}(\mathbb{M})$ in $M_E$, *viz.* $\partial\mathbb{M} \doteq i^+ \cup \mathfrak{I}^+ \cup i^0 \cup \mathfrak{I}^- \cup i^-$. It becomes visible that $\partial\mathbb{M}$ is comprised of five distinct pieces, namely, *future timelike infinity* $i^+$, *future null infinity* $\mathfrak{I}^+$, *spacelike infinity* $i^0$, *past null infinity* $\mathfrak{I}^-$, and *past timelike infinity* $i^-$. While $i^\pm$ and $i^0$ are two-spheres, $\mathfrak{I}^\pm$ are codimension-one submanifolds diffeomorphic to $I \times \mathbb{S}^2$, with $I$ an open interval in $\mathbb{R}$. Moreover, the interpretation of the boundary pieces is such that all future directed timelike geodesics of $\mathbb{M}$ begin in $i^-$ and end in $i^+$, all future directed null geodesics of $\mathbb{M}$ begin on $\mathfrak{I}^-$ and end on $\mathfrak{I}^+$, and all spacelike geodesics of $\mathbb{M}$ 'begin' in $i^0$ and 'end' in $i^0$. Let us stress that the conformal embedding $\chi^{\mathbb{M}}$ was sufficient to define the conformal boundary $\partial\mathbb{M}$, but, if in addition we would like to do tensor analysis on $\partial\mathbb{M}$, a conformal embedding is not enough since $\chi^{\mathbb{M}}_* \eta$ is still divergent in the limit to $\partial\mathbb{M}$. However, we can obviously apply the conformal transformation $\Omega^2_{\mathbb{M}}$ to $\chi^{\mathbb{M}}_* \eta$ to obtain the metric $g_E$ which can be manifestly extended to $\partial\mathbb{M}$ in a smooth way.

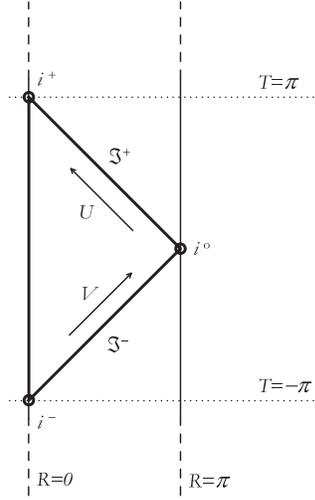

*Figure I.1:* The conformal embedding of Minkowski spacetime $\mathbb{M}$ into the Einstein static universe $M_E$. A point in this picture corresponds to a two-sphere.

Let us now return to NBB spacetimes. As we have already remarked, an NBB spacetime $M_B$ can be conformally embedded into Minkowski spacetime $\mathbb{M}$, which can in turn be conformally embedded into the Einstein static universe. We thus have a chain of conformal embeddings [Pin10]

$$(M_B, g_B) \hookrightarrow (\mathbb{M}, \eta) \hookrightarrow (M_E, g_E)$$

and, if we denote by $\chi^B$ the resulting conformal embedding $\chi^B : (M_B, g_B) \hookrightarrow (M_E, g_E)$, it

satisfies

$$\chi^B_* g_B\big\rvert_{\chi^B(M_B)} = \frac{a^2}{\Omega^2_{\mathbb{M}}} g_E\big\rvert_{\chi^B(M_B)}.$$

Henceforth, we shall omit the restriction of quantities to $\chi^B(M_B)$ whenever it is obvious in order to simplify notation. As the conformal time of an NBB spacetime ranges over $(-\infty, \tau_{max})$, we have

$$\chi^B(M_B) = \chi^{\mathbb{M}}(\mathbb{M})\big\rvert_{\tau < \tau_{max}}$$

and can therefore define the past conformal boundary $\partial^- M_B$ of $M_B$ as the 'past piece' of the conformal boundary $\partial \mathbb{M}$ of $\mathbb{M}$. To perform tensor analysis on $\partial^- M_B$, we have to neutralise the singularity of $\chi^B g_B$ on $\partial^- M_B$. A possibility would be to multiply it with the conformal factor $\Omega^2_{\mathbb{M}} a^{-2}$ in order to obtain the Einstein static universe metric $g_E$ in analogy to the discussion of the Minkowskian boundary $\partial \mathbb{M}$. However, as we are only considering the past boundary of $\mathbb{M}$, a 'weaker' conformal transformation that only counteracts the singularities of $\chi^B g_B$ on this boundary is sufficient. Moreover, the usage of such different conformal transformation yields a simple metric on $\partial^- M_B$, as stated in the following definition.

**Definition I.3.2.1**  *The **past conformal boundary** $\partial^- M_B$ of an NBB spacetime $(M_B, g_B)$ is defined as*

$$\partial^- M_B \doteq \mathfrak{J}^- \cup i^-,$$

*where $\mathfrak{J}^-$ and $i^-$ are **past null infinity** and **past timelike infinity** of Minkowski spacetime respectively. This entails that $i^-$ is a two-sphere $\mathbb{S}^2$, corresponding to the subset of the Einstein static universe $M_E$ specified as*

$$\frac{T-R}{2} = U = \tan^{-1}(u) = \tan^{-1}(\tau - r) = -\frac{\pi}{2}$$

$$and \qquad \frac{T+R}{2} = V = \tan^{-1}(v) = \tan^{-1}(\tau + r) = -\frac{\pi}{2},$$

*while $\mathfrak{J}^-$ is the Cartesian product of an open interval and $\mathbb{S}^2$, corresponding to the subset of $M_E$ specified as*

$$\frac{T-R}{2} = U = \tan^{-1}(u) = \tan^{-1}(\tau - r) = -\frac{\pi}{2}$$

$$and \qquad -\frac{\pi}{2} < \frac{T+R}{2} = V = \tan^{-1}(v) = \tan^{-1}(\tau + r) < \frac{\pi}{2}.$$

*Setting*

$$\Omega_B \doteq 2\cos\left(\frac{T-R}{2}\right) = 2\cos(U) = \frac{2}{\sqrt{1+u^2}} = \frac{2}{\sqrt{1+(\tau - r)^2}},$$

*we shall regard $\mathfrak{J}^-$ as the null manifold $\mathbb{R} \times \mathbb{S}^2$ described by coordinates $v \in \mathbb{R}$ and $(\theta, \varphi) \in \mathbb{S}^2$ and endowed with the metric*

$$h \doteq \frac{\Omega^2_B}{a^2} \chi^B_* g_B\big\rvert_{\mathfrak{J}^-} = (1+v^2) g_E\big\rvert_{\mathfrak{J}^-}$$

which, by a straightforward calculation, corresponds to the line element of the degenerate **Bondi metric**

$$ds^2 = 2d\Omega_B dv + d\mathbb{S}^2(\theta, \varphi).$$

In the definition of $h$, it is understood that the metric $\frac{\Omega_B^2}{a^2}\chi_*^B g_B$ is smoothly extended to $\mathfrak{I}^-$ before restricting it there.

Note that the above definition of $h$ is valid, since the differential $d\Omega_B$ of $\Omega_B$ is non-vanishing on $\mathfrak{I}^-$, although $\Omega_B$ itself is vanishing there.

As already briefly anticipated, we will be interested in exploiting the symmetry of $\mathfrak{I}^-$ and of quantum field theories defined on $\mathfrak{I}^-$ in order to obtain distinguished ground states and thermal states for quantum field theories in $M_B$. It is obvious that $\mathfrak{I}^-$ endowed with the Bondi metric is invariant under translations in the $v$ directions. In fact, we will use $v$ as a 'time coordinate' on $\mathfrak{I}^-$ and define ground states and thermal states with respect to these 'time translations'. The physical picture behind this is that the 'time-translation symmetry' with respect to $v$ on $\mathfrak{I}^-$ is an 'asymptotic conformal time-translation symmetry' of both $M_B$ and the field theories defined on it; this is underlined by the following lemma which in addition shows how the spatial translations extend to $\mathfrak{I}^-$.

**Lemma** I.3.2.2   The conformal Killing vector field $\partial_\tau$ of $(M_B, g_B)$ can be smoothly extended to $\chi^B(M_B) \cup \mathfrak{I}^-$ and its restriction to $\mathfrak{I}^-$ coincides with $\partial_v$.

Moreover, the Killing vector fields of $(M_B, g_B)$ generating spatial translations

$$\partial_{x_1} = \frac{1}{\sin(\theta)\cos(\varphi)}\partial_r + \frac{1}{r\cos(\theta)\cos(\varphi)}\partial_\theta - \frac{1}{r\sin(\theta)\sin(\varphi)}\partial_\varphi,$$

$$\partial_{x_2} = \frac{1}{\sin(\theta)\cos(\varphi)}\partial_r + \frac{1}{r\cos(\theta)\sin(\varphi)}\partial_\theta + \frac{1}{r\sin(\theta)\cos(\varphi)}\partial_\varphi,$$

$$and \quad \partial_{x_3} = \frac{1}{\cos(\theta)}\partial_r - \frac{1}{r\sin(\theta)}\partial_\theta$$

can be smoothly extended to $\chi^B(M_B) \cup \mathfrak{I}^-$ and their restriction to $\mathfrak{I}^-$ coincides with

$$\frac{1}{\sin(\theta)\cos(\varphi)}\partial_v, \quad \frac{1}{\sin(\theta)\sin(\varphi)}\partial_v, \quad and \quad \frac{1}{\cos(\theta)}\partial_v,$$

respectively.

*Proof.* The result follows by a simple computation once we have stressed a possibly obvious, yet important fact: we have used the notion of a conformal embedding and the related symbols to be able to write formulae in a clear and hopefully unambiguous way. However, for all practical purposes, the conformal embeddings discussed in this work are just judicious choices of coordinates in combination with restriction of coordinate ranges. This implies in particular that $\chi_*^{\,B} \partial_\tau \big|_{\chi^B(M_B)} = \partial_\tau$; analogous relations hold for all other vector fields mentioned in the statement of the present lemma.

With this in mind, let us consider $\partial_\tau$ first. Using the various coordinates defined above and the mutual relations among them, we obtain on $\chi^B(M_B)$

$$\partial_\tau = -\partial_u + \partial_v = -\frac{1}{1+\tan^2(U)}\partial_U + \partial_v.$$

Given $\partial_\tau$ in this form, we see that we can smoothly extend it to $\mathfrak{I}^-$ where $U = -\frac{\pi}{2}$ and, hence, $\frac{1}{1+\tan^2(U)} = 0$.

The result for the spatial translations can be obtained in a similar fashion by first recalling that

$$\partial_r = \partial_v + \partial_u,$$

and then realising that $\partial_\theta$ and $\partial_\varphi$ extend to $\mathfrak{I}^-$ in a trivial fashion, while their coefficients are proportional to $r^{-1} = 2(u+v)^{-1}$ and thus vanishing on $\mathfrak{I}^-$. □

The above lemma shows that the spatial translations of $(M_B, g_B)$ become angular dependent $v$-translations on $\mathfrak{I}^-$, so-called *supertranslations*. We can straightforwardly check that supertranslations are isometries of the Bondi metric $h$: in the degenerate coordinates $(v, \Omega_B, \theta, \varphi)$, the Bondi metric reads

$$h_{\mu\nu} = \begin{pmatrix} 0 & 1 & 0 & 0 \\ 1 & 0 & 0 & 0 \\ 0 & 0 & 1 & 0 \\ 0 & 0 & 0 & \sin^2(\theta) \end{pmatrix}. \tag{I.24}$$

Taking into account the coordinate expression for the Lie derivative, see [Wal84, app. 3], we obtain for all vector fields of the form $\alpha \partial_v$ with a smooth function $\alpha(\theta, \varphi)$ depending on the angular coordinates

$$\pounds_{\alpha\partial_v} h_{\mu\nu} = \alpha \partial_v h_{\mu\nu} - 2 h_{\mu\rho} \partial_\nu (\alpha \partial_v)^\rho,$$

the above expression vanishes for all $\mu$, $\nu$, either because $h$ has block diagonal form, or because $\alpha$ and $h$ are independent of $v$. Taking into account the above considerations, lemma I.3.2.2, and the trivial invariance of $h$ under rotations, we obtain the following important result.

**Lemma I.3.2.3** *The conformal isometry of $(M_B, g_B)$ constituted by $\tau$-translations and all isometries of $(M_B, g_B)$, i.e. the spatial translations and rotations, extend to isometries of $(\mathfrak{I}^-, h)$.*

As of now we understand how the conformal $\tau$-translation symmetry of $(M_B, g_B)$ extends to a strict $v$-translation symmetry of $(\mathfrak{I}^-, h)$. However, as already apprehended at several occasions, we will be concerned with mapping *massive* field theories in $M_B$ to $\partial^- M_B$ in such a way that they are invariant under $v$-translations. A reader familiar with conformally invariant theories might wonder how the latter aim can be achieved despite the fact that massive field theories are *not* conformally invariant. Admittedly, conformal transformations act on a massive field by rescaling the mass with the employed conformal factor, as we will discuss in more detail in chapter II. However, the following lemma shows that the additional restrictions we have imposed on $a(t)$ in definition I.3.1.1 entail that the conformal factor appearing in our scenario vanishes smoothly on $\partial^- M_B$, such that a massive field in $M_B$ becomes conformally invariant while approaching $\partial^- M_B$.

**Lemma** I.3.2.4

a) *The scale factor $a(\tau)$ of an NBB spacetime given in terms of the conformal time $\tau$ fulfils: for all $n > 0$ there exist constants $C_n \in (0, \infty)$, $\tau_n \in (-\infty, 0)$ such that*

$$a(\tau) \leq \left( \frac{1}{-C_n \tau} \right)^n \quad \forall \tau < \tau_{1,n}.$$

*Moreover, for all $k \in \mathbb{N}$, $a^{-1}(\tau) |\partial_\tau^k a(\tau)|$ is bounded in the limit $\tau \to -\infty$.*

b) *The conformal factors $\frac{a}{\Omega_B}$ and $\frac{a}{\Omega_{\mathbb{M}}}$ appearing in the definition of the Bondi metric and in the conformal embedding $(M_B, g_B) \hookrightarrow (M_E, g_E)$ respectively vanish smoothly towards both $\mathfrak{I}^-$ and $i^-$.*

*Proof.* Let us start by proving b) under the assumption that a) holds. To this avail, we recall

$$\frac{a}{\Omega_B} = \frac{a\left(\frac{u+v}{2}\right)\sqrt{1+u^2}}{2} = \frac{a(\tau)\sqrt{1+(\tau+r)^2}}{2}$$

$$\frac{a}{\Omega_{\mathbb{M}}} = \frac{a\left(\frac{u+v}{2}\right)\sqrt{1+u^2}\sqrt{1+v^2}}{2} = \frac{a(\tau)\sqrt{1+(\tau+r)^2}\sqrt{1+(\tau-r)^2}}{2}$$

and that taking the limit to $\mathfrak{I}^-$ corresponds to taking the limit $u \to -\infty$ while keeping $v$ finite, whereas taking the limit to $i^-$ is equivalent to taking the limit $\tau \to -\infty$ at finite $r$. We thus directly see that both conformal factors vanish in the considered limit on account of the rapid decrease of $a(\tau)$ towards $\partial^- M_B$. To see that they vanish smoothly, we have to analyse all their derivatives with respect to the coordinates $U$ and $V$, which are regular on $\partial^- M_B$. Let us consider $U$-derivatives and $a\Omega_{\mathbb{M}}^{-1}$ for definiteness, the other cases can be treated analogously. From $\partial_U = (1+u^2)\partial_u$ we can infer that $\partial_U^k a\Omega_{\mathbb{M}}^{-1}$ is given in terms of derivatives of $a$ multiplied by terms which are polynomially bounded in $u$ and $v$. Since a) entails that the $\tau$-derivatives of $a$ vanish in the considered limit at least as fast as $a$ itself, $\partial_U^k a\Omega_{\mathbb{M}}^{-1}$ converges to $0$ towards $\partial^- M_B$ for all $k$.

To prove a), let us remember that, on account of definition I.3.1.1, there are positive constants $C$, $C_\varepsilon$, $t_\varepsilon$, $t_0$ such that, for all $\varepsilon \in (0,1)$,

$$C_\varepsilon t^{1+\varepsilon} \leq a(t) \quad \forall t \leq t_\varepsilon, \qquad a(t) \leq C t \quad \forall t \leq t_0.$$

Inserting the lower bound in the definition of the conformal time (I.22), we obtain

$$-\tau = \int\limits_{t}^{t_1} \frac{dt'}{a(t')} = D_1 + \int\limits_{t}^{t_\varepsilon} \frac{dt'}{a(t')} \le D_1 - \frac{1}{D_2 \varepsilon}\left(t_0^{-\varepsilon} - t^{-\varepsilon}\right) = -D_3 + \frac{1}{D_4} t^{-\varepsilon},$$

with suitable constants $D_i$, which we can all assume to be positive without loss of generality ($D_3$ could in principle be negative, but it will always be positive for a sufficiently small $t_\varepsilon$). Solving the above inequality for $t$, we obtain

$$t \le \left(\frac{1}{D_4(D_3 - \tau)}\right)^{\frac{1}{\varepsilon}}$$

and, hence, the wished-for property with $n = 1/\varepsilon$ and $C_n = D_4/C$ by inserting the above inequality in the upper bound for $a(t)$. $\qquad\square$

To close this subsection, let us describe how the above discussion relates to the scenario analysed in the already mentioned works [DMP06, Mor06, Mor08]. As remarked before, $\partial^- M_B$ is invariant under the extensions of translations in $\tau$, spatial translations, and rotations to $\partial^- M_B$, which, considered on Minkowski spacetime $\mathbb{M}$, are a subset of the proper, orthochronous Poincaré transformations. Since $\partial^- M_B$ is part of the conformal boundary of $\mathbb{M}$, one could expect that $\partial^- M_B$ is invariant under the *full* proper, orthochronous Poincaré group suitably extended to $\partial^- M_B$. This can indeed be shown to be true, but the situation is in fact considerably more general: Minkowski spacetime is a special case of an *asymptotically flat spacetime*, see [DMP06, Mor06, Mor08, Wal84] and references therein. Such spacetimes are generally containing the null boundary piece $\mathfrak{J}^-$, and, depending on the definition, the pieces $i^+$, $i^-$, or $i^0$, where $\mathfrak{J}^-$ is viewed as an *intrinsic* object specified by a manifold $\mathbb{R} \times \mathbb{S}^2$, a degenerate metric $h$, and a null vector field $\partial_v$ generating $\mathfrak{J}^-$. Such intrinsically defined $\mathfrak{J}^-$ turns out to have an *infinite dimensional* symmetry group, the so-called *Bondi-Metzner-Sachs* (BMS) group, which contains the proper, orthochronous Poincaré group as a normal subgroup, *cf.* [ArDa03, ArDa04, Dap05, DMP06, Mor06, Mor08, Wal84] and references therein. To understand why the BMS group is larger than the proper, orthochronous Poincaré group, note that, by the calculation preceding lemma I.3.2.3, *all* translations of $v$ depending on $\mathbb{S}^2$-angles are contained in the BMS group, but not all of them can be extensions of Poincaré transformations. For an introduction to these topics, we refer the reader to [Wal84, chap. 11] and to the lecture notes [Dap09]. As a final comment, note that an NBB spacetime is *not* an asymptotically flat spacetime, but only conformally related to one.

### I.3.3 Spinors in Flat Friedmann-Lemaître-Robertson-Walker Spacetimes

Since we would like to analyse Fermionic field theories on NBB spacetimes, we need to discuss how the spinor-related constructions described in the subsections I.2.1 and I.2.2 are realised in such spacetimes. Particularly, we have to understand if and how we can map spinorial quantities on an NBB spacetime to its past null boundary. Some parts of the following discussion do not depend on the functional behaviour of the scale factor $a$ and will therefore encompass all flat FLRW spacetimes.

To start in more general terms, let us consider the interplay between spin structures and conformal transformations. As exhibited in subsection I.2.1, the starting point to obtain a spin structure on a spacetime $(M, g)$ is the bundle of Lorentz frames $LM$ of $M$. Let us recall that a Lorentz frame is specified by $g(e_a, e_b) \equiv \eta_{ab}$. Therefore, if $e_a$ is a Lorentz frame of $(M, g)$, $\Omega^{-1} e_a$ will be a Lorentz frame of the conformally related spacetime $(M, \Omega^2 g)$. This entails that the Lorentz frame bundles of $(M, g)$ and $(M, \Omega^2 g)$ are equivalent, *i.e.* they only differ by the implementation of their local trivialisations by means of a Lorentz frame proper to $(M, g)$ and $(M, \Omega^2 g)$ respectively. The next step in constructing a spin structure on $(M, g)$ is then to lift the Lorentz frame bundle $LM$ to a bundle of spin frames $SM$ by covering the local Lorentz groups, *i.e.* the fibres of $LM$, in a coherent way. As remarked in subsection I.2.1, the obstruction and essential factor in this step is the topology of $M$. Since the latter is manifestly not affected by a conformal transformation, the spin structures on $(M, g)$ and $(M, \Omega^2 g)$ are again equivalent, *viz.* in case of multiple inequivalent spin structures on $(M, g)$, each such structure is mapped to a one on $(M, \Omega^2 g)$ which is unique up to equivalence in the sense of definition I.2.1.2 and vice versa. Finally, the construction of a spin connection on $(M, \Omega^2 g)$ proceeds in the same way as on $(M, g)$ and the relation between the resulting spin connection coefficients can be explicitly computed employing only the relation between Lorentz frames on $(M, g)$ and $(M, \Omega^2 g)$ mentioned at the beginning of this paragraph; this holds since the $\gamma$-matrices $\gamma_a$ and spin frames $E_A$ on $(M, g)$ and $(M, \Omega^2 g)$ can be chosen to be identical on account of the above discussion. While the preceding discussion may seem too general for our rather specific purposes, it leads us to a very simple computation of the spin connection coefficients in flat FLRW spacetimes.

***Lemma*** **I.3.3.1**

a) *Let $\Gamma_{ac}^{\ b}$ denote the connection coefficients of the Levi-Civita connection on an arbitrary spacetime $(M, g)$ and let $\widetilde{\Gamma}_{ac}^{\ b}$ denote the connection coefficients of the Levi-Civita connection on $(M, \Omega^2 g)$, where $\Omega$ is a strictly positive function on $M$ and the connection coefficients are specified in the Lorentz frame bases $e_a$ on $(M, g)$ and $\widetilde{e}_a = \Omega^{-1} e_a$ on $(M, \Omega^2 g)$, respectively. Then, $\widetilde{\Gamma}_{ac}^{\ b}$ and $\Gamma_{ac}^{\ b}$ are related as*

$$\widetilde{\Gamma}_{ac}^{\ b} = \frac{1}{\Omega} \Gamma_{ac}^{\ b} + \frac{1}{\Omega^2} (\delta_a^{\ b} \nabla_c \Omega - \eta_{ac} \nabla^b \Omega),$$

*where $\nabla_a = \nabla_{e_a}$ denotes the covariant derivative with respect to the Levi-Civita connection on $(M, g)$. Moreover, if $(M, g)$ is a four-dimensional, globally hyperbolic spacetime and we denote by $\sigma_a$ and $\widetilde{\sigma}_a$ the spin connection coefficients related to equivalent spin structures on $(M, g)$ and $(M, \Omega^2 g)$ respectively, then $\widetilde{\sigma}_a$ and $\sigma_a$ are related as*

$$\widetilde{\sigma}_a = \frac{1}{\Omega} \sigma_a + \frac{1}{4\Omega^2} [\gamma_a, \gamma_c] \nabla^c \Omega, \qquad \gamma^a \widetilde{\sigma}_a = \frac{1}{\Omega} \gamma^a \sigma_a + \frac{3}{2\Omega^2} \gamma^a \nabla_a \Omega.$$

*b) The spin connection coefficients in a flat FLRW spacetime satisfy*

$$\gamma^a \sigma_a = \frac{3a'}{2a^2}\gamma^0,$$

*where $\cdot'$ denotes taking the derivative with respect to conformal time.*

*Proof.*

Let us start by proving a). If we denote by $\widetilde{\nabla}$ the covariant derivative associated to $\Omega^2 g$, then the connection coefficients $\widetilde{\Gamma}_{ac}^{\ b}$ and $\Gamma_{ac}^{\ b}$ are defined as

$$\widetilde{\Gamma}_{ac}^{\ b} \doteq \widetilde{e}^b\left(\widetilde{\nabla}_{\widetilde{e}_a}\widetilde{e}_c\right), \qquad \Gamma_{ac}^{\ b} \doteq e^b\left(\nabla_{e_a}e_c\right).$$

Inserting $\widetilde{e}_a = \Omega^{-1}e_a$, the identity $\widetilde{e}^a = \Omega e_a$ related to the former by $\widetilde{e}^a(\widetilde{e}_b) \equiv \delta^a_b$, and recalling the properties of a covariant derivative, we obtain

$$\widetilde{\Gamma}_{ac}^{\ b} = e^b\left(\widetilde{\nabla}_{e_a}\frac{1}{\Omega}e_c\right).$$

Now we have to use the fact that $\widetilde{\nabla}$ and $\nabla$ must agree on scalar functions, whereas the difference of their actions on frames can be obtained from their metric compatibility as [Wal84, app. D]

$$\widetilde{\nabla}_{e_a}e_c = \nabla_{e_a}e_c + \left(\delta^d_a(\nabla_{e_c}\ln\Omega) + \delta^d_c(\nabla_{e_a}\ln\Omega) - \eta_{ac}\eta^{df}(\nabla_{e_f}\ln\Omega)\right)e_d.$$

These considerations combined with some elementary algebraic steps lead to

$$\widetilde{\Gamma}_{ac}^{\ b} = \frac{1}{\Omega}\Gamma_{ac}^{\ b} + \frac{1}{\Omega^2}(\delta^b_a\nabla_c\Omega - \eta_{ac}\nabla^b\Omega).$$

The identities for the spin connection coefficients follow by a direct computation upon recalling

$$\widetilde{\sigma}_a = \frac{1}{4}\widetilde{\Gamma}_{ac}^{\ b}\gamma_b\gamma^c, \qquad \sigma_a = \frac{1}{4}\Gamma_{ac}^{\ b}\gamma_b\gamma^c,$$

and the anticommutation relations of the $\gamma$-matrices, *cf.* lemma I.2.2.9.

To prove b), let us realise and recall the following facts: a flat FLRW spacetime $M$ is conformally related to the topologically trivial Minkowski spacetime $\mathbb{M}$ via $g_B = a^2\eta$. Hence, there exist unique spin structures on both spacetimes, and, on the portion of $\mathbb{M}$ corresponding to the conformal embedding of $M$, they are related by the same conformal transformation mapping $M$ into $\mathbb{M}$. We can therefore apply part a) of the present lemma. To this avail, we have to realise that the canonical frame on $\mathbb{M}$ is provided by

$$e_0 = \partial_\tau, \qquad e_i = \partial_i,$$

with $\partial_i$ denoting the partial derivative with respect to the Cartesian coordinates of $\mathbb{R}^3$. If we insert this and $\Omega = a$ into the identity for $\gamma^a\widetilde{\sigma}_a$ proved in a) and recall that Minkowski spacetime is flat and has vanishing (spin) connection coefficients, we obtain the wished-for result. $\qquad\square$

It is necessary to know how certain quantities behave approaching the past null boundary $\partial^- M_B$ of an NBB spacetime $M_B$ in order to be able to map field theories on $M_B$ to $\partial^- M_B$ in a controlled way. More in detail, as we would like to perform several operations with spinor-tensors on $\partial^- M_B$, we have employed the conformal transformation $\Omega_B^2 a^{-2}$ to endow $\partial^- M_B$ with a non-degenerate, but regular Bondi metric $\mathring{h}$, see definition I.3.2.1. Therefore, 'mapping a quantity $Q$ on $M_B$ to $\partial^- M_B$' has to be read as 'transforming $\chi_*^B Q$ according to the conformal transformation $\chi_*^B g_B \to \Omega_B^2 a^{-2} \chi_*^B g_B$ and then taking the limit to $\partial^- M_B$'. We will of course specify the latter statement precisely whenever it will be necessary, but for the moment it should serve as a motivation to understand why we are interested in whether mapping spinors on $M_B$ to $\partial^- M_B$ is possible and, if this is the case, how the result of such mapping looks like. In fact, it turns out that the canonical spin frame on $M_B$ induced by its canonical Cartesian Lorentz frame is not well suited for this purpose as it is vanishing on $\partial^- M_B$. Hence, as we shall discuss in the following, the conformal transformation $\chi_*^B g_B \to \Omega_B^2 a^{-2} \chi_*^B g_B$ has to be supplemented with a spin frame transformation in order to map spinorial quantities from $M_B$ to $\partial^- M_B$ in a well-defined way.

To this avail, we need some preparatory considerations regarding the possibility to extend the spin structure on $\chi^B(M_B)$ to a spin structure on $\chi^B(M_B) \cup \partial^- M_B$. As $(\chi^B(M_B), \chi_*^B g_B)$ is per definition isometric to $(M_B, g_B)$, all constructions related to spin structures on $M_B$ can be trivially extended to $\chi^B(M_B)$. Moreover, since $\chi_*^B g_B$ is a conformal transformation of the Einstein static universe metric $g_E$, we can immediately relate the spin structures on $(\chi^B(M_B), \chi_*^B g_B)$ and $(\chi^B(M_B), g_E \!\restriction_{\chi^B(M_B)})$ on account of the discussion at the beginning of this subsection. $M_B$ and, hence, $\chi^B(M_B) \subset M_E$ are topologically trivial, while the full manifold $M_E$ is not, but it is at least simply connected, since $\mathbb{S}^3$ is. As a result, all manifolds under consideration carry unique spin structures; the spin structure of $(M_E, g_E)$ restricted to $\chi^B(M_B)$ must therefore be conformally equivalent to the one on $(\chi^B(M_B), \chi_*^B g_B)$, which trivially implies that the spin structure on $(\chi^B(M_B), \chi_*^B g_B)$ can be extended to one on $(M_E, g_E)$ via a conformal transformation. As already remarked at the beginning of this subsection, we can without loss of generality choose the 'same' spin frame on $(\chi^B(M_B), \chi_*^B g_B)$ and $(\chi^B(M_B), g_E \!\restriction_{\chi^B(M_B)})$. However, we can not extend the convenient spin frame employed in part b) of lemma I.3.3.1 from $(\chi^B(M_B), g_E \!\restriction_{\chi^B(M_B)})$ to the full Einstein static universe $(M_E, g_E)$ in a meaningful way, as already the associated Lorentz frame is vanishing on $\partial^- M_B$. To see this, let us introduce the canonical Lorentz frame $e_a^E$ on $(M_E, g_E)$ defined as

$$e_0^E \doteq \partial_T, \quad e_1^E \doteq \partial_R, \quad e_2^E \doteq \frac{1}{\sin(R)} \partial_\theta, \quad e_3^E \doteq \frac{1}{\sin(R)\sin(\theta)} \partial_\phi, \tag{I.25}$$

and let us denote the Lorentz frame on $(\chi^B(M_B), g_E \!\restriction_{\chi^B(M_B)})$ obtained by the conformal transformation $\chi_{\mathbb{M}}^B g_B \to \Omega_{\mathbb{M}}^2 a^{-2} \chi_*^B g_B$ from the canonical Lorentz frame on $(M_B, g_B)$ by

$$\widetilde{e}_a \doteq a \Omega_{\mathbb{M}}^{-1} e_a^B, \quad e_0^B \doteq \frac{1}{a} \partial_\tau, \quad e_i^B \doteq \frac{1}{a} \partial_i. \tag{I.26}$$

Recalling the various relations between the coordinates $(T, R)$, $(U, V)$, $(u, v)$, and $(\tau, r)$, one can for instance straightforwardly compute

$$e_0^E = \Omega_{\mathbb{M}} \left( \frac{2 + u^2 + v^2}{4} \widetilde{e}_0 + \frac{v^2 - u^2}{4} \sum_{i=1}^{3} \frac{\partial x_i}{\partial r} \widetilde{e}_i \right).$$

As $\Omega_{\mathbb{M}} = 2\sqrt{(1 + u^2)(1 + v^2)}^{-1}$, whereas $e_0^E$ is finite on $\partial^- M_B$, which is characterised by $\lim_{u \to -\infty}$, we see that the frame $\widetilde{e}_a$ is vanishing on $\partial^- M_B$. Hence, as already anticipated, in order to map spinorial quantities in the bulk NBB spacetime $M_B$ to its boundary $\partial^- M_B$ in a finite way, we have to perform first a conformal transformation, and then a spin frame transformation, namely, the one related to the Lorentz transformation $\widetilde{e}_a \to e_a^E$. For later reference, we collect these insights and related details in the following lemma.

**Lemma** I.3.3.2 *Let $e_a^E$ denote the canonical global Lorentz frame on $(M_E, g_E)$ defined in (I.25) and let $\widetilde{e}_a$ denote the Lorentz frame on $\left(\chi^B(M_B), g_E\!\restriction_{\chi^B(M_B)}\right)$ obtained by a conformal transformation from the canonical global Lorentz frame on $(M_B, g_B)$, cf. (I.26).*

*a)* $e_a^E$ *and $\widetilde{e}_a$ are related by the Lorentz transformation*

$$e_a^E = \widetilde{e}_b (\Lambda^B)^b{}_a, \quad (\Lambda^B)^b{}_a \doteq \begin{pmatrix} \Omega_{\mathbb{M}} \frac{2 + u^2 + v^2}{4} & \Omega_{\mathbb{M}} \frac{v^2 - u^2}{4} & 0 & 0 \\ \Omega_{\mathbb{M}} \frac{v^2 - u^2}{4} \vec{e}_r & \Omega_{\mathbb{M}} \frac{2 + u^2 + v^2}{4} \vec{e}_r & \vec{e}_\theta & \vec{e}_\phi \end{pmatrix}$$

*which is divergent on $\partial^- M_B$. Here, $\vec{e}_r$, $\vec{e}_\theta$, and $\vec{e}_\phi$ denote the unit vectors of $\mathbb{R}_3$ w.r.t. polar coordinates.*

*b)* *Let $E_A^E$ and $\widetilde{E}_A$ denote the unique (up to a sign) spin frames associated to $e_a^E$ and $\widetilde{e}_a$ respectively via the unique spin structure on $(M_E, g_E)$. Moreover, let $\widetilde{\Lambda}^B$ denote the matrix-valued function on $\chi^B(M_B)$ constituting the transformation $\widetilde{E}_A \to E_A^E$, uniquely determined by*

$$\widetilde{\Lambda}^B \gamma_a (\widetilde{\Lambda}^B)^{-1} = \gamma_b (\Lambda^B)^b{}_a$$

*up to a sign. Then, the spinor covariant derivatives $\nabla_a^E$ and $\widetilde{\nabla}_a$ related to $e_a^E$ and $\widetilde{e}_a$ respectively fulfil*

$$\nabla_a^E = \widetilde{\Lambda}^B \widetilde{\nabla}_b (\widetilde{\Lambda}^B)^{-1} \left[ (\Lambda^B)^{-1} \right]^b{}_a.$$

Although the above-mentioned transformation property of the spinor covariant derivatives follows immediately from lemma I.2.2.10 applied to the two different frames considered here, we have written it down explicitly, as one might think that such identity fails to hold for spacetime-dependent frame transformations at first glance. Finally, let us recall that the Bondi metric on $\partial^- M_B$ arises from $g_E$ via an additional conformal transformation not considered in this lemma, *cf.* definition I.3.2.1. However, this conformal transformation is regular on $\partial^- M_B$, and hence neither essential, nor obstructive for the above results.

## I.4    Spacetime Categories

The main topic of the present thesis is the investigation of the backreaction proper to quantum fields in curved spacetimes, *i.e.* the effect of quantum matter-energy on the curvature of spacetime. This of course necessitates the ability to define quantum field theory on a curved spacetime *without knowing the curved spacetime beforehand*. It is therefore advisable to employ only generic properties of spacetimes in the construction of quantum fields, and we shall follow this philosophy throughout this work. This entails that we have to formulate a quantum field theory in a *local* way, *i.e.* only employing local properties of the underlying curved manifold. In addition, we would like to take into account the diffeomorphism-invariance of General Relativity and therefore construct *covariant* quantum fields. This concept of a *locally covariant quantum field theory* goes back to many works, of which the first one chould mention is [Dim82], followed by many others such as [Wal95, chap. 4.6] and [Ver01, HoWa01]. Building on these works, the authors of [BFV03] have given the first complete definition of a locally covariant quantum field theory.

As shown in [BFV03], giving such a definition in precise mathematical terms requires the language of *category theory*, a branch of mathematics which basically aims to unify *all* mathematical structures into one coherent picture. A category is essentially a class $\mathfrak{C}$ of *objects* denoted by $obj(\mathfrak{C})$, with the property that, for each two objects $A$, $B$ in $\mathfrak{C}$ there is (at least) one *morphism* or *arrow* $\phi : A \to B$ relating $A$ and $B$. The collection of all such arrows is denoted by $hom_{\mathfrak{C}}(A,B)$. Morphisms relating a chain of three objects are required to be associative with respect to compositions, and one demands that each object has an *identity morphism* $id_A : A \to A$ which leaves all morphisms $\phi : A \to B$ starting from $A$ invariant upon composition, *i.e.* $\phi \circ id_A = \phi$. An often cited simple example of a category is the category of sets $\mathfrak{Set}$. The objects of $\mathfrak{Set}$ are sets, while the morphism are maps between sets, the identity morphism of an object just being the identity map of a set. Given two categories $\mathfrak{C}_1$ and $\mathfrak{C}_2$, a *functor* $F : \mathfrak{C}_1 \to \mathfrak{C}_2$ is a map between two categories which maps objects to objects and morphisms to morphisms such that identity morphisms in $\mathfrak{C}_1$ are mapped to identity morphism in $\mathfrak{C}_2$ and the composition of morphisms is preserved under the mapping. This paragraph was only a very brief introduction to category theory and we refer the reader to the standard monograph [Mac98] and to the introduction in [Sze04, sec. 1.7] for further details. A locally covariant quantum field theory according to [BFV03] should be a functor from a category of spacetimes to a category of suitable algebras. The first step in understanding such a construction if of course the definition of judicious categories of spacetimes, which shall thus be the topic of the present section.

We have already explained in the previous sections of this chapter that four-dimensional, oriented and time-oriented, globally-hyperbolic spacetimes are the physically sensible class of spacetimes among all curved Lorentzian manifolds. It is therefore natural to take them as the objects of a potential category of spacetimes. Regarding the morphisms, one could think of various possibilities to select them among all possible maps between the spacetimes under consideration. However, to be able to emphasise the local nature of a quantum field theory, we shall take embeddings between spacetimes. This will allow us to require locality by asking that a quantum field theory on a 'small' spacetime can be easily embedded into a larger spacetime without 'knowing anything' about the remainder of the larger spacetime. Moreover, a sensible quantum field theory will depend on the orientation and time-orientation and the causal structure of the underlying manifold, we should therefore only consider embeddings that preserve these structures. To this avail, the authors in [BFV03] have chosen isometric embeddings with causally convex range (see section I.1 regarding an explanation of these notions), but since the causal structure of a spacetime is left invariant by conformal transformations, one could also choose conformal embeddings, as done in [Pin09]. We will nevertheless follow the choice of [BFV03], since it will be sufficient for our purposes. Let us now subsume the above considerations in a definition.

**Definition I.4.1**  *The **category of spacetimes** $\mathfrak{Man}$ is the category having as its class of objects obj($\mathfrak{Man}$) the globally hyperbolic, four-dimensional, oriented and time-oriented spacetimes $(M, g)$. Given two spacetimes $(M_1, g_1)$ and $(M_2, g_2)$ in obj($\mathfrak{Man}$), the considered morphisms $\hom_{\mathfrak{Man}}((M_1, g_1), (M_2, g_2))$ are isometric embeddings $\chi : (M_1, g_1) \hookrightarrow (M_2, g_2)$ preserving the orientation and time-orientation and having causally convex range $\chi(M_1)$. Moreover, the identity morphism $id_{(M,g)}$ of a spacetime in obj($\mathfrak{Man}$) is just the identity map of $M$ and the composition of morphisms is defined as the usual composition of embeddings.*

The just defined category is sufficient to discuss locally covariant Bosonic quantum field theories. However, for Fermionic quantum field theories, we need a category which incorporates spin structures both on the level of objects and on the level of morphisms, as defined in [Ver01, San08]. Before we state the definition of this enlarged category, let us briefly remark that our usage of the words 'Boson' and 'Fermion' for integer and half-integer spin fields respectively is allowed on account of the spin-statistics theorem in curved spacetimes proved in [Ver01].

**Definition I.4.2**  *The **category of spin spacetimes** $\mathfrak{SMan}$ is the category having as its class of objects obj($\mathfrak{SMan}$) globally hyperbolic, four-dimensional, oriented and time-oriented spacetimes endowed with a spin structure $(M, g, SM, \rho)$. Given two spin spacetimes $(M_1, g_1, SM_1, \rho_1)$ and $(M_2, g_2, SM_2, \rho_2)$ in obj($\mathfrak{SMan}$), the considered morphisms $\hom_{\mathfrak{SMan}}((M_1, g_1, S_1, \rho_1), (M_2, g_2, S_1, \rho_1))$ are pairs $X = (\chi, \widetilde{\chi})$ with $\chi$ an element of $\hom_{\mathfrak{Man}}((M_1, g_1), (M_2, g_2))$ and $\widetilde{\chi} : SM_1 \to SM_2$ being a smooth map which satisfies $\widetilde{\chi} \circ R_S^1 = R_S^2 \circ \widetilde{\chi}$ and $\rho_2 \circ \widetilde{\chi} = \chi_* \circ \rho_1$. Here, $R^1$ and $R^2$ are the right actions of $SM_1$ and $SM_2$ respectively, and $\chi_* : LM_1 \to LM_2$ denotes the canonical push-forward of $\chi$ to the Lorentz frame bundles of the spacetimes under consideration. Identity morphisms and the composition of morphisms are defined in analogy to the definition of $\mathfrak{Man}$.*

The just defined categories are the last spacetime structure we need in order to define field theories on curved spacetimes.

# II

# Classical Fields in Curved Spacetimes

Having set up the geometry and the necessary structures of the spacetimes we shall consider, we will now proceed to discuss fields propagating on these spacetimes. Although a quantum field is arguably a more fundamental object than a classical field, we shall follow the successful approach to discuss classical fields first and quantize them afterwards.

## II.1 The Free Scalar Field in General Curved Spacetimes

We shall start our discussion of classical fields on curved spacetimes with the potentially most exhausted and maybe physically least relevant model, namely, the free scalar field. However, the treatment of this simple case will undoubtedly give us a clear understanding of the essential structures in classical field theory necessary for quantization.

### II.1.1 The Klein-Gordon Equation and its Fundamental Solutions

In physics, we are used to describe dynamics by (partial) differential equations and initial conditions. To apply this philosophy to the scalar field, we therefore need a partial differential equation to start with. In Minkowski spacetime, Poincaré symmetry entails that the *Klein-Gordon equation* $(-\Box + m^2)\phi = (-\partial_\mu \partial^\mu + m^2)\phi = 0$ is the unique relativistic equation for a scalar field $\phi$, with only the mass $m$ as a free parameter. In curved spacetimes, we have only local Lorentz symmetry and global diffeomorphism symmetry at our disposal, see subsection I.2.1. We can therefore only require the equation of motion for a free scalar field on curved spacetimes to be a scalar operator with respect to these two notions. In addition, one may require that such equation of motion reduces to the Klein-Gordon one in the case of vanishing curvature. In the masslass case, this determines the equation to be

$$(-\Box + f)\phi \doteq (-\nabla_\mu \nabla^\mu + f)\phi = 0$$

with the *d'Alembert operator* $\Box$ and some scalar function $f$ of mass dimension 2 constructed out of the metric and vanishing in flat spacetime. One can now, depending on ones taste, either invoke *Occam's razor* or require analytic metric dependence of $f$ like in [HoWa01, sec. 5.1] to restrict the possible freedom to $f = \xi R$ with a free constant parameter $\xi \in \mathbb{R}$. Note that analyticity arguments rule out $f$ such as

$$f = R \frac{R_{\alpha\beta} R^{\alpha\beta}}{R^2},$$

which would certainly vanish in 'some' spacetimes and with respect to 'some' limits towards vanishing curvature. We shall therefore regard

$$(-\Box + \xi R)\phi = 0$$

as the possible equation of motion for massless free scalar fields in curved spacetimes. Note that this implies a coupling of $\phi$ to the curvature of the spacetime both via $\Box$ and via $R$. The case $\xi = 0$ is usually called *minimal coupling*, whereas, in four dimensions, the case $\xi = 1/6$ is called *conformal coupling*. While the motivation for the former name is obvious, the reason for the latter is rooted in the transformation of the covariant derivative under conformal transformations. Namely, if we consider the conformally related metrics $g$ and $\widetilde{g} \doteq \Omega^2 g$ with a strictly positive smooth function $\Omega$, denote by $\nabla$, $\Box$, and $R$ the quantities associated to $g$ and by $\widetilde{\nabla}$, $\widetilde{\Box}$, and $\widetilde{R}$ the quantities associated to $\widetilde{g}$, then the respective metric compatibility of the covariant derivatives $\nabla$ and $\widetilde{\nabla}$ and their agreement on scalar functions imply [Wal84, app. D]

$$\left(-\Box + \frac{1}{6}\widetilde{R}\right)\frac{1}{\Omega} = \frac{1}{\Omega^3}\left(-\Box + \frac{1}{6}R\right). \tag{II.27}$$

This entails that a function $\phi$ solving $(-\Box + \frac{1}{6}R)\phi = 0$ can be mapped to a solution $\widetilde{\phi}$ of $(-\widetilde{\Box} + \frac{1}{6}\widetilde{R})\widetilde{\phi} = 0$ by multiplying it with the conformal factor $\Omega$ to the power of the *conformal weight* $-1$, *i.e.* $\widetilde{\phi} = \Omega^{-1}\phi$. We shall therefore call a scalar field $\phi$ with an equation of motion $(-\Box + \frac{1}{6}R)\phi = 0$ *conformally invariant*. In other spacetimes dimensions $d \neq 4$, the conformal weight and the magnitude of the conformal coupling are different, see [Wal84, app. D], but conformally invariant scalar fields exist in all spacetime dimensions, at least mathematically.

If we consider massive free scalar fields, a potential new freedom in choosing an equation of motion appears in comparison to the massless case due to the introduction of a new scale, which would allow couplings like $f = \widetilde{f}(R/m^2)R$, with some analytic function $\widetilde{f}$. However, one can again either argue with analyticity in $m$ or Occam's razor to view

$$P\phi \doteq \left(-\Box + \xi R + m^2\right)\phi = 0 \tag{II.28}$$

as the natural generalisation of the massive Klein-Gordon equation on Minkowski spacetime to a curved manifold. The general treatment of the subject is independent on the magnitude of either $m$ or $\xi$, but for actual calculations we shall often restrict to conformal coupling for simplicity.

Having a partial differential equation for a free scalar field at hand, one would expect that giving sufficient initial data would determine a unique solution on all $M$. However, this is, in case of the Klein-Gordon operator at hand, in general only true for globally hyperbolic spacetimes. To see a simple counterexample, let us consider Minkowski spacetime with a compactified time direction and the massless case, *i.e.* the equation $(-\partial_t^2 + \partial_x^2 + \partial_y^2 + \partial_z^2)\phi = 0$. Giving initial conditions $\phi|_{t=0} = 0$, $\partial_t\phi|_{t=0} = 1$, a possible *local* solution is $\phi \equiv t$. But this can of course never be a global solution, since one would run into contradictions after a full revolution around the compactified time direction.

In what follows, the *fundamental solutions* or *Green's functions* of the Klein-Gordon equation shall play a distinguished role. Before stating their existence, as well as the existence of general solutions, let us define the function spaces we shall be working with in the following, as well as their topological duals, see [CDD77, chap. VI] for an introduction.

**Definition II.1.1.1** *By* $\mathscr{E}(M) \doteq C^\infty(M, \mathbb{C})$ *we denote the* **smooth, complex-valued functions** *on* $M$ *equipped with the usual* **locally convex topology**, *i.e. a sequence of functions* $f_n \in \mathscr{E}(M)$ *is said to converge to* $f \in \mathscr{E}(M)$ *if all derivatives of* $f_n$ *converge to the ones of* $f$ *uniformly on all compact subsets of* $M$.

*The space* $\mathscr{D}(M) \doteq C_0^\infty(M, \mathbb{C})$ *is the subset of* $\mathscr{E}(M)$ *constituted by the* **smooth, complex-valued functions with compact support**. *We equip* $\mathscr{D}(M)$ *with the locally convex topology determined by* saying that a sequence of functions $f_n \in \mathcal{D}(M)$ converges to $f \in \mathcal{D}(M)$ if there is a compact subset $K \subset M$ such that all $f_n$ and $f$ are supported in $K$ and all derivatives of $f_n$ converge to the ones of $f$ uniformly in $K$.

By $\mathcal{D}(M, \mathbb{R}) \subset \mathcal{D}(M)$, $\mathcal{E}(M, \mathbb{R}) \subset \mathcal{E}(M)$ we denote the real-valued subspaces of $\mathcal{D}(M)$ and $\mathcal{E}(M)$ respectively.

By $\mathcal{D}'(M)$ we denote the space of **distributions**, i.e. the topological dual of $\mathcal{D}(M)$ provided by continuous, linear functionals $\mathcal{D}(M) \to \mathbb{C}$, whereas $\mathcal{E}'(M)$ denotes the topological dual of $\mathcal{E}(M)$, i.e. the space of **distributions with compact support**.

For $f \in \mathcal{D}(M)$, $u \in \mathcal{D}'(M) \supset \mathcal{E}(M) \supset \mathcal{D}(M)$, we shall denote the **dual pairing** of $f$ and $u$ by

$$\langle u, f \rangle \doteq \int\limits_M d_g x \, u(x) f(x).$$

The physical relevance of the above spaces is that functions in $\mathcal{D}(M)$, so-called *test functions*, should henceforth essentially be viewed as encoding the localisation of some observable in space and time. From the point of view of dynamics, initial data for a partial differential equation will always be encoded by some test function, whereas solutions of hyperbolic partial differential equations like the Klein-Gordon one are typically distributions or smooth functions which do *not* have compact support on account of the causal propagation of initial data; having a solution with compact support would entail that data 'is lost somewhere'. Moreover, fundamental solutions of differential equations will always be singular distributions, as can be expected from the fact that they are solutions with a singular $\delta$-distribution as source. Finally, since (anti)commutation relations of quantum fields are usually formulated in terms of fundamental solutions, the quantum fields and their expectation values will also turn out to be singular distributions quite generically.

Let us now state the theorem which guarantees us existence and properties of solutions and fundamental solutions of the Klein-Gordon operator $P$. We refer to the monograph [BGP07] for the proofs.

**Theorem II.1.1** *Let $P$ be a **normally hyperbolic operator** on a globally hyperbolic spacetime $(M, g)$, i.e. in each coordinate patch of $M$, $P$ can be expressed as*

$$P = -g^{\mu\nu} \partial_\mu \partial_\nu + A^\mu \partial_\mu + B$$

*with smooth functions $A^\mu$, $B$ and the **metric principal symbol** $-g^{\mu\nu} \partial_\mu \partial_\nu$. Then, the following results hold.*

a) *Let $f \in \mathcal{D}(M)$, let $\Sigma$ be a smooth Cauchy surface of $M$, let $(u_0, u_1) \in \mathcal{D}(\Sigma) \times \mathcal{D}(\Sigma)$, and let $N$ be the future directed timelike unit normal vector field of $\Sigma$. Then, the **Cauchy problem***

$$P u = f, \qquad u\restriction_\Sigma \equiv u_0, \qquad \nabla_N u \restriction_\Sigma \equiv u_1$$

*has a unique solution $u \in \mathcal{E}(M)$. Moreover,*

$$supp \, u \subset J \left( supp \, f \cup supp \, u_0 \cup supp \, u_1, M \right).$$

b) There exist unique **retarded** $G^+$ and **advanced** $G^-$ fundamental solutions of $P$. Namely, there are unique continuous maps $G^\pm : \mathscr{D}(M) \to \mathscr{E}(M)$ satisfying $P G^\pm = G^\pm P = id_{\mathscr{D}(M)}$ and supp $G^\pm f \subset J^\pm(supp\, f, M)$ for all $f \in \mathscr{D}(M)$.

c) Let $f,\, g \in \mathscr{D}(M)$. If $P$ is formally self-adjoint, i.e. $\langle f, Pg \rangle = \langle Pf, g \rangle$, then $G^+$ and $G^-$ are the formal adjoints of one another, namely, $\langle f, G^\pm g \rangle = \langle G^\mp f, g \rangle$.

d) The **causal propagator** of $P$ defined as $\Delta \doteq G^- - G^+$ is a continuous map $\mathscr{D}(M) \to \mathscr{E}(M)$ satisfying: for all solutions $u$ of $Pu = 0$ with compactly supported initial conditions on a Cauchy surface there is an $f \in \mathscr{D}(M)$ such that $u = \Delta f$. Moreover, for every $f \in \mathscr{D}(M)$ satisfying $\Delta f = 0$ there is a $g \in \mathscr{D}(M)$ such that $f = Pg$.

The Klein-Gordon operator $P$ is manifestly normally hyperbolic. Moreover, one can check by partial integration that $P$ is also formally self-adjoint. Hence, all above-mentioned results hold for $P$.

To close this subsection, let us examine a little the causal propagator $\Delta$. The continuity of $\Delta$ together with the fact that $\mathscr{E}(M) \subset \mathscr{D}'(M)$ guarantees that

$$\Delta(f, g) \doteq \langle f, \Delta g \rangle$$

defines a distribution $\Delta \in \mathscr{D}'(M^2)$ which we shall denote by the same symbol. Moreover, from the properties of $G^\pm$ one obtains that the formal adjoint of $\Delta$ is $-\Delta$ such that $\Delta$ viewed as a distribution is antisymmetric, i.e. $\Delta(f, g) = -\Delta(g, f)$. Additionally, the support properties of $G^\pm$ entail that $\Delta(f, g)$ vanishes if the supports of $f$ and $g$ are spacelike separated. On the level of distribution kernels, this implies that $\Delta(x, y)$ vanishes for spacelike separated $x$ and $y$. In anticipation of the quantization of the free Klein-Gordon field, this qualifies $\Delta(x, y)$ as a *commutator function*.

## II.1.2 The Symplectic Space of Solutions

In the past literature on quantum field theory in curved spacetimes, (at least) two ways how to obtain a quantized field from a classical one have been proposed. Although they are ultimately related, they initially emphasise different aspects of quantization. The one we shall describe in this subsection highlights the *'canonical quantization'* aspect, whereas the one introduced in the next subsection will underline the *'algebraic'* aspect of quantization, namely, that classical fields have an Abelian algebra of observables, while quantum fields have a non-Abelian one.

To perform canonical quantization, we need a *symplectic structure* on the space of solutions of the Klein-Gordon equation, i.e. we need a *Poisson bracket*. In standard treatments on scalar field theory, one usually defines Poisson brackets at 'equal times', but as realised by Peierls in [Pei52], one can give a covariant version of the Poisson bracket which does not depend on a splitting of spacetime into space and time, and we shall directly give such definition.

**Definition** II.1.2.1   *By* $(\mathscr{S}(M), \varsigma_M)$ *we denote the* **symplectic space of real-valued solutions of the Klein-Gordon equation**. *Here,* $\mathscr{S}(M) \subset \mathscr{E}(M, \mathbb{R})$ *is the space of smooth real-valued solutions* $\phi$ *of $P\phi = 0$ with compactly supported (real-valued) initial data on a Cauchy surface and $\varsigma_M : \mathscr{S}(M) \times \mathscr{S}(M) \to \mathbb{R}$ is a* **strongly non-degenerate symplectic form** *defined by*

$$\mathscr{S}(M) \times \mathscr{S}(M) \ni (\phi_1, \phi_2) \mapsto \varsigma_M(\phi_1, \phi_2) \doteq \int\limits_{\Sigma} d\Sigma \; \phi_2 \nabla_N \phi_1 - \phi_1 \nabla_N \phi_2.$$

*Above, the strong non-degeneracy of $\varsigma_M$ entails that $\varsigma_M(\phi_1, \phi_2) = 0$ for all $\phi_1 \in \mathscr{S}(M)$ implies $\phi_2 \equiv 0$ and $\Sigma$ is an arbitrary Cauchy surface of $(M, g)$, whereas $N$ is the forward pointing unit normal vector field on $\Sigma$ and $d\Sigma$ denotes the canonical volume measure on $\Sigma$ obtained by restriction of $d_g x$ to $\Sigma$.*

Definition II.1.2.1 employs an arbitrary Cauchy surface $\Sigma$ of $M$, and to show that it is independent of such $\Sigma$, we have to understand if and why the integration appearing in $\varsigma_M(\phi_1, \phi_2)$ yields the same result on all Cauchy surfaces. An alert reader might have already noticed that $\varsigma_M(\phi_1, \phi_2)$ is independent of $\Sigma$ since it is the 'charge' corresponding to the current $j^\mu(\phi_1, \phi_2) \doteq \phi_2 \nabla^\mu \phi_1 - \phi_1 \nabla^\mu \phi_2$ which is conserved, *i.e.* $\nabla_\mu j^\mu(\phi_1, \phi_2) = 0$, on account of the solution nature of $\phi_1$ and $\phi_2$. However, it is still interesting to compute $\varsigma_M(\phi_1, \phi_2)$ explicitly in a manifestly Cauchy surface-independent way since this will give us an alternative definition of $(\mathscr{S}(M), \varsigma_M)$ in terms of test functions and the causal propagator $\Delta$.

To this avail, let us recall the relation between test functions $\mathscr{D}(M)$ and solutions $\mathscr{S}(M)$. On account of theorem II.1.1, all $f \in \mathscr{D}(M)$ can be promoted to a solution $\Phi_f \in \mathscr{S}(M)$ by setting $\Phi_f \doteq \Delta f$ and every element of $\mathscr{S}(M)$ is of this form. However, while $\Delta : \mathscr{D}(M) \to \mathscr{S}(M)$ is surjective, we know that it is not injective, since all $g \in \mathscr{D}(M)$ of the form $g = Pf$ for some $f \in \mathscr{D}(M)$ are in the kernel of $\Delta$. In fact, the kernel of $\Delta$ is so 'large' that, given a $\phi \in \mathscr{S}(M)$, one can find a suitable $f$ with $\phi = \Delta f$ 'at any time', as the following well-known result shows.

**Lemma** II.1.2.2 *Let $\phi \in \mathscr{S}(M)$ be an arbitrary real-valued solution of $P\phi = 0$ with compactly supported initial data on a Cauchy surface and let $\Sigma$ be any Cauchy surface of $M$. Then, for any bounded neighbourhood $\mathcal{O}(\Sigma)$ of $\Sigma$, we can find a $g \in \mathscr{D}(M, \mathbb{R})$ with supp $g \in \mathcal{O}(\Sigma)$ and $\phi = \Delta g$.*

*Proof.* On account of theorem II.1.1 we now that there exists an $f \in \mathscr{D}(M)$ with $\phi = \Delta f$. Let us assume that $\mathcal{O}(\Sigma)$ lies in the future of supp $f$, *i.e.* $J^-(\text{supp } f, M) \cap \mathcal{O}(\Sigma) = \emptyset$, the case $J^+(\text{supp } f, M) \cap \mathcal{O}(\Sigma) = \emptyset$ can be treated analogously.

With the aforementioned conditions in mind, let us consider two auxiliary Cauchy surfaces $\Sigma_1$ and $\Sigma_2$ which are both contained in $\mathcal{O}(\Sigma)$ and which are chosen such that $\Sigma_2$ lies in the future of $\Sigma$ whereas $\Sigma_1$ lies in the past of $\Sigma$. Moreover, let us take a smooth function $\Lambda \in \mathscr{E}(M)$ which is identically vanishing in the future of $\Sigma_2$ and fulfils $\Lambda \equiv 1$ in the past of $\Sigma_1$ and let us define $g \doteq f - P\Lambda G^+ f$. By construction and on account of the properties of both a globally hyperbolic spacetime $M$ and a retarded fundamental solution $G^+$ on $M$, $\Delta g = \Delta f = \phi$ and supp $g$ is contained in a compact subset of $J^+(\text{supp } f, M) \cap \mathcal{O}(\Sigma)$. $\qquad\square$

The above result may seem surprising at first glance, but it is strongly desirable on physical grounds. Namely, as already anticipated, we shall interpret test functions as localisations of observables. In this regard, lemma II.1.2.2 entails that observables solving the equation of motion can be 'predicted' at all times on the basis of knowing them in an arbitrarily small time interval. We shall come back to this at later stages of the thesis.

As promised, we shall now provide an alternative characterisation of $(\mathscr{S}(M), \varsigma_M)$ based on the properties of $\Delta$.

**Lemma** **II.1.2.3** *Let* $\Phi_f$, $\Phi_g$ *be elements of* $\mathscr{S}(M)$ *given as* $\Phi_f = \Delta f$, $\Phi_g = \Delta g$ *for suitable test functions* $f$ *and* $g$.

*a) The symplectic form* $\varsigma_M(\Phi_f, \Phi_g)$ *can be computed in terms of the causal propagator as*

$$\varsigma_M(\Phi_f, \Phi_g) = \Delta(f, g).$$

*b)* $(\mathscr{S}(M), \varsigma_M)$ *is equivalent to the symplectic space* $(\mathscr{D}(M, \mathbb{R})/\mathrm{Ker}\Delta, \varsigma_M')$, *where* $\mathscr{D}(M, \mathbb{R})/\mathrm{Ker}\Delta$ *is the quotient space obtained from* $\mathscr{D}(M, \mathbb{R})$ *by dividing out the kernel of* $\Delta$, *and, given two equivalence classes* $[f]$, $[g]$ *in* $\mathscr{D}(M, \mathbb{R})/\mathrm{Ker}\Delta$, *the symplectic form* $\varsigma_M'$ *is defined by*

$$\varsigma_M'([f], [g]) \doteq \Delta(f, g).$$

*Proof.* b) follows immediately from a) and the fact that the map from $\mathscr{D}(M, \mathbb{R})/\mathrm{Ker}\Delta$ to $\mathscr{S}(M)$ induced by $\Delta : \mathscr{D}(M) \to \mathscr{S}(M)$ is bijective by construction. The proof of a) can be for instance found in [Dim80]: *Green's identity*, or, in more general terms, *Stokes' theorem* entails that

$$\int_V d_g x \, vPu - uPv = \int_V d_g x \, u \square v - v \square u = \int_{\partial V} d(\partial V) \, u \nabla_{\widetilde{N}} v - v \nabla_{\widetilde{N}} u,$$

where $v$, $u$ are smooth functions with support properties ensuring that all integrals converge, $V$ is a submanifold of $M$ with smooth boundary $\partial V$, $\widetilde{N}$ is the outward pointing unit normal vector field on $\partial V$ and $d(\partial V)$ is the volume measure on $\partial V$ induced by $d_g x$. We can apply the above identity to the two cases

$$V = \Sigma^+ \doteq J^+(\Sigma, M) \setminus \Sigma, \quad \partial V = \Sigma, \quad v = \Phi_f, \quad u = G^- g, \quad \widetilde{N} = -N$$

$$\text{and} \quad V = \Sigma^- \doteq J^-(\Sigma, M) \setminus \Sigma, \quad \partial V = \Sigma, \quad v = \Phi_f, \quad u = G^+ g, \quad \widetilde{N} = N,$$

where again $\Sigma$ is an arbitrary Cauchy surface of $M$ and $N$ is the forward pointing unit normal vector field on $\Sigma$. Note that all integrals we are interested in converge and that we are allowed to view $\Sigma$ as the only relevant boundary of $\Sigma^\pm$ because $V \cap \mathrm{supp}\, u$ is a compact set in both cases on account of the global hyperbolicity of $M$. With this in mind, we can compute

$$\Delta(f, g) = \int_M d_g x \, f \Delta g = \int_{\Sigma^+} d_g x \, f \Phi_g + \int_{\Sigma^-} d_g x \, f \Phi_g = \int_{\Sigma^+} d_g x \, \Phi_g P G^- f + \int_{\Sigma^-} d_g x \, \Phi_g P G^+ f$$

$$= \int_\Sigma d\Sigma \, \Phi_g \nabla_N (G^- f) - G^- f \nabla_N \Phi_g - \int_\Sigma d\Sigma \, \Phi_g \nabla_N (G^+ f) - G^+ f \nabla_N \Phi_g$$

$$= \int_\Sigma d\Sigma \, \Phi_g \nabla_N (\Delta f) - \Delta f \nabla_N \Phi_g = \varsigma_M(\Phi_f, \Phi_g).$$

□

To see that the symplectic structure we have just defined corresponds to 'equal-time Poisson brackets', we need a further well-known result, which will also prove helpful in the discussion of FLRW spacetimes. As a preparation, let $u \in \mathscr{E}'(M)$ and $f \in \mathscr{D}(M)$. By $\langle G^{\mp} u, f \rangle \doteq \langle u, G^{\pm} f \rangle$ we can extend $G^{\pm}$ and, hence, $\Delta$ to maps from $\mathscr{E}'(M)$ to $\mathscr{D}'(M)$. Moreover, given a test function $f \in \mathscr{D}(\Sigma)$, we can naturally interpret it as a compactly supported distribution and therefore apply $\Delta$ to it. With these considerations in mind, one can prove the following result [Dim80, cor. 1.2].

**Lemma II.1.2.4** *Let $\Sigma$ be an arbitrary Cauchy surface of $M$ and $N$ its future pointing unit normal vectorfield. For all $f \in \mathscr{D}(\Sigma)$ it holds*

$$\nabla_N \Delta f \upharpoonright_\Sigma = f, \qquad \Delta f \upharpoonright_\Sigma = 0.$$

*On the level of distribution kernels, this entails that*

$$\nabla_N \Delta(x, y) \upharpoonright_{\Sigma \times \Sigma} = \delta_\Sigma(x, y), \qquad \Delta(x, y) \upharpoonright_{\Sigma \times \Sigma} \equiv 0,$$

*where $\delta_\Sigma$ is the $\delta$-distribution with respect to the canonical measure on $\Sigma$.*

In order to close this subsection and to relate the symplectic form $\varsigma_M$ to equal-time Poisson brackets, we shall only make a few brief and formal considerations and we refer the interested reader to [Wal95, chap. 3] for an exhaustive and precise discussion. First, let us recall that we can consider a Cauchy surface $\Sigma$ as an 'equal-time' surface because each such $\Sigma$ can be obtained a a pre-image of a global time function $t$, *i.e.*, $\Sigma = \{x \in M \mid t(x) = t_0\}$. Then, let us remember $\varsigma_M(\Phi_f, \Phi_g) = \Delta(f, g)$ and let us consider $\Phi_f$ and $\Phi_g$ as the outcome of an application of a distribution $\Phi$ to either $f$ or $g$. Doing this, we can relate the distribution kernels of such $\Phi$ and of $\Delta$ as

$$\varsigma_M(\Phi(x), \Phi(y)) = \Delta(x, y).$$

We can regard $\varsigma_M$ as constant with respect to the time evolution generated by a unit normal vector field $N$ of a Cauchy surface $\Sigma$ and therefore compute

$$\varsigma_M(\nabla_N \Phi(x) \upharpoonright_\Sigma, \Phi(y) \upharpoonright_\Sigma) = \nabla_N \Delta(x, y) \upharpoonright_{\Sigma \times \Sigma} = \delta_\Sigma(x, y),$$

$$\varsigma_M(\Phi(x) \upharpoonright_\Sigma, \Phi(y) \upharpoonright_\Sigma) = \Delta(x, y) \upharpoonright_{\Sigma \times \Sigma} = 0.$$

The above equations can now finally be interpreted as equal-time Poisson brackets of the 'field' $\Phi(x)$ and its 'canonical momentum' $\nabla_N \Phi(x)$.

## II.1.3 The Algebra of Classical Observables

As already anticipated at the beginning of the preceding section, we shall now discuss a different and rather modern viewpoint of quantizing a classical field which puts more emphasis on the algebraic structure of both classical and quantum field theory and allows for a unified description of both. This approach goes under the name of *deformation quantization* and, although not being the first paper to treat this subject, [DüFr01a] can be named as the earliest reference in which deformation quantization has been discussed in the language we shall use here. In this approach to quantization, one essentially treats classical and quantum observables as elements of the same vector space, but endowed with different products and therefore constituting different algebras. In more detail, one starts with the algebra of classical observables, which is endowed with a *commutative* product, and then 'deforms' this product in such a way that one obtains an algebra of quantum observables with a *non-commutative* product. In this process, *Planck's constant* $\hbar$ serves as a 'deformation parameter' and one regains the classical theory in the limit $\hbar \to 0$. Apart from providing a clear viewpoint of quantization, deformation quantization is a very powerful tool to treat perturbative quantum field theory in the algebraic setting, as demonstrated in [BDF09], see also [BrFr09, Kel10] for an introduction. It allows for the treatment of non-polynomial interactions and observables and encodes the combinatorial structure of perturbative quantum field theory in a very efficient way. In the present thesis, we shall only consider free fields and polynomial observables and therefore present a simplified version of the framework displayed in [BDF09]. However, we will nevertheless profit extensively from the deformation quantization approach when discussing *normal ordering* of quantum fields in chapter III. To lay the foundation for the deformation quantization of the free scalar field, we shall introduce its algebra of classical observables in the following.

One usually takes the point of view that states map observables to real numbers. To define the algebra of classical observables, we reverse this way of thinking and regard observables as maps from states to the real numbers. In classical field theory, we shall regard states as being smooth functions on $M$, *i.e.* elements of $\mathscr{E}(M)$. Observables are then maps from $\mathscr{E}(M)$ to the complex numbers[2] and therefore compactly supported distributions in $\mathscr{E}'(M)$. However, for the moment we shall restrict to smooth observables in $\mathscr{D}(M)$ and consider more singular ones in the later treatment of quantum field theory. A reader who may feel uncomfortable with the just described 'definition' of states and observables in classical field theory should have in mind the following example: let us consider a classical field describing the temperature of the universe. The state of this field is just a smooth function $T$ giving the value of the temperature at every point $x \in M$. An observable $f \in \mathscr{D}(M)$ then corresponds to measuring the mean value of the temperature in the support of $f$ and weighted according to the functional behaviour of $f$, *i.e.* the expectation value of $f$ in the state $T$ is just $\langle f, T \rangle$. However, one can have more complicated observables, for instance the correlation of the temperature in $n$ different regions of $M$. Such observables would be represented by sums of elements $f_1 \otimes \cdots \otimes f_n$ in $\mathscr{D}(M^n)$; in fact, we can consider only symmetric tensor products, as the order of the tensor factors in $f_1 \otimes \cdots \otimes f_n$ should not matter in classical field theory on account of the commutativity of classical observables. With this example in mind, we shall proceed to state the definition of the algebra of classical observables.

**Definition II.1.3.1**   *By* $(\mathscr{C}_0(M), \cdot_s)$ *we denote the* **off-shell algebra of classical observables of the**

---

[2]Of course proper observables have real expectation values. However, we shall allow for complex expectation values since we also want to discuss unobservable elements of the field algebra.

*scalar field*. Here, $\mathscr{C}_0(M)$ is the vector space

$$\mathscr{C}_0(M) \doteq \bigoplus_{n=0}^{\infty} \mathscr{D}^s(M^n),$$

with $\mathscr{D}^s(M^n)$ denoting the **symmetric test functions** in $n$ variables, $\mathscr{D}^s(M^0) \doteq \mathbb{C}$, and elements of $\mathscr{C}_0(M)$ have to be understood as **finite** sums $f = \oplus_n \frac{1}{n!} f^{(n)}$ with $f^{(n)} \in \mathscr{D}^s(M^n)$. We equip $\mathscr{C}_0(M)$ with the **symmetric product** $\cdot_s$ defined as

$$\frac{1}{n!}(f \cdot_s g)^{(n)} \doteq \sum_{k+l=n} \mathfrak{Sym} \left( \frac{1}{k!} f^{(k)} \otimes \frac{1}{l!} g^{(l)} \right),$$

where $\mathfrak{Sym} : \mathscr{D}(M^n) \to \mathscr{D}^s(M^n)$ is the **total symmetrisation projector**, i.e. $\mathfrak{Sym}$ is surjective and $\mathfrak{Sym}^2 = \mathfrak{Sym}$. Moreover, we equip $\mathscr{C}_0(M)$ with the following topology: a sequence $\{f_k\}_k = \{\oplus_n f_k^{(n)}\}_k$ in $\mathscr{C}_0(M)$ is said to converge to $f = \oplus_n f^{(n)}$ if $f_k^{(n)}$ converges to $f^{(n)}$ for all $n$ and there exists an $N$ such that $f_k^{(n)} = 0$ for all $n > N$ and all $k$.

We define the **on-shell algebra of classical observables of the scalar field** $(\mathscr{C}(M), \cdot_s)$ by replacing $(\mathscr{C}_0(M), \cdot_s)$ with the quotient space $\mathscr{C}_0(M)/\mathscr{I}_p$ endowed with the induced product and topology. Here, $\mathscr{I}_p$ is the closed ideal generated by elements $Pf$, $f \in \mathscr{D}(M)$.

We would like to close this subsection by providing a few remarks on the above definition. The alert reader might wonder why we have implemented the various factorials in the definition of $\mathscr{C}(M)$. The reason for this is that we want to regard $f \in \mathscr{C}(M)$ as *functionals* $f : \mathscr{E}(M) \ni \phi \mapsto f(\phi) \in \mathbb{C}$, $f^{(n)}$ as their functional derivatives at $\phi = 0$, and $\cdot_s$ as the pointwise product of functionals, i.e. $(f \cdot_s g)(\phi) = f(\phi)g(\phi)$; this is the picture described in [BDF09]. The factorials are then necessary to understand $\cdot_s$ as coming from the Leibniz rule for functional differentiation and we have written them in the way displayed above to underline the isomorphy of $(\mathscr{C}(M), \cdot_s)$ to a subalgebra of the so-called *Borchers-Uhlmann algebra*, an object we shall introduce at the later stages of this thesis. Finally, we refer the reader to [BDF09, BrFr09, Kel10] for a discussion on how to implement the symplectic structure introduced in the preceding subsection in the algebra $(\mathscr{C}(M), \cdot_s)$. Note however, that the quotient appearing in the definition of the on-shell algebra of classical observables corresponds to the quotient with respect to $\text{Ker}\Delta$ taken in lemma II.1.2.3 and therefore all elements $[f] = \oplus_n [f^{(n)}]$ of $\mathscr{C}(M)$ satisfy the Klein-Gordon equation in the sense $P[f] = \oplus_n P[f^{(n)}] = \oplus_n [Pf^{(n)}] = 0$.

## II.2 The Free Scalar Field in Null Big Bang Spacetimes

Following the general organisation of this thesis, we shall now specialise the structures related to a classical free scalar field in an arbitrary spacetime described in the previous subsection to the case of an NBB spacetime. Moreover, in preparation of the wished-for construction of distinguished states in NBB spacetimes, we will display in which sense it is possible to map the classical scalar field in the bulk of such spacetime to its boundary. As conformal transformations play an essential role in the definition of the boundary of NBB spacetimes, we shall restrict attention to a Klein-Gordon field with conformal coupling to curvature, *i.e.* we will set $\xi = \frac{1}{6}$ in (II.28).

### II.2.1   The Mode Expansion of Solutions and the Causal Propagator

As is well-known, the spatial translation and rotation invariance of FLRW spacetimes allows to give a Fourier decomposition of the on-shell scalar field in terms of so-called *modes*. Such mode expansion is not strictly necessary for the constructions we wish to perform on the boundary of an NBB spacetime, but it will allow us to understand these constructions in simpler terms and to investigate their cosmological implications; we shall therefore discuss it in this subsection.

To this avail, let us start by analysing the general behaviour of the Klein-Gordon equation and its solutions under conformal transformations. Building on the discussion of the conformally invariant version of the Klein-Gordon equation at the beginning of this section, we can obtain the following well-known and simple, but very powerful results.

***Lemma*** **II.2.1.1**   *Let $\Omega : M \to (0, \infty)$ be the conformal factor related to a conformal transformation $g \mapsto \widetilde{g} \doteq \Omega^2 g$ and let $\widetilde{\nabla}$, $\widetilde{\Box}$, and $\widetilde{R}$ denote the covariant derivative, d'Alembert operator, and scalar curvature associated to $\widetilde{g}$ respectively. Moreover, let*

$$P \doteq -\Box + \frac{1}{6}R + m^2, \qquad \widetilde{P} \doteq -\widetilde{\Box} + \frac{1}{6}\widetilde{R} + \frac{1}{\Omega^2}m^2.$$

*a)   The two Klein-Gordon operators $P$ and $\widetilde{P}$ are related via*

$$\widetilde{P} = \frac{1}{\Omega^3}P\Omega.$$

*b)   Let $\widetilde{G}^\pm$ and $\widetilde{\Delta}$ denote the retarded and advanced fundamental solutions and the causal propagator associated to $\widetilde{P}$ respectively. The aforementioned propagators are related to the ones proper to $P$ by*

$$\widetilde{G}^\pm = \frac{1}{\Omega}G^\pm\Omega^3, \qquad \widetilde{\Delta} = \frac{1}{\Omega}\Delta\Omega^3.$$

*On the level of distribution kernels, this entails that*

$$\widetilde{G}^\pm(x,y) = \frac{1}{\Omega(x)}G^\pm(x,y)\Omega^3(y), \qquad \widetilde{\Delta}(x,y) = \frac{1}{\Omega(x)}\Delta(x,y)\Omega^3(y).$$

*c)   Let $\mathscr{S}(M)$ and $\widetilde{\mathscr{S}}(M)$ denote the spaces of smooth solutions with compactly supported initial conditions associated to $P$ and $\widetilde{P}$ respectively. Every $u \in \mathscr{S}(M)$ induces an element $\widetilde{u}$ of $\widetilde{\mathscr{S}}(M)$ via*

$$\widetilde{u} \doteq \frac{1}{\Omega}u$$

*and every element of $\widetilde{\mathcal{F}}(M)$ is of this form. Moreover, if $f \in \mathscr{D}(M)$ is such that $u = \Delta f$, then*

$$\widetilde{u} = \widetilde{\Delta} \frac{1}{\Omega^3} f \,.$$

*Proof.* a) follows directly from (II.27) which in turn ensues from the behaviour of the covariant derivative under conformal transformations as displayed in [Wal84, app. D]. Furthermore, $\widetilde{G}^\pm$ manifestly satisfy $\widetilde{P}\widetilde{G}^\pm = id_{\mathscr{D}(M)}$ and have the causal support properties of retarded and advanced fundamental solutions as conformal transformations do not affect the causal structure of a spacetime; this entails b). To prove c), let $\Sigma$ be a Cauchy surface of $(M, g)$ with future pointing unit normal vector field $N$. Then, by the invariance of causal relations under conformal transformations, $\Sigma$ is a Cauchy surface for $(M, \widetilde{g})$ as well. Moreover, since $\widetilde{N} = \Omega^{-1}N$ is the forward poiting unit normal vector field of $\Sigma$ with respect to $\widetilde{g}$, and $\widetilde{\nabla}$ and $\nabla$ agree on scalar functions, $\widetilde{u} = \Omega^{-1}u$ is manifestly the unique solution of $\widetilde{P}\widetilde{u} = 0$ with compactly supported initial conditions $\widetilde{u}\restriction_\Sigma = \Omega^{-1}u\restriction_\Sigma$, $\widetilde{\nabla}_{\widetilde{N}}\widetilde{u}\restriction_\Sigma = (\Omega^{-2}\nabla_N u + \Omega^{-1}u\nabla_N \Omega^{-1})\restriction_\Sigma$. Finally, one can directly compute $\widetilde{u} = \widetilde{\Delta}\Omega^{-3}f = \Omega^{-1}u$, which closes the proof. $\qquad\square$

To obtain the modes proper to a conformally coupled massive scalar field in a FLRW spacetime $(M, g)$, let us recall that the metric $g$ of such a spacetime is conformally related to the Minkowski metric $\eta$ via $g = a^2\eta$. We can therefore apply lemma II.2.1.1 to the case $\widetilde{g} = \eta$, $\Omega = a^{-1}$ to obtain

$$P = -\square + \frac{1}{6}R + m^2 = \frac{1}{a^3}\left(\partial_\tau^2 - \vec{\nabla}^2 + a^2 m^2\right)a \,,$$

where $\vec{\nabla}$ denotes the gradient with respect to the (comoving) spatial coordinates. We thus see that the conformally coupled massive scalar field in a FLRW spacetime is conformally equivalent to a scalar field with time-varying mass in Minkowski spacetimes. This leads us to define a mode solution $\phi_{\vec{k}}$ of $P\phi = 0$ as

$$\phi_{\vec{k}}(\tau, \vec{x}) \doteq \frac{T_k(\tau)e^{i\vec{k}\vec{x}}}{(2\pi)^{\frac{3}{2}}a(\tau)} \,, \tag{II.29}$$

where $k \doteq |\vec{k}|$ and $T_k(\tau)$ is a solution of the ordinary differential equation

$$\left(\partial_\tau^2 + k^2 + a^2(\tau)m^2\right)T_k(\tau) = 0 \,, \tag{II.30}$$

which can be taken to depend only on $k$ and not on $\vec{k}$. To obtain a normalisation for $T_k(\tau)$, let us note that the *Wronskian* $T_k(\tau)\partial_\tau\overline{T_k(\tau)} - \overline{T_k(\tau)}\partial_\tau T_k(\tau)$ is constant in $\tau$ on account of (II.30) and furthermore purely imaginary. We can therefore demand that

$$T_k(\tau)\partial_\tau\overline{T_k(\tau)} - \overline{T_k(\tau)}\partial_\tau T_k(\tau) \equiv i \,. \tag{II.31}$$

This choice of normalisation may seem to be arbitrary at the moment, but we shall see that it leads to concise expressions for the quantities we are interested in.

Note that $\phi_{\vec{k}}$ is *not* an element of $\mathscr{S}(M)$ as it has manifestly non-compact support on any Cauchy surface. However, we can still hope to express any $\phi \in \mathscr{S}(M)$ via a Fourier integral over $\phi_{\vec{k}}$ with suitable coefficient functions. Following [DMP09a, DMP09b], we can show that this is indeed the case, provided the $T_k(\tau)$ fulfil certain regularity conditions as a function of $k$ for all times $\tau$.

**Lemma II.2.1.2** *Let us assume that the time parts $T_k(\tau)$ of the modes $\phi_{\vec{k}}(\tau, \vec{x})$ and their $\tau$-derivatives are continuous functions of $k$ for $k > 0$ which are square-integrable with respect to the measure $k^2 dk$ in any compact neighbourhood of $k = 0$, i.e. $T_k(\tau)$, $\partial_\tau T_k(\tau) \in L^2([0, k_0], k^2 dk)$ for all $k_0 < \infty$ and all $\tau$. Moreover, let us assume that $T_k(\tau)$ and $\partial_\tau T_k(\tau)$ are polynomially bounded in $k$ for large $k$ and all $\tau$. Then, we can express every $\phi \in \mathscr{S}(M)$ via*

$$\phi(\tau, \vec{x}) = \int_{\mathbb{R}^3} d\vec{k} \; \widetilde{\phi}(\vec{k}) \phi_{\vec{k}}(\tau, \vec{x}) + \overline{\widetilde{\phi}(\vec{k})} \phi_{\vec{k}}(\tau, \vec{x}),$$

*where the coefficient functions $\widetilde{\phi}(\vec{k})$ can be obtained as*

$$\widetilde{\phi}(\vec{k}) = -i \int_{\mathbb{R}^3} d\vec{x} \, a^2(\tau) \left( \phi(\tau, \vec{x}) \partial_\tau \overline{\phi_{\vec{k}}(\tau, \vec{x})} - \overline{\phi_{\vec{k}}(\tau, \vec{x})} \partial_\tau \phi(\tau, \vec{x}) \right).$$

*Proof.* Let us start by noting that

$$\widetilde{\phi}(\vec{k}) \doteq -i \int_{\mathbb{R}^3} d\vec{x} \, a^2(\tau) \left( \phi(\tau, \vec{x}) \partial_\tau \overline{\phi_{\vec{k}}(\tau, \vec{x})} - \overline{\phi_{\vec{k}}(\tau, \vec{x})} \partial_\tau \phi(\tau, \vec{x}) \right)$$

is well-defined, as $\phi \in \mathscr{S}(M)$ and its $\tau$-derivative have compact support in $\vec{x}$ for all $\tau$. Our notation suggests that $\widetilde{\phi}(\vec{k})$ is independent of $\tau$ and this is indeed the case as one can see either by deriving the integrand with respect to $\tau$ or by noting that, with the notation of definition II.1.2.1, $d\Sigma \nabla_N = d\vec{x} a^2(\tau) \partial_\tau$ in the case of $\Sigma$ being a hypersurface of constant $\tau$. The latter equality follows from $d\Sigma \nabla_N = \frac{1}{6} \sqrt{|\det g|} g^{\alpha\beta} \varepsilon_{\alpha\mu\nu\rho} dx^\mu \wedge dx^\nu \wedge dx^\rho \partial_\beta$, with $\varepsilon_{\alpha\mu\nu\rho}$ denoting the totally antisymmetric *Levi-Civita symbol* and entails that, barring an abuse of notation as $\phi_{\vec{k}} \notin \mathscr{S}(M)$, $\widetilde{\phi}(\vec{k}) = i\varsigma_M(\overline{\phi_{\vec{k}}}, \phi)$.

Having assured ourselves that $\widetilde{\phi}(\vec{k})$ is well-defined, let us see if the wished-for mode expansion

$$\phi(\tau, \vec{x}) = \int_{\mathbb{R}^3} d\vec{k} \; \widetilde{\phi}(\vec{k}) \phi_{\vec{k}}(\tau, \vec{x}) + \overline{\widetilde{\phi}(\vec{k}) \phi_{\vec{k}}(\tau, \vec{x})}$$

converges and equals $\phi$. To this avail, let us note that $\widetilde{\phi}(\vec{k})$ is both smooth and rapidly decreasing in $\vec{k}$ since it is given by (the derivative of) $\overline{T_k(\tau)}$ times the Fourier transform of a compactly supported smooth function as one can see by inserting the explicit expression for $\phi_{\vec{k}}$ given in (II.29) in the definition of $\widetilde{\phi}(\vec{k})$. Moreover, by our assumptions, $T_k(\tau)$ is square-integrable with respect to $k^2 dk$ for small $k$ and polynomially bounded for large $k$. These data entail that $\widetilde{\phi}(\vec{k})T_k(\tau)$ is an element of $L^2(\mathbb{R}^3, d\vec{k})$ such that the above integral can be interpreted as the well-defined Fourier-Plancherel transform of a square-integrable function (in fact, the remainder of the proof will show that it is the Fourier transform of a smooth and rapidly decreasing function). To see that it really equals $\phi$, let us recall that $\phi \in \mathscr{S}(M)$ is real-valued and insert the definition for $\widetilde{\phi}(\vec{k})$ to obtain

$$\int\limits_{\mathbb{R}^3} d\vec{k}\, \widetilde{\phi}(\vec{k})\phi_{\vec{k}}(\tau, \vec{x}) + \overline{\widetilde{\phi}(\vec{k})\phi_{\vec{k}}(\tau, \vec{x})}$$

$$= i \int\limits_{\mathbb{R}^3 \times \mathbb{R}^3} d\vec{k}\, d\vec{y}\, \phi(\tau, \vec{y}) \left( \overline{\phi_{\vec{k}}(\tau, \vec{x})} \partial_\tau \phi_{\vec{k}}(\tau, \vec{y}) - \phi_{\vec{k}}(\tau, \vec{x}) \partial_\tau \overline{\phi_{\vec{k}}(\tau, \vec{y})} \right).$$

Realising that the integral and the $T_k$ are invariant under $\vec{k} \mapsto -\vec{k}$, we can flip the sign of $\vec{k}$ in the first of the above two summands and then employ the mode normalisation condition (II.31). The resulting integral is the composition of the Fourier transform and inverse Fourier transform of $\phi$ and therefore equals $\phi$. $\qquad\square$

We know that any element $\phi$ of $\mathscr{S}(M)$ can be written as $\phi = \Delta f$ for some $f \in \mathscr{D}(M)$. In view of the just obtained mode expansion for such $\phi$, a natural question one could ask is how the expansion coefficients $\widetilde{\phi}(\vec{k})$ are related to $f$. As we will need the answer to this question for the already anticipated interpretation of the boundary states we wish to construct, we shall discuss this issue in the following. Unsurprisingly, this requires the knowledge of the causal propagator $\Delta$ in terms of modes and the first step we have to take is thus to obtain a mode expansion for $\Delta$. In Minkowski spacetime and for the Klein-Gordon field (with constant mass), we know that $\Delta(x, y)$ is a translationally invariant distribution on rapidly decreasing functions and we can therefore Fourier transform it in $x - y$. In view of the symmetries of a FLRW spacetime, one could expect that such property holds at least in the spatial variables. If this were true, we could directly Fourier transform $\Delta(\tau_x, \vec{x}, \tau_y, \vec{y})$ with respect to $\vec{x} - \vec{y}$ and give the result in terms of a suitable linear combination of $T_k$. However, as we are not aware of a proof of such properties of $\Delta$ which does not require the knowledge of the looked-for mode expansion, we shall follow the way displayed in [LuRo90]: we just 'guess' the correct mode decomposition of $\Delta(x, y)$ and then prove that it coincides with the abstractly defined $\Delta$.

***Lemma* II.2.1.3**

*a) Let us assume that the time parts $T_k(\tau)$ of the modes and their derivatives $\partial_\tau T_k(\tau)$ are continuous in $k$ for $k > 0$, polynomially bounded in $k$ for large $k$, and fulfil $T_k(\tau) = O(k^{-1/2})$, $\partial_\tau T_k(\tau) =*

$O(k^{-1/2})$ *for small* $k$, *where all these regularity conditions are assumed to hold for all* $\tau$. *Then, the distribution kernel of the causal propagator* $\Delta$ *can be expressed as*

$$\Delta(\tau_x, \vec{x}, \tau_y, \vec{y}) = -i \int_{\mathbb{R}^3} d\vec{k} \, \overline{\phi_{\vec{k}}(\tau_x, \vec{x})} \phi_{\vec{k}}(\tau_y, \vec{y}) - \phi_{\vec{k}}(\tau_x, \vec{x}) \overline{\phi_{\vec{k}}(\tau_y, \vec{y})},$$

*where the integral has to be understood as the limit obtained by multiplication of the integrand with* $e^{-\varepsilon k}$, $\varepsilon > 0$ *and then taking* $\varepsilon \to 0$ *after smearing the result with (at least one) test function in* $\mathscr{D}(M)$.

b) *Let* $f \in \mathscr{D}(M, \mathbb{R})$ *be a test function related to a* $\phi \in \mathscr{S}(M)$ *by* $\phi = \Delta f$. *The Fourier coefficients* $\widetilde{\phi}(\vec{k})$ *of* $\phi$ *defined as in lemma II.2.1.2 fulfil*

$$\widetilde{\phi}(\vec{k}) = i \overline{\phi_{\vec{k}}}(f) \doteq i \langle \overline{\phi_{\vec{k}}}, f \rangle.$$

*Proof.* b) follows immediately from a) if we compare $\phi(\tau, \vec{x}) = (\Delta f)(\tau, \vec{x}) = \Delta(\tau, \vec{x}, f)$ with the expansion of $\phi(\tau, \vec{x})$ given in lemma II.2.1.2. To prove a), let us define

$$\widetilde{\Delta}(\tau_x, \vec{x}, \tau_y, \vec{y}) = -i \int_{\mathbb{R}^3} d\vec{k} \, \overline{\phi_{\vec{k}}(\tau_x, \vec{x})} \phi_{\vec{k}}(\tau_y, \vec{y}) - \phi_{\vec{k}}(\tau_x, \vec{x}) \overline{\phi_{\vec{k}}(\tau_y, \vec{y})}$$

$$= \frac{1}{(2\pi)^3} \int_{\mathbb{R}^3} d\vec{k} \, \frac{\Delta_k(\tau_x, \tau_y)}{a(\tau_x) a(\tau_y)} e^{i\vec{k}(\vec{x}-\vec{y})},$$

where

$$\Delta_k(\tau_x, \tau_y) \doteq -i \left( \overline{T_k(\tau_x)} T_k(\tau_y) - T_k(\tau_x) \overline{T_k(\tau_y)} \right).$$

Following [LuRo90], we will first prove that $\widetilde{\Delta} f$ is a smooth solution of the Klein-Gordon equation and then show that $\widetilde{\Delta} f$ and $\Delta f$ have the same initial conditions on a Cauchy surface. The uniqueness of solutions of the Klein-Gordon equation with compactly supported initial conditions then entails that $\widetilde{\Delta} \equiv \Delta$.

To prove that $\widetilde{\Delta}$ maps all $f \in \mathscr{D}(M, \mathbb{R})$ (and therefore all elements in $\mathscr{D}(M)$ because $\Delta$ is real-linear) to smooth solutions, let us analyse

$$\left[ \widetilde{\Delta} f \right](\tau_x, \vec{x}) = \widetilde{\Delta}(\tau_x, \vec{x}, f) = -i \int_{\mathbb{R}^3} d\vec{k} \, \overline{\phi_{\vec{k}}(\tau_x, \vec{x})} \phi_{\vec{k}}(f) - \phi_{\vec{k}}(\tau_x, \vec{x}) \overline{\phi_{\vec{k}}(f)}.$$

By our assumptions on the $k$-regularity of $T_k$ and arguments analogous to the ones given in the proof of lemma II.2.1.2, $T_k(\tau) \phi_{\vec{k}}(f)$ is an element of $L^2(\mathbb{R}^3, d\vec{k})$ for all $\tau$ and the above integral is therefore a well-defined Fourier-Plancherel transform of a square-integrable function. To see that $\widetilde{\Delta}(\tau_x, \vec{x}, f)$ is actually a smooth function, we have to show that all derivatives of its integrand are integrable. Regarding the spatial derivatives, this follows from the regularity of $T_k(\tau)\hat{\phi}_{\vec{k}}(f)$ for small $k$ and its rapid decrease for large $k$, whereas the wished-for property of $\tau$-derivatives follows from the $k$-regularity of $T_k(\tau)$ and $\partial_\tau T_k(\tau)$ and the fact that one can relate higher $\tau$-derivatives of $T_k(\tau)$ to lower ones by the ordinary differential equation (II.30). These considerations entail in particular that $\Delta(\tau_x, \vec{x}, f)$ is a solution of the Klein-Gordon equation since $\hat{\phi}_{\vec{k}}(\tau_x, \vec{x})$ already possesses this property.

In order to show that $\widetilde{\Delta}f = \Delta f$ for all $f$, we need a mode expansion for an arbitrary solution in terms of its initial datum on a Cauchy surface. To wit, let $\Sigma$ denote a Cauchy surface of constant $\tau = \tau_y$ with future pointing unit normal vector field $N = a^{-1}\partial_\tau$ and volume measure $d\Sigma$, let $u \in \mathscr{S}(M)$ with $u\!\restriction_\Sigma = u_0$ and $\nabla_N u\!\restriction_\Sigma = u_1$, and let us denote the Fourier transform of a function $f$ on $\mathbb{R}^3$ by $\mathscr{F}[f]$ and its inverse Fourier transform by $\mathscr{F}^{-1}[f]$. Then, we can express $u$ by

$$u(\tau_x, \vec{x}) = \mathscr{F}^{-1}\left[a^2(\tau_y)\left\{\mathscr{F}[u_0]\partial_{\tau_y}\frac{\Delta_k(\tau_y, \tau_x)}{a(\tau_x)a(\tau_y)} - \mathscr{F}[u_1]\frac{\Delta_k(\tau_y, \tau_x)}{a(\tau_x)}\right\}\right](\vec{x}), \qquad \text{(II.32)}$$

as follows either by direct computation on account of the properties of $\Delta_k$ inherited from the mode normalisation condition (II.31), or from the last identity displayed in the proof of lemma II.2.1.2. Recalling $d\Sigma\nabla_N = d\vec{y}a^2(\tau_y)\partial_{\tau_y}$, we can compute

$$\int_\Sigma d\Sigma\, u_0\nabla_N\widetilde{\Delta}f - \widetilde{\Delta}f\, u_1 = \int_{\mathbb{R}^3} d\vec{y}\, a^2(\tau_y)\left\{u_0(\vec{y})\partial_{\tau_y}(\widetilde{\Delta}f)(\tau_y, \vec{y}) - a(\tau_y)u_1(\vec{y})(\widetilde{\Delta}f)(\tau_y, \vec{y})\right\}$$

$$= \int_{\mathbb{R}} d\tau_x\, a^4(\tau_x)\int_{\mathbb{R}^3} d\vec{k}\, a^2(\tau_y)\, \mathscr{F}^{-1}[f](\tau_x, \vec{k})\left\{\mathscr{F}[u_0](\vec{k})\partial_{\tau_y}\frac{\Delta_k(\tau_y, \tau_x)}{a(\tau_x)a(\tau_y)} - \mathscr{F}[u_1](\vec{k})\frac{\Delta_k(\tau_y, \tau_x)}{a(\tau_x)}\right\}$$

$$= \int_M d_g x\, u(x)f(x),$$

where the second identity follows by insertion of $\widetilde{\Delta}$ and $d_g x = d\tau_x d\vec{x}a^4(\tau_x)$ and the last one follows from Parseval's theorem and (II.32). To close the proof, let us recall that $\Delta$ also fulfils $\int_\Sigma d\Sigma\, u_0\nabla_N\Delta f - \Delta f\, u_1 = \int_M d_g x\, u(x)f(x)$ as proven in lemma II.1.2.3. This entails that $\Delta f\!\restriction_\Sigma = \widetilde{\Delta}f\!\restriction_\Sigma$ and $\nabla_N\Delta f\!\restriction_\Sigma = \nabla_N\widetilde{\Delta}f\!\restriction_\Sigma$ which in turn implicates $\widetilde{\Delta} \equiv \Delta$ due to the uniqueness of solutions of the Klein-Gordon equation with compactly supported initial conditions. $\quad\square$

Note that the just obtained result allows us to give an 'interpretation' of the mode normalisation condition (II.31): by lemma II.1.2.4, we know that the normal derivative of the causal propagator $\Delta$ restricted to a Cauchy surface equals the $\delta$-distribution on that Cauchy surface. In view of lemma II.2.1.3, (II.31) entails exactly such 'normalisation condition' for $\Delta$ in the special case of constant-$\tau$ Cauchy surfaces in FLRW spacetimes.

The alert reader might have already noticed that we lack one ingredient in order to assure that the results described in this subsection are really meaningful. Namely, we need to show that there really exist solutions $T_k(\tau)$ of (II.30) fulfilling the $k$-regularity conditions we have assumed in lemma II.2.1.2 and II.2.1.3. To close both this gap and this subsection, we shall now provide an example for such $T_k$ by showing that, on NBB spacetimes, 'asymptotic positive frequency modes at early times' fulfil the aforementioned conditions. Moreover, as realised in [Pin10], such modes will play an important role in the later discussion of scalar quantum fields on NBB spacetimes. To briefly anticipate the following result, let us recall that the scale factor $a(\tau)$ of an NBB spacetime vanishes in the limit $\tau \to -\infty$. This implies that (II.30) tends to the massless Minkowskian Klein-Gordon equation at early times, and we could therefore expect that the solutions of (II.30) come arbitrarily close to massless Minkowskian Klein-Gordon modes while approaching the past boundary of an NBB spacetime. Following [DMP09a, DMP09b, Pin10], we will show that this is indeed the case by providing a solution of (II.30) in terms of a convergent perturbative expansion around the massless Minkowskian case.

**Lemma II.2.1.4** *Let a be the scale factor of an NBB spacetime, let $V(\tau) \doteq a^2(\tau)m^2$, let us denote the normalised positive frequency solutions of* (II.30) *with $V \equiv 0$ by $T_{0,k}(\tau) \doteq \frac{1}{\sqrt{2k}}e^{-ik\tau}$, and let finally*

$$\Delta_{0,k}(\tau_x, \tau_y) \doteq -i \left( \overline{T_{k,0}(\tau_x)}T_{k,0}(\tau_y) - T_{k,0}(\tau_x)\overline{T_{k,0}(\tau_y)} \right).$$

*There exists a $\tau_0 \leq -1$ such that the series*

$$T_k(\tau) \doteq T_{0,k}(\tau) + \sum_{n=1}^{\infty} T_{n,k}(\tau)$$

*with*

$$T_{n,k}(\tau) \doteq (-1)^n \int_{-\infty}^{\tau} d\tau_1 \cdots \int_{-\infty}^{\tau_{n-1}} d\tau_n \, \Delta_{0,k}(\tau, \tau_1) \cdots \Delta_{0,k}(\tau_{n-1}, \tau_n) V(\tau_1) \cdots V(\tau_n) T_{0,k}(\tau_n)$$

*converges for all $\tau < \tau_0$ and defines a solution of* (II.30) *which fulfils*

$$\lim_{\tau \to -\infty} e^{ik\tau} T_k(\tau) = \frac{1}{\sqrt{2k}} \qquad \lim_{\tau \to -\infty} e^{ik\tau} \partial_\tau T_k(\tau) = -i\sqrt{\frac{k}{2}}$$

*and*

$$T_k(\tau) = O\left(k^{-\frac{1}{2}}\right)$$

*for all $k$ and for all $\tau$, whereas*

$$\partial_\tau T_k(\tau) = \begin{cases} O\left(k^{-\frac{1}{2}}\right) \text{ for } k \to 0 \text{ and} \\ \\ O\left(\sqrt{k}\right) \text{ for } k \to \infty \end{cases}$$

*uniformly in $\tau$.*

*Proof.* Essentially, we want to solve the inhomogeneous differential equation

$$(\partial_\tau^2 + k^2) T_k(\tau) = -V(\tau) T_k(\tau)$$

with the initial conditions

$$\lim_{\tau \to -\infty} \left( T_k(\tau) - T_{0,k}(\tau) \right) = 0, \qquad \lim_{\tau \to -\infty} \left( \partial_\tau T_k(\tau) - \partial_\tau T_{0,k}(\tau) \right) = 0.$$

As $(\partial_{\tau_x}^2 + k^2) \Delta_{0,k}(\tau_x, \tau_y) = 0$ and $\partial_{\tau_x} \Delta_{0,k}(\tau_x, \tau_y)|_{\tau_x = \tau_y} = 1$ by construction, one can straightforwardly compute that the series under consideration is a formal solution of the mentioned differential equation with the correct initial conditions. To show that it is a proper solution which is also smooth in $\tau$, we need to prove that both the series itself and the series of arbitrary $\tau$-derivatives converge. In order to achieve this, it is sufficient to only consider the series of a single $\tau$-derivative, as higher derivatives are related to the ones of zeroth and first order via the homogeneous differential equation fulfilled by $T_{0,k}$ and $\Delta_{0,k}$.

To prove the wished-for convergence of the series itself, let us note that, for $0 \geq \tau_x \geq \tau_y$

$$\Delta_{0,k}(\tau_x, \tau_y) = \frac{\sin k(\tau_x - \tau_y)}{k} = (\tau_x - \tau_y) \frac{\sin k(\tau_x - \tau_y)}{k(\tau_x - \tau_y)} \leq (\tau_x - \tau_y) \leq -\tau_y$$

as $x^{-1} \sin x$ is bounded by 1. Moreover, lemma I.3.2.4 entails that there is a $\tau_0 \leq -1$ and a positive constant $C$ such that $V(\tau) \leq C \tau^{-3}$ for all $\tau < \tau_0$. This in combination with the bound for $\Delta_{0,k}$ and the fact that $|T_{0,k}(\tau)| \equiv (2k)^{-1/2}$ implies that the $\tau_n$-integral in the definition of $T_{n,k}(\tau)$ is bounded by $C(-\tau_{n-1})^{-1} |T_{0,k}|$. Iterating this argument, we obtain

$$|T_{n,k}(\tau)| \leq \frac{C^n}{n!(-\tau)^n} |T_{0,k}|$$

and we see that the series defining $T_k(\tau)$ converges absolutely and fulfils

$$|T_k(\tau)| \leq e^{\frac{C}{-\tau}} |T_{0,k}|,$$

which in particular implies that $T_k(\tau) = O(k^{-1/2})$ for all $\tau$.

In order to analyse the convergence of $\partial_\tau T_k(\tau)$, let us first note that

$$\partial_{\tau_x} \Delta_{0,k}(\tau_x, \tau_y) = \cos k(\tau_x - \tau_y) \leq 1$$

and let us then realise that, by the antisymmetry of $\Delta_{0,k}$ and the absolute convergence of all appearing integrals for $\tau \leq \tau_0$,

$$\partial_\tau T_{n,k}(\tau) = (-1)^n \int_{-\infty}^{\tau} d\tau_1 \cdots \int_{-\infty}^{\tau_{n-1}} d\tau_n \, \partial_\tau \Delta_{0,k}(\tau, \tau_1) \cdots \Delta_{0,k}(\tau_{n-1}, \tau_n) V(\tau_1) \cdots V(\tau_n) T_{0,k}(\tau_n).$$

More in detail, to see the abovementioned absolute convergence and to achieve our goals, we can repeat the above considerations regarding the bound satisfied by $T_{n,k}(\tau)$ to obtain

$$|\partial_\tau T_{n,k}(\tau)| \leq \frac{C^n}{(n+1)(n-1)!(-\tau)^{n+1}}|T_{0,k}| \leq \frac{C^n}{n!(-\tau)^{n+1}}|T_{0,k}|.$$

This bound entails that the series corresponding to $\partial_\tau T_k(\tau)$ converges absolutely and fulfils

$$|\partial_\tau T_k(\tau)| \leq |\partial_\tau T_{0,k}(\tau)| + \frac{1}{-\tau}(e^{\frac{C}{-\tau}}-1)|T_{0,k}|.$$

This implies the wanted $k$-regularity of $\partial_\tau T_k(\tau)$. □

## II.2.2 The Bulk and Boundary Symplectic Spaces of Solutions

As already discussed at the beginning of the present section, one possible approach to field quantization is to promote a symplectic space, *i.e.* a space of classical fields with a Poisson bracket, to an algebra of quantum fields via canonical quantization. As we are interested in defining a quantum field theory both on the bulk and on the boundary of an NBB spacetime, we need a symplectic space on the boundary. Moreover, since we would like to have a sensible relation of the bulk and boundary quantum field theory to be able to interpret boundary results in the bulk, we need a symplectic space on the boundary which corresponds to the one in the bulk in a meaningful way. In fact, we will now, following [DMP06, DMP09a, DMP09b, DMP09c, Mor06, Mor08] , show that we can project all elements of the bulk solution space $\mathscr{S}(M)$ to a certain boundary function space, which turns out to be constituted by the smooth, square-integrable functions with square-integrable $v$-derivatives on past null infinity $\Im^-$ with respect to the integration measure $dv d\mathbb{S}^2$ induced by the Bondi metric $h$. In a second step, we will then show that such $L^2$ space carries a symplectic form and that the bulk symplectic form of any two elements of $\mathscr{S}(M)$ is mapped to the boundary symplectic form of their boundary values. The bulk-to-boundary mapping under consideration is therefore preserving the relevant symplectic structures and therefore a so-called *symplectomorphism*.

Following the just-described route, we shall start by discussing the mapping of bulk solutions to the boundary.

**Proposition II.2.2.1** *Employing the notation introduced in section I.3.2, let*

$$P_B \doteq -\Box_B + \frac{1}{6}R_B + m^2,$$

*where the index* $\cdot_B$ *denotes quantities derived from the metric* $g_B$, *let* $\mathscr{S}(M_B)$ *denote the space of real-valued solutions associated to* $P_B$ *in an NBB spacetime* $(M_B, g_B)$, *and let* $\phi$ *be an arbitrary element of* $\mathscr{S}(M_B)$. *The conformal rescaling*

$$\widetilde{\phi} \doteq \frac{a}{\Omega_B}\phi$$

*of $\phi$ can be extended to a smooth function on $\chi^B(M_B) \cup \partial^- M_B$ and its restriction $\widetilde{\phi}\restriction_{\mathfrak{I}^-}$ to $\mathfrak{I}^-$ is therefore well-defined. Moreover, $\widetilde{\phi}\restriction_{\mathfrak{I}^-}$ and its $v$-derivative are elements of $L^2(\mathfrak{I}^-, dv\,d\mathbb{S}^2)$ and we can thus define a mapping of bulk solutions to the boundary as*

$$\Gamma : \mathscr{S}(M_B) \to S(\mathfrak{I}^-) \doteq \{\Phi \in \mathscr{E}(\mathfrak{I}^-, \mathbb{R}) \mid \Phi \in L^2(\mathfrak{I}^-, dv\,d\mathbb{S}^2),\ \partial_v \Phi \in L^2(\mathfrak{I}^-, dv\,d\mathbb{S}^2)\},$$

$$\Gamma \phi \doteq \widetilde{\phi}\restriction_{\mathfrak{I}^-}.$$

*Proof.* We shall prove the thesis by making once more a detour via the Einstein static universe metric $g_E$, since this is regular on the full spacetime $M_E$, in contrast to the metric corresponding to the conformal transformation $g_B \to \Omega_B^2 a^{-2} g_B = (1 + v^2) g_E \restriction_{\chi^B(M_B)}$ and yielding the Bondi metric $h$ on $\mathfrak{I}^-$. To wit, we can apply lemma II.2.1.1 to the case $g = g_B$, $\widetilde{g} = g_E \restriction_{\chi^B(M_B)} = \Omega_{\mathbb{M}}^2 a^{-2} g$ to find that $\phi_E \doteq a \Omega_{\mathbb{M}}^{-1} \phi$ is a smooth solution of

$$P_E \phi_E \doteq \left( -\Box_E + \frac{1}{6} R_E + \frac{a^2}{\Omega_{\mathbb{M}}^2} m^2 \right) \phi_E = 0$$

with compactly supported initial conditions, where the index $\cdot_E$ denotes quantities derived from the metric $g_E$ and it is understood the the above equation only holds on $\chi^B(M_B) \subset M_E$. Lemma I.3.2.4 entails that the coefficient of the mass term in $P_E$ and all its derivatives vanish on $\partial^- M_B = \mathfrak{I}^- \cup i^-$, we can therefore smoothly extend $P_E$ 'beyond' $\partial^- M_B$ by setting the mass term to zero. More in detail, let us recall the coordinates $T$ and $R$ of $M_E$ introduced in section I.3.2, let us denote by $\mathbb{R} \times \{\pi\} \times \mathbb{S}^2 \supset i^0$ the codimension-one submanifold of $M_E$ specified by $R = \pi$, and let us restrict attention to the 'lower, slit half' of $M_E$, i.e. to $M_E^- \doteq M_E \restriction_{T < 0} \setminus (\mathbb{R} \times \{\pi\} \times \mathbb{S}^2)$ to avoid having to worry both about the behaviour of the scale factor $a$ towards the future and about possible issues regarding spatial infinity. Moreover, let us assume without loss of generality that $\chi^B(M_B)$ is *not* completely contained in $M_E^-$, even though $\chi^B(M_B) \subset M_E \setminus (\mathbb{R} \times \{\pi\} \times \mathbb{S}^2)$. With this in mind, lemma I.3.2.4 entails that we can smoothly extend $P_E$ from $M_E^- \cap \chi^B(M_B)$ to $M_E^-$ by just setting

$$P_E \restriction_{M_E^- \setminus \chi^B(M_B)} \equiv -\Box_E + \frac{1}{6} R_E.$$

Let now $f \in \mathscr{D}(M_B)$ be such that $\phi = \Delta_B f$, where $\Delta_B$ is the causal propagator of $P_B$. On account of lemma II.2.1.1, we know that

$$\Delta_E = \frac{a}{\Omega_{\mathbb{M}}} \Delta_B \frac{\Omega_{\mathbb{M}}^3}{a^3}$$

is the causal propagator of $P_E$ on $(\chi^B(M_B), g_E \restriction_{\chi^B(M_B)})$ and that $\phi_E = \Delta_E \Omega_{\mathbb{M}}^3 a^{-3} f$. Let us now agree to omit denoting the restriction of $g_E$ in order to avoid a notational collapse and let us realise a few simple causal relations between the spacetimes under consideration: $(M_E^-, g_E)$, $(\chi^B(M_B), g_E)$, and $(\chi^B(M_B) \cap M_E^-, g_E)$ are globally hyperbolic spacetimes, as the discussion of the conformal embeddings in section I.3.2 entails that all these spacetimes contain a Cauchy surface individuated by a constant-$\tau$ hypersurface. These considerations entail in particular that $\chi^B(M_B) \cap M_E^-$ is a causally convex region both as a subset of $(\chi^B(M_B), g_E)$ and as a subset of $(M_E^-, g_E)$, which in turn implicates that the restriction of $\Delta_E$ to $\chi^B(M_B) \cap M_E^-$ is the causal propagator of $P_E$ on that subspacetime and that the causal propagator of the extended $P_E$ on $M_E^-$ coincides with such $\Delta_E$ upon restriction to $\chi^B(M_B) \cap M_E^-$. These admittedly rather tiresome considerations lead us to the following intermediate result: by lemma II.1.2.2 and the above discussion, we can choose the $f$ in $\phi = \Delta_B f$ to fulfil (omitting the push-forward) supp $f \subset \chi^B(M_B) \cap M_E^-$. Moreover, if we denote the introduced extension of $\Delta_E$ from $\chi^B(M_B) \cap M_E^-$ to $M_E^-$ by the same symbol, then $\phi_E = \Delta_E \Omega_{\mathbb{M}}^3 a^{-3} f$ is the unique smooth extension[3] of $\phi_E$ on $\chi^B(M_B) \cap M_E^-$ fulfilling $P_E \phi_E = 0$ on the full $M_E^-$ and we can straightforwardly restrict it to $\mathfrak{I}^- \subset M_E^-$. To finally obtained the sought extension of $\widetilde{\phi} = a\Omega_B^{-1}\phi$ to $\mathfrak{I}^-$, let us note that, on $\chi^B(M_B) \cap M_E^-$,

$$\widetilde{\phi} = \frac{\Omega_{\mathbb{M}}}{\Omega_B}\phi_E = \frac{1}{\sqrt{1+v^2}}\phi_E.$$

Since $(1+v^2)^{-1/2}$ is smooth on $\mathfrak{I}^- \cup i^-$, $(1+v^2)^{-1/2}\phi_E$ is the wanted smooth extension of $\widetilde{\phi}$ to the past boundary of $M_B$.

In order to finish the proof, we have to show that the found mapping $\Gamma\phi$ of $\phi$ to a smooth function on $\mathfrak{I}^-$ is such that both $\Gamma\phi$ and $\partial_v\Gamma\phi$ are square-integrable with respect to $dv d\mathbb{S}^2$. As $\mathbb{S}^2$ has finite volume with respect to $d\mathbb{S}^2$, it is enough to show that $\Gamma\phi$ and $\partial_v\Gamma\phi$ fall off sufficiently rapidly in the limit $v \to \pm\infty$. Let us consider the limit $v \to -\infty$ and therefore the limit to past timelike infinity $i^-$ first. The wanted fall-off behaviour of $\Gamma\phi$ in this limit follows immediately from the decay properties of $(1+v^2)^{-1/2}$ and from the fact that $\phi_E$ is smooth on $M_E^-$ and therefore bounded in the limit $v \to -\infty$. We can apply the same reasoning for $\partial_v\Gamma\phi$ if we know that $\partial_v\phi_E$ is bounded for large negative $v$. To assure the latter, let us switch from $v$ to the coordinate $V = \tan^{-1}v$ which is non-singular on $i^-$. We can compute

$$\partial_v\phi_E = \frac{1}{1+v^2}\partial_V\phi_E$$

which entails the boundedness of $\partial_v\phi_E$ in the limit $v \to -\infty$ and therefore closes the discussion of such limit.

To analyse $\Gamma\phi$ and $\partial_v\Gamma\phi$ in the limit $v \to \infty$, we can not repeat the above steps, as this limit corresponds to an evaluation at spacelike infinity $i^0$ which lies outside of $M_E^-$ by construction. We can nevertheless show that $\phi_E$ and therefore $\Gamma\phi$ and $\partial_v\Gamma\phi$ actually vanish identically for large but finite $v$ by the causal support properties of solutions of the Klein-Gordon equation with compactly supported initial conditions. As a matter of fact, the definition of $\phi_E$ entails that supp $\phi_E \subset J(\text{supp } f, M_E^-)$, which in turn implicates that $(\text{supp } \Gamma\phi \cup \text{supp } \partial_v\Gamma\phi) \subset J^-(\text{supp } f, M_E^-) \cap \mathfrak{J}^-$. Let us show that this leads to the fact that there is a $v_0 < \infty$ such that[4]

$$\left(\text{supp } \Gamma\phi \cup \text{supp } \partial_v\Gamma\phi\right) \subset (-\infty, v_0] \times \mathbb{S}^2. \tag{II.33}$$

---

[3]Of course, $\Omega_{\mathbb{M}}^3 a^{-3}$ is not defined on the full $M_E^-$, but only on $\chi^B(M_B) \cap M_E^-$. The definition of the extended $\phi_E$ is nevertheless valid since, due to the support property of $f$, $\Omega_{\mathbb{M}}^3 a^{-3}$ is only evaluated where it is defined.

As supp $f$ is compact, there are finite $u_{\min}$, $u_{\max}$, $v_{\min}$, $v_{\max}$ such that, in terms of the retarded and advanced coordinates $u = \tau - r$ and $v = \tau + r$,

$$\text{supp } f \subset \Diamond \doteq [u_{\min}, u_{\max}] \times [v_{\min}, v_{\max}] \times \mathbb{S}^2$$

which entails that

$$\left(\text{supp } \Gamma\phi \cup \text{supp } \partial_v\Gamma\phi\right) \subset J^-(\text{supp } f, M_E^-) \cap \mathfrak{J}^- \subset J^-(\Diamond, M_E^-) \cap \mathfrak{J}^-.$$

To pursue our goals, let us note that the boundary of $J^-(\Diamond, M_E^-)$ is generated by past directed null curves with constant $\mathbb{S}^2$ angles. In $\chi^B(M_B)$, such curves can be parametrised as $r(\tau) = \pm\tau + C_r$, or, equivalently as $u(\tau) = 2\tau + C_u$ or $v(\tau) = 2\tau + C_v$ with suitable constants $C_r$, $C_u$, $C_v$. Note that we do not have to specify to which metric we refer by the invariance of null curves under conformal transformations. As we are considering past directed curves, we have either the case of monotonically decreasing $u$ and constant $v$, or the case of monotonically decreasing $v$ and constant $u$. To avoid the singularity of $u$ and $v$ while approaching the boundary of $\chi^B(M_B)$, let us switch from $u$ to $U = \tan^{-1} u$ and from $v$ to $V = \tan^{-1} v$; this does not hamper the discussion of the behaviour proper to null curves, as $\tan^{-1}$ is monotonous. In these coordinates, the above considerations entail that

$$J^-(\Diamond, M_E^-) \subset (-\infty, \tan^{-1} u_{\max}] \times (-\infty, \tan^{-1} v_{\max}] \times \mathbb{S}^2$$

which implies (II.33) with $v_0 = v_{\max}$. □

The above proposition contains essentially everything we need in order to define the sought boundary symplectic space, as we can expect that the correct symplectic form on the boundary is already fixed by the requirement that it is the result of mapping the bulk symplectic form via $\Gamma$. In fact, we have introduced the bulk symplectic form in definition II.1.2.1 as an integral of suitable normal derivatives on a Cauchy surface and we have shown that it is independent of the chosen Cauchy surface employing Stokes' theorem and the Klein-Gordon equation. In FLRW spacetimes in general and in NBB spacetimes in particular, we can think of the bulk symplectic form as an integral of $\tau$-derivatives over a constant-$\tau$ surface. Moreover, our current understanding of past infinity $\partial^- M_B = \mathfrak{J}^- \cup i^-$, allows us to view it heuristically as a limit of constant-$\tau$ surfaces towards $\tau \to -\infty$ and lemma I.3.2.2 tells us that $\tau$-derivatives tend to $v$-derivatives in the limit to $\mathfrak{J}^-$. We can therefore expect the correct symplectic form on the boundary to be an integral of suitable $v$-derivatives, barring a small caveat: there could be potential problems viewing $\partial^- M_B$ as a limit of constant-$\tau$ surfaces due to $i^-$ being somehow a 'conic singularity' of $\partial^- M_B$, but proposition II.2.2.1 tells us that such problems do not appear on account of the good decay properties of the mapped solutions towards $i^-$. With this in mind, let us state and prove the anticipated results in the following lemma.

---

[4] The reader is invited to draw a picture based on figure I.1.

**Lemma** II.2.2.2  Let $\Phi_1$, $\Phi_2$ be arbitrary elements of $S(\mathfrak{I}^-)$ and let us define a strongly non-degenerate symplectic form on $S(\mathfrak{I}^-)$ by

$$s(\Phi_1, \Phi_2) \doteq \int\limits_{\mathfrak{I}^-} dv\, d\mathbb{S}^2\, \Phi_2 \partial_v \Phi_1 - \Phi_1 \partial_v \Phi_2\,.$$

In addition, let us denote by $\varsigma_B$ the symplectic form on $\mathscr{S}(M_B)$ defined as in in definition II.1.2.1. For all $\phi_1$, $\phi_2$ in $\mathscr{S}(M_B)$, it holds

$$s(\Gamma\phi_1, \Gamma\phi_2) = \varsigma_B(\phi_1, \phi_2)\,.$$

*Proof.*  Since the bulk-to-boundary map $\Gamma$ involves both a conformal rescaling and a limit towards $\mathfrak{I}^-$, we have to prove that both such operations preserve the relevant symplectic forms. To analyse the first one, let us recall that the definition of a symplectic form on general spacetimes provided in definition II.1.2.1 reads

$$\varsigma_M(u_1, u_2) = \int\limits_{\Sigma} d\Sigma\, u_2 \nabla_N u_1 - u_1 \nabla_N u_2$$

and that

$$d\Sigma\,(u_2 \nabla_N u_1 - u_1 \nabla_N u_2) = \frac{1}{6}\sqrt{|\det g|}\, g^{\alpha\beta}\,(u_2 \partial_\alpha u_1 - u_1 \partial_\alpha u_2)\, \varepsilon_{\beta\gamma\mu\nu} dx^\gamma \wedge dx^\mu \wedge dx^\nu \quad \text{(II.34)}$$

with the totally antisymmetric Levi-Civita symbol $\varepsilon_{\beta\gamma\mu\nu}$. Having written the integrand of the bulk symplectic form in this way, we can manifestly see that it is invariant under conformal transformations $g \mapsto \Omega^2 g$, as such transformations induce $g^{-1} \mapsto \Omega^{-2} g^{-1}$ and $u_i \mapsto \Omega^{-1} u_i$. If we denote by $P_B$ the conformally coupled, massive Klein-Gordon operator on $(M_B, g_B)$ (*cf.* proposition II.2.2.1), set

$$\widetilde{P} \doteq \frac{a^3}{\Omega_B^3} P_B \frac{\Omega_B}{a}\,,$$

and denote the symplectic form on the solutions $\widetilde{\phi}$ of $\widetilde{P}\widetilde{\phi} = 0$ on $(M_B, \Omega_B^2 a^{-2} g_B)$ by $\widetilde{\varsigma}_B$, then the above considerations entail

$$\varsigma_B(\phi_1, \phi_2) = \widetilde{\varsigma}_B\left(\frac{a}{\Omega_B}\phi_1, \frac{a}{\Omega_B}\phi_2\right)\,.$$

The statement and proof of proposition II.2.2.1 now allow us to apply Stokes' theorem in a region of $M_E$ which is bounded by a subset $\mathscr{O}$ of a Cauchy surface $\Sigma$ containing the supports of $\phi_1$, $\phi_2$ and their normal derivatives, by $J^-(\mathscr{O}, M_E) \cap \mathfrak{I}^-$, and by a suitable part of the boundary of $J^-(\mathscr{O}, M_E)$. As a consequence, the integration of (II.34) with $g = \Omega_B^2 a^{-2} g_B$, $u_i = a\Omega_B^{-1}\phi_i$ on either $\Sigma$, or $\mathfrak{I}^-$ gives the same result, and, recalling that $\Omega_B^2 a^{-2} g_B$ restricted to $\mathfrak{I}^-$ equals the Bondi metric (I.24), we can straightforwardly compute that (II.34) with $g = \Omega_B^2 a^{-2} g_B$, $u_i = a\Omega_B^{-1}\phi_i$ evaluated on $\mathfrak{I}^-$ yields the integrand of $s(\Gamma\phi_1, \Gamma\phi_2)$, which closes the proof.  $\square$

Having discussed a decent part of the bulk-to-boundary mapping in NBB spacetimes, we are now in the position to say a bit more about the relation of the treatment in this thesis to the works [DMP06, DMP09a, DMP09b, DMP09c, Mor06, Mor08] , where such bulk-to-boundary machinery has been introduced for the first time. The mentioned works treat three different types of spacetimes, *i.e.* asymptotically flat spacetimes [DMP06, Mor06, Mor08], asymptotically de Sitter spacetimes [DMP09a, DMP09b], and the Schwarzschild spacetime [DMP09c]. Let us say a few words about the first two cases, since these are more closely related to NBB spacetimes. While the null boundary used in asymptotically flat spacetimes is, as already remarked in section I.3.2, the past (or future) conformal boundary and therefore not a part of the asymptotically flat spacetime itself, the null 'boundary' utilised in the asymptotically de Sitter case is the *horizon* of such spacetime (see for instance [HaEl73]) which in particular entails that it is a proper part of a spacetime obtained by enlarging the asymptotically de Sitter spacetime *without* introducing a conformal rescaling. This seems to make the asymptotically de Sitter case an easier one, but such conclusion is not the full truth, as the authors of the above-mentioned works find that, while the null boundary of an asymptotically flat spacetime can contain both timelike and null infinity, the horizon of an asymptotically de Sitter spacetime can only be considered as either $\mathfrak{I}^+$ or $\mathfrak{I}^-$, and an analogue of timelike infinity $i^\pm$ does not exist in this case. This entails a difference in the treatment of the bulk-to-boundary projection map $\Gamma$, namely, in the asymptotically flat case one can use general properties of solutions of the Klein-Gordon equation alone to define $\Gamma$ on a more abstract level, while in the asymptotically de Sitter case one has to take a more hands-on approach and introduce $\Gamma$ via a mode decomposition of solutions. The discussion in section I.3.2 entails that we are closer to the asymptotically flat case in that respect; we have thus been able to introduce $\Gamma$ in the mentioned general way in the present subsection. However, as an NBB spacetime is a flat FLRW spacetime, we could have also used the mode-decomposition approach to $\Gamma$, and we have in fact already collected most results relevant in this respect in the previous subsection, though not to use them for the definition of $\Gamma$, but to employ them for the later explicit interpretation of our boundary constructions. Let us stress that, although both NBB and asymptotically de Sitter spacetimes are FLRW spacetimes, they differ in the behaviour of the scalefactor $a(\tau)$ towards $\tau \to -\infty$, which is of the form $a(\tau) = O(\tau^{-1})$ in the asymptotically de Sitter case, whereas it is vanishing faster than any inverse power of $\tau$ in the NBB case[5]. This strong decay of the NBB scale factor for large negative $\tau$ is the reason why the locus $\tau = -\infty$ in the NBB case is a proper boundary and not a horizon. Finally, let us remark that, although there are similarities between NBB and asymptotically flat spacetimes, there is also major difference. To wit, one can not map massive field theories to the boundary of general asymptotically flat spacetimes, even though it is possible in the special case of Minkowski spacetime [Dap07]. As noted in [Pin10], this situation arises because the mass term $m^2$ in the initial Klein-Gordon operator, which in our case is conformally rescaled as $a^2 \Omega_B^{-2} m^2$ and can therefore be projected to the boundary on account of the decay properties of $a$, becomes 'only' $\Omega_B^{-2} m^2$ in the asymptotically flat case as no scale factor $a$ is present, and therefore diverges on the boundary.

---

[5]In fact, one can weaken the definition I.3.1.1 of NBB spacetimes by allowing all power laws of $a(t)$ yielding an infinite particle horizon and not only 'almost linear' ones. This leads to $a(\tau) = O(\tau^{-1-\varepsilon})$, $\varepsilon > 0$, and such cases can be treated with the mentioned mode-decomposition approach of [DMP09a, DMP09b].

To close the discussion of the classical scalar field in NBB spacetimes, let us note that we have not provided a boundary counterpart of the algebra of classical observables defined in subsection II.1.3, as in this work we will not make use of its main potential benefit for free boundary field theories, namely, the concise formulation of normal ordering. However, we would like to stress that a definition of boundary quantization via a deformation of the classical field algebra is certainly possible, as has been shown in [DPP10].

## II.3   The Free Dirac Field in General Curved Spacetimes

After discussing the classical free scalar field, we shall proceed to discuss the classical free Dirac spinor field. Apart from the fact that we have to deal with a multi-component field whose solution space has a 'symmetric' inner product space structure rather than an antisymmetric symplectic structure, our treatment of the classical Dirac field shall follow the same logic as the one of the classical scalar field.

### II.3.1   The Dirac Equation and its Fundamental Solutions

Following the path outlined in the introductory paragraph of the present section, we shall start our analysis of the classical Dirac field by discussing its equation of motion. As already mentioned in the discussion of spinors on curved spacetime, in Minkowski spacetime, the Dirac equation is uniquely determined by Poincaré invariance and the requirement that the reducible representation of (the double cover of) the Poincaré group constituted by a four-component spinor becomes irreducible upon restriction to the space of solutions [Fol63]. In curved spacetime we have no translation invariance, but only local Lorentz invariance and therefore more freedom in choosing a potential Dirac equation. However, we are still limited to differential equations which transform covariantly under the representation $\pi_\oplus$ of $SL(2,\mathbb{C}) = Spin_0(3,1)$ which entered the definition of spinor fields in chapter I[6] and, if we would like the curved-spacetime Dirac equation to reduce to the usual one $(-\partial\!\!\!/ + m\mathbb{I}_4)\psi = (-\partial_a\gamma^a + m\mathbb{I}_4)\psi = 0$ in the flat case, we are limited to a differential operator which is a scalar with respect to both local Lorentz transformations and diffeomorphisms and has mass-dimension 1. This still leaves us with infinitely many possibilities, but Occam's razor or the analyticity arguments mentioned in the discussion of the Klein-Gordon equation lead us to choose

$$D\psi \doteq (-\overline{\nabla}\!\!\!\!/ + m\mathbb{I}_4)\psi = 0 \qquad (\text{II.35})$$

as the canonical covariant generalisation of the Dirac equation to curved spacetimes and therefore $D$ as the canonical *Dirac operator*, where here $\psi$ is taken as a smooth section of $DM$. Note

---

[6]We would like to stress that initially, any statement about transformation properties of spinors depends on the chosen spinor frame. However, in the long run we are always interested in observables – these are 'scalars' in spinor space and a statement that an observable is invariant under spinor transformations is therefore frame-independent and well-defined.

that analyticity arguments rule out adding terms such as

$$\frac{\slashed{\nabla} R}{m^2}$$

to $D$. In addition to the just defined Dirac equation for spinor fields, *i.e.* sections of $DM$, we define the Dirac equation for cospinor fields as

$$D^*\psi^* = (\slashed{\nabla} + m\mathbb{I}_4)\,\psi^* = 0,$$

where $\psi^*$ is taken as a smooth section of $D^*M$ and the adjoint notation for $D^*$ will be motivated soon. To close the discussion of the Dirac equation, let us stress three things: first, the Dirac equation usually involves an imaginary unit $i$ in standard treatments. This is not the case here, as we are working with the sign convention $(-,+,+,+)$ and our $\gamma$-matrices therefore differ from the usual ones by a factor $i$. Secondly, we have up to now defined $D$ only on spinors and $D^*$ only on cospinors, but we shall understand these operators as acting on both spinors and cospinors in the following, where one should always keep in mind that a cospinor field is a row vector and therefore any matrix-valued operator is understood to act on it *from the right*. Finally, the alert reader might have noticed that the only free parameter in the Dirac equation is the mass $m$, which particularly implies that there is no freedom to choose a coupling to curvature like in the case of scalar fields and the Klein-Gordon equation. In fact, the coupling of an on-shell Dirac field to curvature is *always* conformal, as we shall discuss in more detail in the next section.

Before discussing the solutions of the Dirac equations, let us define the appropriate section spaces and distribution spaces we shall deal with in the following. We refer the reader interested in the physical relevance of these spaces to the comments after definition II.1.1.1.

**Definition II.3.1.1** *By $\mathcal{E}(DM) \doteq C^\infty(M, DM)$ we denote the **space of smooth sections** of $DM$ equipped with the usual **locally convex topology**, i.e. a sequence of sections $f_n \in \mathcal{E}(DM)$ is said to converge to $f \in \mathcal{E}(DM)$ if all derivatives of $f_n$ converge to the ones of $f$ uniformly on all compact subsets of $M$. In analogy, we denote the locally convex space of smooth sections of $D^*M$ by $\mathcal{E}(D^*M)$.*

*The space $\mathcal{D}(DM) \doteq C_0^\infty(M, DM)$ is the subspace of $\mathcal{E}(DM)$ constituted by the **smooth sections with compact support**. We equip $\mathcal{D}(DM)$ with the locally convex topology determined by saying that a sequence of sections $f_n \in \mathcal{D}(DM)$ converges to $f \in \mathcal{D}(DM)$ if there is a compact subset $K \subset M$ such that all $f_n$ and $f$ are supported in $K$ and all derivatives of $f_n$ converge to the ones of $f$ uniformly in $K$. We define the locally convex space $\mathcal{D}(D^*M) \doteq C_0^\infty(M, D^*M)$ analogously.*

*By $\mathcal{D}'(D^*M)$ we denote the space of **distributions on** $\mathcal{D}(DM)$, i.e. the topological dual of $\mathcal{D}(DM)$ provided by continuous, linear functionals $\mathcal{D}(DM) \to \mathbb{C}$, whereas $\mathcal{E}'(DM)$ denotes the topological dual of $\mathcal{E}(D^*M)$, i.e. the space of **distributions with compact support**. Similarly, we define $\mathcal{D}'(DM)$ as the topological dual of $\mathcal{D}(D^*M)$, and $\mathcal{E}'(DM)$ as the one of $\mathcal{E}(D^*M)$. Our notation is chosen such that $\mathcal{D}'(DM) \supset \mathcal{E}(DM) \supset \mathcal{D}(DM)$ and $\mathcal{D}'(D^*M) \supset \mathcal{E}(D^*M) \supset \mathcal{D}(D^*M)$.*

*For $f \in \mathcal{D}(DM)$, $u \in \mathcal{D}'(D^*M)$ we shall denote the **dual pairing** of $f$ and $u$ by*

$$\langle u, f \rangle \doteq \int_M d_g x\, u(x) f(x),$$

where $u(x)f(x) \doteq [u(x)]_A[f(x)]^A$ means the fibrewise dual pairing of $DM$ and $D^*M$ inherited from the Hermitean product on $\mathbb{C}^4$.

With the above-defined dual pairing at hand, we can now motivate the adjoint notation for $D^*$. Namely, if either $f \in \mathscr{D}(D^*M)$ and $g \in \mathscr{E}(DM)$, or $f \in \mathscr{E}(D^*M)$ and $g \in \mathscr{D}(DM)$, we can compute by a partial integration:

$$\langle Df, g \rangle = \langle f, D^*g \rangle, \qquad \langle D^*f, g \rangle = \langle f, Dg \rangle.$$

Having introduced the relevant section spaces, we can now employ the Dirac conjugation matrix $\beta$ and the charge conjugation matrix $\mathscr{C}$ introduced in definition I.2.2.3 to define two associated conjugations on the spaces of smooth sections.

**Definition** II.3.1.2  Let $f \in \mathscr{E}(DM)$ and $g \in \mathscr{E}(D^*M)$ be arbitrary smooth sections. We define **Dirac conjugation** maps $\cdot^\dagger : \mathscr{E}(DM) \to \mathscr{E}(D^*M)$ and $\cdot^\dagger : \mathscr{E}(D^*M) \to \mathscr{E}(DM)$ by setting

$$f^\dagger \doteq f^*\beta \qquad g^\dagger \doteq \beta^{-1}g^*,$$

where $\cdot^*$ denotes the fibrewise Hermitean adjoint inherited from $\mathbb{C}^4$. Additionally, we define **charge conjugation** maps $\cdot^c : \mathscr{E}(DM) \to \mathscr{E}(DM)$ and $\cdot^c : \mathscr{E}(D^*M) \to \mathscr{E}(D^*M)$ via

$$f^c \doteq \mathscr{C}^{-1}\overline{f} \qquad g^c \doteq \overline{g}\,\mathscr{C}.$$

Here, the appearing matrix products are meant to be performed in an arbitrary but fixed spinor frame.

Let us stress once more that, although the above definition manifestly depends on a chosen spin frame, as the matrices $\beta$ and $\mathscr{C}$ are given in a frame-independent manner, such frame-dependence of $\cdot^c$ and $\cdot^\dagger$ will disappear when we restrict attention to observable quantities. Moreover, let us recall that, in the representation of the Clifford algebra chosen by us in theorem I.2.2.2, $\beta = -i\gamma_0 = \beta^{-1}$ and $\mathscr{C} = -i\gamma_2 = -\mathscr{C}^{-1}$. In the follwing, we will need the mutual relations between $\cdot^c$ and $\cdot^\dagger$ and also their interplay with $D$ and $D^*$. As shown in [San08], one can straightforwardly compute for any $f \in \mathscr{E}(DM)$, $g \in \mathscr{E}(D^*M)$:

$$f^{\dagger\dagger} = f, \qquad g^{\dagger\dagger} = g, \qquad f^{cc} = f, \qquad g^{cc} = g,$$
$$(Df)^\dagger = D^*f^\dagger, \qquad (D^*g)^\dagger = Dg^\dagger, \qquad (Df)^c = Df^c, \qquad (D^*g)^c = D^*g^c, \qquad \text{(II.36)}$$
$$f^{\dagger c} = -f^{c\dagger}, \qquad g^{\dagger c} = -g^{c\dagger}.$$

We shall now turn to the problem of existence and uniqueness of solutions and fundamental solutions of the Dirac equations. To attack this problem, one has to make use of the fact, well-known from Minkowski spacetime, that the Dirac operator squares to an operator of Klein-Gordon type. In curved spacetimes, this result is called *Lichnerowicz identity*. In fact, following [Lic64], we can compute the following result.

**Lemma II.3.1.3**  *The Dirac operators $D$ and $D^*$ acting on sections of either $DM$ or $D^*M$ fulfil*

$$DD^* = D^*D = \mathfrak{P} \doteq -\nabla_a \nabla^a + \left( \frac{R}{4} + m^2 \right) \mathbb{I}_4.$$

*Proof.* We shall compute the identity on a section $f \in \mathscr{E}(DM)$, as the computation for a cospinor section is analogous. Let us therefore recall that the $\gamma$-matrices are covariantly constant and compute

$$D^*Df = DD^*f = \left( -\gamma^a \gamma^b \nabla_a \nabla_b + m^2 \mathbb{I}_4 \right) f = \left( -\frac{1}{2} \{ \gamma^a, \gamma^b \} \nabla_a \nabla_b - \frac{1}{2} \gamma^a \gamma^b [\nabla_a, \nabla_b] + m^2 \mathbb{I}_4 \right) f.$$

We can now use the Clifford relations, as well as the identities fulfilled by the spin curvature tensor (*cf.* lemma I.2.2.9) to obtain

$$\left( -\frac{1}{2} \{ \gamma^a, \gamma^b \} \nabla_a \nabla_b - \frac{1}{2} \gamma^a \gamma^b [\nabla_a, \nabla_b] + m^2 \mathbb{I}_4 \right) f = \left( -\nabla_a \nabla^a - \frac{1}{2} \gamma^a \gamma^b \mathfrak{R}_{ab} + m^2 \mathbb{I}_4 \right) f$$

$$= \left( -\nabla_a \nabla^a + \frac{1}{4} R \mathbb{I}_4 + m^2 \mathbb{I}_4 \right) f.$$

□

The identity we have just proved shows somehow more explicitly how the definition of the Dirac equation determines uniquely the coupling to (scalar) curvature. It is important to note that, in contrast to the Minkowskian case, the differential operator $\mathfrak{P}$ is *not* diagonal, which can be seen by expanding the covariant derivative in terms of partial derivatives and the spin connection coefficients. However, such expansion shows that $\mathfrak{P}$ has still (diagonal and) metric principal part, and is therefore a normally hyperbolic operator. Although we have defined such operators in theorem II.1.1 only on scalar functions, this definition can be straightforwardly generalised to sections of vector bundles such as $DM$ and $D^*M$, and the same results on existence and uniqueness of solutions of the differential equation induced by $\mathfrak{P}$ hold also in this more general setting, as shown in [BGP07]. To wit, as proved in [Dim82], one can use this fact to assure existence and uniqueness of solutions of the Dirac equation by reducing the problem to the analysis of $\mathfrak{P}$. Before stating this result, let us stress an additional difference between the curved setting and the Minkowskian case: the identity $D^*D = DD^* = \mathfrak{P}$ holds only for sections of $DM$ and $D^*M$, *i.e.* for quantities with a single spinor index. In case of spinor-tensors of higher rank, additional curvature tensors appear in the computation of $D^*D$ and therefore lead to a different result. Let us now provide the promised theorem proved in [Dim82].

**Theorem II.3.1.4**

    *a) Let $\Sigma$ be an arbitrary smooth Cauchy surface of $M$ with forward pointing unit normal vector field $N$, let $D\Sigma$ be the natural Dirac spinor bundle on $\Sigma$ induced by $DM$ and let $u_0 \in \mathscr{D}(D\Sigma)$. The* **Cauchy problem**

$$Du = 0, \qquad u\!\restriction_\Sigma = u_0$$

*has a unique solution $u \in \mathscr{E}(DM)$ which has the causal support property supp $u \subset J(\text{supp } u_0, M)$. Such $u$ has the form $u = D^* \widetilde{u}$, where $\widetilde{u} \in \mathscr{E}(DM)$ is the unique solution of the Cauchy problem*

$$\mathfrak{P}\widetilde{u} = 0, \qquad \widetilde{u}\!\restriction_\Sigma \equiv 0, \qquad \nabla_N \widetilde{u}\!\restriction_\Sigma \equiv -\slashed{N} u_0.$$

b) *The operators $D$ and $D^*$ possess unique* **fundamental solutions**. *Namely, there exist unique continuous operators*

$$S^\pm : \mathscr{D}(DM) \to \mathscr{E}(EM), \qquad S^\pm_* : \mathscr{D}(D^*M) \to \mathscr{E}(E^*M)$$

*which fulfil*

$$D S^\pm = S^\pm D = id_{\mathscr{D}(DM)}, \qquad D^* S^\pm_* = S^\pm_* D^* = id_{\mathscr{D}(D^*M)},$$

$$\text{supp } S^\pm f \subset J^\pm(\text{supp } f, M), \qquad \text{supp } S^\pm_* g \subset J^\pm(\text{supp } g, M)$$

*for all $f \in \mathscr{D}(DM)$, $g \in \mathscr{D}(D^*M)$. The operators $S^\pm$, $S^\pm_*$ have the form $S^\pm = D^* \mathfrak{G}^\pm = \mathfrak{G}^\pm D^*$, $S^\pm_* = D \mathfrak{G}^\pm_* = \mathfrak{G}^\pm_* D$, where*

$$\mathfrak{G}^\pm : \mathscr{D}(DM) \to \mathscr{E}(EM), \qquad \mathfrak{G}^\pm_* : \mathscr{D}(D^*M) \to \mathscr{E}(E^*M)$$

*are the unique fundamental solutions of $\mathfrak{P}$.*

c) *Let us define the* **causal propagators** *of $D$ and $D^*$ by*

$$S \doteq S^- - S^+ \qquad and \qquad S_* \doteq S^-_* - S^+_*,$$

*respectively. $S$ and $S_*$ are related as*

$$\langle S_* g, f \rangle = -\langle g, Sf \rangle$$

*for all $f \in \mathscr{D}(DM)$, $g \in \mathscr{D}(D^*M)$. Moreover, every solution $u \in \mathscr{E}(DM)$ of $Du = 0$ with compactly supported initial conditions is of the form $u = Sf$ for some $f \in \mathscr{D}(DM)$, and for all $f \in \mathscr{D}(DM)$ satisfying $Sf = 0$ there is an $\widetilde{f} \in \mathscr{D}(DM)$ with $f = D\widetilde{f}$.*

The attentive reader might have noticed that we have provided more statements related to $D$ than properties of $D^*$ in the above theorem. However, the 'missing' statements related to $D^*$ can be obtained from the ones proper to $D$ by Dirac conjugation and the fact that $-S_*$ is the formal adjoint of $S$.

In analogy to the scalar case, we can understand $S$ as a distribution on $\mathscr{D}(D^*M) \otimes \mathscr{D}(DM)$ by setting

$$S(g, f) \doteq \langle g, Sf \rangle$$

for all $f \in \mathscr{D}(DM)$, $g \in \mathscr{D}(D^*M)$. $S(g, f)$ vanishes if the supports of $f$ and $g$ are spacelike separated, which entails that the distribution kernel $S(x, y)$ of $S$ vanishes for spacelike separated points. This property will enable us to use is as an *anticommutator function* for the quantized Dirac field. We would like to stress an important point: $S$ manifestly fulfils $DS = SD = 0$, but, when expressing this identity in terms of the distribution kernel $S(x, y)$ of $S$, one has to be careful and take into account that, in $SD$, $D$ is meant to act on a test section from the left, and we have thus to take the adjoint of $D$ if we want to understand it as acting on $S$ from the right. Using canonical notation, this entails that $DS = SD = 0$ translates into

$$D_x S(x, y) = D_y^* S(x, y) = 0. \tag{II.37}$$

To close this subsection, let us say how $S$ is related to Dirac conjugation and charge conjugation. Namely, using the relation of $\cdot^c$ and $\cdot^\dagger$ to $D$ and $D^*$, one can show for arbitrary test sections $f \in \mathscr{D}(DM)$, $g \in \mathscr{D}(D^*M)$ that

$$(S^\pm f)^\dagger = S_*^\pm f^\dagger, \qquad (S^\pm f)^c = S^\pm f^c, \qquad (S_*^\pm g)^\dagger = S^\pm g^\dagger, \qquad (S_*^\pm g)^c = S_*^\pm g^c \tag{II.38}$$

and therefore [San08]

$$S\left(f^{\dagger c}, g^{\dagger c}\right) = S(g, f) = -\overline{S\left(f^\dagger, g^\dagger\right)} = -\overline{S\left(g^c, f^c\right)}. \tag{II.39}$$

As we shall see later, the first equality in (II.39) entails that the anticommutator of a Dirac quantum field and its adjoint is invariant under charge conjugation.

## II.3.2   The Inner Product Space of Solutions

We have already briefly mentioned at the beginning of this section that the space of solutions of the Dirac equation has a 'symmetric' inner product space structure based on the causal propagator $S$ of the Dirac operator $D$, in contrast to the solution space of the scalar field, which has an antisymmetric symplectic structure constructed out of the causal propagator $\Delta$ of the Klein-Gordon operator $P$. In analogy to the scalar case, we shall first define this inner product structure in a way which *a priori* seems to be independent of the causal propagator, and relate it to the causal propagator afterwards.

To this avail, let us recall that, per definition I.2.2.3, $-i\beta\slashed{N}$ is a positive definite matrix for all future-pointing, timelike vector fields $N$[7]. Hence, $-i\langle u^*, \beta\slashed{N}u\rangle = -i\langle u^\dagger, \slashed{N}u\rangle \geq 0$ for all $u \in \mathscr{D}(DM)$ and $-i\langle u^\dagger, \slashed{N}u\rangle$ vanishes only when $u \equiv 0$. Following [Dim82], this prompts the ensuing definition.

**Definition II.3.2.1**  *By* $(\mathfrak{S}(M), \iota_M)$ *we denote the **inner product space of solutions** of the Dirac equation. Here,* $\mathfrak{S}(M) \subset \mathscr{E}(DM)$ *is the space of smooth solutions* $\psi$ *of the Dirac equation* $D\psi = 0$ *with compactly supported initial conditions on a Cauchy surface and* $\iota_M : \mathfrak{S}(M) \times \mathfrak{S}(M) \to \mathbb{C}$ *is an **inner product** defined by*

$$\mathfrak{S}(M) \times \mathfrak{S}(M) \ni (\psi_1, \psi_2) \mapsto \iota_M(\psi_1, \psi_2) \doteq -i \int_\Sigma d\Sigma\, \psi_1^\dagger \slashed{N} \psi_2.$$

*Above,* $\Sigma$ *is an arbitrary Cauchy surface of* $(M, g)$, *$N$ is the forward pointing unit normal vector field on* $\Sigma$, *and* $d\Sigma$ *denotes the canonical volume measure on* $\Sigma$ *obtained by restriction of* $d_g x$ *to* $\Sigma$.

---

[7]As an example, consider $N^a = (1, 0, 0, 0)^T$. Then, $-i\beta\slashed{N} = \beta^2 = \mathbb{I}_4$.

In analogy to the definition of $(\mathscr{S}(M), \varsigma_M)$ in subsection II.1.2, we have utilised a Cauchy surface $\Sigma$ to define $\iota_M$, and we shall explore the relation between the causal propagator $S$ and $\iota_M$ to show that the latter is independent of the chosen $\Sigma$. To this avail, the starting point are the facts that $S$ maps all test sections $f$ in $\mathscr{D}(DM)$ to $\mathfrak{S}(M)$ and that every element of $\psi \in \mathfrak{S}(M)$ is of the form $\psi = Sf$ for some $f \in \mathscr{D}(DM)$. In fact, we can repeat the proof of lemma II.1.2.2 by replacing the Klein-Gordon operator $P$ and its fundamental solutions $G^\pm$ with $D$ and $S^\pm$ respectively to obtain the following result on the nature of such $f$.

**Lemma II.3.2.2** *Let $\psi \in \mathfrak{S}(M)$ be an arbitrary solution of $D\psi = 0$ with compactly supported initial data on a Cauchy surface and let $\Sigma$ be any Cauchy surface of $M$. Then, for any bounded neighbourhood $\mathscr{O}(\Sigma)$ of $\Sigma$, we can find an $f \in \mathscr{D}(DM)$ with $\mathrm{supp}\, f \subset \mathscr{O}(\Sigma)$ and $\psi = Sf$.*

With the above considerations in mind, the wanted relation between $\iota_M$ and $S$ follows essentially from the application of Stokes' theorem to the current $j^\mu(\psi_1, \psi_2) \doteq \psi_1^\dagger \gamma^\mu \psi_2$, which fulfils $\nabla_\mu j^\mu = 0$ if $\psi_1$ and $\psi_2$ are elements of $\mathfrak{S}(M)$. The following result is taken from [Dim82].

**Lemma II.3.2.3** *Let $\Psi_f$, $\Psi_g$ be elements of $\mathfrak{S}(M)$ given as $\Psi_f = Sf$, $\Psi_g = Sg$ for suitable test sections $f$ and $g$.*

*a) The inner product $\iota_M(\Psi_f, \Psi_g)$ can be computed in terms of the causal propagator as*

$$\iota_M(\Psi_f, \Psi_g) = i\, S(f^\dagger, g).$$

*b) $(\mathfrak{S}(M), \iota_M)$ is equivalent to the inner product space $(\mathscr{D}(DM)/\mathrm{Ker}\, S, \iota'_M)$, where $\mathscr{D}(DM)/\mathrm{Ker}\, S$ is the quotient space obtained from $\mathscr{D}(DM)$ by dividing out the kernel of $S$, and, given two equivalence classes $[f]$, $[g]$ in $\mathfrak{S}(M)/\mathrm{Ker}\, S$, the inner product $\iota'_M$ is defined by*

$$\iota'_M([f], [g]) \doteq i\, S(f^\dagger, g).$$

*Proof.* b) follows from a) and the fact that the map from $\mathscr{D}(DM)/\mathrm{Ker}\, S$ to $\mathfrak{S}(M)$ induced by $S : \mathscr{D}(DM) \to \mathfrak{S}(M)$ is bijective by construction. To prove a), let us note that Stokes' theorem implies

$$\int_V d_g x\, (D^* v)u - v Du = \int_V d_g x\, (\slashed{\nabla} v)u + v\slashed{\nabla} u = \int_{\partial V} d(\partial V)\, v \slashed{N} u,$$

where $v \in \mathscr{E}(D^* M)$, $u \in \mathscr{E}(DM)$ are smooth sections with support properties ensuring that all integrals converge, $V$ is a submanifold of $M$ with smooth boundary $\partial V$, $\tilde{N}$ is the outward pointing unit normal vector field on $\partial V$ and $d(\partial V)$ is the volume measure on $\partial V$ induced by $d_g x$. We apply the above identity to the two cases

$$V = \Sigma^+ \doteq J^+(\Sigma, M) \setminus \Sigma, \quad \partial V = \Sigma, \quad v = (S^- f)^\dagger, \quad u = \Psi_g, \quad \tilde{N} = -N$$

and $\quad V = \Sigma^- \doteq J^-(\Sigma, M) \setminus \Sigma, \quad \partial V = \Sigma, \quad v = (S^+ f)^\dagger, \quad u = \Psi_g, \quad \widetilde{N} = N,$

where $\Sigma$ is an arbitrary Cauchy surface of $M$ and $N$ is its forward pointing unit normal vector field. Moreover, all integrals we are interested in converge and viewing $\Sigma$ as the only relevant boundary of $\Sigma^\pm$ is well-defined because $V \cap \operatorname{supp} v$ is a compact set in both cases due to the causal support properties of $S^\pm f$. Taking into account (II.38), we can therefore compute

$$
\begin{aligned}
i S(f^\dagger, g) = i \int_M d_g x \, f^\dagger S g &= i \int_{\Sigma^+} d_g x \, D^*(S^- f)^\dagger \Psi_g + i \int_{\Sigma^-} d_g x \, D^*(S^+ f)^\dagger \Psi_g \\
&= i \int_\Sigma d\Sigma \, (S^+ f)^\dagger \slashed{N} \Psi_g - i \int_\Sigma d\Sigma \, (S^- f)^\dagger \slashed{N} \Psi_g \\
&= -i \int_\Sigma d\Sigma \, (S f)^\dagger \slashed{N} \Psi_g = \iota_M(\Psi_f, \Psi_g).
\end{aligned}
$$

$\square$

As it holds for every inner product space, we can complete $\mathfrak{G}(M)$ with respect to the norm induced by $\iota_M$ and extend $\iota_M$ to the completion of $\mathfrak{G}(M)$ by continuity to obtain a Hilbert space $(\overline{\mathfrak{G}(M)}, \iota_M)$. The inner product space $(\mathfrak{G}(M), \iota_M)$ will be sufficient for all constructions we shall perform, but we will comment on possible applications of $(\overline{\mathfrak{G}(M)}, \iota_M)$ at the later stages of this work.

In Minkowski spacetime, $\mathfrak{P} = P \mathbb{I}_4$ and the causal propagators of the Dirac and the Klein-Gordon equation are therefore related as $S = D^* \Delta$. Moreover, we know from lemma II.1.2.4 that (the normal derivative of) $\Delta$ restricted to a Cauchy surface $\Sigma$ can be given in terms of the $\delta$-distribution on $\Sigma$. On account of $S = D^* \Delta$, similar 'initial conditions' hold for $S$ and one might wonder if this is also true in general curved spacetimes. The following result shows that this is indeed the case. In preparation for its statement, let us note that, if $u$ is an arbitrary compactly supported distribution in $\mathscr{E}'(DM)$ and $g$ is a test section in $\mathscr{D}(D^*M)$, we can define the action of $S$ on $u$ by

$$
\langle g, S u \rangle \doteq -\langle S_* g, u \rangle
$$

which entails that we can extend $S$ from $\mathscr{D}(DM)$ to $\mathscr{E}'(DM)$. Moreover, we can view every test section in $\mathscr{D}(D\Sigma)$ on a Cauchy surface $\Sigma$ as a compactly supported distribution in $\mathscr{E}'(DM)$ and therefore define $S$ on $\mathscr{D}(D\Sigma)$. This allows to prove the following statement [Dim82, prop 2.4.(c)].

**Lemma** II.3.2.4 *Let $\Sigma$ be an arbitrary Cauchy surface of $M$ and $N$ its future pointing unit normal vectorfield. For all $f \in \mathscr{D}(D\Sigma)$ it holds*

$$
S f \restriction_\Sigma = \slashed{N} f.
$$

*On the level of distribution kernels, this entails that*

$$
S(x, y) \restriction_{\Sigma \times \Sigma} = \slashed{N} \delta_\Sigma(x, y),
$$

*where $\delta_\Sigma$ is the $\delta$-distribution with respect to the canonical volume measure on $\Sigma$.*

As already anticipated, we will later employ $S(x, y)$ to postulate covariant anticommutation relations for the quantized Dirac field. The above result entails that this is equivalent to the often postulated equal-time anticommutation relations.

### II.3.3  The Algebra of Classical Observables

In analogy to the treatment of the Klein-Gordon field, we will now provide the algebra of classical observables of the Dirac field as a preparation for the deformation quantization we wish to perform at the later stages of this work. Recall that, for our purposes, the strength of the deformation quantization approach lies in the efficient formulation of the combinatorics of normal ordering. While the algebra of classical observables of the scalar field has already been given by Brunetti, Dütsch, and Fredenhagen in [BDF09] and in preceding works of the mentioned authors, the formulation of the same algebra for the case of Dirac fields is new. As in the scalar case, we will only discuss polynomial observables. For a treatment of the subject which includes non-polynomial observables and is closer to the spirit of [BDF09], we refer the reader to the recent work [Rej10].

To formulate the wanted algebra of classical observables, it is advantageous to combine a Dirac spinor field and a Dirac cospinor field into a single object (with eight components), a path already pursued by previous works on Dirac fields, *e.g.* [San08]. To wit, we consider the *double spinor bundle* $D^\oplus M$, *i.e.* the direct sum $D^\oplus M \doteq DM \oplus D^*M$ of the Dirac spinor and cospinor bundle, and compactly supported smooth sections $\mathscr{D}(D^\oplus M)$ on this bundle. $\mathscr{D}(D^\oplus M)$ is endowed with the usual locally convex topology and naturally contains $\mathscr{D}(DM)$ and $\mathscr{D}(D^*M)$. Namely, arbitrary $f \in \mathscr{D}(DM)$, $g \in \mathscr{D}(D^*M)$ can be promoted to elements $\widetilde{f}$, $\widetilde{g}$ of $\mathscr{D}(D^\oplus M)$ by setting

$$\widetilde{f} \doteq f \oplus 0, \qquad \widetilde{g} \doteq 0 \oplus g\,.$$

While the algebra of classical observables of the scalar field is an Abelian algebra, observables of the classical Dirac field should constitute an algebra of anticommuting objects. This anticommutativity of classical Dirac fields is often encoded by demanding that the classical Dirac field has values in the *Grassmann numbers*, see for instance [IZ80], but we shall achieve it in simpler terms by employing antisymmetric tensor products. Note that such antisymmetry must take into account both spinor indices and spacetime dependencies. To specify this explicitly, let us define a frame $\mathfrak{E}^\Theta$ of the double spinor bundle as

$$\mathfrak{E}^\Theta \doteq \begin{cases} E_A \oplus 0 & \text{if } \Theta = A \text{ and} \\[2mm] 0 \oplus E^B & \text{if } \Theta = 4 + B, \end{cases} \tag{II.40}$$

where $E_A$ and $E^B$ are frames of $DM$ and $D^*M$ respectively, see definition I.2.1.6. We can decompose every test section $f^{(n)} \in \mathscr{D}(D^\oplus M^n)$ as

$$f^{(n)}(x_1, \cdots, x_n) = f^{(n)}_{\Theta_1 \cdots \Theta_n}(x_1, \cdots, x_n) \, \mathfrak{E}^{\Theta_1}(x_1) \otimes \cdots \otimes \mathfrak{E}^{\Theta_n}(x_n),$$

where the component functions $f^{(n)}_{\Theta_1 \cdots \Theta_n}(x_1, \cdots, x_n)$ are elements of $\mathscr{D}(M^n)$, and we define the antisymmetric subspace $\mathscr{D}^a(D^\oplus M^n)$ of $\mathscr{D}(D^\oplus M^n)$ by requiring that every element $f^{(n)}$ of $\mathscr{D}^a(D^\oplus M^n)$ fulfils

$$f^{(n)}_{\Theta_1 \cdots \Theta_k \Theta_{k+1} \cdots \Theta_n}(x_1, \cdots, x_k, x_{k+1}, \cdots, x_n) = -f^{(n)}_{\Theta_1 \cdots \Theta_{k+1} \Theta_k \cdots \Theta_n}(x_1, \cdots, x_{k+1}, x_k, \cdots, x_n)$$

for all $k \in \{1, \cdots, n-1\}$.

Let us now define the general algebra of classical Dirac fields, and discuss its observable elements afterwards. As in the scalar case, we shall first formulate a version which does not contain any dynamical information and then encode the dynamics by dividing out a suitable ideal, a procedure incorporating the relation between test sections and solutions stemming from the properties of the causal propagator $S$.

**Definition II.3.3.1** *By $(\mathscr{C}_0(DM), \cdot_a)$ we denote the **off-shell algebra of the classical Dirac field**. Here, $\mathscr{C}_0(DM)$ is the vector space*

$$\mathscr{C}_0(DM) \doteq \bigoplus_{n=0}^{\infty} \mathscr{D}^a(D^\oplus M^n),$$

*with $\mathscr{D}^a(D^\oplus M^n)$ denoting the **antisymmetric test sections** in n variables, $\mathscr{D}^a(D^\oplus M^0) \doteq \mathbb{C}$, and elements of $\mathscr{C}_0(DM)$ have to be understood as **finite sums** $f = \oplus_n \frac{1}{n!} f^{(n)}$ with $f^{(n)} \in \mathscr{D}^a(D^\oplus M^n)$. We equip $\mathscr{C}_0(DM)$ with the **antisymmetric product** $\cdot_a$ defined as*

$$\frac{1}{n!}(f \cdot_a g)^{(n)} \doteq \sum_{k+l=n} \mathfrak{As}\left(\frac{1}{k!} f^{(k)} \otimes \frac{1}{l!} g^{(l)}\right),$$

*where $\mathfrak{As} : \mathscr{D}(D^\oplus M^n) \to \mathscr{D}^a(D^\oplus M^n)$ is the **total antisymmetrisation projector**, i.e. $\mathfrak{As}$ is surjective and $\mathfrak{As}^2 = \mathfrak{As}$. Moreover, we equip $\mathscr{C}_0(DM)$ with the following topology: a sequence $\{f_k\}_k = \{\oplus_n f^{(n)}_k\}_k$ in $\mathscr{C}_0(DM)$ is said to converge to $f = \oplus_n f^{(n)}$ if $f^{(n)}_k$ converges to $f^{(n)}$ for all n and there exists an N such that $f^{(n)}_k = 0$ for all $n > N$ and all k.*

*We define the **on-shell algebra of the classical Dirac field** $(\mathscr{C}(DM), \cdot_s)$ by replacing $(\mathscr{C}_0(DM), \cdot_s)$ with the quotient space $\mathscr{C}_0(DM)/\mathscr{I}_D$ endowed with the induced product and topology. Here, $\mathscr{I}_D$ is the closed ideal generated by elements $D^\oplus f$, $f \in \mathscr{D}(D^\oplus M)$ and $D^\oplus \doteq D \oplus D^*$ is the **double Dirac operator**.*

As already briefly remarked on several occasions, observables of the Dirac fields should be *gauge-invariant*, i.e. invariant under spinor transformations. To state this in precise terms, let us define a frame-dependent 'action' of $Spin_0(3,1)$ on $\mathscr{D}(D^\oplus M^n)$. Namely, let $\widetilde{\Lambda} \in Spin_0(3,1)$ and $f^{(n)} \in \mathscr{D}(D^{\oplus}M^n)$ be arbitrary and let $\mathfrak{E}^{\Theta}$ be a frame of $D^{\oplus}M$. We define the pointwise action $A_{\mathfrak{E}}(\widetilde{\Lambda})$ of $\widetilde{\Lambda}$ on the level of components with respect to $\mathfrak{E}^{\Theta}$ by setting

$$\left[ A_{\mathfrak{E}}(\widetilde{\Lambda}) f^{(n)} \right]_{\Theta_1 \cdots \Theta_n} (x_1, \cdots, x_n) = \mathfrak{L}(\widetilde{\Lambda})_{\Theta_1}^{\Theta_1'} \cdots \mathfrak{L}(\widetilde{\Lambda})_{\Theta_n}^{\Theta_n'} f^{(n)}_{\Theta_1' \cdots \Theta_n'} (x_1, \cdots, x_n),$$

where, in accord with (I.5),

$$\mathfrak{L}(\widetilde{\Lambda})_{\Theta}^{\Theta'} \doteq \begin{cases} \left( \widetilde{\Lambda}^{-1} \right)_{A}^{B} & \text{if } (\Theta', \Theta) = (A, B) \\ \\ \widetilde{\Lambda}^{A}_{B} & \text{if } (\Theta', \Theta) = (4+A, 4+B) \\ \\ 0 & \text{otherwise} \end{cases},$$

and the representation $\pi_{\oplus}$ of $Spin_0(3,1)$ is implicit. The above action depends on the chosen frame, see the footnote on [San08, p. 74]. However, a statement like $A_{\mathfrak{E}}(\widetilde{\Lambda}) f^{(n)} = f^{(n)} \; \forall \widetilde{\Lambda} \in Spin_0(3,1)$ is independent of $\mathfrak{E}^{\Theta}$ as any two frames $\mathfrak{E}^{\Theta}$, $\mathfrak{E}'^{\Theta}$ are related by a local $Spin_0(3,1)$ transformation, *i.e.* at every point $x \in M$ there is a $\widetilde{\Lambda} \in Spin_0(3,1)$ such that

$$\mathfrak{E}'^{\Theta'}(x) = \mathfrak{L}(\widetilde{\Lambda})_{\Theta}^{\Theta'} \mathfrak{E}^{\Theta}(x).$$

If $M$ is not simply connected and $\mathfrak{E}$ therefore not a global frame, we interpret $\mathfrak{E}$ as a collection of arbitrary frames supported in local patches covering $M$. Let us therefore extend $A_{\mathfrak{E}}(\widetilde{\Lambda})$ by linearity from $\mathscr{D}(D^{\oplus}M^n)$ to $\mathscr{C}_0(DM)$ and, noting that invariance under $A_{\mathfrak{E}}(\widetilde{\Lambda})$ is compatible with tensor products and therefore also with $\cdot_a$, let us state the following definition.

**Definition** II.3.3.2 *By* $(\mathscr{C}_0^{obs}(DM), \cdot_a)$ *we denote the* **off-shell algebra of observables of the classical Dirac field**. *Here,* $\mathscr{C}_0^{obs}(DM)$ *is defined as*

$$\mathscr{C}_0^{obs}(DM) \doteq \left\{ f \in \mathscr{C}_0(DM) \,|\, A_{\mathfrak{E}}(\widetilde{\Lambda}) f = f \; \forall \widetilde{\Lambda} \in Spin_0(3,1) \right\}.$$

*The* **on-shell algebra of observable of classical Dirac field** $(\mathscr{C}^{obs}(DM), \cdot_a)$ *is then defined by replacing* $(\mathscr{C}_0^{obs}(DM), \cdot_a)$ *with the quotient space* $\mathscr{C}_0^{obs}(DM) / \mathscr{I}_D'$ *endowed with the induced product and topology. Here,* $\mathscr{I}_D'$ *is the closed ideal* $\mathscr{I}_D \cap \mathscr{C}_0^{obs}(DM)$.

As $A_{\mathfrak{E}}(\widetilde{\Lambda})$ essentially transforms spinors by means of the left action of $\widetilde{\Lambda}^{-1}$ and cospinors by virtue of the right action of $\widetilde{\Lambda}$, the above definition implicates the well-known postulate that observables of the Dirac field should be traces of an equal number of spinor and cospinor fields.

## II.4 The Free Dirac Field in Null Big Bang Spacetimes

In this section, we shall treat classical Dirac fields in flat FLRW spacetimes in general and NBB spacetimes in particular. We will discuss how the general structures that we have already analysed in the last section can be re-casted in more concrete terms and we shall introduce new constructions that are only possible on NBB spacetimes. Barring a slight increase in computational work due to the vector-valued nature of the Dirac field, we will be able to repeat all steps taken in the analysis of the Klein-Gordon field in NBB spacetimes.

### II.4.1 The Mode Expansion of Solutions and the Causal Propagator

Let us start the treatment of the Dirac field on flat FLRW spacetimes by analysing the transformation properties of solutions of the Dirac equation with respect to conformal transformations. As already discussed in detail in subsection I.3.3, a conformal transformation $g \mapsto \Omega^2 g$ transforms a Lorentz frame as $e_a \mapsto \Omega^{-1} e_a$, but leaves the related spin frame $E_A$ and matrices $\gamma^a$ invariant. Based on these facts, we can prove the following important result.

***Lemma*** **II.4.1.1**    *Let $\Omega : M \to (0, \infty)$ be the conformal factor related to a conformal transformation $g \mapsto \widetilde{g} \doteq \Omega^2 g$ and let $\widetilde{\nabla}$ denote the covariant derivative associated to $\widetilde{g}$. Moreover, let $e_a$ be a global Lorentz frame of $(M, g)$, let $\widetilde{e}_a = \Omega^{-1} e_a$, and let*

$$D = -\slashed{\nabla} + m = -\gamma^a \nabla_{e_a} + m\,, \qquad \widetilde{D} \doteq -\gamma^a \widetilde{\nabla}_{\widetilde{e}_a} + \frac{1}{\Omega} m\,.$$

*a)* *The two Dirac operators $D$ and $\widetilde{D}$ are related via*

$$\widetilde{D} = \Omega^{-\frac{5}{2}} D \Omega^{\frac{3}{2}}\,.$$

*b)* *Let $\widetilde{S}^\pm$ and $\widetilde{S}$ denote the retarded and advanced fundamental solutions and the causal propagator associated to $\widetilde{D}$ respectively. The aforementioned propagators are related to the ones proper to $D$ by*

$$\widetilde{S}^\pm = \Omega^{-\frac{3}{2}} S^\pm \Omega^{\frac{5}{2}}\,, \qquad \widetilde{S} = \Omega^{-\frac{3}{2}} S \Omega^{\frac{5}{2}}\,.$$

*On the level of distribution kernels, this entails that*

$$\widetilde{S}^\pm(x, y) = \Omega^{-\frac{3}{2}}(x) S^\pm(x, y) \Omega^{\frac{5}{2}}(y)\,, \qquad \widetilde{S}(x, y) = \Omega^{-\frac{3}{2}}(x) S(x, y) \Omega^{\frac{5}{2}}(y)\,.$$

*c)* *Let $\mathfrak{S}(M)$ and $\widetilde{\mathfrak{S}}(M)$ denote the spaces of smooth solutions with compactly supported initial conditions associated to $D$ and $\widetilde{D}$ respectively. Every $u \in \mathfrak{S}(M)$ induces an element $\widetilde{u}$ of $\widetilde{\mathfrak{S}}(M)$ via*

$$\widetilde{u} \doteq \Omega^{-\frac{3}{2}} u$$

*and every element of $\widetilde{\mathfrak{S}}(M)$ is of this form. Moreover, if $f \in \mathscr{D}(DM)$ is such that $u = Sf$, then*

$$\widetilde{u} = \widetilde{S} \Omega^{-\frac{5}{2}} f\,.$$

*Proof.* To prove a), we employ the results on the behaviour of the spin connection coefficients $\sigma_a$ under conformal transformations obtained in lemma I.3.3.1 to compute

$$\gamma^a \widetilde{\nabla}_{\tilde{e}_a} \Omega^{-\frac{3}{2}} = \frac{\gamma^a}{\Omega}\left(\partial_a + \sigma_a + \frac{3\nabla_a\Omega}{2\Omega}\right)\Omega^{-\frac{3}{2}} = \Omega^{-\frac{5}{2}}\gamma^a\left(\partial_a + \sigma_a\right) = \Omega^{-\frac{5}{2}}\slashed{\nabla},$$

whence $\widetilde{D} = \Omega^{-\frac{5}{2}}D\Omega^{\frac{3}{2}}$. b) and c) then follow by computation and the properties of (fundamental) solutions in direct analogy to the proof of lemma II.2.1.1. □

To obtain the mode expansion of the Dirac field in flat FLRW spacetimes $(M, g)$, we recall that $g = a^2\eta$ is conformally related to the Minkowski metric $\eta$ and apply the above result to the case $\widetilde{g} = \eta$, $\Omega = a^{-1}$. If we choose $e_a$ as the canonical frame of a flat FLRW spacetime given by $a^{-1}$ times the partial derivatives with respect to the conformal time $\tau$ and the comoving spatial coordinates $x_i$, the result is

$$D = a^{-\frac{5}{2}}\left[-\gamma^0\partial_\tau - \gamma^i\partial_i + am\mathbb{I}_4\right]a^{\frac{3}{2}},$$

where $\partial_i \doteq \partial_{x_i}$. In Minkowski spacetime, one writes down mode expansions of the Dirac field by making use of the fact that $DD^*$ equals the Klein-Gordon operator times the four-dimensional identity matrix, which entails in particular that $DD^*$ is a *diagonal* operator in this case. Unfortunately, $\slashed{\nabla} = DD^*$ is *not* a diagonal operator in flat FLRW spacetimes. This is related to the fact that $\slashed{\nabla}$ does *not* transform in a sensible way under conformal transformations and is therefore not conformally related to a Minkowskian Klein-Gordon operator. One can diagonalise $\slashed{\nabla}$ by 'brute force', but the result is a second order partial differential operator in which spatial derivatives appear both as $\partial_i\partial^i$ and as $(-\partial_i\partial^i)^{1/2}$. Although terms like $(-\partial_i\partial^i)^{1/2}$ are in principle tractable (they correspond to a multiplication by $k$ in momentum space), we would like to avoid them and to choose a diagonalisation procedure which yields a more simple partial differential operator. Fortunately, such a procedure has been given in [BaDu87] and we shall use in the following, though we will be able to formulate it in more concise terms than the authors of [BaDu87] by not going to momentum space until after the diagonalisation procedure.

To start, let us introduce the modified Dirac operator $\mathfrak{D} \doteq \gamma^0 D$. Recalling the representation of the $\gamma_a$-matrices chosen in theorem I.2.2.2 and $\gamma^0 = -\gamma_0$, $\gamma^i = \gamma_i$, we can compute

$$\mathfrak{D} = \gamma^0 D = a^{-\frac{5}{2}}\begin{pmatrix} (\partial_\tau - iam)\mathbb{I}_2 & \vec{\sigma}\vec{\nabla} \\ \vec{\sigma}\vec{\nabla} & (\partial_\tau + iam)\mathbb{I}_2 \end{pmatrix}a^{\frac{3}{2}},$$

where $\vec{\sigma}\vec{\nabla} = \sigma_i\partial^i$. The equations $D\psi = 0$ and $\mathfrak{D}\psi = 0$ are equivalent, as $\gamma^0$ is invertible. If we define $\mathfrak{D}^* \doteq -aD^*\gamma^0$ (the factor $a$ counteracts the bad behaviour of $\slashed{\nabla}$ under conformal transformations) and recall that $D^*$ differs from $D$ only by the sign in front of $\slashed{\nabla}$, then a computation yields

$$\mathfrak{D}\mathfrak{D}^* = a^{-\frac{5}{2}}\begin{pmatrix} \left[(\partial_\tau - iam)(\partial_\tau + iam) - \vec{\nabla}^2\right]\mathbb{I}_2 & 0 \\ 0 & \left[(\partial_\tau + iam)(\partial_\tau - iam) - \vec{\nabla}^2\right]\mathbb{I}_2 \end{pmatrix}a^{\frac{3}{2}}.$$

The just described procedure is, in the case of Minkowski spacetime, equivalent to 'diagonalising' $DD^*$, as $\mathfrak{D}\mathfrak{D}^* = -\gamma^0 DD^*\gamma^0 = -(\gamma^0)^2 DD^* = DD^*$ in that case.

The outcome of the above calculation is that we can write every solution $\psi$ of $D\psi = 0$, as $\psi = \mathfrak{D}^*\widetilde{\psi}$, with $\widetilde{\psi}$ being a solution of the diagonal partial differential equation $\mathfrak{D}\mathfrak{D}^*\widetilde{\psi} = 0$. This leads us to define a mode basis for solutions of the Dirac equations as

$$\psi_{\vec{k},l}(\tau,\vec{x}) \doteq \frac{\hat{\mathfrak{D}}^* u_{k,l}(\tau)e^{i\vec{k}\vec{x}}}{(2\pi a(\tau))^{\frac{3}{2}}}, \tag{II.41}$$

where

$$\hat{\mathfrak{D}}^* \doteq \begin{pmatrix} (\partial_\tau + ia(\tau)m)\mathbb{I}_2 & -i\vec{\sigma}\vec{k} \\ -i\vec{\sigma}\vec{k} & (\partial_\tau - ia(\tau)m)\mathbb{I}_2 \end{pmatrix},$$

and

$$u_{k,1} \doteq \begin{pmatrix} \mathfrak{T}_k \\ 0 \\ 0 \\ 0 \end{pmatrix}, \quad u_{k,2} \doteq \begin{pmatrix} 0 \\ \mathfrak{T}_k \\ 0 \\ 0 \end{pmatrix}, \quad u_{k,3} \doteq \begin{pmatrix} 0 \\ 0 \\ \overline{\mathfrak{T}_k} \\ 0 \end{pmatrix}, \quad u_{k,4} \doteq \begin{pmatrix} 0 \\ 0 \\ 0 \\ \overline{\mathfrak{T}_k} \end{pmatrix}.$$

Here, $\mathfrak{T}_k$ constitutes an arbitrary solution of

$$\mathscr{P}\mathfrak{T}_k \doteq (\partial_\tau^2 + k^2 + a^2m^2 + ia'm)\mathfrak{T}_k = 0, \tag{II.42}$$

which we can choose to depend only on $k$ and not on $\vec{k}$. To obtain a normalisation condition for the modes $\mathfrak{T}_k$, we require

$$|(\partial_\tau + iam)\mathfrak{T}_k|^2 + k^2|\mathfrak{T}_k|^2 \equiv 1 \tag{II.43}$$

The left hand side of (II.43) is real and one can compute that is independent of $\tau$ for any solution $\mathfrak{T}_k$ of $\mathscr{P}\mathfrak{T}_k = 0$. The required normalisation condition is therefore well-defined, but one might wonder why we have chosen exactly this right hand side of (II.43). At the same time, it seems natural to ask if an arbitrary $\mathfrak{T}_k$ is enough to span all mode solutions of the Dirac equation, and if it is not necessary to consider two linearly independent solutions of $\mathscr{P}\mathfrak{T}_k = 0$. Indeed, while the latter question can be answered in the affirmative for solutions of $\mathfrak{D}\mathfrak{D}^*\psi = 0$, solutions of $D\psi = 0$ are only a subclass of these. Consequently, it seems reasonable that they have fewer independent degrees of freedom, as the following lemma shows.

**Lemma II.4.1.2** *Let us consider the modes $\psi_{\vec{k},l}(\tau,\vec{x})$ defined as in (II.41) and constructed out of an arbitrary $\mathfrak{T}_k$ fulfilling the normalisation condition (II.43). A straightforward computation shows that they are*

a) *orthonormal, i.e.*

$$\int\limits_{\mathbb{R}^3} d\vec{x}\, a^3(\tau)\, \psi_{\vec{k},l}^*(\tau,\vec{x})\psi_{\vec{p},r}(\tau,\vec{x}) = \delta(\vec{k}-\vec{p})\,\delta_{l,r}$$

*b)  and complete, viz.*

$$\int_{\mathbb{R}^3} d\vec{k}\, a^3(\tau) \sum_{l=1}^{4} \psi_{\vec{k},l}(\tau,\vec{x}) \psi_{\vec{k},l}^{*}(\tau,\vec{y}) = \delta(\vec{x}-\vec{y})\,\mathbb{I}_4 \,.$$

Based on this result and provided that $\mathfrak{T}_k$ fulfil certain regularity conditions as functions of $k$, we can state the following mode expansion of arbitrary solutions of the Dirac equation with compactly supported initial conditions.

**Lemma II.4.1.3**  *Let us assume that $\mathfrak{T}_k(\tau)$ and $\partial_\tau \mathfrak{T}_k(\tau)$ are continuous functions of $k$ for $k > 0$ which are square-integrable with respect to the measure $k^2 dk$ in any compact neighbourhood of $k = 0$, i.e. $\mathfrak{T}_k(\tau) \in L^2([0,k_0], k^2 dk)$ for all $k_0 < \infty$ and all $\tau$. Moreover, let us assume that $\mathfrak{T}_k(\tau)$ and $\partial_\tau \mathfrak{T}_k(\tau)$ are polynomially bounded in $k$ for large $k$ and all $\tau$. Then, we can express every $\psi \in \mathfrak{S}(M)$ as*

$$\psi(\tau,\vec{x}) = \int_{\mathbb{R}^3} d\vec{k} \sum_{l=1}^{4} \widetilde{\psi}_l(\vec{k})\, \psi_{\vec{k},l}(\tau,\vec{x}),$$

*where the coefficient functions $\widetilde{\psi}_l(\vec{k})$ can be obtained as*

$$\widetilde{\psi}_l(\vec{k}) = -i \int_{\mathbb{R}^3} d\vec{x}\, a^3(\tau)\, \psi_{\vec{k},l}^{\dagger}(\tau,\vec{x})\, \gamma_0\, \psi(\tau,\vec{x}).$$

*Proof.* The integral defining $\widetilde{\psi}_l(\vec{k})$ converges because elements of $\mathfrak{S}(M)$ have compact support on any Cauchy surface. Moreover, it is independent of $\tau$ as one can see by direct computation or by realising that $\widetilde{\psi}_l(\vec{k})$ can be expressed via the inner product $\iota_M$ as $\widetilde{\psi}_l(\vec{k}) = \iota_M(\psi_{\vec{k},l}, \psi)$, even though $\psi_{\vec{k},l} \notin \mathfrak{S}(M)$; the latter statement follows from

$$d\Sigma N = \frac{1}{6} \sqrt{|\det g|}\, g^{\alpha\beta} \varepsilon_{\alpha\mu\nu\rho} dx^\mu \wedge dx^\nu \wedge dx^\rho \gamma_\beta = d\vec{x}\, a^3(\tau) \gamma_0,$$

which holds on any Cauchy surface $\Sigma$ with forward pointing unit normal vector field $N$ individuated by a surface of constant $\tau$.

Knowing that $\widetilde{\psi}_l(\vec{k})$ is well-defined, we can insert its integral expression into the postulated expansion for $\psi(\tau,\vec{x})$ to see if the appearing integral converges and equals $\psi(\tau,\vec{x})$. In fact, on account of the explicit form of $\psi_{\vec{k},l}$, $\widetilde{\psi}_l(\vec{k})$ is a Fourier transform of a smooth function with compact support and therefore smooth and rapidly decreasing in $\vec{k}$. In analogy to the proof of lemma II.2.1.2, the required $k$-regularity properties of $\mathfrak{T}_k(\tau)$ and $\partial_\tau \mathfrak{T}_k(\tau)$ ensure that each of the four components of

$$\int_{\mathbb{R}^3} d\vec{k} \sum_{l=1}^{4} \widetilde{\psi}_l(\vec{k})\, \psi_{\vec{k},l}(\tau,\vec{x}) \tag{II.44}$$

is the well-defined Fourier-Plancherel transform of a square-integrable function. In fact, the following steps will imply that it is even the Fourier transform of a smooth and rapidly decreasing function. Namely, inserting the integral expression of $\widetilde{\psi}_l(\vec{k})$ yields

$$\int\limits_{\mathbb{R}^3} d\vec{k} \sum_{l=1}^{4} \widetilde{\psi}_l(\vec{k})\, \psi_{\vec{k},l}(\tau,\vec{x}) = -i \int\limits_{\mathbb{R}^3 \times \mathbb{R}^3} d\vec{y} d\vec{k}\, a^3(\tau) \sum_{l=1}^{4} \psi_{\vec{k},l}(\tau,\vec{x})\, \psi_{\vec{k},l}^{\dagger}(\tau,\vec{y})\, \gamma_0\, \psi(\tau,\vec{y}).$$

An application of the completeness relation proved in lemma II.4.1.2 closes the proof. $\qquad\square$

In order to relate the just obtained mode decomposition of an arbitrary element $\psi$ of $\mathfrak{S}(M)$ to its expression $\psi = Sf$ via a suitable $f \in \mathscr{D}(DM)$, we have to know the mode expansion of the causal propagator $S$. As in the Klein-Gordon case, the best we can do is to make an educated guess of the expansion first and to show that it fulfils all properties uniquely determining $S$ afterwards.

***Lemma*** **II.4.1.4**

a) *Let us assume that $\mathfrak{T}_k(\tau)$ and $\partial_\tau \mathfrak{T}_k(\tau)$ are continuous functions of $k$ for $k > 0$, fulfil $\mathfrak{T}_k(\tau) = O(k^{-1})$ and $\partial_\tau \mathfrak{T}_k(\tau) = O(k^0)$ for small $k$, and are polynomially bounded for large $k$, where all these regularity conditions are assumed to hold for all $\tau$. Then, the distribution kernel of the causal propagator $S$ can be expressed as*

$$S(\tau_x,\vec{x},\tau_y,\vec{y}) = i \int\limits_{\mathbb{R}^3} d\vec{k} \sum_{l=1}^{4} \psi_{\vec{k},l}\left(\tau_x,\vec{x}\right) \psi_{\vec{k},l}^{\dagger}\left(\tau_y,\vec{y}\right),$$

*where the integral has to be understood as the limit obtained by multiplication of the integrand with $e^{-\varepsilon k}$, $\varepsilon > 0$ and then taking $\varepsilon \to 0$ after smearing the result with a test section in $\mathscr{D}(DM)$.*

b) *Let $f \in \mathscr{D}(DM)$ be a test section related to a $\psi \in \mathfrak{S}(M)$ by $\psi = Sf$. The Fourier coefficients $\widetilde{\psi}_l(\vec{k})$ of $\psi$ defined as in lemma II.4.1.3 fulfil*

$$\widetilde{\psi}_l(\vec{k}) = i\, \psi_{\vec{k},l}^{\dagger}(f) \doteq i \langle \psi_{\vec{k},l}^{\dagger}, f \rangle.$$

*Proof.* b) follows immediately from a), $\psi = Sf$, and the mode decomposition proved in lemma II.4.1.3. To show that a) holds, we define

$$\widetilde{S}(\tau_x,\vec{x},\tau_y,\vec{y}) = i \int\limits_{\mathbb{R}^3} d\vec{k} \sum_{l=1}^{4} \psi_{\vec{k},l}\left(\tau_x,\vec{x}\right) \psi_{\vec{k},l}^{\dagger}\left(\tau_y,\vec{y}\right)$$

$$= \frac{1}{(2\pi)^3} \int\limits_{\mathbb{R}^3} d\vec{k}\, \frac{S_k(\tau_x,\tau_y)}{a^{\frac{3}{2}}(\tau_x) a^{\frac{3}{2}}(\tau_y)}\, e^{i\vec{k}(\vec{x}-\vec{y})},$$

where

$$S_k(\tau_x, \tau_y) = i \sum_{l=1}^{4} \hat{\mathfrak{D}}^* u_{k,l}(\tau_x) \left[\hat{\mathfrak{D}}^* u_{k,l}\right]^\dagger(\tau_y).$$

Note that $S_k(\tau_x, \tau_y)$ depends only on $k$ and not on $\vec{k}$ due to the behaviour of $D^*$ under Lorentz transformations in general and spatial rotations in particular. Let $f$ be an arbitrary test section in $\mathscr{D}(DM)$ and let us consider

$$\left[\widetilde{S}f\right](\tau_x, \vec{x}) = \widetilde{S}(\tau_x, \vec{x}, f) = i \int_{\mathbb{R}^3} d\vec{k} \sum_{l=1}^{4} \psi_{\vec{k},l}(\tau_x, \vec{x}) \, \psi_{\vec{k},l}^\dagger(f).$$

By the assumed $k$-regularity of $\mathfrak{T}_k$ and $\partial_\tau \mathfrak{T}_k$ and arguments similar to the ones invoked in the proof of lemma II.2.1.3, the above integral converges to a smooth solution of the Dirac equation. To show that $\widetilde{S}f = Sf$, let $u$ be an arbitrary element of $\mathfrak{S}(M)$, let $\Sigma$ denote a Cauchy surface of constant $\tau = \tau_y$ with future pointing unit normal vector field $N = a^{-1}\partial_\tau$ and volume measure $d\Sigma$, let $u$ be specified by $u\!\restriction_\Sigma = u_0$, and let us denote the component-wise Fourier transform of a $\mathbb{C}^4$-valued function $f$ on $\mathbb{R}^3$ by $\mathscr{F}[f]$ and its inverse Fourier transform by $\mathscr{F}^{-1}[f]$. With these data, $u$ can be specified via

$$u(\tau, \vec{x}) = \mathscr{F}^{-1}\left[a^3(\tau_y)\left\{\frac{-S_k(\tau_x, \tau_y)\gamma_0}{a^{\frac{3}{2}}(\tau_x)a^{\frac{3}{2}}(\tau_y)}\mathscr{F}[u_0]\right\}\right],$$

as follows from $S_k(\tau_x, \tau_x) \equiv \gamma_0$, which in turn is equivalent to the completeness relation proven in lemma II.4.1.2. Based on this, we recall $d\Sigma \slashed{N} = d\vec{y}\,a^3(\tau_y)\gamma_0$ and compute in analogy to the proof of lemma II.2.1.3

$$\int_\Sigma d\Sigma\, u_0^\dagger \slashed{N}\, \widetilde{S}f = \int_{\mathbb{R}^3} d\vec{y}\,a^3(\tau_y)\left\{u_0^\dagger(\vec{y})\,\gamma_0\,\left[\widetilde{S}f\right](\tau_y, \vec{y})\right\}$$

$$= \int_\mathbb{R} d\tau_x\,a^4(\tau_x)\int_{\mathbb{R}^3} d\vec{k}\,a^3(\tau_y)\left\{\frac{-S_k(\tau_x,\tau_y)\gamma_0}{a^{\frac{3}{2}}(\tau_x)a^{\frac{3}{2}}(\tau_y)}\mathscr{F}[u_0](\vec{k})\right\}^\dagger \mathscr{F}^{-1}[f](\tau_x,\vec{k}) = \int_M d_g x\, u^\dagger f.$$

Finally, let us recall that, as proven in lemma II.3.2.3, $S$ also fulfils

$$\int_\Sigma d\Sigma\, u_0^\dagger \slashed{N}\, Sf = \int_M d_g x\, u^\dagger f.$$

This entails $\widetilde{S}f\!\restriction_\Sigma = Sf\!\restriction_\Sigma$ and, hence, $\widetilde{S} \equiv S$ by the uniqueness of solutions of the Dirac equation with compactly supported initial data. □

To close the treatment of the Dirac field in flat FLRW spacetimes and to prepare the ground for the analysis of Dirac fields in NBB spacetimes, we shall now provide an example for $\mathfrak{T}_k$ fulfilling the $k$-regularity conditions assumed in the above two lemmata. Namely, lemma I.3.2.4 entails that the differential equation (II.42) for $\mathfrak{T}_k$ tends to the one of massless Minkowskian scalar field modes in the limit $\tau \to -\infty$. In addition, the normalisation condition (II.43) becomes

$$|\partial_\tau \mathfrak{T}_k|^2 + k^2|\mathfrak{T}_k|^2 = 1$$

in that limit. It therefore seems meaningful to require that $\mathfrak{T}_k$ tends asymptotically to $(\sqrt{2}k)^{-1}e^{-ik\tau}$. The following result makes this precise.

**Lemma II.4.1.5** *Let a be the scale factor of an NBB spacetime, let $V(\tau) \doteq ia'(\tau)m + a^2(\tau)m^2$, let $T_{0,k}(\tau) \doteq \frac{1}{\sqrt{2k}}e^{-ik\tau}$, and let finally*

$$\Delta_{0,k}(\tau_x, \tau_y) \doteq -i\left(\overline{T_{k,0}(\tau_x)}T_{k,0}(\tau_y) - T_{k,0}(\tau_x)\overline{T_{k,0}(\tau_y)}\right).$$

*There exists a finite $\tau_0 \leq -1$ such that the series*

$$\mathfrak{T}_k(\tau) \doteq \frac{1}{\sqrt{k}}\left(T_{0,k}(\tau) + \sum_{n=1}^{\infty}\mathfrak{T}_{n,k}(\tau)\right)$$

*with*

$$\mathfrak{T}_{n,k}(\tau) \doteq (-1)^n\int_{-\infty}^{\tau}d\tau_1\cdots\int_{-\infty}^{\tau_{n-1}}d\tau_n\,\Delta_{0,k}(\tau,\tau_1)\cdots\Delta_{0,k}(\tau_{n-1},\tau_n)V(\tau_1)\cdots V(\tau_n)T_{0,k}(\tau_n),$$

*converges for all $\tau < \tau_0$ and defines a solution of (II.42) which fulfils*

$$\lim_{\tau \to -\infty} e^{ik\tau}\mathfrak{T}_k(\tau) = \frac{1}{\sqrt{2k}} \qquad \lim_{\tau \to -\infty} e^{ik\tau}\partial_\tau\mathfrak{T}_k(\tau) = -\frac{i}{\sqrt{2}},$$

*and*

$$\mathfrak{T}_k(\tau) = O\left(k^{-1}\right)$$

*for all $k$ and for all $\tau$, whereas*

$$\partial_\tau\mathfrak{T}_k(\tau) = \begin{cases} O\left(k^{-1}\right) \text{ for } k \to 0 \text{ and} \\[2mm] O\left(k^0\right) \text{ for } k \to \infty \end{cases}$$

*uniformly in $\tau$.*

*Proof.* The proof proceeds largely as the one of lemma II.4.1.5, we therefore only mention the essential steps. As $(\partial_{\tau_x}^2 + k^2)S_{0,k}(\tau_x, \tau_y) = 0$ and $\partial_{\tau_x}S_{0,k}(\tau_x, \tau_y)|_{\tau_x = \tau_y} = 1$, the series is a formal solution of the considered differential equation with the correct initial conditions. In order to prove that the series itself converges to a smooth solution, it is sufficient to consider the original series and its first $\tau$-derivatives.

The starting point is

$$\Delta_{0,k}(\tau_x, \tau_y) \leq -\tau_y \qquad \text{for} \qquad 0 \geq \tau_x \geq \tau_y$$

and the fact that lemma I.3.2.4 entails the existence of a finite $\tau_0 \leq -1$ and a constant $C$ such that $|V(\tau)| \leq C\tau^{-3}$ for all $\tau < \tau_0$. Namely, these data entail

$$|\mathfrak{T}_{n,k}(\tau)| \leq \frac{C^n}{n!(-\tau)^n}|T_{0,k}|,$$

and, hence, both the absolute convergence of the series and

$$|\mathfrak{T}_k(\tau)| \leq \frac{1}{\sqrt{2}k}e^{\frac{C}{-\tau}}$$

uniformly in $\tau$ for all $\tau < \tau_0$. Similar considerations lead to

$$|\partial_\tau \mathfrak{T}_{n,k}(\tau)| \leq \frac{C^n}{n!(-\tau)^{n+1}}|T_{0,k}|,$$

which in turn entails the wanted convergence of the series of first $\tau$-derivatives and finally

$$|\partial_\tau \mathfrak{T}_k(\tau)| \leq \frac{1}{\sqrt{2}} + \frac{1}{-\sqrt{2}k\tau}(e^{\frac{C}{-\tau}} - 1)$$

for all $\tau < \tau_0$. $\qquad\qquad\square$

### II.4.2 The Bulk and Boundary Inner Product Spaces of Solutions

In subsection II.2.2, we have seen that it is possible to map the symplectic space of solutions of the Klein-Gordon equation in an NBB spacetime $M_B$ to a suitable counterpart on the past null boundary $\partial^- M_B$ in a way which preserves the symplectic form. The purpose of the present subsection is to show that a similar procedure is possible also for the inner product space $(\mathfrak{S}(M_B), \iota_B)$ of solutions of the Dirac equation on $M_B$. This will pave the way for the construction of preferred ground and thermal states of a quantized Dirac field on $M_B$ by means of a quantum Dirac field theory on $\partial^- M_B$ in the next chapter. The results we present here constitute the first application of the methods developed in [DMP06, DMP09a, DMP09b, DMP09c, Mor06, Mor08] to the case of Dirac fields.

In contrast to the Klein-Gordon case, the analysis of the bulk-to-boundary correspondence for Dirac fields is complicated by the *a priori* necessity to include a frame transformation in the bulk-to-boundary map, see lemma I.3.3.2 and the preceding discussion. Namely, although the existence of a well-defined map from $(\mathfrak{S}(M_B), \iota_B)$ to a suitable boundary space can be proven in analogy to the scalar case and in a straightforward manner, the inclusion of the mentioned frame transformation makes it difficult to interpret the constructions we shall perform on $\partial^- M_B$ in terms of bulk Fourier modes. This problem arises because, although we have been able to compute this transformation in lemma I.3.3.2 on the level of Lorentz frames, it is hard to give an explicit expression for the related spin frame transformation. Hence, we shall proceed in the following way: we first anticipate a suitable boundary inner product space $(\mathsf{S}(\mathfrak{I}^-), \mathsf{i})$ by mapping the bulk inner product $\iota_B$ to the boundary with the help of Stokes' theorem. Then, we show that a combination of a conformal transformation and a frame transformation indeed maps $(\mathfrak{S}(M_B), \iota_B)$ to $(\mathsf{S}(\mathfrak{I}^-), \mathsf{i})$, were we shall use the fact that solutions of the Dirac equation on the Einstein static universe can be trivially restricted to $\partial^- M_B$. Finally, we introduce a second bulk-to-boundary map, which is effectively equivalent to the one already known and, hence, well-defined, but can be given in explicit terms without having recourse to a spin frame transformation. Altogether, the natural but unwieldy bulk-to-boundary map is used to prove the well-posedness of an effective and simple alternative, which shall be the one used in the constructions performed in the next chapter.

***Theorem*** II.4.2.1    *Let*

$$D_B \doteq -\overline{\nabla}^B + m \,,$$

*where $\overline{\nabla}^B$ denotes the contracted covariant derivative associated to $g_B$ and its canonical Lorentz frame $e_a^B$ defined in (I.26), and let $\mathfrak{S}(M_B)$ denote the space of solutions associated to $D_B$ in an NBB spacetime $(M_B, g_B)$. Moreover, let $e_B^a$ denote the frame dual to $e_a^B$, let $\psi_i$, $i = 1, 2$ be arbitrary elements of $\mathfrak{S}(M_B)$, and let*

$$\widetilde{e}_B^a \doteq \frac{\Omega_B}{a} e_B^a \,, \qquad \widetilde{\psi}_i \doteq \frac{a^{\frac{3}{2}}}{\Omega_B^{\frac{3}{2}}} \psi_i \,.$$

*Finally, let us denote by $\iota_B$ the inner product on $\mathfrak{S}(M_B)$, cf. definition II.3.2.1, and let*

$$K(\vec{n}) \doteq \frac{1}{\sqrt{2}} \left( \begin{array}{cc} \mathbb{I}_2 & \vec{n} \cdot \vec{\sigma} \\ \vec{n} \cdot \vec{\sigma} & \mathbb{I}_2 \end{array} \right) \,,$$

*where $\vec{n}$ is an arbitrary vector in $\mathbb{R}^3$ and $\vec{\sigma}$ denotes the vector of Pauli matrices.*

a)  *Let us assume that*

$$\lim_{u \to -\infty} \widetilde{\psi}_1^\dagger \gamma_a \, \widetilde{\psi}_2 \, \widetilde{e}_B^a(\partial_v)$$

*is smooth and integrable with respect to $d \upsilon d\mathbb{S}^2$. Then,*

$$\iota_B(\psi_1, \psi_2) = -i \int\limits_{\mathfrak{I}^-} d\upsilon d\mathbb{S}^2 \lim_{u \to -\infty} \widetilde{\psi}_1^\dagger \gamma_a \, \widetilde{\psi}_2 \, \widetilde{e}_B^a(\partial_\upsilon) \,.$$

*b)* *Let us define a map* $\widetilde{\mathfrak{G}}$ *as*

$$\widetilde{\mathfrak{G}}\psi_i \doteq \lim_{u \to -\infty} \frac{K(\vec{e}_1)\widetilde{\Lambda}^B\,\widetilde{\psi}_i}{\sqrt[4]{1+v^2}},$$

*where* $\vec{e}_1 \doteq (1,0,0)^T$ *and* $\widetilde{\Lambda}^B$ *is the spin frame transformation introduced in lemma I.3.3.2.* $\widetilde{\mathfrak{G}}$ *is a well-defined map from* $\mathfrak{G}(M_B)$ *to*

$$\mathsf{S}(\mathfrak{I}^-) \doteq \mathscr{C}^\infty(\mathfrak{I}^-, \mathbb{C}^4) \cap L^2(\mathfrak{I}^-, dv\,d\mathbb{S}^2)$$

*and*

$$\widetilde{\mathfrak{G}}(\psi_1)^* \widetilde{\mathfrak{G}}(\psi_2) = \lim_{u \to -\infty} -i\,\widetilde{\psi}_1^\dagger \gamma_a\, \widetilde{\psi}_2\, \widetilde{e}_B^a(\partial_v).$$

*c)* *Let* $\mathfrak{G}$ *be defined as*

$$\mathfrak{G}\psi_i \doteq \lim_{u \to -\infty} \frac{K(\vec{e}_r)\widetilde{\psi}_i}{\sqrt[4]{1+u^2}},$$

*where* $\vec{e}_r \doteq \partial_r \vec{x}$, *and let us define an inner product* $\mathfrak{i}$ *on* $\mathsf{S}(\mathfrak{I}^-)$ *by*

$$\mathfrak{i}(u_1, u_2) \doteq \int\limits_{\mathfrak{I}^-} dv\,d\mathbb{S}^2\, u_1^*\, u_2$$

*for arbitrary* $u_1,\ u_2 \in \mathsf{S}(\mathfrak{I}^-)$. *Then,* $\mathfrak{G}$ *is a well-defined map between the inner product spaces* $(\mathfrak{G}(M_B), \iota_B)$ *and* $(\mathsf{S}(\mathfrak{I}^-), \mathfrak{i})$, *particularly,*

$$\iota_B(\psi_1, \psi_2) = \mathfrak{i}(\mathfrak{G}\psi_1, \mathfrak{G}\psi_2).$$

*d)* $\mathfrak{G}$ *fulfils*

$$\mathfrak{G}\psi_i = \lim_{u \to -\infty} \frac{\sqrt{2}\,\widetilde{\psi}_i}{\sqrt[4]{1+u^2}},$$

*particularly,* $\mathfrak{G}$ *is injective.*

*Proof.* a) On account of definition II.3.2.1, the bulk inner product of $\psi_1, \psi_2 \in \mathfrak{G}(M_B)$ reads

$$\iota_B(\psi_1, \psi_2) = -i \int\limits_{\Sigma} d\Sigma\; \psi_1^\dagger \slashed{N} \psi_2,$$

where $\Sigma$ is an arbitrary Cauchy surface of $(M_B, g_B)$ with future pointing unit normal $N$. Recalling that

$$d\Sigma\, \psi_1^\dagger \slashed{N} \psi_2 = \frac{1}{6}\sqrt{|\det g|}\, g^{\alpha\beta}\, \psi_1^\dagger \gamma_\alpha\, \psi_2\, \varepsilon_{\beta\gamma\mu\nu}\, dx^\gamma \wedge dx^\mu \wedge dx^\nu$$

and that conformal transformations $g \mapsto \Omega^2 g$ entail $g^{-1} \mapsto \Omega^{-2} g^{-1}$, $\gamma_\alpha \mapsto \Omega \gamma_\alpha$, $\psi_i \mapsto \Omega^{-3/2} \psi_i$, we see that $\iota_B(\psi_1, \psi_2)$ is invariant under conformal transformations. If we set

$$\widetilde{D} \doteq \frac{a^{\frac{5}{2}}}{\Omega_B^{\frac{5}{2}}} D_B \frac{\Omega_B^{\frac{3}{2}}}{a^{\frac{3}{2}}},$$

and denote the inner product on the solutions $\widetilde{\psi}$ of $\widetilde{D}\widetilde{\psi} = 0$ on $(M_B, \Omega_B^2 a^{-2} g_B)$ by $\widetilde{\iota}_B$, the above considerations entail

$$\iota_B(\psi_1, \psi_2) = \widetilde{\iota}_B \left( \frac{a^{\frac{3}{2}}}{\Omega_B^{\frac{3}{2}}} \psi_1, \frac{a^{\frac{3}{2}}}{\Omega_B^{\frac{3}{2}}} \psi_2 \right). \tag{II.45}$$

We would now like to apply Stokes' theorem to reach the conclusion that the integration of the integrand proper to (II.45) on either $\Sigma$ or $\Im^-$ yields the same result. Assuming that this is possible, and recalling that $\Omega_B^2 a^{-2} g_B$ restricted to $\Im^-$ equals the Bondi metric (I.24), one can straightforward compute that the integrand of (II.45) evaluated on $\Im^-$ equals

$$\lim_{u \to -\infty} -i \widetilde{\psi}_1^\dagger \gamma_a \widetilde{\psi}_2 \, \widetilde{e}_B^\alpha(\partial_v).$$

By our assumptions on the properties of the above expression and the causal support properties of solutions of the Dirac equation, we can indeed apply Stokes' theorem to prove the thesis.

b) Barring the appearance of a frame transformation, the first part of the thesis can be proved in close analogy to the proof of proposition II.2.2.1, and we only sketch the essential arguments. An application of lemma II.4.1.1 to the case $g = g_B$, $\widetilde{g} = g_E \!\restriction_{\chi^B(M_B)} = \Omega_\mathbb{M}^2 a^{-2} g$ and the results of lemma I.3.3.2 b) imply that $\psi_i^E \doteq \widetilde{\Lambda}^B a^{3/2} \Omega_\mathbb{M}^{-3/2} \psi_i$ are smooth solutions of

$$D_E \psi_i^E \doteq \left( -\overline{\nabla}^E + \frac{a}{\Omega_\mathbb{M}} m \right) \psi_i^E = 0$$

with compactly supported initial conditions, where $\overline{\nabla}^E$ denotes the contracted covariant derivative related to the metric $g_E$ and its canonical Lorentz frame $e_a^E$ (cf. (I.25)) and it is understood the the above equation holds only on $\chi^B(M_B) \subset M_E$. Let now $M_E^-$ denote the 'lower, slit half' of $M_E$ defined in the proof of proposition II.2.2.1. The lemmata I.3.2.4 and I.3.3.2 entail that we can smoothly extend $D_E$ from $M_E^- \cap \chi^B(M_B)$ to $M_E^-$ by setting

$$D_E \!\restriction_{M_E^- \setminus \chi^B(M_B)} \equiv -\overline{\nabla}^E.$$

By lemma II.3.2.2, we can find $f_i \in \mathscr{D}(DM_B)$ fulfilling $\psi_i = S_B f_i$ and supp $f_i \subset \chi^B(M_B) \cap M_E^-$, where $S_B$ denotes the causal propagator of $D_B$. Moreover, on account of lemma

II.4.1.1 and lemma I.3.3.2, we know that

$$S_E = \frac{a^{\frac{3}{2}}}{\Omega_{\mathbb{M}}^{\frac{3}{2}}} \widetilde{\Lambda}^B \, S_B \, (\widetilde{\Lambda}^B)^{-1} \frac{\Omega_{\mathbb{M}}^{\frac{5}{2}}}{a^{\frac{5}{2}}}$$

is the causal propagator of $D_E$ on $(\chi^B(M_B), g_E\!\restriction_{\chi^B(M_B)})$ and that $\psi_i^E = S_E \Omega_{\mathbb{M}}^{\frac{5}{2}} a^{-\frac{5}{2}} f_i$. We can extend $S_E$ in a unique way to the causal propagator of $D_E$ on the full $M_E^-$ and therefore smoothly extend $\psi_i^E$ from $M_E^- \cap \chi^B(M_B)$ to $M_E^-$ by inserting the extension of $S_E$ in the expression

$$\psi_i^E = S_E \Omega_{\mathbb{M}}^{\frac{5}{2}} a^{-\frac{5}{2}} f_i .$$

The result can be restricted to $\mathfrak{I}^-$ and we find that

$$\widetilde{\mathfrak{G}} \psi_i \doteq \lim_{u \to -\infty} \frac{K(\vec{e}_1) \widetilde{\Lambda}^B \widetilde{\psi}_i}{\sqrt[4]{1 + v^2}} = \lim_{u \to -\infty} \frac{K(\vec{e}_1) \psi_i^E}{(1 + v^2)^{\frac{7}{4}}}$$

is a smooth, $\mathbb{C}^4$-valued function on $\mathfrak{I}^-$.

Let us now show that the mapping $\widetilde{\mathfrak{G}}$ is such that $\widetilde{\mathfrak{G}} \psi_i$ is not only smooth, but even square-integrable with respect to $dv\,d\mathbb{S}^2$. As $\psi_i^E$ is smooth on $i^-$, $\|\widetilde{\mathfrak{G}} \psi_i\|_{\mathbb{C}^4}^2$ decays as $(1 + v^2)^{7/2}$ for large negative $v$. Furthermore, the causal support properties of $\psi_i^E$ entail that there is a finite $v_0$ such that $\|\widetilde{\mathfrak{G}} \psi_i\|_{\mathbb{C}^4}^2 \equiv 0$ for all $v \geq v_0$. This implies the wished-for square-integrability. The last asserted property of $\widetilde{\mathfrak{G}}$, namely,

$$\widetilde{\mathfrak{G}}(\psi_1)^* \widetilde{\mathfrak{G}}(\psi_2) = \lim_{u \to -\infty} -i \widetilde{\psi}_1^\dagger \gamma_a \widetilde{\psi}_2 \widetilde{e}_B^a (\partial_v) ,$$

follows straightforwardly from

$$\frac{(\widetilde{\Lambda}^B)^* K(\vec{e}_1)^* K(\vec{e}_1) \widetilde{\Lambda}^B}{\sqrt{1 + v^2}} = -i \frac{(\widetilde{\Lambda}^B)^* \beta (\gamma_0 + \gamma_1) \widetilde{\Lambda}^B}{\sqrt{1 + v^2}} = -i \frac{\beta (\widetilde{\Lambda}^B)^{-1} (\gamma_0 + \gamma_1) \widetilde{\Lambda}^B}{\sqrt{1 + v^2}}$$

$$= -i \beta (\widetilde{\Lambda}^B)^{-1} \gamma_a \widetilde{\Lambda}^B e_E^a (\partial_v) = -i \beta \gamma_a \widetilde{e}_E^a (\partial_v) .$$

Here, the identity $(\widetilde{\Lambda}^B)^* \beta = \beta (\widetilde{\Lambda})^{-1}$, which holds for all $\widetilde{\Lambda} \in Spin_0(3, 1)$ and follows from the defining properties of $\beta$ and the relation of $Spin_0(3, 1)$ to $Cl(3, 1)$ has been used.

c) We start by proving $\iota_B(\psi_1, \psi_2) = i(\mathfrak{G} \psi_1, \mathfrak{G} \psi_2)$. The results obtained in a) and b) imply

$$\iota_B(\psi_1, \psi_2) = -i \int_{\mathfrak{I}^-} dv\,d\mathbb{S}^2 \lim_{u \to -\infty} \widetilde{\psi}_1^\dagger \gamma_a \widetilde{\psi}_2 \widetilde{e}_B^a (\partial_v) = \int_{\mathfrak{I}^-} dv\,d\mathbb{S}^2 (\widetilde{\mathfrak{G}} \psi_1)^* \widetilde{\mathfrak{G}} \psi_2 = i(\widetilde{\mathfrak{G}} \psi_1, \widetilde{\mathfrak{G}} \psi_2) .$$

Hence, the compatibility of $\mathfrak{G}$ with the inner product structures is shown if

$$\mathfrak{G}(\psi_1)^*\mathfrak{G}(\psi_2) = \lim_{u \to -\infty} -i\,\widetilde{\psi}_1^{\dagger}\gamma_a\,\widetilde{\psi}_2\,\widetilde{e}_B^a(\partial_v).$$

This, however, follows straightforwardly from

$$\frac{K(\vec{e}_r)^*K(\vec{e}_r)}{\sqrt{1+u^2}} = -i\,\frac{\beta(\gamma_0 + \gamma_i\partial_r x^i)}{\sqrt{1+u^2}} = -i\beta\gamma_a\widetilde{e}_B^a(\partial_v).$$

By comparing $\mathfrak{G}$ with $\widetilde{\mathfrak{G}}$, we have proven that $\mathfrak{G}$ maps solutions of the Dirac equation in the bulk to square-integrable functions on $\mathfrak{I}^-$. However, it is not yet clear if these functions are actually smooth. To see this, we notice that

$$\frac{\widetilde{\psi}_i}{\sqrt[4]{1+u^2}} = \frac{(\widetilde{\Lambda}^B)^{-1}}{\sqrt[4]{1+u^2}}\,\frac{\psi_i^E}{(1+v^2)^{\frac{3}{2}}} = \frac{\beta(\widetilde{\Lambda}^B)^*\beta}{\sqrt[4]{1+u^2}}\,\frac{\psi_i^E}{(1+v^2)^{\frac{3}{2}}} \doteq \beta\widehat{\Lambda}^*\beta\,\frac{\psi_i^E}{(1+v^2)^{\frac{3}{2}}}$$

Since $\beta$ is invertible, the smoothness of $\mathfrak{G}\psi_i$ is assured once we know that $\lim_{u \to -\infty} \widehat{\Lambda}$ is smooth. As already remarked, it is difficult to compute $\widetilde{\Lambda}^B$, and, hence, $\widehat{\Lambda}$ explicitly. Therefore, we recall the relation between Lorentz and spinor transformations and compute

$$\widehat{\Lambda}\gamma_a\beta\widehat{\Lambda}^*\beta = \frac{1}{\sqrt{1+u^2}}\,\widetilde{\Lambda}^B\gamma_a(\widetilde{\Lambda}^B)^{-1} = \frac{1}{\sqrt{1+u^2}}\,\gamma_b(\Lambda^B)^b{}_a \doteq \widehat{\Lambda}_a.$$

$\widehat{\Lambda}_a$ is a smooth matrix in the limit $u \to -\infty$ as can be computed from the explicit form of $\Lambda^B$ given in lemma I.3.3.2. Additionally, $\widehat{\Lambda}$ is determined uniquely up to a sign in terms of $\widehat{\Lambda}_a$. More in detail, the matrix elements of $\widehat{\Lambda}$ can be obtained from the ones of $\widehat{\Lambda}_a$ by algebraic expressions which map smooth functions to smooth functions. Hence, $\widehat{\Lambda}$ is a smooth matrix in the limit $u \to -\infty$.

d) At this point, one might be worried about the fact that the map $\mathfrak{G}$ contains the non-invertible matrix $K(\vec{e}_r)$, *i.e.* one might think that 'some information gets lost' in the bulk-to-boundary mapping. However, if we prove that the multiplication $K(\vec{e}_r)$ amounts to a multiplication with $\sqrt{2}$ in the case at hand, then the injectivity of $\mathfrak{G}$, being essentially the limit of a smooth section, is assured.

To see this property of $K(\vec{e}_r)$, we recall that $\mathfrak{G}$ is defined only on solutions $\widetilde{\psi}$ of $\widetilde{D}\widetilde{\psi} = 0$. However, any such solution $\widetilde{\psi}$ can be written as

$$\widetilde{\psi} = \widetilde{D}^*\widetilde{u}_\psi, \qquad \widetilde{D}\widetilde{D}^*\widetilde{\phi}_\psi = 0,$$

see theorem II.3.1.4. This is the starting point of the following straightforward chain of identities

$$\lim_{u \to -\infty} \frac{\widetilde{\psi}}{\sqrt[4]{1+u^2}} = \lim_{u \to -\infty} \frac{\widetilde{D}^*\widetilde{\phi}_\psi}{\sqrt[4]{1+u^2}} = \lim_{u \to -\infty} \frac{1}{\sqrt{2}}\,\frac{\sqrt{1+u^2}^{\frac{5}{2}}}{\sqrt[4]{1+u^2}}\,(\slashed{\partial} + am)\,\frac{\widetilde{\phi}_\psi}{\sqrt{1+u^2}^{\frac{3}{2}}}$$

$$= \lim_{u \to -\infty} \frac{1}{2^{\frac{3}{2}}} u^2 \left( \gamma^0 + \gamma^i \partial_r x_i \right) \partial_v \frac{\widetilde{\phi}_\psi}{(-u)^{\frac{3}{2}}} = -K(\vec{e}_r) \frac{i\beta}{2} \lim_{u \to -\infty} \sqrt{-u} \, \partial_v \widetilde{\phi}_\psi,$$

where lemma I.3.2.2 has been used in computing the limit of $\widetilde{\partial}$. The thesis now follows from $K(\vec{e}_r)^2 = \sqrt{2} K(\vec{e}_r)$.

<div align="right">□</div>

Note that the injectivity of $\widetilde{\mathfrak{G}}$ can be proved in quite the same way the one of $\mathfrak{G}$ has been proven. Moreover, the proof of the last item in the above theorem shows once more that the solutions of the Dirac equation constitute only a subclass of the solutions of the 'squared' Dirac equation, as already discussed in subsection II.4.1.

To close the discussion of the classical Dirac field on NBB spacetimes, let us remark that it is possible to complete both the bulk and the boundary inner product spaces to Hilbert spaces and to extend $\mathfrak{G}$ by continuity such that the resulting bulk Hilbert space is mapped to a sub-Hilbert space of the boundary Hilbert space in a unitary manner.

# III

# Quantized Fields in Curved Spacetimes

Building on the concepts and structures introduced in the previous chapter, we shall now discuss the formulation of quantum field theories on curved spacetimes. We will see that conceptual differences to standard treatments of Minkowskian quantum field theories arise which, however, should not be taken as an obstruction, but rather as a guide to the essential ingredients of a quantum field theory.

## III.1 The Free Scalar Field in General Curved Spacetimes

We start our treatment of quantum fields on curved spacetimes by discussing the free neutral scalar field. In analogy to the case of classical fields, this will enable us to concentrate on the essential concepts which appear in the quantum theory of a field of an arbitrary spin.

### III.1.1 The Algebras of Observables

The first conceptual difference of quantum field theories in curved spacetimes and such theories in Minkowski spacetime already arises at the beginning of our discussion. Namely, while in Minkowski spacetime Poincaré invariance and the spectrum condition select a unique vacuum state (see for instance [Ara99]), such phenomenon does not occur in generic curved spacetimes. Admittedly, there are curved spacetimes which have 'enough' symmetries to allow for preferred states, *e.g.* de Sitter spacetime [Al85]. Moreover, we shall provide a construction of preferred states in NBB spacetimes at a later time. Nevertheless, a proper understanding of quantum fields on curved spacetimes requires to construct a quantum field theory without having recourse to states and, hence, Hilbert spaces. Instead, one has to focus solely on the algebraic relations between fields and observable quantities first, and introduce states and related Hilbert space representations as secondary objects. This is achieved by the *algebraic approach* introduced by Rudolf Haag and it has lead to a deeper understanding of quantum field theory already in the context of Minkowski spacetime. A canonical monograph in this respect is the book [Haa92] written by Haag himself, whereas a survey of the algebraic approach to quantum field theory on curved spacetimes is found in Wald's book [Wal95]. Many progresses in algebraic quantum field theory on curved spacetimes have been attained after the appearance of [Wal95], an introduction to these new achievements can be found in [BrFr09].

Let us start by introducing the *Borchers-Uhlmann algebra* $\mathscr{A}(M)$ of the free scalar field. This is the most simple algebra of the free quantum scalar field, it does not contain *normal ordered quantities*, *i.e.* Wick polynomials, yet. The enlargement of $\mathscr{A}(M)$ to an algebra containing Wick polynomials is non-trivial and will be postponed to a later subsection. If $f \in \mathscr{D}(M)$ is a test function, we interpret it as $\phi(f)$, *i.e.* a 'smeared field', which we can formally write as $\phi(f) = \langle \phi, f \rangle$. The product of two smeared fields $\phi(f_1)\phi(f_2)$ is then just represented by the tensor product $f_1 \otimes f_2$. On the long run, we would like to represent $\phi(f)$ as an operator on a Hilbert space, we therefore need an operation which encodes 'taking the adjoint with respect to a Hilbert space inner product' on the abstract algebraic level. We define such a $*$-operation by setting $[\phi(f)]^* = \phi(\overline{f})$ and $[\phi(f_1) \cdots \phi(f_n)]^* = [\phi(f_n)]^* \cdots [\phi(f_1)]^*$. Observables would then be polynomials $\mathscr{P}$ of smeared fields which fulfil $\mathscr{P}^* = \mathscr{P}$. To promote $\phi(f)$ to a proper quantum field, we define

$$[\phi(f), \phi(g)] \doteq \phi(f)\phi(g) - \phi(g)\phi(f) = i\Delta(f, g)\mathbb{I}, \qquad \text{(III.46)}$$

where $\Delta$ is the causal propagator of the Klein-Gordon operator and $\mathbb{I}$ is the identity. Recall that $\Delta(f, g)$ vanishes if the supports of $f$ and $g$ are spacelike separated, the above *canonical commutation relations* (CCR) therefore assure that observables commute at spacelike separations.

One can write the CCR formally as

$$[\phi(x), \phi(y)] = i\Delta(x, y)\mathbb{I}.$$

Recall that this is nothing but the covariant version of the well-known equal-time CCR, see the discussion after lemma II.1.2.4. To encode dynamics, we would like the just defined quantum field $\phi(x)$ to fulfil the Klein-Gordon equation $P\phi(x) = 0$. On the algebraic level, this is encoded by demanding that $\phi(g) = 0$ for all $g \in \mathscr{D}(M)$ of the form $g = Pf$ with $f \in \mathscr{D}(M)$. More in detail, this entails dividing the test functions by the ideal generated by elements of the form $Pf$. Recall that this ideal is nothing but the kernel of $\Delta$, the just described procedure is therefore compatible with the CCR. We subsume the above discussion in the following definition.

**Definition III.1.1.1**  *The **Borchers-Uhlmann algebra** $\mathscr{A}(M)$ of the free scalar field is defined as*

$$\mathscr{A}(M) \doteq \mathscr{A}_0(M)/\mathscr{I},$$

*where $\mathscr{A}_0(M)$ is the direct sum*

$$\mathscr{A}_0(M) \doteq \bigoplus_{n=0}^{\infty} \mathscr{D}(M^n)$$

*($\mathscr{D}(M^0) \doteq \mathbb{C}$) equipped with a product defined by the linear extension of the tensor product of $\mathscr{D}(M^n)$, a $*$-operation defined by the antilinear extension of $[f^*](x_1, \cdots, x_n) = \overline{f}(x_n, \cdots, x_1)$, and it is required that elements of $\mathscr{A}_0(M)$ are finite linear combinations of multi-component test functions. Additionally, we equip $\mathscr{A}_0(M)$ with the topology defined by saying that a sequence $\{f_k\}_k = \{\oplus_n f_k^{(n)}\}_k$ in $\mathscr{A}_0(M)$ converges to $f = \oplus_n f^{(n)}$ if $f_k^{(n)}$ converges to $f^{(n)}$ for all $n$ in the locally convex topology of $\mathscr{D}(M^n)$ and there exists an $N$ such that $f_k^{(n)} = 0$ for all $n > N$ and all $k$. Moreover, $\mathscr{I}$ is the closed $*$-ideal generated by elements of the form $-i\Delta(f, g) \oplus (f \otimes g - g \otimes f)$ and $Pf$, and $\mathscr{A}(M)$ is thought to be equipped with the product, $*$-operation, and topology descending from $\mathscr{A}_0(M)$.*

*If $\mathscr{O}$ is an open subset of $M$, $\mathscr{A}(\mathscr{O})$ denotes the algebra obtained by allowing only test functions with support in $\mathscr{O}$.*

A comparison with definition II.1.3.1 reveals that, barring the CCR, the algebra of classical observables $(\mathscr{C}_0(M), \cdot_s)$ is isomorphic to the subset of $\mathscr{A}(M)$ constituted by (linear combinations of) symmetric testfunctions. We shall examine this relation in more detail when discussion the enlargement of $\mathscr{A}(M)$ to an algebra including Wick polynomials. Moreover, we stress that $\mathscr{A}(M)$, in contrast to $\mathscr{A}_0(M)$, depends explicitly on the metric $g$ of a spacetime $(M, g)$ via the causal propagator and the equation of motion. However, now and in the following we shall omit this dependence in favour of notational simplicity.

In what follows, we will denote the smeared fields generating $\mathscr{A}(M)$ by $\phi(f)$ for an $f \in \mathscr{D}(M)$ and not make the equivalence class corresponding to the quotient $\mathscr{A}_0(M)/\mathscr{I}$ explicit. In fact, let us recall that, by lemma II.1.2.2 and the properties of $\Delta$, the equivalence class of such $f$ is so large that it contains elements with support in an arbitrarily small neighbourhood of any Cauchy surface of $M$. This implies the following well-known result, which on physical grounds entails the predictability of observables.

**Lemma III.1.1.2** *The Borchers-Uhlmann algebra $\mathscr{A}(M)$ of the free scalar field fulfils the **time-slice axiom**. Namely, let $\Sigma$ be a Cauchy surface of $(M, g)$ and let $\mathcal{O}$ be an arbitrary neighbourhood of $\Sigma$. Then $\mathscr{A}(\mathcal{O}){=}\mathscr{A}(M)$.*

The above discussion implicates in particular that we can understand the Borchers-Uhlmann algebra $\mathscr{A}(M)$ equivalently as generated by solutions of the Klein-Gordon equations with compactly supported initial conditions, *i.e.* by tensor products of (the complexification of) $\mathscr{S}(M)$. Given an element $u \in \mathscr{S}(M)$ with $u = \Delta f$, this correspondence is constituted by identifying the smeared field $\phi(f)$ with a field $\Phi(u)$ fulfilling

$$[\Phi(u_1), \Phi(u_2)] = i\varsigma_M(u_1, u_2),$$

where $\varsigma_M$ is the symplectic form on $\mathscr{S}(M)$, see definition II.1.2.1. The field $\Phi(u)$ can be understood as a 'symplectically smeared field' $\Phi(u) = \varsigma_M(\Phi, u)$, see [Wal95] for details.

We now turn our attention to states. Let $\mathfrak{A}$ be a topological, unital $*$-algebra, *i.e.* $\mathfrak{A}$ is endowed with an operation $*$ which fulfils $(AB)^* = B^* A^*$ and $(A^*)^* = A$ for all elements $A, B$ in $\mathfrak{A}$. A *state* $\omega$ on $\mathfrak{A}$ is defined to be a continuous linear functional $\mathfrak{A} \to \mathbb{C}$ which is normalised, *i.e.* $\omega(\mathbb{I}) = 1$ and positive, namely, $\omega(A^*A) \geq 0$ must hold for any $A \in \mathfrak{A}$. Considering the special topological and unital $*$-algebra $\mathscr{A}(M)$, a state on $\mathscr{A}(M)$ is determined by its $n$-point functions

$$\omega_n(f_1, \cdots, f_n) \doteq \omega\left(\phi(f_1) \cdots \phi(f_n)\right).$$

These are distributions in $\mathscr{D}'(M^n)$ by the continuity of $\omega$ and the Schwartz kernel theorem. Given a state $\omega$ on $\mathfrak{A}$, we can represent $\mathfrak{A}$ on a Hilbert space $\mathscr{H}_\omega$ by the so-called *GNS construction* (after Gel'fand, Naimark, and Segal), see for instance [Haa92, Ara99]. By this construction, algebra elements are represented as operators on a common dense and invariant subspace of $\mathscr{H}_\omega$, while $\omega$ is represented as a vector of $\mathscr{H}_\omega$. The GNS construction arises out of the observation that $\omega(A^*A) \geq 0$ entails

$$\overline{\omega(A^*B)} = \omega(B^*A)$$

for arbitrary elements $A, B$ in $\mathfrak{A}$; $\omega$ therefore gives a positive semidefinite Hermitean form on $\mathfrak{A}$. If we take the quotient $\mathfrak{A}/\text{Ker}_\omega$ of $\mathfrak{A}$ and the kernel of $\omega$, we obtain a proper inner product on the resulting quotient space and can complete it to obtain $\mathscr{H}_\omega$. The kernel of $\omega$ is a left ideal of $\mathfrak{A}$ by the Cauchy-Schwarz inequality, elements of $\mathscr{A}(M)$ therefore act on the equivalence classes in $\mathscr{A}(M)/\text{Ker}_\omega$ by left multiplication, *i.e.*

$$B[A] = [BA]$$

for arbitrary $A, B$ in $\mathfrak{A}$. We shall provide a more explicit incarnation of the GNS representation below. The GNS representation of $\mathscr{A}(M)$ yields in general unbounded operators, which are technically more difficult to handle than bounded operators. One therefore often considers the *Weyl algebra* instead of the Borchers-Uhlmann algebra, see for instance [BrRo96v2, BGP07]. This approach heuristically speaking amounts to consider $e^{i\phi(f)}$ instead of $\phi(f)$ and therefore leads to a GNS representation by bounded operators which are defined on the full GNS Hilbert space.

For the purpose of this thesis, it will be sufficient to work on the level of the Borchers-Uhlmann algebra and its extensions.

Among the possible states on $\mathscr{A}(M)$ there are a couple of special classes, which we collect in the following definition. Some of the definitions are sensible for general $*$-algebras, as we point out explicitly.

**Definition** III.1.1.3    *Let $\mathfrak{A}$ denote a general $*$-algebra and let $\mathscr{A}(M)$ denote the Borchers-Uhlmann algebra of the Klein-Gordon field.*

a) *A state $\omega$ on $\mathfrak{A}$ is called* **mixed**, *if it is a convex linear combination of states, i.e. $\omega = \lambda \omega_1 + (1-\lambda)\omega_2$, where $\lambda < 1$ and $\omega_i \neq \omega$ are states on $\mathfrak{A}$. A state is called* **pure** *if it is not mixed.*

b) *A state $\omega$ on $\mathscr{A}(M)$ is called* **even**, *if it is invariant under $\phi(f) \mapsto -\phi(f)$, i.e. it has vanishing n-point functions for all odd n.*

d) *An even state on $\mathscr{A}(M)$ is called* **quasifree** *or* **Gaussian** *if, for all even n,*

$$\omega_n(f_1, \cdots, f_n) = \sum_{\pi_n \in S'_n} \prod_{i=1}^{n/2} \omega_2\left(f_{\pi_n(2i-1)}, f_{\pi_n(2i)}\right).$$

*Here, $S'_n$ denotes the set of ordered permutations of n elements, namely, the following two conditions are satisfied for $\pi_n \in S'_n$:*

$$\pi_n(2i-1) < \pi_n(2i) \quad \text{for} \quad 1 \leq i \leq n/2, \qquad \pi_n(2i-1) < \pi_n(2i+1) \quad \text{for} \quad 1 \leq i < n/2.$$

e) *Let $\alpha_t$ denote a one-parameter group of $*$-automorphisms on $\mathfrak{A}$, i.e. for arbitrary elements A, B of $\mathfrak{A}$,*

$$\alpha_t(A^*B) = \left(\alpha_t(A)\right)^* \alpha_t(B), \qquad \alpha_t\left(\alpha_s(A)\right) = \alpha_{t+s}(A), \qquad \alpha_0(A) = A.$$

*A state $\omega$ on $\mathfrak{A}$ is called $\alpha_t$-**invariant** if $\omega(\alpha_t(A)) = \omega(A)$ for all $A \in \mathfrak{A}$.*

e) *An $\alpha_t$-invariant state $\omega$ on $\mathfrak{A}$ is said to satisfy the* **KMS condition** *for an inverse temperature $\beta = T^{-1} > 0$ if, for arbitrary elements A, B of $\mathfrak{A}$, the two functions*

$$F_{AB}(t) \doteq \omega\left(B\alpha_t(A)\right), \qquad G_{AB}(t) \doteq \omega\left(\alpha_t(A)B\right)$$

*extend to functions $F_{AB}(z)$ and $G_{AB}(z)$ on the complex plane which are analytic in the strips $0 < Im\, z < \beta$ and $-\beta < Im\, z < 0$ respectively, continuous on the boundaries $Im\, z \in \{0, \beta\}$, and fulfil*

$$F_{AB}(t + i\beta) = G_{AB}(t).$$

The KMS condition (after Kubo, Martin, and Schwinger) holds naturally for Gibbs states of *finite* systems in quantum statistical mechanics, *i.e.* for states that are given as $\omega_\beta(A) = Tr\, \rho A$ with a density matrix $\rho = \exp(-\beta H)(Tr \exp(-\beta H))^{-1}$, $H$ the Hamiltonian operator of the system, and $Tr$ denoting the trace over the respective Hilbert space. This follows by setting

$$\alpha_t(A) = e^{itH}Ae^{-itH},$$

making use of the cyclicity of the trace, and considering that $\exp(-\beta H)$ is bounded and has finite trace in the case of a finite system. In the thermodynamic limit, $\exp(-\beta H)$ does not possess these properties any more, but the authors of [HHW76] have shown that the KMS condition is still a reasonable condition in this infinite-volume limit. Physically, KMS states are states which are in (thermal) equilibrium with respect to the time evolution encoded in the automorphism $\alpha_t$. In general curved spacetimes, there is no 'time evolution' which acts as an automorphism on $\mathscr{A}(M)$. One could be tempted to introduce a time evolution by a canonical time-translation with respect to some time function of a globally hyperbolic spacetime. However, the causal propagator $\Delta$ will in general not be invariant under this time translation if the latter does not correspond to an isometry of $(M, g)$. Hence, such time-translation would not result in an automorphism of $\mathscr{A}(M)$. There have been various proposals to overcome this problem and to define generalised notions of thermal equilibrium in curved spacetimes. We will comment on this when constructing KMS states on the boundary of NBB spacetimes.

The authors of [HHW76] have provided an equivalent formulation of the KMS condition on the level of Fourier transforms of $F_{AB}(t)$ and $G_{AB}(t)$. As we wish to construct KMS states on the boundary of NBB spacetimes directly in Fourier space, this equivalent condition will be very convenient.

**Lemma** III.1.1.4 *An $\alpha_t$-invariant state $\omega$ on $\mathscr{A}(M)$ fulfils the KMS condition at the inverse temperature $\beta$ if the functions $F_{AB}(t)$ and $G_{AB}(t)$ introduced in definition III.1.1.3 are bounded, continuous, and fulfil the following condition: denoting by $\hat{F}(E)$ and $\hat{G}(E)$ the Fourier transforms of $F_{AB}(t)$ and $G_{AB}(t)$ in the sense of distributions on rapidly decreasing smooth functions,*

$$\hat{F}(E) \equiv e^{\beta E}\hat{G}(E).$$

*Here, the multiplication $e^{\beta E}\hat{G}(E)$ has to be understood as first restricting $\hat{G}(E)$ to a distribution on $\mathscr{D}(\mathbb{R})$, then multiplying it with $e^{\beta E}$, and finally considering the unique extension of the result to a distribution on rapidly decreasing smooth functions.*

The KMS condition seems difficult to check, as it suggests that one has to assure it for arbitrary elements $A$ and $B$ of $\mathscr{A}(M)$. However, on account of the linearity of a state and the fact that $\alpha_t$ is an automorphism, the KMS condition already holds for all $A$ and $B$ if it holds for $A = \phi(f)$. In the case of a quasifree state $\omega$, it is therefore sufficient to check the KMS condition for the two-point function $\omega_2$. We refer the reader interested in further properties of KMS states to the standard monographs [BrRo96v2, Haa92] and to [DMP09c, app. B].

Let us now consider quasifree states. These states are closely related to a Fock space picture. In fact, one can show that pure states can be represented as Fock vacua and therefore define a particle picture. To understand this in more detail, we shall introduce the notion of a *one-particle Hilbert space structure*. To this avail, we need to consider what the general requirement of positivity means in the special case of a quasifree state. Namely, positivity of a state entails that $\omega_2(\overline{f}, f) \geq 0$ for all $f \in \mathscr{D}(M)$. However, to obtain the reversed implication, namely, that a positive two-point function yields a quasifree state, one has to require that the symmetric part

$$\omega_s(f, g) \doteq \frac{1}{2}\left(\omega(f, g) + \omega(g, f)\right)$$

of $\omega_2$ (which is a real and positive distribution by the positivity of $\omega$) fulfils a particular lower bound, namely [KaWa91],

$$\frac{1}{4}|\Delta(\overline{f}, g)| \leq \omega_s(\overline{f}, f)\omega_s(\overline{g}, g). \tag{III.47}$$

The quasifree state defined by such a two-point function is pure if and only if [KaWa91]

$$\omega_s(\overline{f}, f) = \sup_{g \in \mathscr{D}, \, \omega_s(\overline{g}, g) \neq 0} \frac{|\Delta(\overline{f}, g)|}{4\omega_s(\overline{g}, g)}.$$

Note that (III.47) entails that the kernel of $\omega_s$ is contained in the kernel of $\Delta$. If we recall that this kernel has been identified with zero in the construction of $\mathscr{A}(M)$, we find that $\omega_s$ is already strictly positive definite on $\mathscr{A}(M)$. On account of the relation between $\mathscr{D}(M)$ and $\mathscr{S}(M)$, this entails that $\omega_s$ defines an inner product $\mu$ on $\mathscr{S}(M)$. Namely, if $u_1 = \Delta f_1$, $u_2 = \Delta f_2$ are arbitrary elements of $\mathscr{S}(M)$ with suitable $f_i \in \mathscr{D}(M, \mathbb{R})$, then we can define such $\mu$ as

$$\mu(u_1, u_2) \doteq \omega_s(f_1, f_2).$$

The resulting $\mu$ fulfils

$$\frac{1}{4}|\varsigma_M(u_1, u_2)| \leq \mu(u_1, u_1)\mu(u_2, u_2). \tag{III.48}$$

and we can interpret it as the expectation value of the product of the symplectically smeared field operators $\Phi(u_1)\Phi(u_2)$. This shows that viewing $\mathscr{A}(M)$ as constructed out of solutions or test functions is equivalent also on the level of states (once $\mathscr{S}(M)$ is complexified). Moreover, we see explicitly that the specification of a state on $\mathscr{A}(M)$ amounts to providing a real and symmetric distribution fulfilling (III.47), or, equivalently, an inner product on $\mathscr{S}(M)$ fulfilling (III.48). This is clarified further by the following proposition, taken from [KaWa91].

**Proposition** III.1.1.5 *Let $\mathscr{S}$ be a real vector space equipped with a strongly non-degenerate symplectic form $\varsigma$ and an inner product $\mu$. Under these circumstances, one can always find a complex Hilbert space $\mathscr{H}$ with inner product $(\cdot, \cdot)$ and a real-linear map $K : \mathscr{S} \to \mathscr{H}$ satisfying the following properties.*

*a) The complexified range $K\mathscr{S} + iK\mathscr{S}$ is dense in $\mathscr{H}$.*

*b) $\mu(u_1, u_2) = Re(Ku_1, Ku_2).$*

*a)* $\varsigma(u_1, u_2) = 2 Im (K u_1, K u_2)$.

*Moreover, the pair $(K, \mathscr{H})$ is unique up to unitary equivalence.*

We refer to [KaWa91, app. A] for the proof. The above result gives a very explicit realisation of a representation of the Borchers-Uhlmann algebra on a Hilbert space. Namely, let us consider the Fock space $\mathscr{F}$ built out of symmetric tensor products of $\mathscr{H}$, let $\psi_1$, $\psi_2 \in \mathscr{H}$, and let $a^\dagger(\psi_2)$, $a(\psi_1)$ denote the usual creation and annihilation operators on $\mathscr{F}$ fulfilling

$$\left[a(\psi_1), a^\dagger(\psi_2)\right] = (\psi_1, \psi_2)\mathbb{I}, \qquad \left[a(\psi_1), a(\psi_2)\right] = \left[a^\dagger(\psi_1), a^\dagger(\psi_2)\right] = 0,$$

see for instance [ReSi75, BrRo96v2]. Then, the symplectically smeared field $\Phi(u)$ can be represented on $\mathscr{F}$ as

$$\Phi(u) = i a^\dagger(K u) - i a(K u)$$

which is equivalent to viewing the smeared field $\phi(f)$ as being represented on $\mathscr{F}$ as

$$\phi(f) = i a^\dagger(K \Delta f) - i a(K \Delta f).$$

Note that, by the properties of $K$, both $\phi(f)$ and $\Phi(u)$ are real-linear, even though the annihilation operator $a(\psi)$ is complex antilinear. The map $K$ can be viewed as selecting a 'positive frequency subspace' of the (complexified) space of solutions $\mathscr{S}(M)$, see the discussion in [KaWa91, sec. 3.2]; this corresponds to the well-known mode decomposition of a quantum field in Minkowski spacetime. Moreover, in [KaWa91] it is also shown that the range $K\mathscr{S}$ of $K$ is already dense in $\mathscr{H}$ if the considered $\mu$ satisfies (III.48) and therefore stems from a pure state.

### III.1.2 Hadamard States

The Borchers-Uhlmann algebra $\mathscr{A}(M)$ contains only very basic observables, namely, linear combinations of products of free fields at separate points, *e.g.* $\phi(x)\phi(y)$. However, if one wants to treat interacting fields in perturbation theory, or the backreaction of quantum fields on curved spacetimes via their stress-energy tensor, ones needs a notion of normal ordering, *i.e.* a way to define field monomials like $\phi^2(x)$ at the same point. To see that this requires some work, let us consider the massless scalar field in Minkowski spacetime. Its two point function reads

$$\omega_2(x,y) = \omega(\phi(x)\phi(y)) = \lim_{\varepsilon \downarrow 0} \frac{1}{4\pi^2} \frac{1}{(x-y)^2 + i\varepsilon(x_0 - y_0) + \varepsilon^2}, \qquad \text{(III.49)}$$

where $(x-y)^2$ denotes the Minkowskian product induced by the metric $\eta$ and the limit has to be understood as being performed after integrating $\omega_2$ with at least one test function. This is a smooth function if $x$ and $y$ are spacelike or timelike separated. It is singular at $(x-y)^2 = 0$, but the singularity is 'good enough' to give a finite result when smearing $\omega_2(x,y)$ with two test functions. Hence, $\omega_2$ is a well-defined (tempered) distribution. Loosely speaking, this shows once more that the product of fields $\phi(x)\phi(y)$ is 'well-defined' at non-null related points. However, if we were to define $\phi^2(x)$ by some 'limit' like

$$\phi^2(x) \doteq \lim_{x \to y} \phi(x)\phi(y),$$

the expectation value of the resulting object would 'blow up' and would not be any meaningful object. The well-known solution to this apparent problem is to define field monomials by appropriate regularising subtractions. For the squared field, this is achieved by setting

$$:\phi^2(x): \doteq \lim_{x \to y}(\phi(x)\phi(y) - \omega_2(x, y)\mathbb{1}),$$

where of course one would have to specify in which sense the limit should be taken. Omitting the details of this procedure, it seems still clear that the *Wick square* $:\phi^2(x):$ is a meaningful object, as it has a sensible expectation value, *i.e.* $\omega(:\phi^2(x):) = 0$. In the standard Fock space picture, one heuristically writes the field (operator) in terms of creation and annihilation operators in momentum space, *i.e.*

$$\phi(x) = \frac{1}{\sqrt{2\pi}^3} \int \frac{d\vec{k}}{\sqrt{2k_0}} a_{\vec{k}}^\dagger e^{ikx} + a_{\vec{k}} e^{-ikx},$$

and defines $:\phi^2(x):$ by writing the mode expansion of the product $\phi(x)\phi(y)$, 'normal ordering' the appearing products of creation and annihilation operators such that the creation operators are standing on the left hand side of the annihilation operators, and then finally taking the limit $x \to y$. It is easy to see that this procedure is equivalent to the above defined subtraction of the vacuum expectation value. However, having defined the Wick polynomials is not enough. We would also like to multiply them, *i.e.*, we would like them to constitute an algebra. Using the mode-expansion picture, one can straightforwardly compute

$$:\phi^2(x)::\phi^2(y): = :\phi^2(x)\phi^2(y): + 4 :\phi(x)\phi(y): \omega_2(x, y) + 2(\omega_2(x, y))^2,$$

which is a special case of the well-known *Wick theorem*, see for instance [IZ80]. The right hand side of the above equation is a sensible object if the appearing square of of the two-point function $\omega_2(x, y)$ is well-defined. In more detail, we know that $\omega_2(x, y)$ has singularities, and that these are integrable with test functions. Obviously, $(\omega_2(x, y))^2$ has singularities as well, and the question is whether the singularities are still good enough to be integrable with test functions. In terms of a mode decomposition, one could equivalently wonder whether the momentum space integrals appearing in the definition of $:\phi^2(x)\phi^2(y):$ via normal ordering creation and annihilation operators converge in a sensible way. The answer to these questions is 'yes' because of the energy positivity property of the Minkowskian vacuum state, and this is the reason why one usually never worries about whether normal ordering is well-defined in quantum field theory on Minkowski spacetime. In more detail, the Fourier decomposition of the massless two-point function $\omega_2$ reads

$$\omega_2(x, y) = \lim_{\varepsilon \downarrow 0} \frac{1}{(2\pi)^3} \int dk\, \Theta(k_0)\delta(k^2)\, e^{ik(x-y)}e^{-\varepsilon k_0}, \tag{III.50}$$

where $\Theta(k_0)$ denotes the Heaviside step function and we omit the necessary $\varepsilon$-regularisation for simplicity. We see that the Fourier transform of $\omega_2$ has only support on the forward lightcone (or the positive mass shell in the massive case); this corresponds to the fact that we have associated the positive frequency modes to the creation operator in the above mode expansion of the quantum field. This insight allows to determine (or rather, define) the square of $\omega_2(x, y)$ by a convolution in Fourier space

$$
\begin{aligned}
(\omega_2(x, y))^2 &= \lim_{\varepsilon \downarrow 0} \frac{1}{(2\pi)^6} \int dq \int dp \, \Theta(q_0) \, \delta(q^2) \, \Theta(p_0) \, \delta(p^2) \, e^{i(q+p)(x-y)} e^{-\varepsilon q_0} \\
&= \lim_{\varepsilon \downarrow 0} \frac{1}{(2\pi)^6} \int dk \int dq \, \Theta(q_0) \, \delta(q^2) \, \Theta(k_0 - q_0) \, \delta((k-q)^2) \, e^{ik(x-y)} e^{-\varepsilon q_0}.
\end{aligned}
$$

Without going too much into details here, let us observe that the above expression can only give a sensible result (a distribution) if the integral over $q$ converges, *i.e.* if the integrand is rapidly decreasing in $q$. To see that this is the case, note that for an arbitrary but fixed $k$ and large $q$ where here 'large' is meant in the Euclidean norm on $\mathbb{R}^4$, the integrand is vanishing on account of $\delta(q^2)$ and $\Theta(k_0 - q_0)$ as $k_0 - q_0 < 0$ for large $q$. Loosely speaking, we observe the following: by the form of a convolution, the Fourier transform of $\omega_2$ is multiplied by the same Fourier transform, but with negative momentum. Since the $\omega_2$ has only Fourier support in one 'energy direction', namely the positive one, the intersection of its Fourier support and the same support evaluated with negative momentum is compact, and the convolution therefore well-defined. Moreover, as this statement only relies on the large momentum behaviour of Fourier transforms, it holds equally in the case of massive fields, as the mass shell approaches the light cone for large momenta.

The outcome of the above considerations is the insight that, if we want to define a sensible generalisation of normal ordering in curved spacetimes, we have to select states whose two-point functions are singular, but regular enough to allow for pointwise multiplication. Even though general curved spacetimes are not translationally invariant and therefore do not allow to define a global Fourier transform and a related global energy positivity condition, one could think that this task can be achieved by some kind of a 'local Fourier transform' and a related 'local energy positivity condition'. In fact, as showed in the pioneering work of Radzikowski [Rad96a, Rad96b], this heuristic idea can be made precise in terms of *microlocal analysis*, a modern branch of Mathematics. In the aforementioned works [Rad96a, Rad96b], it has been shown that so-called *Hadamard states*, which have already been known to allow for a sensible renormalisation of the stress-energy tensor [Wal77, Wal78a, Wal95], indeed fulfil a local energy positivity condition in the sense that their two-point function has a specific *wave front set*. Based on this, Brunetti, Fredenhagen, Köhler, Hollands, and Wald [BFK95, BrFr00, HoWa01, HoWa02, HoWa05] have been able to show that one can as a matter of fact define normal ordering and perturbative interacting quantum field theories based on Hadamard states essentially in the same way as on Minkowski spacetimes. Though, it turned out that there is a big conceptual difference to flat spacetime quantum field theories, namely, new regularisation freedoms in terms of curvature terms appear. Although these are finitely many, and therefore lead to the result that theories which are perturbatively renormalisable in Minkowski spacetime retain this property in curved spacetimes [BrFr00], the appearance of these additional renormalisation freedoms has a profound impact on the backreaction of quantum fields on curved spacetimes, as we will discuss in the next chapter.

We have already anticipated that Hadamard states can be approached from two angles. One way to discuss them is to look at the concrete realisation of their two-point function. This treatment has lead to the insight that Hadamard states are the sensible starting point for the definition of a regularised stress-energy tensor [Wal77, Wal78a, Wal95], and it is well-suited for actual calculations in particular. On the other hand, the rather abstract study of Hadamard states based on microlocal analysis is well-suited to tackle and solve conceptual problems. Following our discussion of the obstructions in the definition of normal ordering, we shall start our treatment by considering the microlocal aspects of Hadamard states. A standard monograph on microlocal analysis is the book of Hörmander [Hör90], who has also contributed a large part to this field of Mathematics [Hör71, DuHo72]. Introductory treatments can be found in [ReSi75, BrFr00, Kra00, Str09].

Let us start be introducing the notion of a wave front set. To motivate it, let us recall that a smooth function on $\mathbb{R}^m$ with compact support has a rapidly decreasing Fourier transform. If we take an distribution $u$ in $\mathscr{D}'(\mathbb{R}^m)$ and multiply it by an $f \in \mathscr{D}(\mathbb{R}^m)$ with $f(x_0) \neq 0$, then $uf$ is an element of $\mathscr{E}'(\mathbb{R}^m)$, *i.e.*, a distribution with compact support. If $fu$ were smooth, then its Fourier transform $\widehat{fu}$ would be smooth and rapidly decreasing. The failure of $fu$ to be smooth in a neighbourhood of $x_0$ can therefore be quantitatively described by the set of directions in Fourier space where $\widehat{fu}$ is not rapidly decreasing. Of course it could happen that we choose $f$ badly and therefore 'cut' some of the singularities of $u$ at $x_0$. To see the full singularity structure of $u$ at $x_0$, we therefore need to consider all test functions which are non-vanishing at $x_0$. With this in mind, one first defines the wave front set of distributions on $\mathbb{R}^m$ and then extends it to curved manifolds in a second step.

**Definition III.1.2.1** *A neighbourhood $\Gamma$ of $k_0 \in \mathbb{R}^m$ is called **conic** if $k \in \Gamma$ implies $\lambda k \in \Gamma$ for all $\lambda \in (0, \infty)$. Let $u \in \mathscr{D}'(\mathbb{R}^m)$. A point $(x_0, k_0) \in \mathbb{R}^m \times (\mathbb{R}^m \setminus \{0\})$ is called a regular directed point of $u$ if there is an $f \in \mathscr{D}(\mathbb{R}^m)$ with $f(x_0) \neq 0$ such that, for every $n \in \mathbb{N}$, there is a constant $C_n \in \mathbb{R}$ fulfilling*

$$|\widehat{fu}(k)| \leq C_n (1 + |k|)^{-n}$$

*for all $k$ in a conic neighbourhood of $k_0$. The **wave front set** $WF(u)$ is the complement in $\mathbb{R}^m \times (\mathbb{R}^m \setminus \{0\})$ of the set of all regular directed points of $u$.*

Let us immediately state a few important properties of wave front sets, the proofs of which can be found in [Hör90] (see also [Str09]).

**Theorem III.1.2.2** *Let $u \in \mathscr{D}'(\mathbb{R}^m)$.*

*a) If $u$ is smooth, then $WF(u)$ is empty.*

b) *Let $P$ be an arbitrary partial differential operator. It holds*

$$WF(Pu) \subset WF(u).$$

c) *Let $U$, $V \subset \mathbb{R}^m$, let $u \in \mathscr{D}'(V)$, and let $\chi : U \to V$ be a diffeomorphism. The pull-back $\chi^*(u)$ of $u$ defined by $\chi^* u(f) = u(\chi_* f)$ for all $f \in \mathscr{D}(U)$ fulfils*

$$WF(\chi^* u) = \chi^* WF(u) \doteq \left\{ (\chi^{-1}(x), \chi^* k) \,|\, (x,k) \in WF(u) \right\},$$

*where $\chi^* k$ denotes the push-forward of $\chi$ in the sense of **cotangent vectors**. Hence, the wave front set transforms covariantly under diffeomorphisms as an element of $T^*\mathbb{R}^m$, and we can extend its definition to distributions on general curved manifolds $M$ by patching together wave front sets in different coordinate patches of $M$. As a result, for $u \in \mathscr{D}'(M)$, $WF(u) \subset T^*M \setminus \{0\}$, where $0$ denotes the zero section of $T^*M$.*

d) *Let $u_1$, $u_2 \in \mathscr{D}'(M)$ and let*

$$WF(u_1) \oplus WF(u_2) \doteq \left\{ (x, k_1 + k_2) \,|\, (x, k_1) \in WF(u_1), \, (x, k_2) \in WF(u_2) \right\}.$$

*If $WF(u_1) \oplus WF(u_2)$ does not intersect the zero section, then one can define the product $u_1 u_2$ in such a way that it yields a well-defined distribution in $\mathscr{D}'(M)$ and that it reduces to the standard pointwise product of smooth functions if $u_1$ and $u_2$ are smooth. Moreover, the wave front set of such product is bounded in the following way*

$$WF(u_1 u_2) \subset WF(u_1) \cup WF(u_2) \cup \left( WF(u_1) \oplus WF(u_2) \right).$$

Note that the wave front set transforms as a subset of the cotangent bundle on account of the covector nature of $k$ in $\exp(ikx)$. The last of the above statements is exactly the criterion for pointwise multiplication of distributions we have been looking for. Namely, from (III.50) and (III.49) one can infer that the wave front set of the Minkowskian two-point function (for $m \geq 0$) is [ReSi75]

$$WF(\omega_2) = \left\{ (x, y, k, -k) \in T^*\mathbb{M}^2 \,|\, x \neq y, \, (x-y)^2 = 0, \, k \| (x-y), \, k_0 > 0 \right\} \qquad \text{(III.51)}$$

$$\cup \left\{ (x, x, k, -k) \in T^*\mathbb{M}^2 \,|\, k^2 = 0, \, k_0 > 0 \right\},$$

particularly, it is the condition $k_0 > 0$ which encodes the energy positivity of the Minkowskian vacuum state. We can now rephrase our observation that the pointwise square of $\omega_2(x,y)$ is a well-defined distribution by noting that $WF(\omega_2) \oplus WF(\omega_2)$ does not contain the zero section. In contrast, we know that the $\delta$-distribution $\delta(x)$ is singular at $x = 0$ and that its Fourier transform is a constant. Hence, its wave front set reads

$$WF(\delta) = \left\{ (0, k) \,|\, k \in \mathbb{R} \setminus \{0\} \right\},$$

and we see that the $\delta$-distribution does not have a 'one-sided' wave front set and, hence, can not be squared. The same holds if we view $\delta$ as a distribution $\delta(x,y)$ on $\mathscr{D}(\mathbb{R}^2)$. Then

$$WF(\delta(x,y)) = \{(x,x,k,-k) \mid k \in \mathbb{R} \setminus \{0\}\}.$$

The previous discussion suggests that a generalisation of (III.51) to curved spacetimes is the sensible requirement to select states which allow for the construction of Wick polynomials. We shall now define such a generalisation.

**Definition III.1.2.3**  *Let $\omega$ be a state on $\mathscr{A}(M)$. We say that $\omega$ fulfils the **Hadamard condition** and is therefore a **Hadamard state** if its two-point function $\omega_2$ fulfils*

$$WF(\omega_2) = \left\{(x,y,k_x,-k_y) \in T^*M^2 \setminus \{0\} \mid (x,k_x) \sim (y,k_y),\, k_x \triangleright 0\right\}.$$

*Here, $(x,k_x) \sim (y,k_y)$ implies that there exists a null geodesic $c$ connecting $x$ to $y$ such that $k_x$ is coparallel and cotangent to $c$ at $x$ and $k_y$ is the parallel transport of $k_x$ from $x$ to $y$ along $c$. Finally, $k_x \triangleright 0$ means that the covector $k_x$ is future-directed.*

Having discussed the rather abstract aspect of Hadamard states, let us now turn to their more concrete realisations. To this avail, let us consider a geodesically convex set $\mathcal{O}$ in $M$, see section I.1. By definition, there are open subsets $\mathcal{O}'_x \subset T_x M$ such that the exponential map $\exp_x : \mathcal{O}'_x \to \mathcal{O}$ is well-defined for all $x \in \mathcal{O}$, *i.e.* we can introduce Riemannian normal coordinates on $\mathcal{O}$. For any two points $x,\, y \in \mathcal{O}$, we can therefore define the *half squared geodesic distance* $\sigma(x,y)$ as

$$\sigma(x,y) \doteq \frac{1}{2} g\left(\exp_x^{-1}(y), \exp_x^{-1}(y)\right).$$

This entity is sometimes also called *Synge's world function* and is both smooth and symmetric on $\mathcal{O} \times \mathcal{O}$. Moreover, one can show that it fulfils the following identity

$$\sigma_{;\mu}\sigma_{;}{}^{\mu} = 2\sigma, \tag{III.52}$$

where the covariant derivatives are taken with respect to $x$ (even though this does not matter by the symmetry of $\sigma$), see for instance [Fri75, Poi03]. Let us introduce a couple of standard notations related to objects on $\mathcal{O} \times \mathcal{O}$ such as $\sigma$. If $VM$ and $WM$ are vector bundles over $M$ with typical fibers constituted by the vector spaces $V$ and $W$ respectively, then we denote by $VM \boxtimes WM$ the *exterior tensor product* of $VM$ and $WM$. $VM \boxtimes WM$ is defined as the vector bundle over $M \times M$ with typical fibre $V \otimes W$. The more familiar notion of the tensor product bundle $VM \otimes WM$ is obtained by considering the pull-back bundle of $VM \boxtimes WM$ with respect to the map $M \ni x \mapsto (x,x) \in M^2$. Typical exterior product bundles are for instance the tangent bundles of Cartesian products of $M$, *e.g.* $T^*M \boxtimes T^*M = T^*M^2$. A section of $VM \boxtimes WM$ is called a *bitensor*. We introduce the *Synge bracket notation* for the coinciding point limits of a bitensor. Namely, let $B$ be a smooth section of $VM \boxtimes WM$. We define

$$\left[B(x,y)\right] \doteq \lim_{y \to x} B(x,y).$$

With this definition, $[B(x,y)]$ is a section of $VM \otimes WM$. In the following, we shall denote by unprimed indices tensorial quantities at $x$, while primed indices denote tensorial quantities at $y$. As an example, let us state the well-known *Synge rule*, proved for instance in [Chr76, Poi03].

**Lemma III.1.2.4** *Let $B$ be an arbitrary smooth bitensor. Its covariant derivatives at $x$ and $y$ are related by **Synge's rule**. Namely,*

$$[B_{;\mu'}] = [B]_{\mu} - [B_{;\mu}].$$

*Particularly, let $VM$ be a vector bundle, let $f_a$ be a local frame of $VM$ defined on $\mathcal{O} \subset M$ and let $x$, $y \in \mathcal{O}$. If $B$ is **symmetric**, i.e. the coefficients $B_{ab'}(x,y)$ of*

$$B(x,y) \doteq B^{ab'}(x,y) f_a(x) \otimes f_{b'}(y)$$

*fulfil*

$$B^{ab'}(x,y) = B^{b'a}(y,x),$$

*then*

$$[B_{;\mu'}] = [B_{;\mu}] = \frac{1}{2}[B]_{;\mu}.$$

The half squared geodesic distance is a prototype of a class of bitensors of which we shall encounter many in the following. Namely, $\sigma$ fulfils a partial differential equation (III.52) which relates its higher order derivatives to lower order ones. Hence, given the initial conditions

$$[\sigma] = 0, \qquad [\sigma_{;\mu}] = 0, \qquad [\sigma_{;\mu\nu}] = g_{\mu\nu}$$

which follow from the very definition of $\sigma$, one can compute the coinciding point limits of its higher derivatives by means of an inductive procedure, see for instance [DeWBr60, Chr76, Ful89, Poi03]. As an example, in the case of $[\sigma_{;\mu\nu\rho}]$, one differentiates (III.52) three times and then takes the coinciding point limit. Together with the already known relations, one obtains

$$[\sigma_{;\mu\nu\rho}] = 0.$$

At a level of fourth derivative, the same procedure yields a linear combination of three coinciding fourth derivatives, though with different index orders. To relate those, one has to commute derivatives to rearrange the indices in the looked-for fashion, and this ultimately leads to the appearance of Riemann curvature tensors and therefore to

$$[\sigma_{;\mu\nu\varrho\tau}] = -\frac{1}{3}(R_{\mu\varrho\nu\tau} + R_{\mu\tau\nu\varrho}).$$

A different bitensor of the abovementioned kind we shall need in the following is the *bitensor of parallel transport* $g_{\rho'}^{\mu}(x,y)$. Namely, given a geodesically convex set $\mathcal{O}$, $x$, $y \in \mathcal{O}$, and a vector $v = v^{\mu'}\partial_{\mu'}$ in $T_yM$, the parallel transport of $v$ from $y$ to $x$ along the unique geodesic in $\mathcal{O}$ connecting $x$ and $y$ is given by the vector $\tilde{v}$ in $T_xM$ with components

$$\tilde{v}^{\mu} = g_{\rho'}^{\mu}v^{\rho'}.$$

This definition of the bitensor of parallel transport entails

$$[g^{\mu}_{\ \rho'}] = \delta^{\mu}_{\ \rho}, \qquad g^{\mu}_{\ \rho';\alpha}\sigma^{;\alpha} = 0, \qquad g^{\mu}_{\ \rho'}\sigma^{;\rho'} = -\sigma^{;\mu}.$$

In fact, the first two identities can be taken as the defining partial differential equation of $g^{\mu}_{\ \rho'}$ and its initial condition (one can even show that the mentioned partial differential equation is an ordinary one). Out of these, one can obtain by the inductive procedure outlined above

$$[g^{\mu}_{\ \rho';\alpha}] = 0, \qquad [g^{\mu}_{\ \rho';\alpha\beta}] = \frac{1}{2}R^{\mu}_{\ \nu\alpha\beta}.$$

With these preparations at hand, let us now provide the explicit form of Hadamard states.

**Definition** III.1.2.5   Let $\omega_2$ be the two-point function of a state on $\mathcal{A}(M)$, let $t$ be a time function on $(M, g)$, let

$$\sigma_{\varepsilon}(x, y) \doteq \sigma(x, y) + 2i\varepsilon(t(x) - t(y)) + \varepsilon^2,$$

and let $\lambda$ be an arbitrary length scale. We say that $\omega_2$ if of **local Hadamard form** if, for every $x_0 \in M$ there exists a geodesically convex neighbourhood $\mathcal{O}$ of $x_0$ such that $\omega_2(x, y)$ on $\mathcal{O} \times \mathcal{O}$ is of the form

$$\omega_2(x, y) = \lim_{\varepsilon \downarrow 0} \frac{1}{8\pi^2} \left( \frac{u(x, y)}{\sigma_{\varepsilon}(x, y)} + v(x, y) \log\left( \frac{\sigma_{\varepsilon}(x, y)}{\lambda^2} \right) + w(x, y) \right)$$

$$\doteq \lim_{\varepsilon \downarrow 0} \frac{1}{8\pi^2} (h_{\varepsilon}(x, y) + w(x, y)).$$

Here, the **Hadamard coefficients** $u$, $v$, and $w$ are smooth, real-valued biscalars, where $v$ is given by a series expansion in $\sigma$ as

$$v = \sum_{n=0}^{\infty} v_n \sigma^n$$

with smooth biscalar coefficients $v_n$. The bidistribution $h_{\varepsilon}$ shall be called **Hadamard parametrix**, indicating that it solves the Klein-Gordon equation up to smooth terms.

Note that the above series expansion of $v$ does not necessarily converge on general smooth spacetimes, however, it is known to converge on analytic spacetimes [Gar64]. One therefore often truncates the series at a finite order $n$ and asks for the $w$ coefficient to be only of regularity $C^n$, see [KaWa91]. Moreover, the local Hadamard form is special case of the *global Hadamard form* defined for the first time in [KaWa91]. The definition of the global Hadamard form in [KaWa91] assures that there are no (spacelike) singularities in addition to the lightlike ones visible in the local form and, moreover, that the whole concept is independent of the chosen time function $t$. However, as proven by Radzikowski in [Rad96b] employing the microlocal version of the Hadamard condition, the local Hadamard form already implies the global Hadamard form on account of the fact that $\omega_2$ must be positive, have the causal propagator $\Delta$ as its antisymmetric part, and fulfil the Klein-Gordon equation in both arguments. It is exactly this last fact which serves to determine the Hadamard coefficients $u$, $v$, and $w$ by a recursive procedure.

To see this, let us omit the subscript $\varepsilon$ and the scale $\lambda$ in the following, since they do not influence the result of the follwing calculations, and let us denote by $P_x$ the Klein-Gordon operator action on the $x$-variable. Applying $P_x$ to $h$, we obtain potentially singular terms proportional to $\sigma^{-n}$ for $n = 1, 2, 3$ and to $\log \sigma$, as well as smooth terms proportional to positive powers of $\sigma$. We know, however, that the total result is smooth because $P_x(h + w) = 0$ since $\omega_2$ is a bisolution of the Klein-Gordon equation and $w$ is smooth. One possible way to achieve the smoothness of the so calculated $P_x h$ is to demand that the coefficients of the potentially singular terms are identically vanishing. Let us stress that, since we do *a priori* not know if $u$ contains positive powers of $\sigma$, the terms proportional to negative powers of $\sigma$ could in principle cancel each other to yield a smooth result. It is therefore a choice and not a necessity to require the coefficients of the inverse powers of $\sigma$ to vanish. The afore laid down line of argument does, however, not hold for the coefficients of $P_x h$ proportional to $\log \sigma$; since $u$ and $v$ are required to be smooth, they can not contain a logarithmic dependence on $\sigma$ and the terms proportional to $\log \sigma$ have to vanish necessarily.

The result of the previously described procedure are the well-known *Hadamard recursive relations*, and we shall now see how they arise explicitly. Let us therefore examine the terms $u/\sigma$ and $v \log \sigma$ individually. Starting with the latter, we have

$$P_x(v \log \sigma) = (P_x v) \log \sigma + \sum_{n=0}^{\infty} \left( v_n(\Box_x \sigma - 2 + 4n) + 2\sigma_{;}{}^{\mu} v_{n;\mu} \right) \sigma^{n-1}, \qquad \text{(III.53)}$$

where we have employed the identity (III.52). Remembering our previous discussion, we can now extract our first differential equation by requiring the coefficient of $\log \sigma$ to vanish, *i.e.*

$$P_x v = 0. \qquad \text{(III.54)}$$

To obtain further differential equations, we need to look at the terms involving $u$, *viz.*,

$$P_x \left( \frac{u}{\sigma} \right) = \frac{(P_x u)}{\sigma} - \frac{2\sigma_{;\mu} u^{\mu} + (\Box_x \sigma - 4) u}{\sigma^2},$$

which, combined with the $\sigma^{-1}$ coefficient coming from the series obtained out of differentiating $v \log \sigma$, leads us to the following two identities:

$$P_x u + 2v_{0;\mu} \sigma^{\mu} + (\Box_x \sigma - 2) v_0 = 0, \qquad \text{(III.55)}$$

$$2u_{;\mu} \sigma^{\mu} + (\Box_x \sigma - 4) u = 0, \qquad \text{(III.56)}$$

referring to the $\sigma^{-1}$ and $\sigma^{-2}$ coefficients, respectively. To obtain differential equations for the $v_n$, one observes that (III.54) implies

$$\sum_{n=0}^{\infty} (P_x v_n) \sigma^n + \sum_{l=1}^{\infty} \left( 2l v_{l;\mu} \sigma_{;}{}^{\mu} + (l\Box \sigma + 2l(l-1)) v_l \right) \sigma^{l-1} = 0,$$

and, if we require this identity to hold true at each order in $\sigma$, we get

$$P_x v_0 + 2v_{1;\mu}\sigma_;^{\ \mu} + (\Box_x \sigma) v_1 = 0, \qquad \text{(III.57)}$$

$$P_x v_n + 2(n+1)v_{n+1;\mu}\sigma_;^{\ \mu} + \left((n+1)\Box_x\sigma + 2n(n+1)\right)(v_{n+1}) = 0. \quad \forall n \geq 1 \qquad \text{(III.58)}$$

To solve these recursive partial differential equations, let us now focus on (III.56). Since the only derivative appearing in this equation is the derivative along the geodesic connecting $x$ and $y$, (III.56) is in fact an ordinary differential equation with respect to the affine parameter of the mentioned geodesic. $u$ is therefore uniquely determined once a suitable initial condition is given. Comparing the Hadamard form with the Minkowskian two-point function (III.49), the initial condition is usually chosen as

$$[u] = 1,$$

which leads to the well-known result that $u$ is given by the square root of the so-called *Van Vleck-Morette* determinant, see for instance [DeWBr60, Chr76, Ful89, Poi03]. Similarly, given $u$, the differential equation (III.55) is again an ordinary one with respect to the geodesic affine parameter, and it can be immediately integrated since taking the coinciding point limit of (III.55) and inserting the properties of $\sigma$ yield the initial condition

$$[v_0] = \frac{1}{2}[P_x u].$$

It is clear how this procedure can be iterated to obtain solutions for all $v_n$. Particularly, one obtains the initial conditions

$$[v_{n+1}] = -\frac{1}{2(n+1)(n+2)}[P_x v_n]$$

for all $n > 0$. Moreover, one finds that $u$ depends only on the local geometry of the spacetime, while the $v_n$ and, hence, $v$ depend only on the local geometry and the parameters appearing in the Klein-Gordon operator $P$, namely, the mass $m$ and the coupling to the scalar curvature $\xi$. These observations entail that the state dependence of $\omega_2$ is encoded in the smooth biscalar $w$, which furthermore has to be symmetric because it is bound to vanish in the difference of two-point functions yielding the antisymmetric causal propagator $\Delta$, *viz*.

$$\omega_2(x,y) - \overline{\omega_2(x,y)} = \omega_2(x,y) - \omega_2(y,x) = i\,\Delta(x,y).$$

More precisely, this observation ensues from the following important result obtained in [Mor99, Mor00].

**Theorem** III.1.2.6   *The Hadamard coefficients $v_n$ are symmetric biscalars.*

As easy as this result sounds, as difficult has its proof been. Namely, it required the introduction of both a local approximation of smooth Lorentzian metrics by analytic Lorentzian metrics and a local *Wick rotation* of the approximating analytic metrics to ones of Euclidean signature. By assuring that in these transformations a common geodesic neighbourhood can be maintained, the symmetry of the coefficients in the Euclidean regime, which follows by functional analytic methods, could be transported to the smooth Lorentzian case. The above theorem proves the folklore knowledge that the causal propagator $\Delta$ is locally given by

$$i\Delta = \lim_{\varepsilon \downarrow 0} \frac{1}{8\pi^2}(h_\varepsilon - h_{-\varepsilon}).$$

Even though we can in principle obtain the $v_n$ as unique solutions of ordinary differential equations, we shall only need their coinciding point limits and coinciding points limit of their derivatives in what follows. In this respect, the symmetry of the $v_n$ will prove very valuable in combination with lemma III.1.2.4. In fact, employing the Hadamard recursion relations, we find the following results [Mor03].

**Lemma III.1.2.7** *The following identities hold for the Hadamard parametrix $h(x, y)$*

$$[P_x h] = [P_y h] = 6[v_1], \quad [(P_x h)_{;\mu'}] = [(P_y h)_{;\mu'}] = 4[v_1]_{;\mu}, \quad [(P_x h)_{;\mu'}] = [(P_y h)_{;\mu}] = 2[v_1]_{;\mu}.$$

*Proof.* Let us recall our previous derivation of the Hadamard recursion relations. After inserting the found partial differential equations for $u$ and $v_0$, we find based on (III.53) that

$$P_x h = \sum_{n=1}^{\infty} \left( v_n(\Box_x \sigma - 2 + 4n) + 2\sigma_{;}{}^{\mu} v_{n;\mu} \right) \sigma^{n-1}.$$

We stress once more that, to our current knowledge, nothing assures that the above sum converges. However, as we are interested in coinciding point limits of finite order, the high summands will always vanish on account of the properties of $\sigma$, and the convergence of the infinite sum is not required for our purposes. With this in mind, let us rewrite the above expression in a more tidy way as

$$P_x h = v_1(\Box_x \sigma + 2) + 2\sigma_{;}{}^{\mu} v_{1;\mu} + O(\sigma).$$

Starting from this expression, the identities involving $P_x$ are easily computed employing the known coinciding point limits of $\sigma$, the symmetry of $v_1$, and lemma III.1.2.4. The identities involving $P_y$ then follow analogously. $\square$

It is remarkable that these rather simple computations will be essentially sufficient for the construction of a conserved stress-energy tensor of a free scalar quantum field [Mor03]. Particularly, the knowledge of the explicit form of, say, $[v_1]$ is not necessary to accomplish such a task. However, if one is interested in computing the actual backreaction of a scalar field on curved spacetimes, one needs the explicit form of $[v_1]$. One can compute this straightforwardly by the inductive procedure already mentioned at several occasions and the result is well-known, see for instance [DeFo08]. However, the necessary computations are quite involved and PhD students equipped with computer algebra systems seem to be the canonical candidates for doing such calculations. Moreover, there is a remarkable relation between the Hadamard coefficients and the so-called *DeWitt-Schwinger* coefficients, see for instance [Mor99, Mor00, DeFo06], which stem from an *a priori* completely different expansion of two-point functions. The latter have been computed for the first time in [Chr76, Chr78][8] and can also be found in many other places like, *e.g.* [DeFo06, Ful89]. We shall provide some new insights on the relation between Hadamard and DeWitt-Schwinger renormalisation in the next chapter. For now, let us state the explicit form of $[v_1]$.

$$[v_1] = \frac{m^4}{8} + \frac{(6\xi - 1)\,m^2 R}{24} + \frac{(6\xi - 1) R^2}{288} + \frac{(1 - 5\xi)\Box R}{120} - \frac{R_{\alpha\beta} R^{\alpha\beta}}{720} + \frac{R_{\alpha\beta\gamma\delta} R^{\alpha\beta\gamma\delta}}{720}$$
$$= \frac{m^4}{8} + \frac{(6\xi - 1)\,m^2 R}{24} + \frac{(6\xi - 1) R^2}{288} + \frac{(1 - 5\xi)\Box R}{120} + \frac{C_{\alpha\beta\gamma\delta} C^{\alpha\beta\gamma\delta} + R_{\alpha\beta} R^{\alpha\beta} - \frac{R^2}{3}}{720}.$$

$$(III.59)$$

Having discussed the Hadamard form to a large extent, let us state the already anticipated equivalence result obtained by Radzikowski in [Rad96a]. See also [SaVe01] for a slightly different proof, which closes a gap in the proof of [Rad96a].

**Theorem** III.1.2.8 *Let $\omega_2$ be the two-point function of a state on $\mathcal{A}(M)$. $\omega_2$ fulfils the Hadamard condition of definition III.1.2.3 if and only if it is of global Hadamard form.*

By the result of [Rad96b], that a state which is locally of Hadamard form is already of global Hadamard form, we can safely replace 'global' by 'local' in the above theorem. Moreover, from the above discussion it should be clear that the two-point functions of two Hadamard states differ by a smooth and symmetric biscalar.

In past works on (algebraic) quantum field theory in curved spacetimes, one has often considered only on quasifree Hadamard states. For non-quasifree states, a more general *microlocal spectrum condition* has been proposed in [BFK95]. Such condition requires certain wave front set properties of the higher order $n$-point functions of a non-quasifree state. However, as shown in [San08, San09a], the Hadamard condition of the two-point function of a non-quasifree state alone determines the singularity structure of all higher order $n$-point functions by the CCR. It is therefore sufficient to specify the singularity structure of $\omega_2$ also in the case of non-quasifree states. Note however, that certain technical results on the structure of Hadamard states have up to now only been proven for the quasifree case [Ver94].

Before closing the discussion of Hadamard states, let us consider the obvious question whether Hadamard states exist at all on generic spacetimes. In [FNW81], it has been proven that Hadamard states exist on ultrastatic spacetimes, these include in particular ground and KMS states. Moreover, in the same work it has been shown that this result can be generalised to arbitrary spacetimes by two steps. First, one shows that it is possible to 'deform' the past of a Cauchy surface proper to a generic spacetime in such a way that it contains a neighbourhood of a Cauchy surface of an ultrastatic spacetime. Then, one picks one of the Hadamard states known to exist in this ultrastatic region. By the time-slice axiom (lemma III.1.1.2), one knows that determining a state in a neighbourhood of a Cauchy surface already determines it on the full spacetime under consideration. Finally, one proves that this globally determined state is locally Hadamard also outside the ultrastatic region. This is achieved employing the result proved in [FSW78] (see also [KaWa91, SaVe01]) that a bidistribution which is of local Hadamard form in the neighbourhood of a Cauchy surface is already of Hadamard form on the full spacetime. The mentioned deformation argument proved to be very useful in other works of quantum field theory on curved spacetimes, and we shall also exploit it in more detail at a later time. Apart from this rather abstract existence result of [FNW81], there have also been more concrete examples of Hadamard states. A well-known example is the *Bunch-Davies state* [BuDa78, Al85] on de Sitter spacetime. Technically, it has only been known to be of local Hadamard form, but this is not a flaw in view of [Rad96b]. In the already mentioned work [DMP09b], the authors have provided Hadamardian generalisations of the Bunch-Davies state on spacetimes which are not strictly de Sitter, but only asymptotically of this form. In [Mor08] it has been proved that a distinguished state on asymptotically flat pastimes (with either $i^+$ or $i^-$) constructed by an already discussed bulk-to-boundary approach in [DMP06] is of Hadamard form and thus provides a generalisation of the Minkowski vacuum (which by now should be clear to be the prototype of a Hadamard state) to asymptotically flat spacetimes. Moreover, with similar holographic methods the Hadamard property of the *Unruh state* on Schwarzschild spacetime has recently been proven in [DMP09c]. Referring to FLRW spacetimes and apart from the already mentioned results of [DMP09b], known Hadamard states are the *states of low energy* constructed in [Olb07] (see also [DeVe09] for applications). While ground states on symmetric spacetimes have 'minimal energy' everywhere, states of low energy have minimal energy with respect to a finite time interval on the worldline of an observer, and are therefore observer-dependent. Generalisation of states of low energy are the equilibrium-like states of low 'free energy' constructed in [Küs08]. In [Pin10], it has been proved that a distinguished asymptotic ground state on NBB spacetimes is Hadamard. We shall add to this result by providing asymptotic equilibrium states of Hadamard type on NBB spacetimes (and both asymptotic ground and equilibrium states on such spacetimes for the case of Dirac fields).

---

[8]On a personal note, we would like to point out how history can repeat itself. Namely, Christensen has performed the computations whose results are stated in [Chr76, Chr78] during his PhD thesis. It seems that he also felt that such computations are an inhuman task and therefore should be done by computers. He thus developed a computer algebra package suitable for computations with bitensors which is by now a commercially successful product. Unfortunately, this package has not been at our disposal, we have therefore chosen to work with Mathematica and the free package [Ricci], suitable for performing calculations with vector bundles. The codes we have used to implement the recursive procedures and coinciding point limits are available upon request from `t.p.hack@gmx.de`.

A class of states related to Hadamard states is constituted by *adiabatic states*. These have been introduced in [Par69] and put on rigorous grounds by [LuRo90]. Effectively, they are states which approximate ground states if the curvature of the background spacetime is only slowly varying. In [JuSch01], the concept of adiabatic states has been generalised to arbitrary curved spacetimes. There, it has also been displayed in a quantitative way how adiabatic states are related to Hadamard states. Namely, an adiabatic state of a specific order $n$ has a certain *Sobolev wave front set* (in contrast to the $C^\infty$ wave front set introduced above), and, hence, loosely speaking differs from a Hadamard state by a biscalar of finite regularity $C^n$. In this sense, Hadamard states are adiabatic states of 'infinite order'.

Finally, let us remark that one can define the Hadamard form also in spacetimes with dimensions differing from 4, see for instance [SaVe01, Mor03]. Moreover, the proof of the equivalence of the concrete Hadamard form and the microlocal Hadamard condition also holds in arbitrary spacetime dimensions, as shown in [SaVe01].

### III.1.3 The Enlarged Algebra of Observables

We shall now proceed to discuss how one can apply Hadamard states to introduce a notion of normal ordering in generic curved spacetimes, *i.e.* to enlarge the algebra $\mathscr{A}(M)$ in such a way that it includes Wick polynomials of the field. It is here where the power of the deformation quantization approach of [BDF09] becomes visible. Namely, one is able to introduce a generalisation of normal ordering *without* having recourse to a state, or a related Hilbert space picture with associated creation and annihilation operators. This is achieved by encoding Wick's theorem directly as a judicious choice of a product on the algebra of observables. We stress once more that our review of the construction of [BDF09] is restricted to the simpler case of polynomial expression of the field, while [BDF09] treat also non-polynomial Wick ordered quantities.

We start by 'deforming' the algebra of classical observables $(\mathscr{C}(M), \cdot_s)$ introduced in section II.1.3. This is achieved by replacing the commutative product $\cdot_s$ by a non-commutative one encoding the CCR. To this avail, let us consider an arbitrary bidistribution $B \in \mathscr{D}'(M^2)$ and $n \geq 2$. Given an arbitrary test function $f^{(n)} \in \mathscr{D}(M^n)$, we define the *contraction operator* $\Gamma_B : \mathscr{D}(M^n) \to \mathscr{D}(M^{n-2})$ related to $B$ as

$$[\Gamma_B f^{(n)}](x_1, \cdots, x_{n-2}) \doteq \sum_{i=1}^{n-1} \sum_{j=i+1}^{n} \int_{M^2} d_g y_1 d_g y_2 \, B(y_1, y_2) \quad \times \tag{III.60}$$
$$\times \quad f^{(n)}(x_1, \cdots, y_1, x_{i+1}, \cdots, x_{j-1}, y_2, \cdots, x_{n-2}).$$

For $n < 2$ and $f^{(n)} \in \mathscr{D}(M^n)$, we set $\Gamma_B f^{(n)} = 0$. Given two elements $f = \oplus_n \frac{1}{n!} f^{(n)}$, $g = \oplus_m \frac{1}{m!} g^{(m)} \in \mathscr{C}(M)$, we define a *star product* $\star_\Delta$ on $\mathscr{C}(M)$ by setting

$$\frac{1}{n!}(f \star_\Delta g)^{(n)} \doteq \mathfrak{Sym} \left[ e^{\frac{i}{2}\Gamma_\Delta}(f \otimes g) \right]^{(n)}, \tag{III.61}$$

where $\Delta$ is the causal propagator of the Klein-Gordon operator, $\mathfrak{Sym}$ is the total symmetrisation projector, and $\left[ e^{\frac{i}{2}\Gamma_\Delta}(f \otimes g) \right]^{(n)}$ means taking the $\mathscr{D}(M^n)$-component of $e^{\frac{i}{2}\Gamma_\Delta}(f \otimes g)$. Note that, by the symmetry of elements in $\mathscr{C}(M)$, only mutual contractions between $f$ and $g$ appear. To see that the introduced product is a non-commutative extension of $\cdot_s$, let us consider the case of $f, g \in \mathscr{D}(M) \subset \mathscr{C}(M)$. We find

$$\frac{1}{2} f \star_\Delta g = f \cdot_s g \oplus \frac{i}{2} \Delta(f, g)$$

and, hence,

$$\frac{1}{2}(f \star_\Delta g - g \star_\Delta f) = i\,\Delta(f,g).$$

Furthermore, note that the star product $\star_\Delta$ is compatible with the dynamic nature of elements in $\mathscr{C}(M)$ that we have implemented by taking an appropriate quotient, see definition II.1.3.1. Altogether, it is not difficult to see that the map $\mathscr{T} : \mathscr{A}(M) \to (\mathscr{C}(M), \star_\Delta)$ defined as

$$\mathscr{T}\left[\bigoplus_n f^{(n)}\right] \doteq \bigoplus_n \frac{1}{n!} \mathfrak{Sym} f^{(n)}$$

is an algebra homomorphism. While obtaining $(\mathscr{C}(M), \star_\Delta)$ from $\mathscr{A}(M)$ is straightforward, recovering $\mathscr{A}(M)$ from $(\mathscr{C}(M), \star_\Delta)$ seems a little more difficult. Namely, $\mathscr{A}(M)$ contains arbitrary test functions and not only symmetric ones and one may wonder how to obtain non-symmetric test functions from $(\mathscr{C}(M), \star_\Delta)$. Let us briefly sketch how this can be achieved. To this avail, let $f$, $g \in \mathscr{D}(M)$ and consider $f \otimes g \in \mathscr{D}(M^2) \subset \mathscr{A}(M)$. We identify $f \otimes g$ with $f \star_\Delta g \in (\mathscr{C}(M), \star_\Delta)$, here (inverting the $1/n!$ factors in $\mathscr{T}$) viewed as $\frac{1}{2}(f \otimes g + g \otimes f) \oplus i\,\Delta(f,g) \subset \mathscr{A}(M)$. Proceeding like this, we have rearranged $f \otimes g$ in such a way that its antisymmetric part $\frac{1}{2}(f \otimes g - g \otimes f) \in \mathscr{D}(M^2)$ is stored as the 'zero order component' $i\,\Delta(f,g)$. Given a state $\omega$ on $\mathscr{A}(M)$, the evaluation of $\omega$ on either $f \otimes g$ or $f \star_\Delta g$, gives the same result and we therefore regain the full information stored in $f \otimes g$. One can show that it is possible to extend this to arbitrary elements, thus writing every tensor product of $n$ test functions in terms of symmetrised tensor products of up to $n$ functions, by employing the CCR and storing the antisymmetric part in the coefficients[9]. Note however, that this is in principle not necessary if one considers only observable elements. Namely, one would call an element $A$ in $\mathscr{A}(M)$ an observable if $A^* = A$. For $A = \phi(f)\phi(g)$ this means that

$$[\phi(f)\phi(g)]^* = \phi(\overline{g})\phi(\overline{f}) \stackrel{!}{=} \phi(f)\phi(g)$$

must be fulfilled. Hence, the observable elements of $\mathscr{A}(M)$ have to be generated in a real-linear way by real and symmetric test functions.

Let us now consider the wished-for extension of $\mathscr{A}(M)$. We shall obtain it by extending $(\mathscr{C}(M), \star_\Delta)$ and then interpret this extension as an enlargement of $\mathscr{A}(M)$ by the aforementioned equivalence of $\mathscr{A}(M)$ and $(\mathscr{C}(M), \star_\Delta)$. To this avail, we first extend the off-shell space $\mathscr{C}_0(M)$ (see definition II.1.3.1) and consider including dynamic information afterwards. Up to now, $\mathscr{C}_0(M)$ contains only rather regular objects. If we want to consider objects like $:\!\phi^2(f)\!:$, then we must include distributions of the form $f(x)\delta(x,y)$, $f \in \mathscr{D}(M)$ in the wanted extension of $\mathscr{C}_0(M)$. However, we can not include objects which are too singular. The reason to choose the following regularity condition will soon become clear. We define the distribution space

---

[9] I am grateful to Valter Moretti for pointing this out to me.

$\mathcal{E}'_V(M^n) \subset \mathcal{E}'(M^n)$ (recall $\mathscr{D}(M^n) \subset \mathcal{E}'(M^n)$) as

$$\mathcal{E}'_V(M^n) \doteq \left\{ u \in \mathcal{E}'(M^n) \mid WF(u) \subset T^*M^n \setminus \left( \bigcup_{x \in M} \left( V_x^+ \right)^n \cup \bigcup_{x \in M} \left( V_x^- \right)^n \right) \right\},$$

where $V_x^\pm$ denote the closed future and past lightcones in the fibre of the cotangent bundle at a point $x$ in $M$. Moreover, we denote the subspace of $\mathcal{E}'_V(M^n)$ constituted by symmetric elements as $\mathcal{E}'^s_V(M^n)$. We therefore demand that the wave front set of distributions in $\mathcal{E}'_V(M^n)$ does not contain points with all covectors causal and future-pointing, or all covectors causal and past-pointing. Particularly, $f(x)\delta(x,y) \in \mathcal{E}'^s_V(M^2)$, see the previous subsection. We now set

$$\mathscr{C}_{0,\text{ext}}(M) \doteq \bigoplus_{n=0}^{n} \mathcal{E}'^s_V(M^n), \tag{III.62}$$

where it is again understood that we consider only *finite* sequences of distributions. We would like to equip this space with a topology. As we have a restriction on the wave front sets of the considered distributions, we need a topology which keeps these under control. Following [BrFr00, HoWa01, HoRu01], we consider:

**Definition III.1.3.1**  *Let $\Gamma$ be an arbitrary but fixed open cone in $T^*M^n$, let*

$$\mathcal{E}'_\Gamma(M^n) \doteq \left\{ u \in \mathcal{E}'(M^n) \mid WF(u) \subset T^*M^n \setminus \Gamma \right\},$$

*and let $\{u_k\}_k$ be a sequence in $\mathcal{E}'_\Gamma(M^n)$. We say that $u_k$ converges to $u \in \mathcal{E}'_\Gamma(M^n)$, if the following three conditions are fulfilled.*

*a) $\bigcup_k \operatorname{supp} u_k$ is compact.*

*b) $u_k$ converges to $u$ weakly in the sense of distributions.*

*c) For every properly supported pseudodifferential operator $P$ with $Pu_k \in \mathscr{D}(M^n)$ for all $k$ and $Pu \in \mathscr{D}(M^n)$, $Pu_k$ converges to $Pu$ in $\mathscr{D}(M^n)$.*

*This defines the **Hörmander pseudo topology** on $\mathcal{E}'_\Gamma(M^n)$.*

We refer the reader to [Tay81] for the notion of a properly supported pseudodifferential operator and only briefly remark that such operators are generalisations of partial differential operators in the sense that one considers 'functions' of derivatives more general than polynomials. Additionally, let us point out that it is the very last item in the above definition which keeps the wave front set under control. Namely, by $Pu_k \in \mathscr{D}(M^n)$, $P$ contains information on the complete singularity structure of $u_k$. Demanding that $Pu_k \to Pu \in \mathscr{D}(M^n)$ therefore assures that $WF(u) \subset WF(u_k)$. Let us state an important result in the context of the Hörmander pseudo topology [Hör90].

**Proposition III.1.3.2**  *Let $\Gamma$ be an arbitrary but fixed open cone in $T^*M^n$ and let $u$ be an arbitrary distribution in $\mathscr{E}'_\Gamma(M^n)$. Then, there exists a sequence $\{u_k\}_k$ of test functions in $\mathscr{D}(M^n)$ which converges to $u$ in the sense of the Hörmander pseudo topology.*

This result immediately entails that $\mathscr{C}_0(M)$ is dense in $\mathscr{C}_{0,\mathrm{ext}}(M)$.

The wave front set of the causal propagator reads [Rad96a]

$$WF(\Delta) = \left\{ (x, y, k_x, -k_y) \in T^*M^2 \setminus \{0\} \mid (x, k_x) \sim (y, k_y) \right\}, \tag{III.63}$$

there is no restriction on the direction of $k_0$ in particular. Hence, the square of $\Delta$ is not well-defined and we see that the product $\star_\Delta$ is not well-defined on $\mathscr{C}_{0,\mathrm{ext}}(M)$ since the $\star_\Delta$-product of objects like $f(x)\delta(x, y)$ would immediately lead to pointwise powers of $\Delta$. This is the rather technical motivation that leads us to choose a different product than $\star_\Delta$ on $\mathscr{C}_{0,\mathrm{ext}}(M)$. On the more physical ground, we expect that we have to encode some kind of subtraction of singularities into the wanted product, in order to obtain an abstract version of normal ordering. We have introduced the concept of Hadamard states in the previous section in preparation to the here considered definition of the Wick polynomial algebra. However, we would like to avoid considering subtractions involving the two-point function of a Hadamard state because we are interested in *covariant* Wick polynomials which are *locally* defined, see the discussion in section I.4 and the next subsection. Yet, a state is an inherently non-local object, as it 'knows' the full spacetime by means of the Klein-Gordon equation [HoWa01]. Therefore, as realised in [HoWa01], one should implement normal ordering by means of the singular part $h(x, y)$ of a Hadamard state alone, as our discussion in the previous subsection had demonstrated that $h(x, y)$ depends only on local curvature terms and the constants $m, \lambda, \xi$. In view of our above presentation of the product $\star_\Delta$, one may want to introduce a $\star$-product by replacing $\Delta$ with $h$. However, here a problem arises[10]. Namely, as $h$ is in *a priori* defined as a local distribution on a geodesically convex set, it may not be possible to unambiguously define it on the full spacetime $M$. Hence, given two test functions $f$ and $g$ whose support is such that it is not contained in a common geodesic set, an expression like $h(f, g)$ is not well-defined. Therefore, if one wants to define a product by means of $h$ on $\mathscr{C}_0(M)$ or even $\mathscr{C}_{0,\mathrm{ext}}(M)$, this is only possible if one restricts *all* algebra elements to have support in a common geodesic set. Note, however, that the above considerations do not hold if we consider the two-point function $\omega_2(x, y)$ of a Hadamard state. Namely, this is a distribution which is already defined on the full spacetime $M$. If one introduces a $\star$-product by means of $\omega_2(x, y)$, this can, barring potential singularities, in principle be defined on the full algebras $\mathscr{C}_0(M)$ and $\mathscr{C}_{0,\mathrm{ext}}(M)$. With this in mind, we choose the following approach to be as close as possible to the locally covariant spirit. We first consider the extended algebra constructed out of $h$ on a single causal domain $N$, see section I.1. Recall that these sets are globally hyperbolic subsets of geodesically convex sets and can therefore be considered as a globally hyperbolic spacetime in their own. After constructing the extended algebra on $(N, g\!\restriction_N)$, we will show the well-known result that two algebras constructed in this way out of two distributions which are both of Hadamard form but differ by a non-zero smooth term are isomorphic. Hence, the construction we have introduced will already hold for $(M, g)$ if we replace $h$ by $\omega_2$, and the resulting algebra on $(M, g)$, once restricted to a causal domain $N$, will be naturally isomorphic to the extended algebra on $(N, g\restriction_N)$ constructed by means of only a local Hadamard distribution $h$.

---

[10]I am very grateful to Valter Moretti for pointing this out.

With the above considerations in mind, let in the following $(N, g\restriction_N)$ be a causal domain in $(M, g)$, a globally hyperbolic subspacetime of $(M, g)$ in particular. We choose the following new $\star$-product [BrFr00, HoWa01]. We pick a bidistribution in $\mathscr{D}'(N^2)$ which is of the Hadamard form $h$, see definition III.1.2.5. Given two elements $f = \oplus_n \frac{1}{n!} f^{(n)}$, $g = \oplus_m \frac{1}{m!} g^{(m)} \in \mathscr{E}'_V(N^n)$, we define in analogy to (III.61)

$$\frac{1}{n!}\, (f \star_h g)^{(n)} \doteq \mathfrak{Sym}\left[ e^{\Gamma_h^{\mathrm{mut}}}(f \otimes g) \right]^{(n)}, \tag{III.64}$$

where $\Gamma_h^{\mathrm{mut}}$ is defined like the contraction operator $\Gamma_h$, but with the difference that only *mutual* contractions between $f$ and $g$ are allowed. To see that the new product is still encoding the CCR, let us again consider $f$, $g \in \mathscr{D}(N) \subset \mathscr{C}_{0,\mathrm{ext}}(N)$. We find

$$\frac{1}{2} f \star_h g = f \cdot_s g \oplus h(f, g)$$

and, hence,

$$\frac{1}{2}(f \star_h g - g \star_h f) = i\Delta(f, g)$$

on account of the fact that the antisymmetric part of $h$ is given by $i\Delta$. More generally, it is not difficult to see that $(\mathscr{C}_0(N), \star_h)$ is a *deformation* of $(\mathscr{C}_0(N), \star_\Delta)$ in the sense that for arbitrary elements $f$, $g$ of $\mathscr{C}_0(N)$,

$$f \star_h g = e^{\Gamma_{h^s}}\left[ (e^{-\Gamma_{h^s}} f) \star_\Delta (e^{-\Gamma_{h^s}} g) \right],$$

where $h^s \doteq h - \frac{i}{2}\Delta$.

It remains to be shown that $\star_h$ is a sensible product on $\mathscr{C}_{0,\mathrm{ext}}(N)$, namely, that $\mathscr{C}_{0,\mathrm{ext}}(N)$ is closed with respect to $\star_h$. To see this, we note that $\left[ e^{\Gamma_h^{\mathrm{mut}}}(f \otimes g) \right]^{(n)}$ is given by sums of elements of the form

$$\int_{N^{2k}} d_g x_1 \cdots d_g x_k\, d_g y_1 \cdots d_g y_k \prod_{i=1}^{k} h(x_i, y_i)\, f^{(n)}(x_1, \cdots, x_n)\, g^{(m)}(y_1, \cdots, y_m),$$

where we stress that, on account of the symmetry of $f^{(n)}$ and $g^{(m)}$, it is sufficient to consider contractions of the first $k$ arguments. At this point, our choice of regularity of elements in $\mathscr{C}_{0,\mathrm{ext}}(N)$ is crucial. Namely, with our knowledge of the wave front set of $h$ and the wave front sets of $f^{(n)} \in \mathscr{E}'^s_V(N^n)$, $g^{(m)} \in \mathscr{E}'^s_V(N^m)$, we find by theorem III.1.2.2 that the pointwise product

$$\prod_{i=1}^{k} h(x_i, y_i) f^{(n)}(x_1, \cdots, x_n) g^{(m)}(y_1, \cdots, y_m)$$

is a well-defined distribution because the sum of the wave front sets of $\prod_{i=1}^{k} h(x_i, y_i)$ and $f^{(n)} \otimes g^{(m)}$ does manifestly not intersect the zero section. In fact, based on known wave front sets of the above factors, an application of theorem 8.2.13 in [Hör90] even yields that the integral of the above product is well-defined and gives an element of $\mathscr{E}'_V(N^{m+n-2k})$. Moreover, one can show that the product $\star_h$ is continuous with respect to the Hörmander pseudo topology, and therefore constitutes a well-defined product on $\mathscr{C}_{0,ext}(N)$ [BrFr00, HoWa01]. We subsume the above discussion in the following definition.

**Definition III.1.3.3**    *Let $N$ be a causal domain in $M$, let $\mathscr{C}_{0,ext}(N)$ be defined as in* (III.62) *with $M$ replaced by $N$ and let $\star_h$ be defined as in* (III.64)*. By*

$$\mathscr{W}_0(N) \doteq (\mathscr{C}_{0,ext}(N), \star_h)$$

*we denote the **off-shell extended algebra of observables of the quantized Klein-Gordon field**.*

Let us note that the above defined algebra can be shown to contain not only Wick polynomials, but also time ordered product of such objects [BrFr00].

Two questions are now in order. First, the Hadamard parametrix $h$ employed in the construction of $\star_h$ depends on the scale $\lambda$ which is necessary to make the argument of the logarithmic singularity dimensionless. One may thus wonder to what extent the definition of $\mathscr{W}_0(N)$ depends on this scale. Let us therefore consider two Hadamard parametrices $h$ and $h'$ constructed with two different scales $\lambda$ and $\lambda'$. It follows that $h' - h$ is given by a constant times $\log(\lambda'/\lambda)v$ and therefore smooth by the smoothness of $v$. As observed in [HoWa01], this entails that the algebras $(\mathscr{C}_{0,ext}(N), \star_h)$ and $(\mathscr{C}_{0,ext}(N), \star_{h'})$ are isomorphic via the relation

$$f \star_{h'} g = e^{\Gamma_d} \left[ (e^{-\Gamma_d} f) \star_h (e^{-\Gamma_d} g) \right],$$

with $d = h' - h$. This obviously hold also if $h'$ and $h$ are *arbitrary* distributions which are of Hadamard form and only differ by a smooth term. The second question is if and how it is possible to encode dynamical information in $\mathscr{W}_0(N)$. As we have explicitly chosen $h$ to be the state-independent part of any Hadamard two-point function $\omega_2$, $h$ does not fulfil the Klein-Gordon equation. Taking the quotient of $\mathscr{W}_0(N)$ with an ideal generated by the Klein-Gordon equation is therefore an operation which is not compatible with the product $\star_h$. However, we know that, from an abstract point of view, defining $\mathscr{W}_0(N)$ by means of $\omega_2$ instead of $h$ gives an isomorphic algebra. In such a concrete realisation, it is possible to encode the Klein-Gordon equation by the above described procedure, and the resulting on-shell algebra can be shown to fulfil the time-slice axiom as a consequence [ChFr08]. Moreover, as discussed shortly above the introduction of $\star_h$, defining $\mathscr{W}_0(N)$ by means of $\omega_2$ allows us to directly extend the definition to the full spacetime $M$.

**Definition III.1.3.4**  *Let $\mathscr{C}_{0,ext}(M)$ be defined as in (III.62), let $\omega_2$ be the two-point function of a Hadamard state, and let $\star_{\omega_2}$ be defined as in (III.64) with $h$ replaced by $\omega_2$. By*

$$\mathscr{W}(M) \doteq \left(\mathscr{C}_{0,ext}(M)/\mathscr{I}, \, \star_{\omega_2}\right)$$

*we denote the **on-shell extended algebra of observables of the quantized Klein-Gordon field**. Here, $\mathscr{I}$ is the ideal in $\left(\mathscr{C}_{0,ext}(M), \star_{\omega_2}\right)$ generated by elements of the form $\mathfrak{Sym}\, Pf$ for $f \in \mathscr{C}_{0,ext}(M)$.*

After these rather technical considerations, it is time to reap the reward and to see how the product $\star_h$ encodes Wick's theorem as promised. To this avail, let us recall that a concrete realisation of a scalar quantum field on a Hilbert space in terms of creation and annihilation operators in combination with the standard definition of normal ordering yields

$$:\phi^2(f)::\phi^2(g): \; = \; :\phi^2(f)\phi^2(g): + 4\langle:\phi(x)\phi(x):\omega_2(x,y),\, f(x)\otimes g(y)\rangle + 2\left(\omega_2(f,g)\right)^2.$$

In comparison, let us consider $\tilde{f} \doteq f(x)\delta(x,y)$, $\tilde{g} \doteq g(x)\delta(x,y) \in \mathscr{E}_V'^{\prime s}(M^2)$ (in case $h$ is not coming from a state, we implicitly assume that everything in this example happens in a causal domain $N$). A computation results in

$$\tilde{f} \star_h \tilde{g} = [\tilde{f} \star_s \tilde{g}](x_1,x_2,x_3,x_4) \oplus 4h(x_1,x_2)f(x_1)g(x_2) \oplus 2h^2(f,g)$$

and shows that the elements of $\mathscr{W}(M)$ have to be interpreted as already regularised Wick polynomials. This entails in particular that, algebraically, we have to interpret $\tilde{f}$ as representing $:\phi^2(f):$. However, it is obvious that, given a Hadamard state $\omega$, the evaluation of $\omega$ on $\tilde{f}$ can not be just the integration of $\omega_2(x,y)$ with $f(x)\delta(x,y)$, as this clearly does not give a meaningful result. Instead, one has to consider the *regularised two-point function* $:\omega_2 \doteq \omega_2 - h$; this is smooth and can be integrated with $f(x)\delta(x,y)$. Hence, in the picture presented here, $\omega(:\phi^2(f):)$ corresponds to $\langle:\omega_2:, \tilde{f}\rangle$. Generalising this, we define the regularised $n$-point functions of a *quasifree state* as

$$:\omega_n(x_1,\cdots,x_n): \doteq \begin{cases} 1 & \text{for } n = 0, \\[2mm] 0 & \text{for odd } n \text{ and} \\[2mm] \displaystyle\sum_{\pi_n \in S_n'} \prod_{i=1}^{n/2} (\omega_2 - h)\left(x_{\pi_n(2i-1)}, x_{\pi_n(2i)}\right) & \text{for even } n. \end{cases},$$

where $S_n'$ denotes the set of ordered permutations introduced in definition III.1.1.3. Given an arbitrary $f = \bigoplus_n \frac{1}{n!}f^{(n)} \in \mathscr{W}(M)$ and a quasifree Hadamard state $\omega$ on $\mathscr{A}(M)$, we then define the expectation value of $f$ in $\omega$ as

$$\omega(f) \doteq \sum_n \left\langle :\omega_n:, \frac{1}{n!}f^{(n)}\right\rangle. \tag{III.65}$$

This defines a complex-valued, normalised, linear functional on $\mathscr{W}(M)$ which extends the action of the quasifree Hadamard state $\omega$ on $\mathscr{A}(M)$ to $\mathscr{W}(M)$. To show that in this way we obtain a proper state on $\mathscr{W}(M)$, we have to show that the just defined functional is positive. But, as proved in [HoRu01], this already follows from the positivity and continuity of $\omega$ on $\mathscr{A}(M)$ as this is a dense subset of $\mathscr{W}(M)$ in the Hörmander pseudo topology. This result is a special case of theorem III.1 in [HoRu01]. There, the authors prove that continuous states on $\mathscr{W}(M)$ are in one-to-one correspondence to (not necessarily quasifree) Hadamard states on $\mathscr{A}(M)$ which in addition have smooth truncated $n$-point functions for $n \neq 2$. Without going into details, let us briefly remark that truncated $n$-point functions are defined as the 'connected parts' of the usual $n$-point functions [HoRu01]. In the case of quasifree states, they are vanishing for $n \neq 2$ and, hence, smooth in particular.

(III.65) fits nicely into the discussion of the algebra of classical observables in chapter I. There, we have mentioned that a state in classical field theory is given by a smooth function, whereas we have just seen that in the quantum case this is replaced by the smooth regularised $n$-point functions of the quantum field.

To close the discussion of the enlarged algebra of observables, let us mention two important related results. One the one hand, we have not explicitly discussed Wick polynomials including derivatives. These are of course of great importance for us, as they include the regularised stress-energy tensor. However, as discussed in [Mor03, HoWa05], the inclusion of polynomials with derivatives in $\mathscr{W}(M)$ is a tractable task, essentially because derivatives do not increase the wave front set of a distribution, see theorem III.1.2.2. On the other hand, as we shall discuss in more detail in the next chapter, the construction of Wick polynomials on curved spacetimes is inherently ambiguous. We have already seen this by realising that choosing different Hadamard forms $h$ in the regularisation procedure leads to different results, although the algebra $\mathscr{W}(M)$ is independent on the chosen $h$ on abstract grounds. A careful analysis performed in [HoWa01, HoWa05] extends this preliminary insight by showing that certain physically sensible requirements, *e.g.* appropriate scaling behaviour and specific commutation relations with the single free field, only determine Wicks products up to polynomials of lower order, where the coefficients of such polynomials are completely determined by local curvature terms, the mass $m$ and the coupling to scalar curvature $\xi$. Particularly, they find that two definitions of the Wick square $:\phi^2(x):$, $:\phi^2(x):'$ allowed by the mentioned requirements are related as

$$:\phi^2(x):' = :\phi^2(x): + \alpha R(x) + \beta m^2,$$

where $\alpha$ and $\beta$ are dimensionless constants depending on $\xi$. This is well in line with the freedom of choosing a scale $\lambda$ in the construction of the Hadamard parametrix. In fact, as already remarked, two such parametrices differ by a constant multiple of the Hadamard coefficient $v$. As the Wick square is essentially defined by subtraction of a Hadamard parametrix and a coinciding point 'limit', defining Wick squares by choosing two different scales $\lambda$ yields two objects which differ by a multiple of $[v] = [v_0]$. However, the latter can be computed by the inductive procedure explained in the previous subsection to be

$$[v_0] = \frac{1}{2}\left[m^2 + \left(\xi - \frac{1}{6}\right)R\right].$$

We thus find that a change of scale $\lambda$ in $h$ is included in the renormalisation freedom found by [HoWa01, HoWa05]. Particularly, in the case of conformally invariant scalar fields, *i.e.* $\xi = \frac{1}{6}$ and $m = 0$, one can definite the Wick square in a way which maintains conformal invariance, see [Pin09]. Finally, let us remark that the form of the renormalisation freedom appearing in the definition of Wick polynomials depends on the dimension of the spacetime, although such freedom appears in all higher spacetime dimensions as well.

### III.1.4  Locality and General Covariance

As already argued in section I.4, taking general relativity and quantum field theory serious, one should always strive to define quantum field theories on curved spacetimes by employing only local properties of the manifold and in a generally covariant way. This is even more important if one is interested in the backreaction of a quantum field on the background spacetime, as this implies that the spacetime is taken to be dynamical and, hence, undetermined in particular. The requirement of locality and general covariance is therefore not only a generalisation of flat spacetime Poincaré-invariance, but a fundamentally new concept. In [BFV03], such concept been formulated for the first time, though based on earlier works, *cf.* the introduction in [BFV03] and section I.4 in this thesis. We shall now review this notion and explain why the quantum scalar field provides and example. However, locally covariant theories can already be formulated at the classical level, see [BrFr09, BFR].

To introduce the notion of a locally covariant quantum field theory and the related concept of a locally covariant quantum field, we need a few categories in addition to the ones introduced in section I.4. By $\mathfrak{TAlg}$ we denote the category of unital topological $*$-algebras, where for two $\mathscr{A}_1$, $\mathscr{A}_2$ in $obj(\mathfrak{TAlg})$, the considered morphisms $hom_{\mathfrak{TAlg}}(\mathscr{A}_1, \mathscr{A}_2)$ are continuous, unit-preserving, injective $*$-homomorphisms. In addition, we introduce the category $\mathfrak{Test}$ of test function spaces $\mathscr{D}(M)$ of objects $(M, g)$ in $\mathfrak{Man}$, where here the morphisms $hom_{\mathfrak{Test}}(\mathscr{D}(M_1), \mathscr{D}(M_2))$ are push-forwards $\chi_*$ of the isometric embeddings $\chi : M_1 \hookrightarrow M_2$. In fact, by $\mathscr{D}$ we shall denote the functor between $\mathfrak{Man}$ and $\mathfrak{Test}$ which assigns to a spacetime $(M, g)$ in $\mathfrak{Man}$ its test function space $\mathscr{D}(M)$ and to a morphism in $\mathfrak{Man}$ its push-forward. For reasons of nomenclature, we consider $\mathfrak{TAlg}$ and $\mathfrak{Test}$ as subcategories of the category $\mathfrak{Top}$ of all topological spaces with morphisms given by continuous maps. Let us now state the first promised definition.

**Definition III.1.4.1**  *A locally covariant quantum field theory is a (covariant) functor $\mathscr{A}$ between the two categories $\mathfrak{Man}$ and $\mathfrak{TAlg}$. Namely, let us denote by $\alpha_\chi$ the mapping $\mathscr{A}(\chi)$ of a morphism $\chi$ in $\mathfrak{Man}$ to a morphism in $\mathfrak{TAlg}$ and by $\mathscr{A}(M, g)$ the mapping of an object in $\mathfrak{Man}$ to an object in*

$\mathfrak{TAlg}$, *see the following diagram.*

$$
\begin{array}{ccc}
(M_1, g_1) & \xrightarrow{\ \chi\ } & (M_2, g_2) \\[4pt]
{\scriptstyle\mathscr{A}}\Big\downarrow & & \Big\downarrow{\scriptstyle\mathscr{A}} \\[4pt]
\mathscr{A}(M_1, g_1) & \xrightarrow{\ \alpha_\chi\ } & \mathscr{A}(M_2, g_2)
\end{array}
$$

*Then, the following relations hold for all morphisms $\chi_{ij} \in hom_{\mathfrak{Man}}((M_i, g_i), (M_j, g_j))$:*

$$
\alpha_{\chi_{23}} \circ \alpha_{\chi_{12}} = \alpha_{\chi_{23} \circ \chi_{12}}, \qquad \alpha_{id_M} = id_{\mathscr{A}(M, g)}.
$$

*A locally covariant quantum field theory is called **causal** if in all cases where $\chi_i \in hom_{\mathfrak{Man}}((M_i, g_i), (M, g))$ are such that the sets $\chi_1(M_1)$ and $\chi_2(M_2)$ are spacelike separated in $(M, g)$,*

$$
\Big[\alpha_{\chi_1}(\mathscr{A}(M_1, g_1)), \alpha_{\chi_2}(\mathscr{A}(M_2, g_2))\Big] = \{0\}
$$

*in the sense that all elements in the two considered algebras are mutually commuting.*

*Finally, one says that a locally covariant quantum field theory fulfils the **time-slice axiom**, if the situation that $\chi \in hom_{\mathfrak{Man}}((M_1, g_1), (M_2, g_2))$ is such that $\chi(M_1, g_1)$ contains a Cauchy surface of $(M_2, g_2)$ entails*

$$
\alpha_\chi(\mathscr{A}(M_1, g_1)) = \mathscr{A}(M_2, g_2).
$$

The authors of [BFV03] also give the definition of a state space of a locally covariant quantum field theory and this turns out to be dual to a functor, by the duality relation between states and algebras. One therefore chooses the notation of *covariant functor* for a functor in the strict sense, and calls such a mentioned dual object a *contravariant functor*. We stress once more that the term 'local' refers to the size of spacetime regions. A locally covariant quantum field theory is such that it can be constructed on arbitrarily small (causally convex) spacetime regions without having any information on the remainder of the spacetime. In more detail, this means that the algebraic relations of observables in such small region are already fully determined by the information on this region alone. This follows by application of the above definition to the special case that $(M_1, g_1)$ is a causally convex subset of $(M_2, g_2)$.

As shown in [BFV03], the quantum field theory given by assigning the Borchers-Uhlmann algebra $\mathscr{A}(M)$ of the free Klein-Gordon field to a spacetime $(M, g)$) is a locally covariant quantum field theory[11] fulfilling causality and the time-slice axiom. This follows from the fact that the construction of $\mathscr{A}(M)$ only employs compactly supported test functions and the causal propagator $\Delta$. The latter is uniquely given on any globally hyperbolic spacetime, particularly, the causal propagator on a causally convex subset $(M_1, g_1)$ of a globally hyperbolic spacetime $(M_2, g_2)$ coincides with the restriction of the same propagator on $(M_2, g_2)$ to $(M_1, g_1)$. Finally, causality follows by the causal support properties of the causal propagator, and the time-slice axiom follows by lemma III.1.1.2.

---

[11] Strictly speaking, the authors in [BFV03] provide this example in terms of Weyl algebras, but the necessary arguments are the same.

Let us now discuss the notion of a *locally covariant quantum field*. These fields are particular observables in a locally covariant quantum field theory which transform covariantly, *i.e.* loosely speaking, as a tensor. In categorical terms, this means that they are *natural transformations* between the functors $\mathscr{D}$ and $\mathscr{A}$. We refer to [Mac98] for the notion of a natural transformation, however, its meaning in our context should be clear from the following definition.

**Definition III.1.4.2** *A **locally covariant quantum field** $\Phi$ is a natural transformation between the functors $\mathscr{D}$ and $\mathscr{A}$. Namely, for every object $(M, g)$ in $\mathfrak{Man}$ there exists a morphism $\Phi_{(M,g)} : \mathscr{D}(M, g) \to \mathscr{A}(M, g)$ in $\mathfrak{Top}$ such that, for each morphism $\chi \in hom_{\mathfrak{Man}}((M_1, g_1), (M_1, g_1))$, the following diagram commutes.*

$$
\begin{array}{ccc}
\mathscr{D}(M_1, g_1) & \xrightarrow{\Phi_{(M_1, g_1)}} & \mathscr{A}(M_1, g_1) \\
\chi_* \downarrow & & \downarrow \alpha_\chi \\
\mathscr{D}(M_2, g_2) & \xrightarrow[\Phi_{(M_2, g_2)}]{} & \mathscr{A}(M_2, g_2)
\end{array}
$$

*Particularly, this entails that*

$$
\alpha_\chi \circ \Phi_{(M_1, g_1)} = \Phi_{(M_2, g_2)} \circ \chi_* .
$$

It is easy to see that the Klein-Gordon field $\phi(f)$ is locally covariant. Namely, the remarks on the local covariance of the quantum field theory given by $\mathscr{A}(M)$ after definition III.1.4.1 entail that an isometric embedding $\chi : (M_1, g_1) \hookrightarrow (M_2, g_2)$ transforms $\phi(f)$ as

$$
\alpha_\chi \left( \phi(f) \right) = \phi(\chi_* f),
$$

or, formally,

$$
\alpha_\chi \left( \phi(x) \right) = \phi \left( \chi(x) \right).
$$

Hence, local covariance of the Klein-Gordon field entails that it transforms as a 'scalar'. While the locality and covariance of the Klein-Gordon field itself are somehow automatic, one has to take care that all extended quantities, like Wick powers and time-ordered products, maintain these good properties. In fact, as already discussed and as shown in [HoWa01], the Wick monomials $:\phi^k(f):$ of the Klein-Gordon field are also locally covariant quantum fields, provided they are defined by means of the local Hadamard parametrix $h$ and, hence, in a state-independent manner. The same holds also for field monomials including derivatives [HoWa01, Mor03] and, hence, the regularised stress-energy tensor.

For more recent works in the framework of locally covariant quantum field theory, see [FePf06, Few06, BPR05, Pin09, San08, San09a, San09b]. In [BrFr06], the concepts introduced in [BFV03] have been applied to suggest a possible background-independent formulation of perturbative quantum gravity.

Finally, let us briefly explain how the requirement of local covariance is interrelated with the freedom in the definition of Wick polynomials. Namely, a Wick polynomial is determined up to a finite number of parameters, *i.e.* constant coefficients multiplying expressions built out of local curvature invariants and the parameters of the quantum field theory appearing in the Langrangean/the equation of motion [HoWa01, HoWa05]. Let us assume we have fixed these coefficients on one spacetime, preferably by some experimental data. Then, the requirement of local covariance already determines these coefficients on *all* globally hyperbolic spacetimes (of the same dimension). Admittedly, in view of the above presentation of locality and general covariance, one might think that this holds only for spacetimes with isometric subregions (or spacetimes with conformally related subregions on account of [Pin09]). However, given two spacetimes $(M_1, g_1)$ and $(M_2, g_2)$ with not necessarily isometric subregions, one can employ the deformation argument of [FSW78] to deform, say, $(M_1, g_1)$ such that it contains a subregion isometric to $(M_2, g_2)$. As the renormalisation freedoms are constants multiplying curvature terms or dimensionful constants which maintain their form under such deformation, one can require that the mentioned constants are the same on $(M_1, g_1)$ and $(M_2, g_2)$ in a meaningful way.

## III.2 The Free Scalar Field in Null Big Bang Spacetimes

Having discussed the general treatment of the quantized Klein-Gordon field in generic spacetimes, we shall now focus on the special case of NBB spacetimes. In this respect, the first main result of our thesis will be provided: the construction of Hadamardian asymptotic equilibrium states on NBB spacetimes.

### III.2.1 The Bulk and Boundary Algebras

We have already anticipated how holographic constructions can be achieved in NBB spacetimes $(M_B, g_B)$. Namely, one first projects a field theory from the bulk $M_B$ to the boundary $\partial^- M_B$. Note that this essentially means a projection to $\mathfrak{J}^-$, as the fields mapped to the boundary decay rapidly towards $i^-$ and the latter subset of $\partial^- M_B$ therefore *a posteriori* plays a passive role, although its existence has *a priori* been necessary to derive such decay properties in the subsections II.2.2 and II.4.2. In the aforementioned two subsections, the projection of the field theories has only been provided on the level of classical fields. Here, we shall extend this to a projection of quantum field theories, where we first analyse the scalar case. Such projection will be obtained as a continuous and unit-preserving $*$-homomorphism $i_\Gamma$ mapping the Borchers-Uhlmann algebra $\mathscr{A}(M_B)$[12] of the bulk NBB spacetime to a subalgebra of a suitable Borchers-Uhlmann algebra $\mathscr{A}(\mathfrak{J}^-)$ on its boundary. In a second step, one then exploits the high symmetry of $\mathfrak{J}^-$ and the fact that $\mathscr{A}(\mathfrak{J}^-)$ is compatible with this symmetry to construct preferred ground and KMS states on $\mathscr{A}(\mathfrak{J}^-)$. These can then be pulled back via $i_\Gamma$ to obtain preferred states in the bulk, which turn out to be Hadamard and can be interpreted as asymptotic ground and thermal states. Following the above line of thought, we shall provide the bulk-to-boundary mapping $i_\Gamma$ in this subsection and discuss the related preferred states in the next one.

To this avail, we shall first define a suitable Borchers-Uhlmann algebra on $\mathfrak{J}^-$.

---

[12]The index $_B$ in $\mathscr{A}(M_B)$ serves to stress that the constructions performed in this subsection are not possible on general curved spacetimes.

**Definition III.2.1.1** *Let* $(S(\mathfrak{I}^-), s)$ *denote the boundary symplectic space, cf. subsection II.2.2 and let* $S_{\mathbb{C}}(\mathfrak{I}^-)$ *denote the complexification of* $S(\mathfrak{I}^-)$. *The* **boundary Borchers-Uhlmann algebra** $\mathscr{A}(\mathfrak{I}^-)$ *of the free scalar field is defined as*

$$\mathscr{A}(\mathfrak{I}^-) \doteq \mathscr{A}_0(\mathfrak{I}^-)/\mathscr{I},$$

*where* $\mathscr{A}_0(\mathfrak{I}^-)$ *is the direct sum*

$$\mathscr{A}_0(\mathfrak{I}^-) \doteq \bigoplus_{n=0}^{\infty} S_{\mathbb{C}}(\mathfrak{I}^-)^n$$

$(S_{\mathbb{C}}(\mathfrak{I}^-)^0 \doteq \mathbb{C})$. *Elements of* $\mathscr{A}_0(\mathfrak{I}^-)$ *are required to be finite linear combinations of tensor powers of elements in* $S_{\mathbb{C}}(\mathfrak{I}^-)$ *and* $\mathscr{A}_0(\mathfrak{I}^-)$ *is equipped with a product defined by the linear extension of the tensor product of* $\mathscr{E}((\mathfrak{I}^-)^n)$, *a* *-operation defined by the antilinear extension of* $[u^*](x_1, \cdots, x_n) = \overline{u}(x_n, \cdots, x_1)$, *and the topology defined by saying that a sequence* $\{u_k\}_k = \{\oplus_n u_k^{(n)}\}_k$ *in* $\mathscr{A}_0(\mathfrak{I}^-)$ *converges to* $u = \oplus_n u^{(n)}$ *if* $u_k^{(n)}$ *converges to* $u^{(n)}$ *for all* $n$ *in the locally convex topology of* $\mathscr{E}((\mathfrak{I}^-)^n)$ *and there exists an* $N$ *such that* $u_k^{(n)} = 0$ *for all* $n > N$ *and all* $k$. *Moreover,* $\mathscr{I}$ *is the closed* *-ideal generated by elements of the form* $-i\,s(u_1, u_2) \oplus (u_1 \otimes u_2 - u_2 \otimes u_1)$, *and* $\mathscr{A}(\mathfrak{I}^-)$ *is thought to be equipped with the product,* *-operation, and topology descending from* $\mathscr{A}_0(\mathfrak{I}^-)$.

Note that the above algebra is not necessarily closed with respect to the topology we have introduced, as it is not clear if a sequence in $S(\mathfrak{I}^-)$ converges to an element of $S(\mathfrak{I}^-)$ with respect to the locally convex topology of $\mathscr{E}(\mathfrak{I}^-)$. However, this does not pose a problem as the topology will be sufficient for our purposes.

We can now immediately state the mapping of the bulk algebra $\mathscr{A}(M_B)$ to $\mathscr{A}(\mathfrak{I}^-)$. Namely, by the discussion in subsection III.1.1, we know that we can regard $\mathscr{A}(M_B)$ either as generated by equivalence classes of test functions or as built out of elements in the symplectic space of solutions $(\mathscr{S}(M_B), \varsigma_B)$. More in detail, given a test function $f \in \mathscr{D}(M_B, \mathbb{R})$, its equivalence class $[f] \in \mathscr{D}(M_B, \mathbb{R})/(P\mathscr{D}(M_B, \mathbb{R})) = \mathscr{D}(M_B, \mathbb{R})/(\ker\Delta_B) \subset \mathscr{A}(M_B)$ corresponds in a one-to-one fashion to an element of $\mathscr{S}(M_B)$ by the properties of $\Delta_B$, namely, to $\Delta_B f$. Recalling the bulk-to-boundary map $\Gamma$, see subsection II.2.2, we can therefore define a bulk-to-boundary algebra map via the tensorialisation and subsequent complexification of

$$i_\Gamma([f]) \doteq \Gamma\Delta_B f. \tag{III.66}$$

It is clear that $i_\Gamma$ is continuous by the continuity of $\Gamma$, the latter being essentially a smooth extension of a conformally rescaled smooth function, see subsection II.2.2. Moreover, by lemma II.2.2.2, we find that the tensorialisation of $i_\Gamma$ is an injective homomorphism from $\mathscr{A}(M_B)$ to $\mathscr{A}(\mathfrak{I}^-)$, as $\Gamma : (\mathscr{S}(M_B), \varsigma_B) \to (S(\mathfrak{I}^-), s)$ preserves the relevant symplectic forms and its kernel must be trivial by the strong non-degeneracy of $s$ on $S(\mathfrak{I}^-)$. Finally, $i_\Gamma$ is obviously unit preserving if we define it as the identity on $\mathbb{C} \subset \mathscr{A}(M)$. We collect the above findings in the following proposition.

**Proposition III.2.1.2** *The map* $i_\Gamma : \mathscr{A}(M_B) \to \mathscr{A}(\mathfrak{I}^-)$ *defined by the tensorialisation of* (III.66) *is a continuous, unit-preserving, injective* *-homomorphism.*

We now recall that $\mathfrak{I}^-$ endowed with the Bondi metric $h$ has a large class of Killing-symmetries, namely, all angular dependent translations in the $v$-direction – the supertranslations – constitute such symmetries, *cf.* subsection I.3.2. As the symplectic form $s$ on $S(\mathfrak{I}^-)$ is given by an integral with respect to the canonical volume measure $dv d\mathbb{S}^2$ derived from $h$ (*cf.* subsection II.2.2), it is invariant under supertranslations and, hence, the following result ensues.

**Lemma** III.2.1.3 *Let* $\zeta : \mathbb{S}^2 \to \mathbb{R}$ *be a smooth function on the two-sphere* $\mathbb{S}^2$, *let* $u$ *be an element of* $S(\mathfrak{I}^-)$, *and let* $\theta$, $\varphi$ *denote a coordinate system on* $\mathbb{S}^2$. *The action*

$$\alpha_{\zeta\,t} : S(\mathfrak{I}^-) \to S(\mathfrak{I}^-), \quad u(v, \theta, \varphi) \mapsto u(v - \zeta(\theta, \varphi)t, \theta, \varphi)$$

*induces a* *-automorphism on* $\mathscr{A}(\mathfrak{I}^-)$. *In the case* $\zeta \equiv 1$, *we shall denote it by* $\alpha_t : \mathscr{A}(\mathfrak{I}^-) \to \mathscr{A}(\mathfrak{I}^-)$.

Note that we have chosen the minus sign in the above definition, as one would usually set

$$\alpha_t \left[ \phi(u) \right] \doteq \phi \left( \alpha_{-t} u \right)$$

in obvious notation.

Let us remark once more that the supertranslations are only a subgroup of the full symmetry group of $\mathfrak{I}^-$, namely, the BMS group, see the discussion at the end of subsection I.3.2 and [DMP06]. Moreover, the full group can be implemented as an an automorphism group on the boundary algebra, and many interesting results can be obtained, see [DMP06]. However, the supertranslations will suffice to define preferred states on $\mathscr{A}(\mathfrak{I}^-)$. This can be understood in the following simple way: in Minkowski spacetime, the vacuum state is invariant under the full Poincaré group, but one only has to require translation invariance to single it out as a preferred state [Ara99]. By lemma I.3.2.2, we know that the generators of 'translational symmetries' of an NBB spacetime, namely, the translations in the comoving spacelike directions and the translations in conformal time, are all mapped to supertranslations on the boundary. Moreover, the same holds in the case of Minkowski spacetime, seen as an asymptotically flat spacetime [DMP06], where the conformal time translation corresponds to a usual time translation. It is therefore not surprising that invariance under supertranslations alone suffices to obtain preferred BMS-invariant states, as shown in [Mor06, Mor08].

Finally, let $\omega$ be a state on $\mathscr{A}(\mathfrak{I}^-)$. We can immediately define a state on $\mathscr{A}(M_B)$ by $\omega \circ i_{\Gamma}$. Given a preferred state on $\mathscr{A}(\mathfrak{I}^-)$, one therefore obtains a preferred state on $\mathscr{A}(M_B)$. We shall now proceed to construct such states.

## III.2.2 Preferred Asymptotic Ground States and Thermal States of Hadamard Type

The discussion of one-particle Hilbert space structures (*cf.* proposition III.1.1.5) in general and the case of Minkowski spacetime in particular suggest what one has to do to obtain a preferred state on $\mathscr{A}(\mathfrak{I}^-)$. One has to select a preferred positive frequency subspace of the space of solutions, which in our case is replaced by $(S(\mathfrak{I}^-), s)$. To obtain a state which is invariant under supertranslations, one needs to consider positive frequencies with respect to $v$-translations. We are therefore lead to consider the Fourier-Plancherel transform with respect to $v$ on $\mathfrak{I}^-$. We refer the reader to [Mor08, app. C] for an account of this operation and immediately state the following definition.

**Definition III.2.2.1**   *Let $u$ be an arbitrary element of $S(\mathfrak{J}^-)$ and let $\Theta(k)$ denote the Heaviside step function, i.e. $\Theta(k) = 0$ for $k < 0$ and $\Theta(k) = 1$ for $k \geq 0$. We define the **positive frequency part** $u^+$ of $u$ as*

$$u^+(v, \theta, \varphi) \doteq \frac{1}{\sqrt{2\pi}} \int\limits_{\mathbb{R}} dk \, \Theta(k) \, \hat{u}(k, \theta, \varphi) \, e^{-ikv},$$

*where $\hat{u}(k, \theta, \varphi)$ is the $v$-Fourier-Plancherel transform of $u$ defined as*

$$\hat{u}(k, \theta, \varphi) \doteq \frac{1}{\sqrt{2\pi}} \int\limits_{\mathbb{R}} dv \, u(v, \theta, \varphi) \, e^{ikv}.$$

*Moreover, by $S(\mathfrak{J}^-)_{\mathbb{C}}^+$ we denote the **complexified space of positive frequency parts**, i.e. the complexification of $S(\mathfrak{J}^-)^+ \doteq \{u^+ \mid u \in S(\mathfrak{J}^-)\}$.*

Note that the above definition implies

$$u = u^+ + \overline{u^+}$$

for all $u \in S(\mathfrak{J}^-)$, we can therefore interpret $\overline{u^+}$ as the *negative frequency part* of $u$.

To define a state on $\mathscr{A}(\mathfrak{J}^-)$, we make the following observation. As $u \in S(\mathfrak{J}^-)$ is smooth and both such $u$ and $\partial_v u$ are square-integrable with respect to $dv\, d\mathbb{S}^2$ and to $dv$ in particular, one finds that both $\Theta(k)\hat{u}(k)$ and $\Theta(k)k\hat{u}(k)$ are square-integrable and, hence, $u^+$, $\partial_v u^+$ are square-integrable as well. Consequently, it is possible to extend the symplectic form $s$ from $S(\mathfrak{J}^-)$ to $S(\mathfrak{J}^-)_{\mathbb{C}}^+$, and we set

$$\left\langle u_1^+ \,, u_2^+ \right\rangle \doteq -i \, s \left( \overline{u_1^+}, u_2^+ \right). \tag{III.67}$$

With this definition, an application of Parseval's theorem and a short computation yield

$$\left\langle u_1^+ \,, u_2^+ \right\rangle = \int\limits_{\mathbb{R} \times \mathbb{S}^2} dk\, d\mathbb{S}^2 \, 2k \, \Theta(k) \overline{\hat{u}_1}(k, \theta, \varphi) \, \hat{u}_2(k, \theta, \varphi), \tag{III.68}$$

particularly, we see that $\langle \cdot\,, \cdot \rangle$ yields an inner product on $S(\mathfrak{J}^-)_{\mathbb{C}}^+$. Actually, one can improve these insights to obtain the following theorem, proved in [DMP06, thm 2.2].

**Theorem III.2.2.2**

*a) The symplectic form $s$ can be extended from $S(\mathfrak{J}^-)$ to $S(\mathfrak{J}^-)_{\mathbb{C}}^+$, and (III.67) defines and inner product on $S(\mathfrak{J}^-)_{\mathbb{C}}^+$.*

b) Let $\mathcal{H}$ be the Hilbert space obtained by completing $S(\mathfrak{J}^-)_{\mathbb{C}}^+$ with respect to $\langle \cdot, \cdot \rangle$. The unique complex-linear, continuous extension of $u^+ \mapsto \widehat{u^+}$ from $S(\mathfrak{J}^-)_{\mathbb{C}}^+$ to $\mathcal{H}$ is a unitary isomorphism $\mathcal{H} \to L^2((0, \infty) \times \mathbb{S}^2, 2k\,dk\,d\mathbb{S}^2)$.

c) The map $K : S(\mathfrak{J}^-) \ni u \mapsto u_+ \in \mathcal{H}$ has dense range in $\mathcal{H}$.

By the convolution theorem, one computes that $2\mathrm{Im}\langle u_1^+, u_2^+ \rangle = -s(u_1, u_2)$ and that $\mu(u_1, u_2) \doteq \mathrm{Re}\langle u_1^+, u_2^+ \rangle$ and $s$ are an inner product and a symplectic form respectively satisfying the relation (III.48). Together with the above theorem, we therefore find that $u \mapsto u_+$ defines a one-particle Hilbert space structure of a pure and quasifree state $\omega^3$ on $\mathcal{A}(\mathfrak{J}^-)$, whose two-point function $\omega_2^3$ is given as

$$\omega_2^3(u_1, u_2) \doteq \left\langle u_1^+, u_2^+ \right\rangle$$

for all $u_1, u_2 \in S(\mathfrak{J}^-)$ [DMP06]. Let us recall the $*$-automorphism $\alpha_t$ on $\mathcal{A}(\mathfrak{J}^-)$, see lemma III.2.1.3. A direct computation yields

$$\widehat{\alpha_t u}(k, \theta, \varphi) = e^{ikt}\hat{u}(k, \theta, \varphi). \tag{III.69}$$

Inserting this into (III.68), we immediately obtain the following result.

**Lemma** III.2.2.3  *The state $\omega^3$ on $\mathcal{A}(\mathfrak{J}^-)$ is invariant under $\alpha_t$, i.e.*

$$\omega^3 \circ \alpha_t = \omega^3$$

*for all $t \in \mathbb{R}$.*

It is somehow clear that the just defined state is preferred and enjoys 'energy positivity' properties that make it a ground state by construction. But one can make this precise, and in fact it has been shown in [Mor06] that the just defined state is the unique state which is invariant under $v$-translations and whose GNS Hilbert space-implementation of the $v$-translations is a strongly continuous one-parameter group with non-negative generator. For details and other related and interesting properties of $\omega^3$, we refer the reader to [Mor06].

As already anticipated, the state $\omega^3$ enjoys a further remarkable property. Namely, if we define its pull-back to $\mathcal{A}(M_B)$ by

$$\omega^B \doteq \omega^3 \circ i_\Gamma,$$

then one can show that the resulting state $\omega^B$ is Hadamard. Initially, this has been proven in [Mor08] for the case of asymptotically flat spacetimes with either $i^+$ or $i^-$. However, as realised in [Pin10], the results of [Mor08] can be directly extended to the case of NBB spacetimes. We shall now review the proof of the Hadamard property of $\omega^B$. It proceeds in several steps, the first of which is the following [Mor08, thm. 4.1.].

**Proposition III.2.2.4**  *Let $f_1$, $f_2 \in \mathscr{D}(M_B)$, let $\Delta_B$ denote the causal propagator of the conformally coupled Klein-Gordon operator (with either $m^2 = 0$ or $m^2 > 0$) on $(M_B, g_B)$, and let $\Gamma : \mathscr{S}(M_B) \to S(\mathfrak{I}^-)$ denote the bulk-to-boundary map (cf. subsection II.2.2). The two-point function $\omega_2^B$ of the quasifree state $\omega^B$ is of the following form*

$$\omega_2^B(f_1, f_2) = \lim_{\varepsilon \downarrow 0} -\frac{1}{\pi} \int\limits_{\mathbb{R}^2 \times \mathbb{S}^2} dv\, dv'\, d\mathbb{S}^2\, \frac{[\Gamma \Delta_B f_1](v, \theta, \varphi)[\Gamma \Delta_B f_2](v', \theta, \varphi)}{(v - v' - i\varepsilon)^2}$$

*and it defines a distribution in $\mathscr{D}'(M_B^2)$.*

The proof of this statement proceeds essentially exactly as in [Mor08, thm. 4.1.], we therefore only sketch the main arguments. The starting point is (III.68) with $u_i = \Gamma \Delta f_i$. One uses Parseval's theorem to transform this integral with respect to $k$ into one with respect to $v$. Thereby, the inverse Fourier-Plancherel transform of

$$2k\Theta(k)\widehat{[\Gamma \Delta_B f_2]}(k)$$

has to be computed. One would like to achieve this by the convolution theorem of the Fourier-Plancherel transform, therefore a regularising factor $\exp(-\varepsilon k)$ is inserted. Denoting the inverse Fourier-Plancherel transformation by $\mathscr{F}^{-1}$, one finds

$$\mathscr{F}^{-1}\left[2k e^{-\varepsilon k}\widehat{\Theta \Gamma \Delta_B f_2}\right](v, \theta, \varphi) = \lim_{\varepsilon \downarrow 0} \frac{1}{\pi} \int\limits_{\mathbb{R}} dv'\, \frac{\partial_{v'}[\Gamma \Delta_B f_2](v', \theta, \varphi)}{v - v' - i\varepsilon}\,.$$

From the proof of proposition II.2.2.1, we know that $\Gamma \Delta_B f_i$ vanish smoothly for large $v$. A continuity argument and a partial integration therefore lead us to

$$\omega_2^B(f_1, f_2) = \lim_{\varepsilon \downarrow 0} -\frac{1}{\pi} \int\limits_{\mathbb{R} \times \mathbb{S}^2} dv\, d\mathbb{S}^2 \int\limits_{\mathbb{R}} dv'\, \frac{[\Gamma \Delta_B f_1](v, \theta, \varphi)[\Gamma \Delta_B f_2](v', \theta, \varphi)}{(v - v' - i\varepsilon)^2}\,.$$

It therefore remains to be shown that the above expression is integrable with respect to the joint measure $dv\, dv'\, d\mathbb{S}^2$ and that the result is continuous with respect to the locally convex topology of $\mathscr{D}(M^2)$. These facts follow by combining the continuity of $\Delta_B$ with the decay properties of $\Gamma \Delta_B f_i$ discussed in the proof of proposition II.2.2.1, as these allow to bound the above integrand by simple and integrable expressions. We know state the wanted result, already found in [Pin10].

**Theorem III.2.2.5**  *The quasifree state $\omega^B$ on $\mathscr{A}(M_B)$ fulfils the Hadamard condition (cf. definition III.1.2.3) and is therefore a Hadamard state.*

The proof of this statement is quite involved, but again closely mimics the one of theorem 4.2. and proposition 4.3. in [Mor08] and we only sketch the main arguments (according to [Mor08], a proof of a similar statement in a different context can be found in [Hol00]). Namely, we recall that, by the proof of proposition II.2.2.1, $\Gamma \Delta_B f_i$ are given as

$$\Gamma \Delta_B f_i = \frac{\Omega_{\mathbb{M}}}{\Omega_B} \Delta_E \frac{\Omega_{\mathbb{M}}^3}{a^3} f_i = \frac{1}{\sqrt{1+v^2}} \Delta_E \frac{\Omega_{\mathbb{M}}^3}{a^3} f_i \doteq \widetilde{\Delta} f_i \,,$$

where $\Delta_E$ is the causal propagator of the conformally rescaled Klein-Gordon operator

$$P_E = -\Box_E + \frac{1}{6} R_E + \frac{a^2}{\Omega_{\mathbb{M}}^2} m^2$$

on the lower, slit half $(M_E^-, g_E)$ of the Einstein static universe $(M_E, g_E)$, and where we recall that the coefficient of $m^2$ is smoothly extended by 0 'beyond' $\mathfrak{J}^-$, see the proof of proposition II.2.2.1. We also refer to section I.3.2 for the explicit form of the various conformal factors, and only recall that the NBB metric $g_B$ and the Einstein static universe metric $g_E$ are related as

$$g_B = \frac{a^2}{\Omega_{\mathbb{M}}^2} g_E \,.$$

Given the above form of $\Gamma \Delta_B f_i$, we see that $\omega_2^B(x, y)$ is constituted by the composition of the distribution

$$T \doteq \lim_{\varepsilon \downarrow 0} \frac{1}{(v - v' - i\varepsilon)^2} \otimes \delta(\vec{\theta}, \vec{\theta}')_{\mathbb{S}^2 \times \mathbb{S}^2}$$

with two copies of $\widetilde{\Delta}$, where the latter have one entry restricted to $\mathfrak{J}^-$ and one restricted on $M_B$. Here, the second tensor factor in $T$ denotes the $\delta$-distribution on $\mathbb{S}^2 \times \mathbb{S}^2$ and we use the abbreviating notation $\vec{\theta} \doteq (\theta, \varphi)$. Roughly speaking, the main idea of the proof is the following. One would like to compute the wave front set of the mentioned composition of $T$ with two copies of $\widetilde{\Delta}$ using Hörmander's theorem of composition of distributions [Hör90, thm 8.2.13.]. Let us for the moment assume that this is possible. The wave front set of $\widetilde{\Delta}$ is essentially the same as the one of $\Delta_E$ because of the smoothness of the conformal factors multiplying $\Delta_E$. It therefore has the symmetric form displayed in (III.63). In contrast, $T$ has the wavefront set [Mor08]

$$WF(T) = \left\{ (v, \vec{\theta}, v', \vec{\theta}', k, k_\theta, k', k_\theta') \in T^*(\mathfrak{J}^-)^2 \mid v = v', \vec{\theta} = \vec{\theta}', 0 < k = -k', k_\theta = -k_\theta' \right\}$$
$$\cup \ \left\{ (v, \vec{\theta}, v', \vec{\theta}', k, k_\theta, k', k_\theta') \in T^*(\mathfrak{J}^-)^2 \mid \vec{\theta} = \vec{\theta}', k = k' = 0, k_\theta = -k_\theta' \right\} \,,$$

which follows from the well-known wave front sets of the $\delta$-distribution and the distribution $\lim_{\varepsilon \to 0} (x \pm i\varepsilon)^{-1}$ and where we have denoted elements of $T^* \mathbb{S}^2$ by $k_\theta$. By the properties of the wave front sets proper to the composition of distributions [Hör90, thm 8.2.13.], we see that, considering $x$ and $y$ in $\omega_2^B(x, y)$ and $WF(\Delta_E)$, potential singularities propagate from $x$ and $y$ to $\mathfrak{J}^-$ (past-pointing covector) and back (future-pointing covector) along null geodesics. On $\mathfrak{J}^-$, they continue propagating along null geodesics by $WF(T)$, but only in one direction, namely, from $v$ to $v'$. Heuristically, we can therefore imagine that the null geodesics relating $x$ to $v$, $v$ to $v'$, and $v'$ to $y$ constitute three sides of a parallelogram, their composition therefore equals the fourth side of the same parallelogram, which is firstly a null geodesic, and secondly one pointing from $x$ to $y$. This heuristic argument therefore entails that $\omega_2^B(x,y)$ is singular for null-related points $x$ and $y$ with future pointing covectors at $x$ and thus has the correct wave front set required for a two-point function of a Hadamard state. To put this heuristic argument on firm grounds, one has to be able to use [Hör90, thm 8.2.13.], which is only possible if $T$ and the two copies of $\widetilde{\Delta}$ are composed on compact subset of $\mathfrak{J}^-$. In order to assure this situation, one firstly restricts $\omega_2^B(x,y)$ to $N \times N \subset (M_B \cap M_E^-)^2$, where $N$ is of the form

$$N \doteq (u_{\min}, u_{\max}) \times (v_{\min}, v_{\max}) \times \mathbb{S}^2.$$

Note that the closure of such set has been denoted by $\diamond$ in the proof of proposition II.2.2.1, but we choose the different notation here to be in accord with the one in [Mor08] for the convenience of the reader[13]. Given such $N$, arguments similar to the one given in the proof of proposition II.2.2.1 entail that all null geodesics emanating from $N$ can only intersect $\mathfrak{J}^-$ in a compact set $[v'_{\min}, v_{\max}] \times \mathbb{S}^2 \subset \mathfrak{J}^-$ with $v'_{\min} \leq v_{\min}$. Next, one has to assure that the distributions $\widetilde{\Delta}$ can be restricted to $\mathfrak{J}^- \times N$. One finds that this is possible via [Hör90, thm 8.2.4.] as $WF(\widetilde{\Delta})$ turns out to be (co)normal to $\mathfrak{J}^-$. We now choose two Cauchy surfaces $S_1$ and $S_2$ of $M_E^-$ with the following properties: $S_1$ lies to the future of $N$, while $S_2$ lies to the past of $N$, but to the future of $i^-$. Moreover, we require that the intersection of $S_2$ with $\mathfrak{J}^-$ lies to the past of $[v'_{\min}, v_{\max}] \times \mathbb{S}^2 \subset \mathfrak{J}^-$. Let $H$ be the compact subset of $J^+(i^-, M_E^-)$ enclosed by the two Cauchy surfaces $S_1$ and $S_2$, see figure III.2. We pick a cut-off function $\chi \in \mathscr{D}(M_E)$ which is such that it is identically 1 on a neighbourhood of $H$ disjoint from $i^-$ and has values in $[0, 1]$ otherwise. By construction, supp $\chi \cap \mathfrak{J}^- \supset [v'_{\min}, v_{\max}] \times \mathbb{S}^2$. We now define

$$\mathbb{E} \doteq \chi \widetilde{\Delta}, \qquad \mathscr{E} \doteq (1-\chi)\widetilde{\Delta},$$

which entails that we can decompose $\widetilde{\Delta}$ as

$$\widetilde{\Delta} = \mathbb{E} + \mathscr{E}. \tag{III.70}$$

This entails that, provided $f_i \in \mathscr{D}(N)$, we can decompose $\omega_2^B(f_1, f_2)$ as

$$\omega_2^B(f_1, f_2) = \omega_{\mathbb{E}\mathbb{E}}(f_1, f_2) + \omega_{\mathscr{E}\mathbb{E}}(f_1, f_2) + \omega_{\mathbb{E}\mathscr{E}}(f_1, f_2) + \omega_{\mathscr{E}\mathscr{E}}(f_1, f_2),$$

$$\omega_{\mathbb{E}\mathbb{E}}(f_1, f_2) \doteq \lim_{\varepsilon \downarrow 0} -\frac{1}{\pi} \int\limits_{\mathbb{R}^2 \times \mathbb{S}^2} dv\,dv'\,d\mathbb{S}^2 \frac{[\mathbb{E}f_1](v, \theta, \varphi)\,[\mathbb{E}f_2](v', \theta, \varphi)}{(v - v' - i\varepsilon)^2},$$

$$\omega_{\mathscr{E}\mathbb{E}}(f_1, f_2) \doteq \lim_{\varepsilon \downarrow 0} -\frac{1}{\pi} \int\limits_{\mathbb{R}^2 \times \mathbb{S}^2} dv\,dv'\,d\mathbb{S}^2 \frac{[\mathscr{E}f_1](v, \theta, \varphi)\,[\mathbb{E}f_2](v', \theta, \varphi)}{(v - v' - i\varepsilon)^2},$$

---

[13]Note that, to compare our presentation with [Mor08], one has to exchange 'future' with 'past', as there the future null boundary $\mathfrak{J}^+$ (of an asymptotically flat spacetime) is considered.

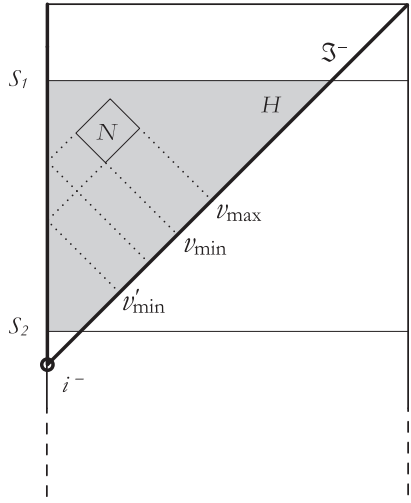

*Figure III.2:* The various geometric elements appearing in the proof of theorem III.2.2.5.

$$\omega_{\mathbb{E}\mathscr{E}}(f_1, f_2) \doteq \lim_{\varepsilon \downarrow 0} -\frac{1}{\pi} \int_{\mathbb{R}^2 \times \mathbb{S}^2} dv\, dv'\, d\mathbb{S}^2 \, \frac{[\mathbb{E}f_1](v, \theta, \varphi)\,[\mathscr{E}f_2](v', \theta, \varphi)}{(v - v' - i\varepsilon)^2},$$

$$\omega_{\mathscr{E}\mathscr{E}}(f_1, f_2) \doteq \lim_{\varepsilon \downarrow 0} -\frac{1}{\pi} \int_{\mathbb{R}^2 \times \mathbb{S}^2} dv\, dv'\, d\mathbb{S}^2 \, \frac{[\mathscr{E}f_1](v, \theta, \varphi)\,[\mathscr{E}f_2](v', \theta, \varphi)}{(v - v' - i\varepsilon)^2}.$$

One can now analyse the four pieces in the above decomposition separately. By construction, $[\mathscr{E}f_i](v, \theta, \varphi)$ are not supported in the subset $[v'_{\min}, v_{\max}] \times \mathbb{S}^2$ of $\mathfrak{I}^-$ related to $N$ by null geodesics. Hence, $\mathscr{E}$ turns out to be a smooth function on $\mathfrak{I}^- \times N$. In contrast, $\mathbb{E}$ is a distribution of compact support on $\mathfrak{I}^- \times N$. We are thus finally in the wished-for situation and can compute the wave front set of $\omega_{\mathbb{E}\mathbb{E}}(x, y)$ by [Hör90, thm 8.2.13.]. According to [Mor08], one finds

$$WF(\omega_{\mathbb{E}\mathbb{E}}) \subset \left\{ (x, y, k_x, -k_y) \in T^* N^2 \setminus \{0\} \mid (x, k_x) \sim (y, k_y),\, k_x \rhd 0 \right\},$$

which already looks very promising. Let us now consider $\omega_{\mathscr{E}\mathscr{E}}$. The fall-off behaviour of $\mathscr{E}(v, \theta, \varphi, x)$ for $x \in N$ is exactly the one in $S(\mathfrak{I}^-)$ by the support property of $(1 - \chi)$. Using this and the mentioned smoothness of $\mathscr{E}$, one finds that $\omega_{\mathscr{E}\mathscr{E}}$ is smooth and does not contribute to $WF(\omega_2^B)$. Finally, let us consider $\omega_{\mathbb{E}\mathscr{E}}$ and $\omega_{\mathscr{E}\mathbb{E}}$. Heuristically, these terms do not add singularities to $WF(\omega_2^B)$, as potential singularities in, *e.g.* $\omega_{\mathbb{E}\mathscr{E}}(x, y)$, propagate from $x$ to $\mathfrak{I}^-$, but can not propagate back to $y$ as $\mathscr{E}$ is smooth. On rigorous grounds, this is found by employing [Hör90, thm 8.2.14.]. Altogether we find

$$WF(\omega_2^B) \subset \left\{ (x, y, k_x, -k_y) \in T^* N^2 \setminus \{0\} \mid (x, k_x) \sim (y, k_y),\, k_x \rhd 0 \right\}.$$

To show equality of the two sets, one uses a proof by contradiction often employed in similar circumstances. Namely, let us assume that there are $(x, y, k_x, -k_y)$ in the set on the right hand side of the above bound which are not contained in $WF(\omega_2^B)$. $N$ is a geodesically convex set, being essentially a 'diamond' of Minkowski spacetime. Hence, given null-related $x$ and $y$ in $N$, there is a unique geodesic connecting them which entails that the covector of $k_x$ of this geodesic at $x$ is determined up to a positive constant. Moreover, as $\omega_2^B$ is the two-point function of a state on $\mathscr{A}(M_B)$, we have $\omega_2^B(x, y) - \omega_2^B(y, x) = i\Delta_B(x, y)$, where we recall

$$WF(\Delta_B) = \left\{ (x, y, k_x, -k_y) \in T^*N^2 \setminus \{0\} \mid (x, k_x) \sim (y, k_y) \right\}.$$

Particularly, $WF(\Delta_B)$ contains both elements $(x, k_x) \sim (y, k_y)$ with $k_x \triangleright 0$ and elements with $k_x \triangleleft 0$. Hence, if $WF(\omega_2^B)$ does not contain $(x, k_x) \sim (y, k_y)$ with $k_x \triangleright 0$, then it must contain $(y, k_y) \sim (x, k_x)$ with $k_y \triangleright 0$, otherwise the relation between $\omega_2^B$ and $\Delta_B$ would not be compatible with $WF(\Delta_B)$. We can now employ the *propagation of singularities* theorem, see [Hör71, Rad96a], to conclude that, since $\omega_2^B$ satisfies the Klein-Gordon equation in both arguments, its wave front set must be a union of sets of the form $B(x, k_x) \times B(y, k_y)$, where $B(x, k_x)$ is the lift of the unique geodesic through $x$ with cotangent vector $k_x$ to $T^*N$. This entails that, if $(y, k_y) \sim (x, k_x)$ with $k_y \triangleright 0$ is contained in $WF(\omega_2^B)$, then $(x, k_x) \sim (y, k_y)$ with $k_x \triangleright 0$ must be contained in the same set as well. We thus reach the wanted contradiction and find the wished-for equality of wave front sets. To extend the found Hadamard property of $WF(\omega_2^B)$ on $N \times N$ to $(M_E^- \cap M_B)^2$, we remark that $N$ has been arbitrary and that diamonds like $N$ form a base of the topology of a spacetime. Combining this with theorem III.1.2.8 and the fact that $(N, g_B\!\restriction_N)$ is a globally hyperbolic spacetime, we thus find that $\omega_2^B$ is of Hadamard form on $N$ and on a geodesically convex neighbourhood of any point in $M_E^- \cap M_B$. Moreover, the local-to-global theorem of [Rad96b] entails that $\omega_2^B$ is globally Hadamard on $M_E^- \cap M_B$. Finally, by a propagation of singularities argument in combination with the propagation of the Hadamard form as devised in the proof of [SaVe01, thm 5.8.], we find that $\omega_2^B$ is globally Hadamard on $M_B$.

We now proceed to define equilibrium states on $\mathscr{A}(\mathfrak{I}^-)$. In order to achieve this, we insert a Bose-Einstein distribution 'by hand' in (III.68). Let us stress that, while the above defined ground state was already known [Pin10], the equilibrium states we present here are new.

**Proposition** III.2.2.6  *Let $\beta \in (0, \infty)$ and $u_1$, $u_2 \in S(\mathfrak{I}^-)$ be arbitrary and let us define a bidistribution on $\omega_{\beta,2}^3$ on $S(\mathfrak{I}^-)$ by*

$$\omega_{\beta,2}^3(u_1, u_2) \doteq \int\limits_{\mathbb{R} \times \mathbb{S}^2} dk\, d\mathbb{S}^2\, 2k\, \Theta(k) \left[ \frac{\overline{\hat{u}_1(k, \theta, \varphi)}\, \hat{u}_2(k, \theta, \varphi)}{1 - e^{-\beta k}} + \frac{\overline{\hat{u}_2(k, \theta, \varphi)}\, \hat{u}_1(k, \theta, \varphi)}{e^{\beta k} - 1} \right].$$

*This bidistribution induces a quasifree, $\alpha_t$-invariant state $\omega_\beta^3$ on $\mathscr{A}(\mathfrak{I}^-)$ which fulfils the KMS condition at inverse temperature $\beta$.*

*Proof.* Let us start by checking if the above integral is well-defined. First, recall that the Fourier-Plancherel transforms $\hat{u}_i$ of $u_i$ are square-integrable. Moreover, the factors $(1-\exp(-\beta k))^{-1}$ and $(\exp(\beta k)-1)^{-1}$ are smooth and bounded everywhere except at $k=0$, but de l'Hospital's rule shows that they they are still bounded at $k=0$ after multiplication with $k$. The above integrand is therefore integrable and the whole integral well-defined. Moreover, it yields a bidistribution on $S(\mathfrak{J}^-)$ by the continuity of the Fourier-Plancherel transform. To see that it defines a quasifree state on $\mathscr{A}(\mathfrak{J}^-)$, we first compute

$$\omega^{\mathfrak{J}}_{\beta,2}(u_1,u_2)-\omega^{\mathfrak{J}}_{\beta,2}(u_2,u_1)=\omega^{\mathfrak{J}}_2(u_1,u_2)-\omega^{\mathfrak{J}}_2(u_2,u_1),$$

which shows that $\omega^{\mathfrak{J}}_{\beta,2}$ fulfils the CCR. A second simple computation proves the inequality (III.47) required for positivity and therefore assures that the two-point function under consideration induces the wanted state. $\alpha_t$-invariance of this state then follows in the same way as the result of lemma III.2.2.3.

To see that the induced state fulfils the KMS condition, we recall that, for a quasifree state on a Borchers-Uhlmann algebra, it is sufficient to check this condition for the two-point function alone. We therefore define

$$F(t)\doteq\omega^{\mathfrak{J}}_{\beta,2}\left(u_2,\alpha_t(u_1)\right),\qquad G(t)\doteq\omega^{\mathfrak{J}}_{\beta,2}\left(\alpha_t(u_1),u_2\right).$$

Using (III.69), we compute

$$F(t)=\int\limits_{\mathbb{R}\times\mathbb{S}^2}dkd\mathbb{S}^2\,2k\,\Theta(k)\left[\frac{e^{ikt}\,\overline{\hat{u}_2}(k,\theta,\varphi)\,\hat{u}_1(k,\theta,\varphi)}{1-e^{-\beta k}}+\frac{e^{-ikt}\,\overline{\hat{u}_1}(k,\theta,\varphi)\,\hat{u}_2(k,\theta,\varphi)}{e^{\beta k}-1}\right],$$

$$G(t)=\int\limits_{\mathbb{R}\times\mathbb{S}^2}dkd\mathbb{S}^2\,2k\,\Theta(k)\left[\frac{e^{-ikt}\,\overline{\hat{u}_1}(k,\theta,\varphi)\,\hat{u}_2(k,\theta,\varphi)}{1-e^{-\beta k}}+\frac{e^{ikt}\,\overline{\hat{u}_2}(k,\theta,\varphi)\,\hat{u}_1(k,\theta,\varphi)}{e^{\beta k}-1}\right].$$

We see that both $F(t)$ and $G(t)$ are continuous functions of $t$ on the full real line, as the integrands are uniformly bounded in $t$ and integrable by the arguments used in the above paragraph. Moreover, the same arguments entail that we can view $F(t)$ and $G(t)$ as Fourier-Plancherel transforms of square-integrable functions. We can therefore directly compute their Fourier-Plancherel transforms as

$$\widehat{F}(E)=\int\limits_{\mathbb{S}^2}d\mathbb{S}^2\,2E\,\Theta(E)\left[\frac{\overline{\hat{u}_2}(-E,\theta,\varphi)\,\hat{u}_1(-E,\theta,\varphi)}{1-e^{\beta E}}+\frac{\overline{\hat{u}_1}(E,\theta,\varphi)\,\hat{u}_2(E,\theta,\varphi)}{e^{\beta E}-1}\right],$$

$$\widehat{G}(E)=\int\limits_{\mathbb{S}^2}d\mathbb{S}^2\,2E\,\Theta(E)\left[\frac{\overline{\hat{u}_1}(E,\theta,\varphi)\,\hat{u}_2(E,\theta,\varphi)}{1-e^{-\beta E}}+\frac{\overline{\hat{u}_2}(-E,\theta,\varphi)\,\hat{u}_1(-E,\theta,\varphi)}{e^{-\beta E}-1}\right].$$

We find

$$\widehat{F}(E) = e^{\beta E} \widehat{G}(E)$$

which, in combination with the continuity of $F(t)$ and $G(t)$, is equivalent to the KMS condition by lemma III.1.1.4. □

Again, it seems clear how the just defined state is a preferred state on $\mathscr{A}(\mathfrak{I}^-)$. Although we shall not prove it here, we expect that it is possible to show rigorously that the just found KMS state is the unique one at inverse temperature $\beta$.

We can pull-back $\omega_\beta^3$ by $i_\Gamma$ to obtain a state

$$\omega_\beta^B \doteq \omega_\beta^3 \circ i_\Gamma$$

on $\mathscr{A}(M_B)$. To assure that this state is physically sensible, we would like it to meet the Hadamard condition already fulfilled by the related state $\omega^B$. We can expect that this is the case, as the modification we have done in the Fourier space with respect to $v$ does not alter the behaviour for large $k$. In fact, looking at the definition of $\omega_\beta^3$ in proposition III.2.2.6, we see that the occurring integrand approaches the one in (III.68) for large $k$. Let us proceed to prove this heuristic expectation on rigorous grounds.

**Theorem** III.2.2.7   *The quasifree state $\omega_\beta^B$ on $\mathscr{A}(M_B)$ fulfils the Hadamard condition and is therefore a Hadamard state.*

*Proof.* We proceed similar as in the proof of theorem III.2.2.5. Let therefore $N$ and $\widetilde{\Delta} = \mathbb{E} + \mathscr{E}$ be defined as in the mentioned proof, and let $f_1, f_2 \in \mathscr{D}(N)$. Using our knowledge of the Hadamard property of $\omega^B$ on $N \times N$, it is sufficient to show that the kernel of the distribution

$$d_\beta(f_1, f_2) \doteq \omega_\beta^B(f_1, f_2) - \omega^B(f_1, f_2)$$

is smooth. We start by computing

$$d_\beta(f_1, f_2) = \int\limits_{\mathbb{R} \times \mathbb{S}^2} dk\, d\mathbb{S}^2\, 2k\, \Theta(k) \left[ \frac{\overline{\widehat{\widetilde{\Delta} f_1}(k, \theta, \varphi)}\, \widehat{\widetilde{\Delta} f_2}(k, \theta, \varphi)}{1 - e^{-\beta k}} + \frac{\overline{\widehat{\widetilde{\Delta} f_2}(k, \theta, \varphi)}\, \widehat{\widetilde{\Delta} f_1}(k, \theta, \varphi)}{e^{\beta k} - 1} \right.$$

$$\left. - \overline{\widehat{\widetilde{\Delta} f_1}(k, \theta, \varphi)}\, \widehat{\widetilde{\Delta} f_2}(k, \theta, \varphi) \right]$$

$$= 2\,\mathrm{Re} \int\limits_{\mathbb{R} \times \mathbb{S}^2} dk\, d\mathbb{S}^2\, 2k\, \Theta(k) \frac{\overline{\widehat{\widetilde{\Delta} f_1}(k, \theta, \varphi)}\, \widehat{\widetilde{\Delta} f_2}(k, \theta, \varphi)}{e^{\beta k} - 1}$$

$$= 2\,\mathrm{Re} \int\limits_{\mathbb{R} \times \mathbb{S}^2} dv\, d\mathbb{S}^2\, \left[ \widetilde{\Delta} f_1 \right](v, \theta, \varphi)\, \mathscr{F}^{-1} \left[ \frac{2k\, \Theta\, \widehat{\widetilde{\Delta} f_2}}{e^{\beta k} - 1} \right](v, \theta, \varphi)$$

$$= \frac{2\,\mathrm{Re}}{\sqrt{2\pi}} \int\limits_{\mathbb{R} \times \mathbb{S}^2} dv\, d\mathbb{S}^2\, \left[ \widetilde{\Delta} f_1 \right](v, \theta, \varphi) \left( \mathscr{F}^{-1} \left[ \frac{2k\, \Theta}{e^{\beta k} - 1} \right] * \left[ \widetilde{\Delta} f_2 \right] \right)(v, \theta, \varphi),$$

where $f * g$ indicates convolution. Let us consider

$$R(v) \doteq \frac{1}{\sqrt{2\pi}}\, \mathscr{F}^{-1} \left[ \frac{2k\, \Theta}{e^{\beta k} - 1} \right](v).$$

A closer look reveals that it is the (inverse) Fourier-Plancherel transform of a square-integrable function. Moreover, all $v$-derivatives of the occurring integrand are again square-integrable, and we therefore find that $R(v)$ is a smooth, square-integrable function and thus falls off at large $v$. Particularly, its convolution with $\widetilde{\Delta} f_2$ is well-defined and the last line in the computation of $d_\beta(f_1, f_2)$ can be rearranged as

$$d_\beta(f_1, f_2) = 2\,\mathrm{Re} \int\limits_{\mathbb{R} \times \mathbb{S}^2} dv\, dv'\, d\mathbb{S}^2\, \left[ \widetilde{\Delta} f_1 \right](v, \theta, \varphi)\, R(v - v') \left[ \widetilde{\Delta} f_2 \right](v', \theta, \varphi).$$

Let us recall that $\widetilde{\Delta}$ can be restricted to $\mathfrak{I}^-$, as found in the proof of theorem III.2.2.5. We find that $d_\beta(x, y)$ is a compostion of the smooth, square-integrable function $R$ and two copies of $\widetilde{\Delta}$. One would now like to use [Hör90, thm 8.2.14.] to compute the wave front set of this composition. To achieve this, one has to restrict $\widetilde{\Delta}$ to a compact subset of $\mathfrak{I}^-$, and we therefore decompose $d_\beta(x, y)$ as

$$d_\beta(x, y) \doteq d_{\mathbb{E}\mathbb{E}}(x, y) + d_{\mathbb{E}\mathscr{E}}(x, y) + d_{\mathscr{E}\mathbb{E}}(x, y) + d_{\mathscr{E}\mathscr{E}}(x, y)$$

in direct analogy to the decomposition of $\omega^B$ in the proof of theorem III.2.2.5. Analysing these four terms one-by-one, we first observe that $d_{\mathscr{E}\mathscr{E}}(x,y)$ is the convolution of three smooth functions which all fall-off rapidly on $\mathfrak{I}^-$ and therefore gives a smooth kernel. We recall that, in the case of $\mathscr{E}$, this follows from the fall-off behaviour of elements in $S(\mathfrak{I}^-)$. To treat $d_{\mathbb{E}\mathscr{E}}(x,y)$ and $d_{\mathscr{E}\mathbb{E}}(x,y)$, one uses the mentioned [Hör90, thm 8.2.14.] which entails that the wave front set of these kernels is empty by the smoothness of $\mathscr{E}$ and $R$ and by the fact that the wave front set of $\widetilde{\Delta}(x,y)$ restricted to $\mathfrak{I}^- \times N$ fulfils

$$WF'(\widetilde{\Delta})_N \doteq \left\{ (y,k_y) \mid (x,y,0,-k_y) \in WF(\widetilde{\Delta}) \right\} = \emptyset,$$

as computed in [Mor08]. Essentially, this holds because all singularities in $WF(\widetilde{\Delta})$ at $x$, being the wave front set of a causal propagator of the Klein-Gordon equation on the Einstein static universe, must have propagated along a null geodesic from $\mathfrak{I}^-$, particularly, $\widetilde{\Delta}$ can not have 'isolated singularities' in $N$. Applying the same argument and in particular [Hör90, thm 8.2.14.] twice to the explicit form of $d_{\mathbb{E}\mathbb{E}}(x,y)$, one finds that this is again a smooth kernel. Altogether, we find that $d_\beta(x,y)$ is smooth on $N \times N$, which in turn implies that $\omega^B_\beta(x,y)$ is Hadamard on this set. The finalising arguments in the proof of theorem III.2.2.5 can now be repeated to find that $\omega^B_\beta(x,y)$ is Hadamard on the full NBB spacetime. □

With the rather technical machinery of microlocal analysis we have been able to prove that the found ground and KMS states on the boundary of an NBB spacetime can be pulled back to Hadamard states on its bulk. However, for actual computations and for a better interpretation of these states in the bulk spacetime, we will now make use of our preparations in section II.2.1 and relate these results to the more explicit mode-decomposition picture. To this avail, let us recall that, by lemma II.2.1.3 and lemma II.2.1.4, we can decompose every element of $\mathscr{S}(M_B)$, say, $\Delta_B f$ with $f \in \mathscr{D}(M_B)$ as

$$[\Delta_B f](\tau,\vec{x}) = i \int_{\mathbb{R}^3} d\vec{k} \, \overline{\phi_{\vec{k}}(f)} \phi_{\vec{k}}(\tau,\vec{x}) - \phi_{\vec{k}}(f) \overline{\phi_{\vec{k}}(\tau,\vec{x})}, \tag{III.71}$$

where the modes $\phi_{\vec{k}}$ are of the form

$$\phi_{\vec{k}}(\tau,\vec{x}) \doteq \frac{T_k(\tau) e^{i\vec{k}\vec{x}}}{(2\pi)^{\frac{3}{2}} a(\tau)},$$

with the $T_k(\tau)$ given by the asymptotic positive frequency modes found in lemma II.2.1.4. Particularly, the decomposition we have is *a priori* only known to be valid for large negative $\tau$, namely, in the region where we know the perturbative series defining $T_k(\tau)$ to converge. Here, the strength of the microlocal proof of the Hadamard condition met by the found preferred boundary-induced states becomes visible. Namely, since we are interested in treating the backreaction of quantum fields on NBB spacetimes, we do not want to specify the scale factor $a(\tau)$ on the full NBB spacetime beforehand, but only its asymptotic behaviour towards $\mathfrak{J}^-$. Hence, we have no *a priori* control of the mode functions $T_k$ on the full NBB spacetime, and the abstract microlocal tools seem really necessary to prove the wanted Hadamard properties by only knowing that $a(\tau)$ is smooth in the non-asymptotic region. *A posteriori*, the abstract knowledge of the Hadamard condition allows to give mode decompositions of the found preferred states on the full NBB spacetime. To this avail, we recall that the bulk-to-boundary map $\Gamma : \mathscr{S}(M_B) \to S(\mathfrak{J}^-)$ is given by a conformal rescaling and a smooth extension to $\mathfrak{J}^-$, *i.e.* a limit $u \to -\infty$. To relate the boundary construction to bulk modes, we therefore need to analyse how the mode decomposition of $[\Delta_B f]$ stated above behaves under such bulk-to-boundary map. To this end, we follow the ideas introduced in [DMP06] in order to relate constructions on the null boundary of Minkowski spacetime to the canonical mode decomposition in its bulk.

**Proposition III.2.2.8** *Let $f, g \in \mathscr{D}(M_B, \mathbb{R})$ be such that their support lies in the region of $M_B$ where the perturbative series defining the modes $T_k(\tau)$ are known to converge by lemma II.2.1.4 and let $\phi_{\vec{k}}$ be the modes resulting from this very lemma.*

*a) The mapping of $\Delta_B f$ to $S(\mathfrak{J}^-)$ is of the form*

$$[\Gamma\Delta_B f](v, \theta, \varphi) = \frac{1}{\sqrt{2\pi}} \int_{\mathbb{R}} dk\, \Theta(k) \left[ \sqrt{\frac{k}{2}}\, \overline{\phi_{-\vec{k}}(f)}\, e^{-ikv} + c.c. \right],$$

*where $\phi_{\vec{k}}(f) = \langle \phi_{\vec{k}}, f \rangle$ and $\vec{k}$ is thought to be given in polar coordinates $(k, \theta, \varphi)$.*

*b) The preferred Hadamard states $\omega^B$ and $\omega^B_\beta$ can be decomposed as*

$$\omega^B(f, g) = \int_{\mathbb{R}^3} d\vec{k}\, \overline{\phi_{\vec{k}}(f)} \phi_{\vec{k}}(g), \qquad \omega^B_\beta(f, g) = \int_{\mathbb{R}^3} d\vec{k}\, \frac{\overline{\phi_{\vec{k}}(f)}\phi_{\vec{k}}(g)}{1 - e^{-\beta k}} + \frac{\overline{\phi_{\vec{k}}(g)}\phi_{\vec{k}}(f)}{e^{\beta k} - 1},$$

*or, equivalently, as*

$$\omega^B(x, y) = \frac{1}{(2\pi)^3 a(\tau_x)a(\tau_y)} \int_{\mathbb{R}^3} d\vec{k}\, \overline{T_k(\tau_x)} T_k(\tau_y)\, e^{i\vec{k}(\vec{x}-\vec{y})},$$

$$\omega^B_\beta(x, y) = \frac{1}{(2\pi)^3 a(\tau_x)a(\tau_y)} \int_{\mathbb{R}^3} d\vec{k} \left[ \frac{\overline{T_k(\tau_x)}T_k(\tau_y)}{1 - e^{-\beta k}} + \frac{T_k(\tau_x)\overline{T_k(\tau_y)}}{e^{\beta k} - 1} \right] e^{i\vec{k}(\vec{x}-\vec{y})}.$$

*Consequently, $\omega^B$ are $\omega^B_\beta$ are homogeneous and isotropic quasifree states, and $\omega^B$ is pure.*

*Proof.* As always, we start by showing how b) emerges from a). Computing the Fourier-Plancherel transform of the result found in a), we find

$$\widehat{[\Gamma\Delta_B g]}(k, \theta, \varphi) = \sqrt{\frac{k}{2}}\, \overline{\phi_{-\vec{k}}(f)}.$$

Next, we recall

$$\omega^B(f,g) = \left\langle [\Gamma \Delta_B f]^+, [\Gamma \Delta_B g]^+ \right\rangle = \int\limits_{\mathbb{R} \times \mathbb{S}^2} dk \, d\mathbb{S}^2 \, 2k \, \Theta(k) \overline{\widehat{[\Gamma \Delta_B f]}}(k,\theta,\varphi) \, \widehat{[\Gamma \Delta_B g]}(k,\theta,\varphi).$$

Inserting the found $\widehat{[\Gamma \Delta_B g]}(k,\theta,\varphi)$ into this expression trivially leads to the asserted form of $\omega^B(f,g)$. The one of $\omega_\beta^B(f,g)$ follows analogously. The assertions that the states are homogeneous and isotropic and that $\omega^B$ is pure follow from [LuRo90, sec. 2] and [LuRo90, thm 2.3.], respectively on account of the $k$-regularity conditions satisfied by the modes $T_k(\tau)$ (see also [Olb07, DeVe09]). Essentially, it should be clear that the states are homogeneous, *i.e.* invariant under translations in the comoving spacial coordinates, as they are given as a (single) integral in 3-momentum space, where the integrands are invariant under rotations.

Let us now prove a). To start, we recall the definition of $\Gamma$ given in proposition II.2.2.1 and apply it to the mode decomposition of $\Delta_B f$ to obtain

$$\Gamma \Delta_B f = \lim_{u \to -\infty} \frac{a}{\Omega_B} \int\limits_{\mathbb{R}^3} d\vec{k} \left[ i \overline{\phi_{\vec{k}}(f)} \phi_{\vec{k}}(\tau,\vec{x}) + c.c. \right]$$

$$= \lim_{u \to -\infty} \frac{1}{\sqrt{2\pi}^3 \, \Omega_B} \int\limits_{\mathbb{R}^3} d\vec{k} \left[ i \overline{\phi_{\vec{k}}(f)} T_k(\tau) e^{i\vec{k}\vec{x}} + c.c. \right].$$

We now remind the reader that

$$\frac{1}{\Omega_B} = \frac{\sqrt{1+u^2}}{2}.$$

As we know that $\Gamma \Delta_B f$ gives a meaningful and finite result, the above integral must vanish in the limit $u \to -\infty$ and this vanishing is perfectly cancelled by the blow-up of $\sqrt{1+u^2}$. Hence, we are free to replace $\sqrt{1+u^2}$ by its large-$u$ limit $-u$. Now, we recall $u = t - r$ and $v = t + r$ with $r$ the Euclidean norm of $\vec{x}$. Moreover, we use that

$$T_k(\tau) \to \frac{1}{\sqrt{2k}} e^{-ik\tau}$$

for large negative $\tau$ and that the $k$-regularity of $T_k(\tau)$ is uniform in $\tau$ in the same $\tau$-regime, as found in lemma II.2.1.4. Therefore, we can safely replace $T_k(\tau)$ in the above integrand by its asymptotic limit. Merging the last considerations, we find

$$[\Gamma \Delta_B f](v,\theta,\varphi) = \lim_{u \to -\infty} \frac{-u}{8\sqrt{\pi}^3} \int\limits_{\mathbb{R} \times \mathbb{S}^2} dk \, d\mathbb{S}^2 \, \sqrt{k}^3 \left[ i \overline{\phi_{\vec{k}}(f)} \, e^{-\frac{i}{2}k[u(1+\cos\alpha)+v(1-\cos\alpha)]} + c.c. \right],$$

where $\alpha$ denotes the angle between $\vec{x}$ (here expressed in terms of $v$, $u$, $\theta$, $\varphi$) and $\vec{k}$. Following [DMP06], we now observe the following: by the Riemann-Lebesgue lemma, the above integral is certainly vanishing in the limit of large $u$ due to infinitely large oscillations, except when $\cos\alpha = -1$, *i.e.* when $\vec{x}$ and $\vec{k}$ are antiparellel. We thus expect that in this case the integral is proportional to $u^{-1}$, whereas it vanishes rapidly in $u$ in all other cases. To make this clear, we momentarily choose coordinates such that $\theta = 0$. Denoting the angular coordinates of $\vec{k}$ by $\theta', \varphi'$, setting $c \doteq \cos\alpha = \cos\theta'$, and inserting the well-known volume measure on the 2-sphere, we find

$$[\Gamma\Delta_B f](v,0,\varphi) = \lim_{u\to-\infty} \frac{1}{4\sqrt{\pi^3}} \int\limits_{-\infty}^{\infty}\int\limits_{0}^{2\pi}\int\limits_{-1}^{1} dk\,d\varphi'\,dc\,\sqrt{k}\left[\overline{\phi_{(k,\theta',\varphi')}(f)}\,e^{-\frac{i}{2}kv(1-c)}\partial_c e^{-\frac{i}{2}ku(1+c)} + c.c.\right].$$

We now perform a partial integration with respect to $c$, thus getting a surface term for $c = -1$, a similar term for $c = 1$, and a remaining integral with the $c$-derivative applied to $\overline{\phi_{(k,\theta',\varphi')}(f)}e^{-\frac{i}{2}kv(1-c)}$. As already remarked, the latter two of these three terms vanish by the Riemann-Lebesgue lemma, this follows particularly from the rapid decrease of $\phi_{(k,\theta',\varphi')}(f)$ in $k$ (and from the compactness of $\mathbb{S}^2$), see section II.2.1. We now remark that in the case $c = \pm 1$, *i.e.* for $\theta' \in \{0,\pi\}$, the integrand is independent of $\varphi'$ and the respective integral gives a factor of $2\pi$. Altogether, we find

$$[\Gamma\Delta_B f](v,0,\varphi) = \frac{1}{\sqrt{4\pi}}\int\limits_{\mathbb{R}} dk\,\sqrt{k}\left[\overline{\phi_{(k,0,\cdot)}(f)}\,e^{-ikv} + c.c.\right],$$

where we have 'omitted' the $\varphi$-dependence of $\phi_{(k,0,\cdot)}(f)$ to indicate the degeneracy we have integrated over. However, this degeneracy is an artifact of the 'bad'-coordinate system we have used, *i.e.* a usual pole-singularity of spherical coordinates. To resolve it, let us remark that, independently from the chose coordinate system, the only contribution of the computed integral surviving by the Riemann-Lebesgue lemma is the one where $\vec{x}$ and $\vec{k}$ are antiparallel. Hence, restoring the $\theta$ dependence in the left hand side of the found expression, the right hand side has an angular dependence obtained from the right hand side by inversion at the centre of the sphere. □

The just obtained explicit expressions for the preferred states enable us to interpret them in physical terms on the bulk spacetime. Namely, let us consider the massless case $m^2 = 0$. Then, the conformally coupled Klein-Gordon operator is a conformal transformation of the massless Minkowskian one. Particularly, the mass term in (II.30) disappears and the modes $T_k(\tau)$ are not only asymptotically of positive frequency type, but strictly and globally. This entails that the states we are considering fulfil

$$\omega^B(x,y) = \frac{\omega^{\mathbb{M}}(x,y)}{a(\tau_x)a(\tau_y)}, \qquad \omega_\beta^B(x,y) = \frac{\omega_\beta^{\mathbb{M}}(x,y)}{a(\tau_x)a(\tau_y)},$$

where $\omega^{\mathbb{M}}$ and $\omega_\beta^{\mathbb{M}}$ denote respectively the vacuum state and the KMS state of the massless scalar field on Minkowski spacetime. In the massless case, our construction therefore yields the *conformal vacuum* and *conformal equilibrium states*, see for instance [BiDa82] for a detailed discussion of these concepts. Particularly, the thermal interpretation of the conformal equilibrium state is the following. If we compute the Wick square $:\phi^2:$ in the state $\omega_\beta^B$ by subtraction of $\omega^B(x,y)$[14], we find by a simple computation (in units where the *Boltzmann constant* $k_B = 1$)

$$\omega_\beta^B(:\phi^2:) \doteq \lim_{x \to y} \left( \omega_\beta^B(x,y) - \omega^B(x,y) \right) = \frac{1}{12a^2(\tau)\beta^2} = \frac{T^2}{12a^2(\tau)}. \qquad (III.72)$$

The usual interpretation of this is that the temperature $T$ scales as $a^{-1}$ and in fact this is exactly the situation assumed in the standard cosmological picture, see for instance [Dod03]. A related interpretation is obtained by formally expanding the (represented) massless scalar field $\phi(x)$ on an NBB spacetime in terms of creation and annihilation operators $a_{\vec{k}}^\dagger$ and $a_{\vec{k}}$ as

$$\phi(\vec{x},\tau) = \frac{1}{a(\tau)} \int d\vec{k} \left( T_k(\tau) e^{i\vec{k}\vec{x}} a_{\vec{k}}^\dagger + \overline{T_k(\tau)} e^{-i\vec{k}\vec{x}} a_{\vec{k}} \right).$$

Defining the *number operator* in momentum space as $n_{\vec{k}} \doteq a_{\vec{k}}^\dagger a_{\vec{k}}$, the mode expansion of $\omega_\beta^B(x,y)$ entails that

$$\omega_\beta^B(n_{\vec{k}}) = \frac{1}{e^{\beta k} - 1}$$

which one can interpret as saying that the modes have a number spectrum which is given by the *Bose-Einstein distribution*. To reach this conclusion, one combines the interpretation of the $a^{-1}$-scaling behaviour of the temperature $T$ with the same scaling behaviour of the 'physical energy of a mode' $k/a$. The latter follows from the observation that the derivatives with respect to conformal and cosmological, *i.e.*, 'physical', time are related as $\partial_t = a^{-1}\partial_\tau$. Let us now consider the massive case. Even in Minkowski spacetime, the interpretation of the Wick square as 'thermometer' fails, because a computation shows that its expectation value computed by regularising the KMS state with the vacuum yields a complicated function of both $T$ and $m$, *cf.* [Küh03] (although the validity of KMS states for massive quantum fields in Minkowski spacetime is beyond any doubt!). We therefore expect that a computation of the Wick square as in (III.72) also yields a complicated expression depending on $T$, $m$, and $a$. However, by the asymptotic behaviour of the modes $T_k(\tau)$ for large negative $\tau$, we can compute that

$$\lim_{\tau \to -\infty} 12a^2(\tau)\omega_\beta^B(:\phi^2:) = T^2.$$

In this sense, the state $\omega_\beta^B$ in the case $m \neq 0$ can be interpreted as an *asymptotic conformal equilibrium state*. Similarly, we interpret $\omega^B$ as an *asymptotic conformal ground state*. We would like to point out that the state $\omega^B$ had already been discussed on the level of modes in [And85, And86].

Let us comment on the relation of our equilibrium-like states to past attempts to define sensible notions of equilibrium (for non-conformal fields) in the non-stationary (and therefore seemingly non-equilibrium) situation provided by cosmological spacetimes. To start with, it seams reasonable to assume that the conformal interpretation can be maintained in the massive case if the background spacetime (*i.e.* $a(\tau)$) is slowly varying [Hu82]. This interpretation is valid up to corrections proportional to curvature terms, namely, higher derivatives of $a$. In contrast, the asymptotic conformal equilibrium states we have introduced come closest to a strict notion of thermal equilibrium in the vicinity of the Big Bang, particularly, for *large* curvatures. A common paradigm related to the adiabatic picture is that, in the treatment of finite temperature effects in the early universe like *Baryogenesis* and *Leptogenesis* (see [Buc07] for a review), the interaction time scales are very large in comparison to the curvature scales; an adiabatic treatment is therefore implicitly assumed to be sensible and one 'forgets' about curvature effects. A somehow related concept is the one of a *local thermal equilibrium*. Heuristically, one expects that a thermal interpretation makes sense only locally anyways, and that a global equilibrium on the full spacetime is only an idealisation. These heuristic ideas have been formulated in conceptually clear terms in [BOR01]. Without going too much into details, let us remark that one tries to implement the idea of a local equilibrium in the sense that *thermodynamic laws* hold everywhere, but the parameters specifying the equilibrium are spacetime dependent. The explicit implementation of these ideas in the treatment of quantum fields on curved spacetimes has been pursued [BuSc06, ScVe08, Sch10], and the topic is currently under active investigation. However, it seems that the task to strictly maintain thermodynamic laws, yet with varying equilibrium parameters is a non-trivial one, as deviations from the usual thermodynamic relations appear even if one allows for varying, say, temperature. We refer to [Sch10] for recent results and note that the problem of formulating a sensible notion of thermal equilibrium in (cosmological) spacetimes seems to lack a conclusive answer to date. We also remark that a somewhat different approach to the definition of local thermal equilibrium has been taken in [Küs08]. In this work, one defines a notion of a 'generalised free energy' and obtains Hadamard states minimising this generalised free energy locally for timelike observers.

---

[14]This definition of a Wick square may seem to violate the requirements of local covariance at first glance, but this is not the case, as the Minkowskian two point function of a massless scalar field has an identically vanishing Hadamard coefficient $w$.

To close, we remark that the conformal coupling we have employed is not strictly necessary for the definition of asymptotic conformal equilibrium and vacuum states. Namely, allowing for non-conformal coupling $\xi \neq \frac{1}{6}$, the ordinary differential equation of the modes $T_k(\tau)$ changes to

$$\left(\partial_\tau^2 + k^2 + (6\xi - 1)\frac{a''}{a} + a^2 m^2\right) T_k(\tau) = 0$$

By our assumptions on the rapid decay of $a(\tau)$ for large negative $\tau$, the appearing $a''$ vanishes towards $\mathfrak{I}^-$ in the same way as $a^2$ does and a similar definition of perturbative solutions around the massless Minkowskian case is possible. Similarly, without giving a detailed proof, we remark that a proof of the Hadamard property of the related preferred states constructed on the boundary should be possible with minor modifications, especially in view of the results found in [DMP09b] in the case of asymptotically de Sitter spacetimes. Finally, by the results of [Pin10], especially theorem 3.1. in that paper, one knows how to obtain a large class of pure Hadamard states starting from $\omega^B$. Namely, it turns out that performing a *Bogoliubov transformation*, *i.e.*

replacing the modes $T_k(\tau)$ in the mode expansion of $\omega^B(x, y)$ by

$$T_k^{AB}(\tau) \doteq A(k)T_k(\tau) + B(k)\overline{T_k(\tau)}$$

where the mode normalisation condition (II.31) entails

$$|A(k)|^2 - |B(k)|^2 = 1$$

yields a Hadamard state if and only if $B(k)$ is rapidly decreasing for large $k$. We strongly expect that an analogous statement holds for the states $\omega_\beta^B$, although we shall not prove it here.

## III.3 The Free Dirac Field in General Curved Spacetimes

The last section on the quantized Klein-Gordon field in curved backgrounds should have given us a good understanding about the essential (and the dispensable) concepts necessary for the formulation and analysis of quantum field theories on curved spacetimes. In the present section we will extend the treatment to the case of quantized Dirac fields and thereby find many new (expected, yet valuable) results.

### III.3.1 The Algebras of Observables

We start our treatment of quantized Dirac fields on curved spacetimes by providing the Borchers-Uhlmann algebra of Dirac fields. As observed in previous works on the subject, see for instance [Köh95, FeVe02, San08], this is best formulated by combining the spinor and cospinor field into a single object as in section II.3.3.

**Definition III.3.1.1** *Let $D^\oplus M = DM \oplus D^*M$ denote the double spinor bundle, let $D^\oplus = D \oplus D^*$ denote the double Dirac operator, and let us introduce a **conjugation** $\varkappa$ on $\mathcal{E}(D^\oplus M)$ by setting*

$$\varkappa(f_1 \oplus f_2) \doteq f_2^\dagger \oplus f_1^\dagger$$

*for a generic element $f = f_1 \oplus f_2 \in \mathcal{E}(D^\oplus M)$ with $f_1 \in \mathcal{E}(DM)$, $f_2 \in \mathcal{E}(D^*M)$. Moreover, we define the **double causal propagator***

$$S^\oplus : \mathscr{D}(D^\oplus M) \to \mathcal{E}(D^\oplus M), \qquad S^\oplus \doteq S \oplus S_*$$

*as the sum of the causal propagators of the Dirac operator on $\mathcal{E}(DM)$ and its adjoint on $\mathcal{E}(D^*M)$ which corresponds to a distribution on $\mathscr{D}(D^\oplus M^2)$ by setting*

$$S^\oplus(f, g) \doteq \langle g_2, Sf_1\rangle - \langle S_*f_2, g_1\rangle = S(g_2, f_1) + S(f_2, g_1).$$

*The **Borchers-Uhlmann algebra** $\mathscr{A}(DM)$ of the free Dirac field is defined as*

$$\mathscr{A}(DM) \doteq \mathscr{A}_0(DM)/\mathscr{I}$$

where $\mathscr{A}_0(DM)$ is the direct sum

$$\mathscr{A}_0(DM) \doteq \bigoplus_{n=0}^{\infty} \mathscr{D}(D^{\oplus}M^n)$$

*($\mathscr{D}(D^{\oplus}M^0) \doteq \mathbb{C}$). Elements of $\mathscr{A}_0(DM)$ are required to be finite linear combinations of multi-component double test sections and $\mathscr{A}_0(DM)$ is equipped with a product defined by the linear extension of the tensor product of $\mathscr{D}(D^{\oplus}M^n)$, a $*$-operation defined by the antilinear extension of*

$$[f^*](x_1, \cdots, x_n) = [\varkappa^{\otimes n} f](x_n, \cdots, x_1),$$

*and the topology defined by saying that a sequence $\{f_k\}_k = \{\oplus_n f_k^{(n)}\}_k$ in $\mathscr{A}_0(DM)$ converges to $f = \oplus_n f^{(n)}$ if $f_k^{(n)}$ converges to $f^{(n)}$ for all $n$ in the locally convex topology of $\mathscr{D}(D^{\oplus}M^n)$ and there exists an $N$ such that $f_k^{(n)} = 0$ for all $n > N$ and all $k$. Moreover, given arbitrary $f$, $g \in \mathscr{D}(D^{\oplus}M)$, $\mathscr{I}$ is the closed $*$-ideal generated by elements of the form $-iS^{\oplus}(f,g) \oplus (f \otimes g + g \otimes f)$ and $D^{\oplus}f$, $\mathscr{A}(DM)$ is taken to be equipped with the product, $*$-operation, and topology descending from $\mathscr{A}_0(DM)$, and the fields generating $\mathscr{A}(DM)$ shall be denoted by $B(f)$.*

*If $\mathscr{O}$ is an open subset of $M$, $\mathscr{A}(D\mathscr{O})$ denotes the algebra obtained by allowing only double test sections with support in $\mathscr{O}$.*

Note that, as in the scalar case, we leave the equivalence class $[f]$ of an $f \in \mathscr{D}(D^{\oplus}M)$ implicit in the notation of the double spinor field $B(f)$. Let us remark how the single Dirac spinor and cospinor fields are elements of $\mathscr{A}(DM)$. Namely, given $f, g \in \mathscr{D}(D^{\oplus}M)$ of the form

$$f = 0 \oplus f_2, \qquad g = g_1 \oplus 0$$

with $g_1 \in \mathscr{D}(DM)$, $f_2 \in \mathscr{D}(D^*M)$, we define

$$\psi(f_2) \doteq B(f), \qquad \psi^{\dagger}(g_1) \doteq B(g).$$

We stress that a cospinor field is smeared with a spinor test section and vice versa. Related to this, one has that by the chosen quotient in the definition of $\mathscr{A}(DM)$, the spinor and cospinor fields fulfil the Dirac equation and its adjoint in the *dual* sense. Namely, $B(D^{\oplus}h) = 0$ for arbitrary $h \in \mathscr{D}(D^{\oplus}M)$ entails that

$$D\psi(f_2) \doteq \psi(D^*f_2) = 0, \qquad D^*\psi^{\dagger}(g_1) \doteq \psi^{\dagger}(Dg_1) = 0.$$

Moreover, on the level of single fields, the imposed $*$-operation on $\mathscr{A}(DM)$ has the form

$$[\psi(f_2)]^* = \psi^{\dagger}(f_2^{\dagger}), \qquad [\psi^{\dagger}(g_1)]^* = \psi(g_1^{\dagger}).$$

Finally, we point out the maybe the most important feature of $\mathscr{A}(DM)$ constituted by the *canonical anticommutation relations* (CAR). On the level of $B(f)$, we have imposed

$$\{B(f), B(g)\} \doteq B(f)B(g) + B(g)B(f) = iS^{\oplus}(f,g)$$

which for the single Dirac fields implies that

$$\left\{ \psi(f_2), \psi^\dagger(g_1) \right\} = i\, S(f_2, g_1)$$

and that all other anticommutators vanish. We recall that, by lemma II.3.2.4, these CAR are equivalent to 'equal-time' CAR on a Cauchy surface. In this respect, we also mention that lemma II.3.2.2 has the following easy consequence.

**Lemma III.3.1.2**   *The Borchers-Uhlmann algebra $\mathscr{A}(DM)$ of the free Dirac field fulfils the **time-slice axiom**. Namely, let $\Sigma$ be a Cauchy surface of $(M, g)$ and let $\mathcal{O}$ be an arbitrary (globally hyperbolic) neighbourhood of $\Sigma$. Then $\mathscr{A}(D\mathcal{O}) = \mathscr{A}(DM)$.*

The double spinor formulation has two main advantages, one of which is maybe already obvious from the definition of $\mathscr{A}(DM)$. Namely, we are interested in an algebra generated by arbitrary (tensor) products of spinor and cospinor fields, mixed ones in particular. The double spinor formulation clearly provides the most simple mean to formulate such requirement. Related to this is the fact that the double spinor formulation has enabled us to state the classical algebra of Dirac fields in section II.3.3 in terms of antisymmetric tensor products and therefore no explicit introduction of Grassmann numbers has been necessary, although this concept is somehow still implicit. To unravel the second advantage of the double spinor formulation, we remark that, on account of lemma II.3.2.3, one can define an inner product on the quotient space $\mathscr{D}(D^\oplus M)/(D^\oplus \mathscr{D}(D^\oplus M)) = \mathscr{D}(D^\oplus M)/\ker S^\oplus$ via

$$\iota^\oplus(f, g) \doteq S^\oplus(\varkappa f, g).$$

Particularly, we can complete $\mathscr{D}(D^\oplus M)$ to a Hilbert space $\mathscr{H}$ and extend $\mathscr{A}(DM)$ to an antisymmetric tensor algebra $\mathfrak{F}$ over $\mathscr{H}$. As is well-known, this turns out to be a $C^*$-algebra [Ar70, Ar70, BrRo96v2, San08], as the fields $B(f)$ can be understood as *bounded* operators acting on the antisymmetric Fock space $\mathfrak{F}$. Note however, that this is a somewhat 'abstract' Fock space, as no state has been introduced. In this sense, a *self-dual*[15] formulation of a CAR algebra, introduced in [Ar70, Ar70] and manifesting itself in $\mathscr{A}(DM)$ here, makes the $C^*$-structure of such algebra easier accessible. Notwithstanding, we stress that we shall not have recourse to these facts in the following. Particularly, we shall not employ the Hilbert space topology induced by $\iota^\oplus$, but only the locally convex topology explicitly mentioned in the above definition.

The Borchers-Uhlmann algebra $\mathscr{A}(DM)$ contains many elements we would not regard as physically observable, but we can single out the observable elements in the same way as we have done it on the level of classical fields.

**Definition III.3.1.3**   *Let $\mathfrak{E}$ be an arbitrary frame of $D^\oplus M$, let $A_{\mathfrak{E}}(\widetilde{\Lambda})$ denote the frame-dependent action of an element $\widetilde{\Lambda}$ in $Spin_0(3, 1)$ introduced in section II.3.3, and let us denote the map induced on the quotient space $\mathscr{A}(DM)$ using the same symbol. If $M$ is not simply connected and $\mathfrak{E}$ therefore not a*

---

[15]This term has been coined in [Ar70] and in our case refers to the fact that $\varkappa$ leaves $\mathscr{D}(D^\oplus M)$ invariant, whereas it does not leave the single spinor section spaces invariant.

*global frame, we interpret* $\mathfrak{E}$ *as a collection of arbitrary frames supported in local patches covering* $M$. *With this setup, we denote by* $\mathscr{A}^{obs}(DM)$ *the* **algebra of observables of the quantized Dirac field** *defined as*

$$\mathscr{A}^{obs}(DM) \doteq \left\{ f \in \mathscr{A}(DM) \,|\, A_{\mathfrak{E}}(\widetilde{\Lambda})f = f \;\forall \widetilde{\Lambda} \in Spin_0(3,1) \right\}.$$

We remark that $\mathscr{A}^{obs}(DM)$ contains elements generated by products of an *even* number of Dirac fields in particular. This is markedly important in physical terms, because general elements of $\mathscr{A}(DM)$ anticommute at spacelike separations on account of the CAR and the causal support properties of $S$. However, we would like physical observables to *commute* in such case. This requirement is fulfilled by all elements in the even subalgebra of $\mathscr{A}(DM)$, as the following result shows.

**Lemma III.3.1.4** *Let* $A_i$, $i \in \{1,2\}$ *be two elements of* $\mathscr{A}(DM)$ *which arise as finite linear combinations of an even numbers of generators* $B(f)$ *as*

$$A_i \doteq \sum_n B(f^i_{n,1}) \cdots B(f^i_{n,2k_n})$$

*such that*

$$\bigcup_{n,j} supp\, f^1_{n,j} \quad and \quad \bigcup_{n,j} supp\, f^2_{n,j}$$

*are spacelike separated. Then*

$$[A_1, A_2] = 0.$$

*Proof.* Let us start by noticing the following simple relation between the commutator and anticommutator of four algebra elements $A, B, C$ and $D$.

$$[AB, CD] = A\{B,C\}D - AC\{B,D\} - C\{A,D\}B + \{A,C\}DB. \tag{III.73}$$

Then, since only products of an even number of fields appear in $A_i$, the properties of the commutator allow to reduce $[A_1, A_2]$ to a linear combination of commutators, all of the form (III.73) with $AB$ and $CD$ of the form $B(f^1_{n,j_1})B(f^1_{n,j_2})$ and $B(f^2_{n,j_1})B(f^2_{n,j_2})$, respectively. The thesis therefore follows by the required support properties and the CAR. $\qquad\square$

In the following, we shall work with $\mathscr{A}(DM)$ for simplicity and just keep in mind how observable elements in $\mathscr{A}(DM)$ and its possible extensions can be singled out.

Let us now discuss states on $\mathscr{A}(DM)$. Some of the definitions stated in definition III.1.1.3 make sense for arbitrary $*$-algebras, and, hence, $\mathscr{A}(DM)$ in particular. These are the notion of pure and mixed states, the notion of automorphism-invariant states, and the notion of KMS states. However, regarding gauge-invariance and quasifreeness, we have to give new definitions suitable for $\mathscr{A}(DM)$. We also define charge-conjugation invariance in the way formulated by [San08].

**Definition III.3.1.5**

a) *A state $\omega$ on $\mathscr{A}(DM)$ is called **even**, if it is invariant under $B(f) \mapsto -B(f)$, i.e. it has vanishing n-point functions for all odd n.*

b) *An even state on $\mathscr{A}(DM)$ is called **quasifree** or **Gaussian** if, for all even n,*

$$\omega_n(f_1, \cdots, f_n) = \sum_{\pi_n \in S'_n} (-1)^{sign[\pi_n]} \prod_{i=1}^{n/2} \omega_2\left(f_{\pi_n(2i-1)}, f_{\pi_n(2i)}\right).$$

*Here, $S'_n$ denotes the set of ordered permutations of n elements already introduced in definition III.1.1.3, namely, the following two conditions are satisfied for $\pi_n \in S'_n$:*

$$\pi_n(2i-1) < \pi_n(2i) \quad for \quad 1 \le i \le n/2, \qquad \pi_n(2i-1) < \pi_n(2i+1) \quad for \quad 1 \le i < n/2.$$

c) *Let $^c$ denote the charge conjugation maps on $\mathscr{E}(DM)$ and $\mathscr{E}(D^*M)$ introduced in definition II.3.1.2. We extend them to $\mathscr{E}(D^{\oplus}M)$ by setting*

$$f^c \doteq f_1^c \oplus f_2^c$$

*for a generic element $f \in \mathscr{E}(D^{\oplus}M)$, and finally define an action $\alpha_C$ on $\mathscr{A}(DM)$ by the tensorialisation of*

$$\mathscr{D}(D^{\oplus}M) \ni f \mapsto \varkappa(f^c)$$

*in the canonical way. We say that a state $\omega$ on $\mathscr{A}(DM)$ is **charge-conjugation invariant** if*

$$\omega \circ \alpha_C = \omega.$$

Recall that, as discussed at the end of subsection II.3.1, the causal propagator $S$ fulfils

$$S(f^{\dagger c}, g^{\dagger c}) = S(g, f)$$

for arbitrary $f \in \mathscr{D}(DM)$, $g \in \mathscr{D}(D^*M)$ which in turn entails the identity

$$S^{\oplus}(\alpha_C f, \alpha_C g) = S^{\oplus}(f, g)$$

for any $f, g \in \mathscr{D}(D^{\oplus}M)$. We thus find that $\alpha_C$ is an automorphism of $\mathscr{A}(DM)$ [San08] and see that the above definition of a charge-conjugation invariant state makes sense. Moreover, note that the property 'even' for states of the Dirac fields is often replaced by the stronger requirement of 'gauge-invariance' formulated by asking that the state is invariant under the action which, given an arbitrary complex number $\lambda$ of unit norm, maps $\psi(f)$ to $\lambda\psi(f)$ and $\psi^{\dagger}(g)$ to $\overline{\lambda}\psi^{\dagger}(g)$. This would be equivalent to saying that the state gives zero when evaluated on the elements like $\psi(f_1)\psi(f_2)$ or $\psi^{\dagger}(g_1)\psi^{\dagger}(g_2)$. However, the given formulation is sufficient for our purposes, and we will see in the next subsection that the Hadamard condition will already force the expectation values of elements like $\psi^{\dagger}(g_1)\psi^{\dagger}(g_2)$ to be 'small' as the corresponding distribution kernels turn out to be smooth. This leads us to single out two distributions. Namely, given a state $\omega$ on $\mathscr{A}(DM)$, we define

$$\omega^+(f,g) \doteq \omega\left(\psi(g)\psi^\dagger(f)\right), \qquad \omega^-(f,g) \doteq \omega\left(\psi^\dagger(f)\psi(g)\right), \qquad \text{(III.74)}$$

where $f \in \mathscr{D}(DM)$ and $g \in \mathscr{D}(D^*M)$. $\omega^\pm$ turn out to be distributions in $\mathscr{D}'(D^*M \boxtimes DM)$ (recall our dual notational convention of distributions, *cf.* definition II.3.1.1) and clearly fulfil

$$\omega^+(f,g) + \omega^-(f,g) = iS(g,f).$$

Moreover, if $\omega$ is invariant under charge conjugations, then

$$\omega^\pm(g^{\dagger c}, f^{\dagger c}) = \omega^\mp(f,g). \qquad \text{(III.75)}$$

Finally, let us remark under which condition a bidistribution $\omega_2$ in $\mathscr{D}'(D^{\oplus}M^2)$ defines a quasifree state $\omega$ on $\mathscr{A}(DM)$. Namely, positivity of a state $\omega$ would imply $\omega_2(\tilde{f}, \varkappa\tilde{f}) \geq 0$ for all $\tilde{f} \in \mathscr{D}(D^{\oplus}M)$ and one can show that this is even a sufficient condition for a bidistribution to induce a quasifree state [BrRo96v2, p. 44]. On the level of single-spinor two-point functions, positivity is therefore assured if and only if

$$\omega^-(f_1, f_1^\dagger) + \omega^+(f_2, f_2^\dagger) + \omega\left(\psi(f_1^\dagger)\psi(f_2^\dagger)\right) + \omega\left(\psi^\dagger(f_1)\psi^\dagger(f_2)\right) \geq 0 \qquad \text{(III.76)}$$

for all $f_1, f_2 \in \mathscr{D}(DM)$.

### III.3.2 Hadamard States

We shall now discuss Hadamard states for Dirac fields. As in the scalar case, it is possible to analyse them both from the perspective of microlocal analysis and in more explicit terms via a local series in the geodesic distance. We start with the former point of view and therefore need to introduce the notion of a wave front set for vector-valued distributions. We illustrate this for the example of $\omega^\pm$. Being distributions in $\mathscr{D}'(D^*M \boxtimes DM)$, we can expand $\omega^\pm$ in a local frame as

$$\omega^\pm(x,y) = \omega^\pm{}_A{}^{B'}(x,y)\, E^A(x) \otimes E_{B'}(y),$$

where the expansion coefficients are *scalar* distributions and we recall our primed index notation of bitensorial quantities. We now define the wave front set of $\omega^\pm$ as

$$WF(\omega^\pm) \doteq \bigcup_{A=1}^4 \bigcup_{B'=1}^4 \omega^\pm{}_A{}^{B'}(x,y).$$

*A priori*, it seems possible that this definition depends on the chosen frame. However, one can show that this is not the case, see [Den82, Kra00, SaVe01]. Moreover, it is possible that some 'directions', *i.e.* components, of $\omega^\pm$ are more singular than others, and with the above definition of the wave front set we clearly loose such information. However, there is a meaningful way to retain this information by using the so-called *polarised wave front sets* introduced in [Den82]. We shall not go into details here, but we remark the following: in the treatment of Hadamard states for the scalar field, we have frequently mentioned the propagation of singularities theorem for scalar distributions which are solutions of a partial differential equation. If we have a vector-valued distribution solving a non-diagonal partial differential equation (this is the case at hand with $\omega^{\pm}$ solving the Dirac equations), then one can expect that the singularities of various components in general get mixed up in the propagation. However, as proven in [Den82], the polarised wave front set propagates meaningfully, and this has been successfully applied in [Kra00, Hol99] to compute the polarised wave front set of Hadamard states. Hence, whenever one makes statements on the wave front set of a vector-valued distribution by means of the propagation of singularities theorem, it is understood that one computes the propagation of the polarised wave front set and then projects down to the usual wave front set. Apart from this, it seems that the polarised wave front set has not found many applications in (mathematical) physics yet, although one may suspect that vector-valued distributions that seem to not allow for pointwise multiplication by means of the wave front set criterion, may very well be multiplied by their polarised wave front set properties, as certain singularities 'cancel' in this way, see [Kra00] for an example of this phenomenon.

In Minkowski spacetime, $\omega_{\mathbb{M}}^{\pm}(x,y)$ are obtained by applying suitable derivatives to the two-point function $\omega_2^{\mathbb{M}}(x,y)$ of the Klein-Gordon field (see, *e.g.* [Sch95]), *i.e.*

$$\omega_{\mathbb{M}}^{+}(x,y) = D_y^* \omega_2^{\mathbb{M}}(y,x), \qquad \omega_{\mathbb{M}}^{-}(x,y) = -D_y^* \omega_2^{\mathbb{M}}(x,y).$$

Note that the reversed order of arguments in the first equation is due to our 'reversed' definition of $\omega^+$ in (III.74). Knowing the wave front set of $\omega_2^{\mathbb{M}}$ and recalling that derivatives do not increase the wavefront set, we thus find that $\omega_{\mathbb{M}}^{-}(x,y)$ has the canonical 'positive energy' wave front set of scalar Hadamard distributions, whereas $\omega_{\mathbb{M}}^{+}(x,y)$ has the 'flipped' negative energy wave front set. It therefore seems meaningful to require the same condition for states of the Dirac field in order to assure that a sensible definition of normal ordering is also possible in this case.

**Definition III.3.2.1**   *Let $\omega$ be a state on $\mathcal{A}(DM)$. We say that $\omega$ fulfils the **Hadamard condition** and is therefore a **Hadamard state** if its two-point function $\omega_2$ fulfils*

$$WF(\omega_2) = \left\{ (x,y,k_x,-k_y) \in T^*M^2 \setminus \{0\} \mid (x,k_x) \sim (y,k_y),\ k_x \triangleright 0 \right\}.$$

*Here, $(x,k_x) \sim (y,k_y)$ implies that there exists a null geodesic $c$ connecting $x$ to $y$ such that $k_x$ is coparallel and cotangent to $c$ at $x$ and $k_y$ is the parallel transport of $k_x$ from $x$ to $y$ along $c$. Finally, $k_x \triangleright 0$ means that the covector $k_x$ is future-directed.*

We have stated the definition on the level of double spinor distributions. However, the next result shows that this is equivalent to a statement in terms of distributions on $\mathcal{D}(DM)$ and $\mathcal{D}(D^*M)$.

**Lemma III.3.2.2** *A state $\omega$ on $\mathcal{A}(DM)$ fulfils the Hadamard condition if and only if*

$$WF(\omega^\pm) = \left\{(x,y,k_x,-k_y) \in T^*M^2 \setminus \{0\} \,|\, (x,k_x) \sim (y,k_y),\, k_x \gtrless 0\right\}$$

*and*

$$\omega\left(\psi(f_2)\psi(g_2)\right), \qquad \omega\left(\psi^\dagger(f_1)\psi^\dagger(g_1)\right)$$

*are distributions with smooth kernels.*

*Proof.* Let us recall that, for a double test section of the form $f = f_1 \oplus f_2$, we have set

$$B(f) \doteq \psi^\dagger(f_1) \oplus \psi(f_2).$$

Starting from this, one computes

$$\omega_2(f,g) \doteq \omega(B(f)B(g)) = \omega^-(f_1,g_2) + \omega^+(g_1,f_2) + \omega(\psi(f_2)\psi(g_2)) + \omega\left(\psi^\dagger(f_1)\psi^\dagger(g_1)\right).$$

The 'if' direction of the thesis follows directly from this expression. To see that the remaining one is valid, we remark that the Hadamard property of $\omega$ implies

$$WF(\omega^\pm) \subset \left\{(x,y,k_x,-k_y) \in T^*M^2 \setminus \{0\} \,|\, (x,k_x) \sim (y,k_y),\, k_x \lessgtr 0\right\}.$$

To see equality of the two sets, one employs a standard argument. Namely, one recalls $\omega^+(f,g) + \omega^-(f,g) = iS(f,g)$. The wave front set of $S(f,g)$ can be computed by the methods of [Rad96a, SaVe01] (recall that $S$ is given as a Dirac derivative applied to the causal propagator of $\mathfrak{P} = D^*D$). One finds that $WF(S) = WF(\Delta)$, with $\Delta$ the causal propagator of the Klein-Gordon field. Particularly, $WF(S)$ contains all elements of the form $(x,k_x) \sim (y,k_y)$ with both $k_x \triangleright 0$ and $k_x \triangleleft 0$. From this it clearly follows that $\omega^\pm$ must have the asserted maximal wave front set in order to be compatible with the CAR. It remains to be shown that the wave front sets of the distributions given by $\omega(\psi(f_2)\psi(g_2))$ and $\omega(\psi^\dagger(f_1)\psi^\dagger(g_1))$ are empty. We show this for the case

$$\widetilde{\omega}(f_2,g_2) \doteq \omega(\psi(f_2)\psi(g_2)),$$

the remaining one follows analogously. To this end, we note that the CAR implicate

$$\widetilde{\omega}(f_2,g_2) = -\widetilde{\omega}(g_2,f_2)$$

and use a second standard argument, see for instance [Rad96a, SaVe01]. Namely, let us assume that $(x,k_x) \sim (y,k_y)$ with $k_x \triangleright 0$ is contained in the wave front set of $\widetilde{\omega}(x,y)$. Then, the above equation entails that $(x,k_x) \sim (y,k_y)$ with $k_x \triangleleft 0$ must be contained in the same set. However, the Hadamard property of $\omega$ implies that

$$WF(\widetilde{\omega}) \subset \left\{(x,y,k_x,-k_y) \in T^*M^2 \setminus \{0\} \,|\, (x,k_x) \sim (y,k_y),\, k_x \triangleright 0\right\}$$

and we reach a contradiction. The wave front set of $\widetilde{\omega}$ must therefore be empty. $\qquad\square$

We see that in general, one has to restrict the wave front set of *both* $\omega^+$ and $\omega^-$ to assure the Hadamard property of $\omega$. However, as already remarked in [Hol99], (III.75) implies that for charge-conjugation invariant states, it is sufficient to restrict only the wave front set of either $\omega^+$ or $\omega^-$.

Let us now turn to the more explicit expression of Hadamard states. Based on the aforementioned relation between the vacuum two-point functions of the Dirac and Klein-Gordon field in Minkowski spacetime, a generalisation of the (global) Hadamard form introduced in [KaWa91] to the case of Dirac fields has been given in [Köh95, Ver96]. In contrast to the scalar case, there does not seem to be a local-to-global proof like the one in [Rad96b]. However, we expect that a similar proof can be given in the case of Dirac fields and shall be content with providing only the local definition of the Hadamard form. We refer the reader to [Köh95, Ver96, SaVe01] for the full global definition.

**Definition** III.3.2.3   *Let $\omega^\pm$ be the two-point functions of a state $\omega$ on $\mathcal{A}(DM)$ defined as in* (III.74), *let $t$ be a time function on $(M, g)$, let*

$$\sigma_\varepsilon(x, y) \doteq \sigma(x, y) + 2i\varepsilon(t(x) - t(y)) + \varepsilon^2,$$

*and let $\lambda$ be an arbitrary length scale. We say that $\omega$ is of **local Hadamard form** if, for every $x_0 \in M$ there exists a geodesically convex neighbourhood $\mathcal{O}$ of $x_0$ such that $\omega^\pm(x, y)$ on $\mathcal{O} \times \mathcal{O}$ are of the form*

$$\omega^\pm(x, y) = \lim_{\varepsilon \downarrow 0} \pm \frac{1}{8\pi^2} D_y^* \left( \frac{U(x, y)}{\sigma_{\mp\varepsilon}(x, y)} + V(x, y) \log\left( \frac{\sigma_{\mp\varepsilon}(x, y)}{\lambda^2} \right) + W(x, y) \right)$$

$$\doteq \lim_{\varepsilon \downarrow 0} \pm \frac{1}{8\pi^2} D_y^* \left( H^\pm(x, y) + W(x, y) \right).$$

*Here, the **Hadamard coefficients** $U$, $V$, and $W$ are smooth bispinors, where $V$ is given by a series expansion in $\sigma$ as*

$$V = \sum_{n=0}^\infty V_n \sigma^n$$

*with smooth bispinorial coefficients $V_n$. We furthermore require $H^\pm$ to be parametrices, namely, bisolutions of the spinorial Klein-Gordon equations up to smooth terms, i.e.*

$$\mathfrak{P}_x H^\pm(x, y) \in \mathcal{E}(D^*M \boxtimes DM), \quad \mathfrak{P}_y H^\pm(x, y) \in \mathcal{E}(D^*M \boxtimes DM) \tag{III.77}$$

*and demand that their difference is specified by the causal propagator of $\mathfrak{P}$, viz.*

$$H^+(f, g) - H^-(f, g) = i\langle g, (\mathfrak{G}^- - \mathfrak{G}^+)f\rangle,$$

*for arbitrary $f \in \mathcal{D}(DM)$ and $g \in \mathcal{D}(D^*M)$.*

It will be clear from what follows that, as in the scalar case, $U$ and $V_n$ are completely determined by local curvature terms and the mass $m$ (recall that there is no choice of coupling to curvature in the Dirac case). It therefore turns out that the bispinor $W$ encodes the complete state dependence of $\omega^{\pm}$. Note that (III.75) enforces a kind of 'symmetry' of $W$ in the case of a charge-conjugation invariant state. As $H^{\pm}$ are the 'same' for all Hadamard states and in particular the charge-conjugation invariant ones, we expect that (III.75) forces $W$ to share the (potential) symmetries of $U$ and $V_n$, though we have not analysed this question in detail.

One may wonder why we have *required* the difference of $H^+$ and $H^-$ to be given by the causal propagator of $\mathfrak{P}$. In fact, this is not guaranteed by the CAR, as this only implies the said statement up to (distributional) bispinors in the kernel of $D_y^*$ (recall that $S(y,x)$ is equal to $D_y^*$ applied to the mentioned causal propagator by theorem II.3.1.4). Similarly, we have demanded $H^{\pm}$ to be bisolutions which is in contrast to the scalar case where this follows automatically. The reason for this is that the definitions (III.74) of $\omega^{\pm}(x,y)$ entail

$$D_x^* \omega^{\pm}(x,y) = D_y \omega^{\pm}(x,y) = 0$$

and, hence,

$$D_x^* D_y^* \left( H^{\pm}(x,y) + W^{\pm}(x,y) \right) = 0$$

$$D_y D_y^* \left( H^{\pm}(x,y) + W^{\pm}(x,y) \right) = \mathfrak{P}_y \left( H^{\pm}(x,y) + W^{\pm}(x,y) \right).$$

These data imply that both $D_x^* D_y^* H^{\pm}$ and $\mathfrak{P}_y H^{\pm}$ are smooth by the smoothness of $W^{\pm}$, but nothing guarantees us that $\mathfrak{P}_x H^{\pm}$ is smooth (the result of [SaVe01, lem. 5.4], which solves a similar problem in the case of Majorana spinor fields, unfortunately does not hold in our more general case of Dirac spinor fields). In Minkowski spacetime, $H^{\pm}$ would be translationally invariant and the action of $D_y^*$ would be equivalent to the one of $D_x$, thus assuring the smoothness of $\mathfrak{P}_x H^{\pm}$. In general curved spacetimes, one can expect that this is true up to smooth terms and we shall soon provide a result which confirms this expectation. In fact, this result will be important in the computation of the Hadamard coefficients $V_n$. However, as it requires the equivalence of the Hadamard condition and the Hadamard form for the present case of Dirac fields, we shall now state this result, obtained in the works [Köh95, Kra00, Hol00, SaVe01, San09b]. Let us remark that these results hold in arbitrary spacetime dimensions, barring the necessary modification of the Hadamard form, see for instance [Hol00].

**Theorem III.3.2.4** *A state $\omega$ on $\mathcal{A}(DM)$ fulfils the Hadamard condition if and only if it is of global Hadamard form.*

We remark that the equivalence of the Hadamard condition and the Hadamard form is also known to hold in the case of non-quasifree states [San08, San09b]. However, certain technical results are up to now only available in the restricted case of quasifree states [DaHo06].

Having assured the above equivalence, we now focus on the mentioned behaviour of $H^{\pm}(x,y)$ with respect to moving Dirac derivatives from $x$ to $y$ and vice versa by proving that $(D_x - D_y^*)H^{\pm}$ is smooth. We stress that this does not follow by the required property of $H^{\pm}$ to be bisolutions of $\mathfrak{P}$ as the latter demand *a priori* only assures the smoothness of $(D_x - D_y^*)H^{\pm}$ up to a (possibly singular) element in the kernel of $D_x^*$. To achieve the wanted smoothness, we recall that a causal domain is a subset of a geodesically convex set which is in addition globally hyperbolic, see section I.1 and [Fri75]. The restriction to the possibly very small sets in the following result is not a disadvantage, as we will ultimately only need it to compute coinciding point limits of certain quantities.

**Proposition** III.3.2.5 *Let $\mathscr{O}$ be a causal domain in $(M, g)$ which has the topology of $\mathbb{R}^4$ (this is permitted as causal domains form the base of the topology of $M$ [Fri75]), and let $\Sigma$ be an arbitrary Cauchy surface of the spacetime $(\mathscr{O}, g \restriction_{\mathscr{O}})$. There exists a neighbourhood $\mathscr{O}_\Sigma$, such that, for every geodesically convex set $N$ in $\mathscr{O}_\Sigma$ and every $x, y \in N$, the Hadamard parametrices $H^\pm(x, y)$ introduced in definition III.3.2.3 fulfil*

$$\left( D_x - D_y^* \right) H^\pm(x, y) = \left( D_y - D_x^* \right) H^\pm(x, y) \in \mathscr{E}(D^*M \boxtimes DM).$$

*Proof.* The idea of the proof is to deform the spacetime $(\mathscr{O}, g \restriction_{\mathscr{O}})$ in such a way that it can be related to Minkowski spacetime. There we know that the wanted identities hold strictly, and a propagation argument then shows that they hold up to smooth terms in $\mathscr{O}$.

To this avail, we use [FNW81, prop. C.1.]. This proposition entails that there is a globally hyperbolic spacetime $(M', g')$ with two Cauchy surfaces $\Sigma_1$ and $\Sigma_2$ and related neighbourhoods $\mathscr{O}_1$ and $\mathscr{O}_2$ such that the following holds: $\mathscr{O}_1$ lies to the future of $\mathscr{O}_2$, $\Sigma_1$ is isometric to $\Sigma$ and there is a neighbourhood $\mathscr{O}_\Sigma$ of $\Sigma$ such that $(\mathscr{O}_1, g')$ is isometric to $(\mathscr{O}_\Sigma, g)$. Moreover, $(\mathscr{O}_1, g')$ is isometric to a subset of a neighbourhood $(\mathscr{O}_\mathbb{M}, \eta)$ of a Cauchy surface $\Sigma_\mathbb{M}$ in Minkowski spacetime $(\mathbb{M}, \eta)$ and $\Sigma_2$ is isometric to the intersection of a Minkowskian Cauchy surface with $\mathscr{O}_\mathbb{M}$. Here, our requirement that $\mathscr{O}$ has the topology of Minkowski spacetime is crucial, as the deformation can only affect the metric, but not the topology of a spacetime.

Let us now consider a geodesically convex subset $N$ of $\mathscr{O}_\Sigma$. We can push-forward $H^\pm$ from $N$ to an isometrically related subset of $\mathscr{O}_1$. We construct $\mathscr{A}(M')$ on $(M', g')$ and consider a Hadamard state on $\mathscr{A}(M')$. Existence of such states follows by the very results of [FNW81] in combination with the propagation of the Hadamard form obtained for vector-valued fields in [SaVe01], but we can also take the pragmatic point of view that we start the proof by assuming that $H^\pm$ stem from a Hadamard state on $\mathscr{A}(M)$, the state on $\mathscr{A}(M')$ is then obtained by a push-forward and the time-slice axiom (cf. lemma III.3.1.2). Particularly, we obtain bidistributions $\widetilde{H}^\pm$ which are of Hadamard form in any geodesically convex set of $(M', g')$ and coincide with the push-forwards of $H^\pm$ on a suitable subset of $\mathscr{O}_1$. On $\mathscr{O}_2$, $\widetilde{H}^\pm$ locally coincide with the Hadamard forms on Minkowski spacetime, and are therefore translationally invariant.

We now define

$$u^\pm(x, y) \doteq \left( D_x - D_y^* \right) H^\pm(x, y) = \left( D_y - D_x^* \right) H^\pm(x, y), \qquad \widetilde{u}^\pm(x, y) \doteq \left( D_x - D_y^* \right) \widetilde{H}^\pm(x, y).$$

$u^\pm$ can be pushed forward from $N$ to $\mathscr{O}_1$, where they coincide with $\widetilde{u}^\pm$. Moreover, we know that $\widetilde{u}^\pm = 0$ on $\mathscr{O}_2$. Realising that $\widetilde{u}^\pm$ fulfil $\mathfrak{P}_x \mathfrak{P}_y \widetilde{u}^\pm$ on the full spacetime $(M', g')$, we can apply the theorem of propagation of singularities [Den82] to find that, as $WF(\widetilde{u}^\pm) = \emptyset$ on $\mathscr{O}_2$, the wave front set of $\widetilde{u}^\pm$ and, hence, $u^\pm$ on $N$ can only contain elements of the form

$$(x, y, k_x, 0) \quad \text{or} \quad (x, y, 0, k_y), \tag{III.78}$$

as these would not propagate to $\mathcal{O}_2$. Following an argument presented in the proof of [SaVe01, thm 5.8.], we can infer that $WF(u^\pm) = \emptyset$ in the following way: by the very definition of $u^\pm$, we have $WF(u^\pm) \subset WF(H^\pm)$. We now need a stronger result than theorem III.3.2.4. Namely, the latter theorem assures that $\omega^\pm$ and, hence, $D_y^* H^\pm$ have the wave front sets as found in lemma III.3.2.2. However, by [San09b, prop. 4.11], we even have that $H^\pm$ have the same wave front sets as $\omega^\pm$, which is *a priori* not clear as derivatives can in principle shrink the wave front set of a distribution. If we now consider the mentioned 'antisymmetric' wavefront set of $H^\pm$, it follows that $WF(u^\pm)$ can not even contain elements of the form (III.78) and are thus empty.    $\square$

With the above proposition at hand, we possess all tools necessary to compute the Hadamard coefficients $U$ and $V_n$. For the convenience of the reader, we recall all partial differential operators that are by now known to give a smooth bispinor once applied to $H^\pm$.

$$\mathfrak{P}_x = D_x^* D_x, \qquad \mathfrak{P}_y = D_y^* D_y, \qquad D_x - D_x^* = D_y - D_y^* \qquad \text{(III.79)}$$

From our presentation of the computation of the Hadamard coefficients $u$ and $v_n$ of the Klein-Gordon field, it is in principle clear how to proceed to obtain recursive differential equations for $U$ and $V_n$. Namely, we apply one of the above differential operators to the explicit form of $H^\pm$ given in definition III.3.2.3, we group the found terms by the appearing powers of the half squared geodesic distance $\sigma$, and we require that the coefficients of the inverse powers of $\sigma$ and of $\log\sigma$ vanish by our knowledge of the smoothness of the total outcome. However, it is important to stress that this procedure requires two explicit choices. On the one hand, it is by no means clear that the differential equations for $U$ and $V_n$ we obtain in this way are independent of the operator chosen from the ones in (III.79). On the other hand, setting the coefficients of the superficially divergent inverse powers $\sigma$ to zero is also only sufficient, but not necessary, as 'hidden' $\sigma$-dependencies of $U$ and $V_n$ could in principle lead to cancellations of terms which are *a priori* of different order in $\sigma$. The latter remark does, however, not extend to the terms proportional to $\log\sigma$. As $U$ and $V_n$ are required to be smooth, they can not contain implicit logarithmic dependencies on $\sigma$, and any term proportional to $\log\sigma$ obtained *after* application of one of the operators listed in (III.79) must necessarily have a vanishing coefficient. These considerations immediately lead to the following partial differential equations for $V = \sum_n V_n \sigma^n$.

$$\mathfrak{P}_x V = \mathfrak{P}_y V = (D_x - D_y^*)V = (D_y - D_x^*)V = 0. \qquad \text{(III.80)}$$

The first two equations have direct analogues in the scalar case, whereas the latter corresponds to $(\partial_x + \partial_y)v = 0$ in the such case, which in turn corresponds to the symmetry of $v_n$ proven in [Mor00, Mor99] (and the fact that $v_n$ depends on $x$ and $y$ via '$x - y$'). There is no symmetry proof available for the Diracian Hadamard coefficients to date, and, although we suspect that a proof can be given with the methods of [Mor00, Mor99], we have not attempted to do this. Hence, our proposition (III.3.2.5) and the latter of the above differential equations for $V$, which results from this proposition, will serve as a 'replacement' for the lacking knowledge of the symmetry of $V$ on computational grounds. To continue the discussion of the Hadamard recursion relations, we have to make the anticipated choice of differential operator. In order to be in close analogy to the scalar case, we choose $\mathfrak{P}_x$ for convenience. We therefore find the following differential equations by asking that the coefficients of $\sigma^{-2}$ and $\sigma^{-1}$ respectively in $\mathfrak{P}_x H$ vanish identically.

$$2U_{;\mu}\sigma_{;}^{\ \mu} + (\Box_x\sigma - 4)U = 0, \tag{III.81}$$

$$\mathfrak{P}_x U + 2V_{0;\mu}\sigma_{;}^{\ \mu} + (\Box_x\sigma - 2)V_0 = 0. \tag{III.82}$$

Let us stress once more that the above equations are ordinary differential equations with respect to the affine parameter along the unique geodesic connecting $x$ and $y$. To obtain similar differential equations for $V_n$, with $n > 1$, one starts from the $\mathfrak{P}_x V = 0$ equation in (III.80), inserts the series expansion of $V$, and demands that $\mathfrak{P}_x V = 0$ holds at each (now positive) order in $\sigma$ separately. One therefore finds

$$\mathfrak{P}_x V_0 + 2V_{1;\mu}\sigma_{;}^{\ \mu} + (\Box_x\sigma)V_1 = 0, \tag{III.83}$$

$$\mathfrak{P}_x V_n + 2(n+1)V_{n+1;\mu}\sigma_{;}^{\ \mu} + \big((n+1)\Box_x\sigma + 2n(n+1)\big)V_{n+1} = 0 \quad \forall n \geq 1. \tag{III.84}$$

It seems that the found Hadamard recursion relations have the same form as in the scalar case. Particularly, we find differential equations for $V_n$ which are of ordinary type once lower orders are known and we realise that the identity $(D_x - D_y^*)V$ is in principle not strictly necessary to determine $V_n$ uniquely. Let us consider how the found Hadamard recursion relations in the Dirac spinor case can be solved. We start with (III.81) and point out that it has a unique solution once an initial condition is given. In the Minkowskian case, $\mathfrak{P} = P\mathbb{I}_4$ with $P$ denoting the Klein-Gordon operator, and $H^\pm$ are therefore just given by the scalar parametrices $h_{\pm\varepsilon}$ times the identity matrix $\mathbb{I}_4$. A sensible initial condition for $U$ in terms of the Synge bracket notation of coinciding point limits is, hence,

$$[U] = \mathbb{I}_4$$

and a sensible ansatz for $U$ is

$$U = u\,\mathscr{I},$$

with $u$ the scalar Hadamard coefficient constituted by the square root of the Van Vleck-Morette determinant and a suitable bispinor $\mathscr{I}$. Inserting this ansatz and the known differential equation (III.56) of $u$ into (III.81), one finds

$$[\mathscr{I}] = \mathbb{I}_4, \qquad \mathscr{I}_{;\mu}\sigma_{;}^{\ \mu} = 0.$$

These ordinary differential equations determine $\mathscr{I}$ to be the *spinor parallel transport* along the unique geodesic connecting $x$ and $y$. Namely, given a spinor $F = F^A(x)E_A(y)$ at $x$, its parallel transport along the mentioned geodesic from $x$ to $y$ has the components

$$F^{B'}(y) = F^A(x)\,\mathscr{I}_A^{\ B'}(x,y),$$

where $\mathscr{I}_A^{\ B'}$ are the components of $\mathscr{I}$ in a the local spin frame $E_A$, namely,

$$\mathscr{I}(x,y) = \mathscr{I}_A^{\ B'}(x,y)\,E^A(x) \otimes E_{B'}(y).$$

Therefore, $\mathscr{I}(x,y)$ can be understood as a 'generalised trace' and it appears naturally in the construction of elements in $\mathscr{A}^{\text{obs}}(DM)$. One could now hope that the remaining coefficients $V_n$ are also proportional to the scalar ones $v_n$ by means of $\mathscr{I}$. However, as is well-known (see for instance [Ful89, Hol00]), this is not the case as terms proportional to the spin curvature tensor $\mathfrak{R}_{\alpha\beta}$ enter the arena via the derivatives of $\mathscr{I}$ present in $\mathfrak{P}_x U$. In fact, following the inductive procedure outlined in subsection III.1.2, one can compute

$$\left[\mathscr{I}_A{}^{B'}{}_{;\alpha\beta}\right] = \frac{1}{2}\mathfrak{R}_A{}^B{}_{\alpha\beta}.$$

We also remark the following relation between the (coordinate basis) $\gamma$-matrices and $\mathscr{I}$. Namely, recalling the vector parallel transport $g_{v'}^{\mu}$ introduced in subsection III.1.2, the fact that the section $\gamma$ is covariantly constant is equivalent to

$$\gamma^{\mu} = g_{v'}^{\mu}\,\mathscr{I}\gamma^{v'}\mathscr{I}^{-1},$$

where we have suppressed spinor indices.

Returning to the solution of the Hadamard recursion relations, we find in analogy to the scalar case

$$[V_0] = -\frac{1}{2}[\mathfrak{P}_x U], \qquad [V_{n+1}] = -\frac{1}{2(n+1)(n+2)}[\mathfrak{P}_x V_n] \qquad \forall n \geq 0.$$

These initial conditions are, depending on ones interest, either the starting point for the determination of coinciding point limits of derivatives, or the initial conditions to integrate the ordinary differential equations (III.82)-(III.84). We remark that the Hadamard coefficients are are essentially equal to the DeWitt-Schwinger coefficients, as in the scalar case. The latter have been computed for Dirac fields in [Chr78], but unfortunately not to the order necessary for our purposes. We therefore had to obtain these results ourselves. Namely, we shall be interested in computing the (normal-ordered) stress-energy tensor of the Dirac field. To assure that it is covariantly conserved – a requirement one has to impose if one wants to take the expectation value of such tensor as a source term in the Einstein equation – one has to compute $[V_{1;\mu}]$, as we shall see in the next chapter. In the scalar case, one knows that $v_1$ is symmetric [Mor99, Mor00] and, hence, $[v_{1;\mu}] = \frac{1}{2}[v_1]_{;\mu}$ holds (see subsection III.1.2). As already mentioned, such symmetry is not known to hold in the Dirac case and $[V_{1;\mu}]$ therefore needs to be computed explicitly. We warn the reader that this task is not as harmless as it may seem. To perform the wanted calculation, one has to compute terms such as

$$[\sigma_{;\alpha\beta\gamma\delta\phi\varepsilon\lambda}] = -\frac{1}{6}R_{\alpha\beta\gamma\delta;\phi\varepsilon\lambda} + 779 \text{ terms}.$$

The following results are thus the outcome of several months of computer-aided work (see the footnote on page 116). Before we state them, we need an additional small lemma which turns out to simplify computations considerably.

**Lemma III.3.2.6**  *Given a smooth bitensor $B(x,y)$ which is such that $\frac{B(x,y)}{\sigma^n(x,y)}$ is a smooth bitensor, it holds*

$$\left[\frac{B}{\sigma^n}\right] = \frac{[\Box^n B]}{[\Box^n(\sigma^n)]}.$$

*Proof.* Since $B$ and $B/\sigma^n$ are smooth, their coinciding point limits depend neither on $y$ nor on the path along which one approaches $x$. Thus, we can apply *de l'Hospital's rule* to our smooth bitensors restricted to arbitrary smooth curves, in particular coordinate curves, thereby expressing coinciding point limits of fractions via such limits of fractions of covariant derivatives with respect to $y$, *e.g.*,

$$\left[\frac{B}{\sigma^n}\right] = \left[\frac{B_{;\mu'}}{(\sigma^n)_{;\mu'}}\right].$$

As the coinciding point limit of $\sigma$ and its first covariant derivative vanish, the coinciding point limit of the $k$-th covariant derivative of $\sigma^n$ vanishes for all $0 \le k \le 2n-1$. Due to the smoothness of $B/\sigma^n$ and de l'Hospital's rule, the same must hold for the $k$-th covariant derivative of $B$ as well. Consequently, a multiple application of de l'Hospital's rule yields

$$\left[\frac{B}{\sigma^n}\right] = \left[\frac{B_{;\mu'_1\cdots\mu'_{2n}}}{(\sigma^n)_{;\mu'_1\cdots\mu'_{2n}}}\right] = \frac{\left[B_{;\mu'_1\cdots\mu'_{2n}}\right]}{\left[(\sigma^n)_{;\mu'_1\cdots\mu'_{2n}}\right]} = \frac{\left[B_{;\mu_1\cdots\mu_{2n}}\right]}{\left[(\sigma^n)_{;\mu_1\cdots\mu_{2n}}\right]},$$

where the third equality holds due to Synge's rule and the vanishing of lower order derivatives of both $B$ and $\sigma^n$ in the coinciding point limit. Since the above equalities do not depend on the choice of $\mu_1\cdots\mu_{2n}$, it holds (even for the vanishing components)

$$\left[B_{;\mu_1\cdots\mu_{2n}}\right] = \left[\frac{B}{\sigma^n}\right]\left[(\sigma^n)_{;\mu_1\cdots\mu_{2n}}\right]$$

and appropriate contractions with the metric yield

$$[\Box^n B] = \left[\frac{B}{\sigma^n}\right][\Box^n(\sigma^n)],$$

which closes the proof. $[\Box^n(\sigma^n)]$ can be expressed solely in terms of traces of the metric and is non-vanishing in particular, further applications of the de l'Hospital rule are thus not necessary. □

We can now finally state our main result in this subsection.

**Theorem III.3.2.7**  *Denoting by $H$ either $H^+$ or $H^-$, the following identities hold.*

*a)* $[\mathfrak{P}_x H] = 6[V_1]$, $\qquad [(\mathfrak{P}_x H)_{;\mu}] = 8[V_{1;\mu}]$, $\qquad [(\mathfrak{P}_x H)_{;\mu'}] = -8[V_{1;\mu}] + 6[V_1]_{;\mu}$,

*b)* $[\mathfrak{P}_y H] = 6[V_1]$, $\qquad [(\mathfrak{P}_y H)_{;\mu}] = 8[V_{1;\mu}] - 2[V_1]_{;\mu}$, $\qquad [(\mathfrak{P}_y H)_{;\mu'}] = -8[V_{1;\mu}] + 8[V_1]_{;\mu}$,

*c)* $Tr\,[D_x' D_y' H] = -Tr\,[\mathfrak{P}_x H]$, $\quad Tr\,[(D_x' D_y' H)_{;\mu}] = -Tr\,[(\mathfrak{P}_x H)_{;\mu}] + [V_1]_{;\mu}$,

$\quad Tr\,[(D_x^* D_y^* H)_{;\mu'}] = -Tr\,[(\mathfrak{P}_x H)_{;\mu'}] - [V_1]_{;\mu}$,

*d)* $Tr\,[(\mathfrak{P}_y H - P_x H)_{;\mu'}]\gamma^\mu \gamma_\nu = 2Tr\,[V_1]_{;\nu}$.

*Proof.* a) We shall employ (III.81)-(III.83). These data entail

$$\mathfrak{P}_x H = 2V_{1;\varrho}\sigma_{;}{}^{\varrho} + V_1(\Box_x \sigma + 2) + O(\sigma), \qquad \text{(III.85)}$$

and thus, taking the coinciding point limit and remembering those of $\sigma$ computed in subsection III.1.2, $[\mathfrak{P}_x H] = 6[V_1]$. Similarly, one gets, deriving (III.85) once and performing the limit, $[(\mathfrak{P}_x H)_{;\mu}] = 8[V_{1;\mu}]$. By means of Synge's rule, we finally have $[(\mathfrak{P}_x H)_{;\mu'}] = -8[V_{1;\mu}] + 6[V_1]_{;\mu}$.

b) We would like to compute $\mathfrak{P}_y H$ directly, but without any knowledge on the symmetries of the Diracian Hadamard coefficients, we have to verify the transport equations for $\mathfrak{P}_y$, which otherwise would follow automatically from those for $\mathfrak{P}_x$ as it happens in the scalar case for the scalar Hadamard coefficients $u$ and $v$ [Mor99, Mor00]. To wit,

$$2U_{;\mu'}\sigma_{;}{}^{\mu'} + U(\Box_y \sigma - 4) = \mathscr{I}\left(2u_{;\mu'}\sigma_{;}{}^{\mu'} + u(\Box_y \sigma - 4)\right) = 0,$$

where the first equality holds since the derivative of $\mathscr{I}$ vanishes along the geodesic connecting $x$ and $y$ and the second one holds since $u(x,y) = u(y,x)$[16] and $u$ is thus subject to transport equations for both $\mathfrak{P}_x$ and $\mathfrak{P}_y$. Since $\mathfrak{P}_y H$ is smooth by our definition of $H$, we now know that

$$Z_1 \doteq \frac{Y_1}{\sigma} \doteq \frac{\mathfrak{P}_y U + 2V_{0;\mu'}\sigma_{;}{}^{\mu'} + V_0(\Box_y \sigma - 2)}{\sigma}$$

must be smooth as well. Alas, it does not factorise into a term only involving the scalar coefficients $u$ and $v$ times $\mathscr{I}$ and, up to now, we are unaware of a way to prove that it is identically vanishing. But we can try to compute whether it vanishes up to the derivative order we need for our purposes. To this end, it helps to split $V$ into $v \cdot \mathscr{I} + \tilde{V}$, where $\tilde{V}$ is the non-trivial matrix part of $V$ stemming from the spin curvature. In this way one can separate from $Y_1$ a term which vanishes due to the transport equation for $v$ and has to cope with the remainder only. Involved calculations yield

$$[Y_1] = [Y_{1;\mu}] = [\Box Y_1] = [(\Box Y_1)_{;\mu}] = 0,$$

[16]The symmetry of $u$ does not have to be proved in the same long way as that of $v$ (see [Mor99, Mor00]), but it follows automatically by its explicit form $u(x,y) = \sqrt{\det(\sigma_{;\mu\nu'}(x,y))}\sqrt{|g(x)|^{-1}}\sqrt{|g(y)|^{-1}}$.

and thus, employing lemma III.3.2.6, $[Z_1] = [Z_{1;\mu}] = 0$. Consequently,

$$\mathfrak{P}_y H = 2V_{1;\varrho'}\sigma_{\;;}^{\varrho'} + V_1(\Box_y\sigma + 2) + \text{terms vanishing in the limit}$$

and

$$(\mathfrak{P}_y H)_{;\mu} = 2V_{1;\varrho'}\sigma_{;\mu}^{\varrho'} + V_{1;\mu}(\Box_y\sigma + 2) + \text{terms vanishing in the limit}.$$

One can now straightforwardly obtain $[\mathfrak{P}_y H] = 6[V_1]$, $[(\mathfrak{P}_y H)_{;\mu}] = 8[V_{1;\mu}] - 2[V_1]_{;\mu}$, and $[(\mathfrak{P}_y H)_{;\mu'}] = 8[V_{1;\mu}] - 8[V_1]_{;\mu}$.

c) Let us define

$$Z_2 \doteq (D_x - D_y^*)H = (D_y - D_x^*)H.$$

By direct inspection,

$$D_x^* D_y^* H = -\mathfrak{P}_x H - D_x^* Z_2.$$

We know that $Z_2$ is smooth by proposition III.3.2.5 but, alas, neither this quantity, nor $D_x^* Z_2$ turn out to be vanishing. Luckily enough, we can still extract some useful results at the level of traced coinciding point limits, at an order of derivatives high enough for our purposes. One computes

$$\begin{aligned}
Z_2 &= -\frac{U(D_x - D_y^*)\sigma}{\sigma^2} + \frac{(D_x - D_y^*)U - V(D_x - D_y^*)\sigma}{\sigma} + \log(\sigma)(D_x - D_y^*)V \\
&\doteq -\frac{U(D_x - D_y^*)\sigma}{\sigma^2} + \frac{Y_2}{\sigma} + \log(\sigma)(D_x - D_y^*)V.
\end{aligned} \tag{III.86}$$

As already discussed, the last term vanishes identically and so does the first term on account of

$$U(D_x - D_y^*)\sigma = -\imath(\mathscr{I}\gamma^\mu\sigma_{;\mu} + \gamma^{\mu'}\mathscr{I}\sigma_{;\mu'}) = -\imath(\mathscr{I}\gamma^\mu\sigma_{;\mu} + \mathscr{I}\gamma^\mu g_\mu^{\;\mu'}\sigma_{;\mu'}) = 0.$$

This leaves us with $Z_2 = Y_2/\sigma$. Involved computations, employing $(D_x - D_y^*)V = \mathfrak{P}_x V = 0$ to exchange higher derivative terms with terms of lower derivative order in the appearing commutators with $\gamma$-matrices, yield

$$[Y_2] = [Y_{2;\mu}] = [\Box Y_2] = 0, \qquad [(\Box Y_2)_{;\mu}] = 6\left[[V_1], \gamma_\mu\right].$$

After a few rearrangements and out of lemma III.3.2.6, one gets

$$[Z_2] = 0, \quad [Z_{2;\mu}] = \left[[V_1], \gamma_\mu\right].$$

Hence, $[Z_{2;\mu}]$ is traceless due to the antisymmetry of the commutator. By means of lemma I.2.2.9, one can show that $Tr[V_1]\gamma_\mu\gamma_\nu = Tr[V_1]g_{\mu\nu}$ which entails that even $D_x^* Z_2$ is traceless and thus

$$Tr[D_x^* D_y^* H] = -Tr[\mathfrak{P}_x H].$$

In order to compute $Tr\left[(D_x^*D_y^*H)_{;\mu}\right]$ and $Tr\left[(D_x^*D_y^*H)_{;\mu'}\right]$, let us consider that $D_y^*Z_2 = D_x^*Z_2 + \mathfrak{P}_x H - \mathfrak{P}_y H$. Employing this as well as the previous results and other tricks we have discussed in this proof, one obtains the following chain of identities

$$
\begin{aligned}
Tr\left[(D_x^*Z_2)_{;\mu}\right] &= Tr\,\gamma^\nu[Z_{2;\nu\mu}] = -Tr\,\gamma^\nu[Z_{2;\nu'\mu}] + Tr\,\gamma^\nu[Z_{2;\mu}]_\nu \\
&= Tr\left[(D_y^*Z_2)_{;\mu}\right] = Tr\left[(D_x^*Z_2)_{;\mu}\right] - Tr\left[(\mathfrak{P}_y H - \mathfrak{P}_x H)_{;\mu}\right] \\
&= -\frac{1}{2}Tr\left[(\mathfrak{P}_y H - \mathfrak{P}_x H)_{;\mu}\right] = Tr\left[V_1\right]_{;\mu} \\
&= -Tr\left[(D_y^*Z_2)_{;\mu'}\right].
\end{aligned}
\tag{III.87}
$$

We can finally use this last calculation to obtain

$$
Tr\left[(D_x^*D_y^*H)_{;\mu}\right] = -Tr\left[(\mathfrak{P}_x H)_{;\mu}\right] + [V_1]_{;\mu}, \quad Tr\left[(D_x^*D_y^*H)_{;\mu'}\right] = -Tr\left[(\mathfrak{P}_x H)_{;\mu'}\right] - [V_1]_{;\mu}.
$$

d) Inserting the previous results, we have $\left[(\mathfrak{P}_y H - \mathfrak{P}_x H)_{;\nu}\right] = 2[V_1]_{;\nu}$. As already discussed, due to $\gamma$-matrix identities, tracing $\left[V_1\right]$ with two $\gamma$-matrices amounts to a multiplication with the metric. Since the operations of trace and covariant derivation commute, we have $Tr\left[V_1\right]_{;\nu}\gamma^\nu\gamma_\mu = Tr\left[V_1\right]_{;\mu}$ and thus

$$
Tr\left[(\mathfrak{P}_y H - \mathfrak{P}_x H)_{;\nu}\right]\gamma^\nu\gamma_\mu = 2[V_1]_{;\mu}.
$$

$\square$

We conclude the subsection by stating the coinciding point limit of $V_1$, *viz.*

$$
[V_1] = \left(\frac{m^4}{8} + \frac{m^2R}{48} + \frac{R^2}{1152} + \frac{\Box R}{480} - \frac{R_{\alpha\beta}R^{\alpha\beta}}{720} + \frac{R_{\alpha\beta\gamma\delta}R^{\alpha\beta\gamma\delta}}{720}\right)\mathbb{I}_4 + \frac{\mathfrak{R}_{\alpha\beta}\mathfrak{R}^{\alpha\beta}}{48}.
\tag{III.88}
$$

### III.3.3 The Enlarged Algebra of Observables

In the previous subsection, we have seen that Hadamard states are a sensible and possible concept also in the case of Dirac fields. Let us recall that the necessity of Hadamard states arose out of the question whether and how it is possible to define normal ordering in curved spacetimes. Particularly, we had to assure that we are dealing with distributions whose spectral properties allow for pointwise multiplication, a situation which necessarily arises out of Wick's theorem when dealing with an algebra of Wick polynomials. We shall now proceed to generalise the construction of the scalar Wick polynomial algebra provided in [BFK95, BrFr00, HoWa01, Mor03, HoWa05] to the case of Dirac fields, and we stress that the result is expected, but new. As in the scalar case, the power of the following construction will be the ability to encode Wick's theorem already on the algebraic level, without having recourse to a state or any related representation in terms of creation and annihilation operators.

Again, we start by promoting the antisymmetric algebra $(\mathscr{C}(DM), \cdot_a)$ introduced in subsection II.3.3 to a proper CAR algebra by exchanging $\cdot_a$ with a suitable product encoding the CAR. To this end, we need a contraction operator like the one in (III.60), but suitable for our case of vector-valued, anticommuting fields. Hence, let us recall our notation $\mathfrak{E}^\Theta$ for frames of $D^\oplus M$, let $\mathfrak{E}_\Theta$ denote the dual frame of $\mathfrak{E}^\Theta$, and let us consider an arbitrary bidistribution $\mathfrak{B} \in \mathscr{D}'(D^\oplus M^2)$, whose kernel can be decomposed as

$$\mathfrak{B}(x_1, x_2) = \mathfrak{B}^{\Theta_1 \Theta_2}(x, y) \, \mathfrak{E}_{\Theta_1}(x_1) \otimes \mathfrak{E}_{\Theta_2}(x_2).$$

Let now $n \geq 2$. Given an arbitrary test section $f^{(n)} \in \mathscr{D}(D^\oplus M^n)$, we define the *contraction operator* $\mathfrak{G}_\mathfrak{B} : \mathscr{D}(D^\oplus M^n) \to \mathscr{D}(D^\oplus M^{n-2})$ related to $\mathfrak{B}$ on the level of frame coefficients as

$$[\mathfrak{G}_\mathfrak{B} f^{(n)}]_{\Theta_1 \cdots \Theta_{n-2}}(x_1, \cdots, x_{n-2}) \doteq \sum_{i=1}^{n-1} \sum_{j=i+1}^{n} \int_{M^2} d_g y_1 d_g y_2 \, (-1)^{j-i+1} \, \mathfrak{B}^{\Theta \Theta'}(y_1, y_2) \quad \times \tag{III.89}$$

$$\times \quad f^{(n)}_{\Theta_1 \cdots \Theta \Theta_{i+1} \cdots \Theta_{j-1} \Theta' \cdots \Theta_{n-2}}(x_1, \cdots, y_1, x_{i+1}, \cdots, x_{j-1}, y_2, \cdots, x_{n-2}).$$

For $n < 2$, and $f^{(n)} \in \mathscr{D}(D^\oplus M^n)$, we set $\mathfrak{G}_\mathfrak{B} f^{(n)} = 0$. Given two elements $f = \oplus_n \frac{1}{n!} f^{(n)}$, $g = \oplus_m \frac{1}{m!} g^{(m)} \in \mathscr{C}(DM)$, we define a *star product* $\star_S$ on $\mathscr{C}(DM)$ by setting

$$\frac{1}{n!} (f \star_S g)^{(n)} \doteq \mathfrak{A}\mathfrak{s} \left[ e^{\frac{i}{2} \mathfrak{G}_{S^\oplus}} (f \otimes g) \right]^{(n)}, \tag{III.90}$$

where $S^\oplus$ is the 'double causal propagator' introduced in definition III.3.1.1, $\mathfrak{A}\mathfrak{s}$ is the total antisymmetrisation projector, and $[e^{\frac{i}{2} \mathfrak{G}_{S^\oplus}}(f \otimes g)]^{(n)}$ means taking the $\mathscr{D}(D^\oplus M^n)$-component of $e^{\frac{i}{2} \mathfrak{G}_{S^\oplus}}(f \otimes g)$. In analogy to the scalar case, only mutual contractions between $f$ and $g$ appear by the antisymmetry of elements in $\mathscr{C}(DM)$. Let us see how the introduced product is an extension of $\cdot_a$ which encodes the CAR. Considering the case of $f$, $g \in \mathscr{D}(D^\oplus M) \subset \mathscr{C}(DM)$. We find

$$\frac{1}{2} f \star_S g = f \cdot_a g \oplus \frac{i}{2} S^\oplus(f, g)$$

and, hence,

$$\frac{1}{2}(f \star_S g + g \star_S f) = i S^\oplus(f, g).$$

Furthermore, the star product $\star_S$ is compatible with the dynamic nature of elements in $\mathscr{C}(DM)$ that we have implemented by taking an appropriate quotient in definition II.3.3.1, as $S^\oplus$ is per construction the causal propagator of the double Dirac operator $D^\oplus$. Clearly, we have tailored $(\mathscr{C}(DM), \star_S)$ in such a way that the map $\mathfrak{T} : \mathscr{A}(DM) \to (\mathscr{C}(DM), \star_S)$ defined as

$$\mathfrak{T} \left[ \bigoplus_n f^{(n)} \right] \doteq \bigoplus_n \frac{1}{n!} \mathfrak{A}\mathfrak{s} \, f^{(n)}$$

is an algebra homomorphism. To see that $\mathscr{A}(DM)$ and $(\mathscr{C}(DM), \star_S)$ are isomorphic, one realises that the non-antisymmetric parts of elements in $\mathscr{A}(DM)$ can be implemented in $\mathscr{C}(DM)$ by storing their evaluation with (the symmetric) $S^{\oplus}$ in numerical coefficients of antisymmetric test sections, in close analogy to the scalar case (*cf.* subsection III.1.3).

We now extend $(\mathscr{C}(DM), \star_S)$ and interpret the result as an extension of $\mathscr{A}(DM)$ in view of the above discussed relation of these two algebras. To this avail, we first extend the off-shell space $\mathscr{C}_0(DM)$ (see definition II.3.3.1) and include the dynamics in a second step. Therefore, let us recall that Wick monomials correspond essentially to objects derived from $\delta$-distributions and define the distribution space $\mathscr{E}'_V(D^{\oplus}M^n) \subset \mathscr{E}'(D^{\oplus}M^n)$ as

$$\mathscr{E}'_V(D^{\oplus}M^n) \doteq \left\{ u \in \mathscr{E}'(D^{\oplus}M^n) \,|\, WF(u) \subset T^*M^n \setminus \left( \bigcup_{x \in M} \left(V_x^+\right)^n \cup \bigcup_{x \in M} \left(V_x^-\right)^n \right) \right\},$$

where $V_x^{\pm}$ denote the closed future and past lightcones in the fibre of the cotangent bundle at a point $x$ in $M$. Moreover, we denote the subspace of $\mathscr{E}'_V(D^{\oplus}M^n)$ constituted by antisymmetric elements as $\mathscr{E}'^a_V(D^{\oplus}M^n)$. We will soon give a detailed account on how objects like $:\psi^{\dagger}\psi:$ are contained in $\mathscr{E}'^a_V(D^{\oplus}M^n)$. We now set

$$\mathscr{C}_{0,\text{ext}}(DM) \doteq \bigoplus_{n=0}^{n} \mathscr{E}'^s_V(D^{\oplus}M^n), \tag{III.91}$$

where it is again understood that we consider only *finite* sequences of distributions. As in the scalar case [BrFr00, HoWa01, HoRu01], we endow the above set with the Hörmander pseudo topology to have a means to control the wave front set of sequences of distributions (it is understood that we only consider sequences in $\mathscr{C}_{0,\text{ext}}(DM)$ with a finite maximal number of direct sum entries). In definition III.1.3.1, we have stated this concept strictly speaking only for scalar distributions. But with our current understanding of wave front sets of vector-valued distributions, the extension of definition III.1.3.1 to the latter is clear and we shall not repeat it here. Similarly, proposition III.1.3.2 which entails that test functions are dense in the compactly supported distributions (with the wanted restriction on the wavefront set to a closed cone) with respect to the Hörmander pseudo topology extends immediately to the case at hand, thus meaning that $\mathscr{C}_0(DM)$ is dense in $\mathscr{C}_{0,\text{ext}}(DM)$.

Let us now recall that $S^{\oplus}$ has the 'symmetric' wave front set (III.63) of the causal propagator $\Delta$ of the Klein-Gordon field. By our choice of $\mathscr{C}_{0,\text{ext}}(DM)$, the product $\star_S$ is therefore not well-defined on all elements of $\mathscr{C}_{0,\text{ext}}(DM)$, as potential pointwise powers of $S^{\oplus}$ can appear. But these are not sensible objects due to the shape of $WF(S^{\oplus})$ and the criterion of Hörmander, see theorem III.1.2.2. We therefore choose a new product built out of contractions with a bidistribution that has an 'antisymmetric' wave front set. Moreover, we are again interested in constructing locally covariant Wick polynomials. Our treatment of the scalar case in subsection (III.1.3) suggests what we have to do. Namely, we should consider subtractions by means of the local Hadamard forms to obtain locally covariant Wick ordered quantities. However, if we want to define an algebra by means of the local Hadamard parametrices alone, then we need to restrict ourselves to a causal domain $N$ in $M$. But, this is not a flaw, as isomorphy of algebras constructed with distributions differing by a smooth section will also hold in this case. Hence, if we define the extended algebra by means of a two-point function $\omega_2$ of a Hadamard state, a definition on the full $M$ is possible, and the restriction to any causal domain $N$ is isomorphic to any extended algebra constructed on $N$ by means of only the local Hadamard parametrices.

With this in mind, let $N$ be a causal domain in $M$. We define a bidistribution in $\mathscr{D}'(D^{\oplus}N^2)$ by setting

$$\mathfrak{H}(f,g) \;\doteq [D_y^* H^+](g_1, f_2) - [D_y^* H^-](f_1, g_2) \tag{III.92}$$

$$= H^+(g_1, Df_2) - H^-(f_1, g_2) \tag{III.93}$$

for arbitrary $f = f_1 \oplus f_2$, $g = g_1 \oplus g_2 \in \mathscr{D}(D^{\oplus}N^2)$ Based on this, we set

$$\frac{1}{n!}(f \star_H g)^{(n)} \doteq \mathfrak{A}\mathfrak{s} \left[ e^{\mathfrak{G}_{\mathfrak{H}}^{\mathrm{mut}}}(f \otimes g) \right]^{(n)}, \tag{III.94}$$

where $\mathfrak{G}_{\mathfrak{H}}^{\mathrm{mut}}$ is defined like the contraction operator $\mathfrak{G}_{\mathfrak{H}}$, but with the difference that only *mutual* contractions between $f$ and $g$ are allowed. It is easy to see that the new product is still encoding the CAR. Namely, let us again consider $f, g \in \mathscr{D}(D^{\oplus}N) \subset \mathscr{C}_{0,\mathrm{ext}}(DN)$. We find

$$\frac{1}{2}f \star_H g = f \cdot_a g \oplus \mathfrak{H}(f,g)$$

and, hence,

$$\frac{1}{2}(f \star_H g - g \star_H f) = iS^{\oplus}(f,g)$$

on account of the fact that $H^+(f_1, Dg_2) - H^-(f_1, Dg_2) - = iS(g_2, f_1)$. More generally, it follows that $(\mathscr{C}_0(DN), \star_H)$ is a deformation of $(\mathscr{C}_0(DN), \star_S)$ in the sense that, for arbitrary elements $f, g$ of $\mathscr{C}_0(DN)$,

$$f \star_H g = e^{\mathfrak{G}_{\mathfrak{H}^d}} \left[ (e^{-\mathfrak{G}_{\mathfrak{H}^d}} f) \star_S (e^{-\mathfrak{G}_{\mathfrak{H}^d}} g) \right],$$

where $\mathfrak{H}^d \doteq \mathfrak{H} - \frac{i}{2}S^{\oplus}$.

In order to show that $\star_H$ is a sensible product on $\mathscr{C}_{0,\mathrm{ext}}(DN)$, namely, that $\mathscr{C}_{0,\mathrm{ext}}(DN)$ is closed with respect to $\mathscr{C}_{0,\mathrm{ext}}(DN)$, we note that $\left[ e^{\mathfrak{G}_{\mathfrak{H}}^{\mathrm{mut}}}(f \otimes g) \right]^{(n)}$ is, on the level of frame coefficients, given by sums of elements of the form

$$\int_{M^{2k}} d_g x_1 \cdots d_g x_k\, d_g y_1 \cdots d_g y_k \prod_{i=1}^{k} \mathfrak{H}^{\Theta_i \Theta'_i}(x_i, y_i) f^{(n)}_{\Theta_1 \cdots \Theta_n}(x_1, \cdots, x_n) g^{(m)}_{\Theta'_1 \cdots \Theta'_m}(y_1, \cdots, y_m), \tag{III.95}$$

where we stress that, on account of the antisymmetry of $f^{(n)}$ and $g^{(m)}$, it is sufficient to consider contractions of the first $k$ arguments. By our knowledge of the wave front set of $\mathfrak{H}$ and the wave front sets of $f^{(n)} \in \mathcal{E}'^{t_a}_V(D^{\oplus}N^n)$, $g^{(m)} \in \mathcal{E}'^{t_a}_V(D^{\oplus}N^m)$, we find by theorem III.1.2.2 that the pointwise product appearing in (III.95) is a well-defined distribution because the sum of the wave front sets of $\prod_{i=1}^k \mathfrak{H}(x_i, y_i)$ and $f^{(n)} \otimes g^{(m)}$ does manifestly not contain the zero section. In fact, based on known wave front sets of the above factors, an application of [Hör90, thm. 8.2.13] even yields that the integral of the above product is well-defined and gives an element of $\mathcal{E}'_V(D^{\oplus}N^{m+n-2k})$. Moreover, it is not difficult to see that the product $\star_H$ is continuous with respect to the Hörmander pseudo topology, and thus constitutes a well-defined product on $\mathcal{C}_{0,\text{ext}}(DN)$, in analogy to the scalar case. The following definition is thus a sensible one.

**Definition III.3.3.1** *Let $\mathcal{C}_{0,ext}(DN)$ be defined as in* (III.91) *and let $\star_H$ be defined as in* (III.94)*. By*

$$\mathcal{W}_0(DN) \doteq (\mathcal{C}_{0,ext}(DN), \star_H)$$

*we denote the **off-shell extended algebra of observables of the quantized Dirac field**.*

Let us now assume that we have defined a second product $\star_{H'}$ employing a bidistribution $\mathfrak{H}'$ which fulfils $WF(\mathfrak{H} - \mathfrak{H}') = \emptyset$. This case arises in particular if we define $\star_{H'}$ by means of changing the scale $\lambda$ in the logarithm appearing in $\mathfrak{H}$, or if we define $\star_{H'}$ utilising the two-point function $\omega_2$ of a Hadamard state $\omega$ on $\mathcal{A}(DM)$. In this case, one immediately obtains that the two algebras $(\mathcal{C}_{0,\text{ext}}(DN), \star_H)$ and $(\mathcal{C}_{0,\text{ext}}(DN), \star_{H'})$ are isomorphic via the relation

$$f \star_{H'} g = e^{\mathfrak{G}_d} \left[ (e^{-\mathfrak{G}_d} f) \star_H (e^{-\mathfrak{G}_d} g) \right],$$

with $d = H' - H$. In this sense, the algebra $(\mathcal{C}_{0,\text{ext}}(DN), \star_H)$ is *independent* of the chosen $\mathfrak{H}$, as long as it fulfils the correct antisymmetric wave front set condition. It is therefore not a flaw that one has to choose a concrete realisation $\star_{H'}$ of an allowed $\star$-product by means of a $\mathfrak{H}'$ which fulfils the (double) Dirac equation to be able to encode dynamics in $\mathcal{W}_0(DN)$ in a meaningful way by taking a suitable quotient. We expect the on-shell algebra resulting in this case can be shown to fulfil the time-slice axiom with the methods employed in [ChFr08]. Moreover, defining the construction by means of the two-point function $\omega_2$ of a Hadamard state, it can directly be extended to the full spacetime $M$.

**Definition III.3.3.2** *Let $\mathcal{C}_{0,ext}(DM)$ be defined as in* (III.91)*, let $\omega_2$ be the two-point function of a Hadamard state on $\mathcal{A}(DM)$, and let $\star_\omega$ be defined as in* (III.94)*, but with $\mathfrak{H}$ replaced by $\omega_2$. By*

$$\mathcal{W}(DM) \doteq (\mathcal{C}_{0,ext}(DM)/\mathcal{I}, \star_\omega)$$

*we denote the **on-shell extended algebra of observables of the quantized Dirac field**. Here, $\mathcal{I}$ is the ideal in $(\mathcal{C}_{0,ext}(DM), \star_\omega)$ generated by elements of the form $\mathfrak{Als}\, D^{\oplus} f$ for $f \in \mathcal{C}_{0,ext}(DM)$, where $D^{\oplus} = D \oplus D^*$.*

We shall now provide an example showing that the rather abstract algebra we have just defined really contains Wick polynomials and encodes the Wick theorem, which has ultimately been our initial motivation to consider Hadamard states after all. To this end, let us consider the smeared Wick square $:[\psi^\dagger\psi](f):$, where we take $f \in \mathscr{D}(M)$ and therefore assume implicitly that the spinor indices have been contracted. Our discussion of the enlarged algebra of the Klein-Gordon field suggests that such object corresponds to $f(x)\delta(x,y)\mathscr{I}(x,y)$, where $\delta(x,y)$ is the usual $\delta$-distribution on $M^2$ and $\mathscr{I}(x,y)$ is the spinor parallel transport along the geodesic connecting $x$ and $y$ (we implicitly assume that such geodesic exists and is unique). This test distribution is rather symmetric, and at first glance this seems to be in contradiction with the antisymmetry we have required for elements in $\mathscr{W}(DM)$. Notwithstanding, let us recall that we have required antisymmetry on the level of *double spinor* coefficients. Particularly, let us define

$$\mathscr{I}^\oplus_{\Theta\Theta'}(x,y) \doteq \begin{cases} (\mathscr{I}^{-1})^A{}_{B'}(x,y) & \text{if } (\Theta,\Theta') = (A, 4+B') \\[2mm] -\mathscr{I}_A{}^{B'}(x,y) & \text{if } (\Theta,\Theta') = (4+A, B') \\[2mm] 0 & \text{otherwise} \end{cases}.$$

Written in terms of a block matrix, this is equivalent to

$$\mathscr{I}^\oplus_{\Theta\Theta'}(x,y) = \begin{pmatrix} 0 & (\mathscr{I}^{-1})^A{}_{B'}(x,y) \\ -\mathscr{I}_A{}^{B'}(x,y) & 0 \end{pmatrix}.$$

We now propose that

$$\tilde{f} \doteq \frac{1}{2}f(x)\delta(x,y)\mathscr{I}^\oplus_{\Theta\Theta'}(x,y) \tag{III.96}$$

is the correct representative of $:[\psi^\dagger\psi](f):$ in $\mathscr{W}(DM)$. Let us see why this is the case. First, $\tilde{f}$ has the correct wave front set and antisymmetry property. Secondly, as we have the formal identity $B^\Theta(x) = \psi^\dagger_A(x) \oplus \psi^B(x)$, $:B(x)B(y):$ can be written in matrix form as

$$:B^\Theta(x)B^{\Theta'}(y): := \begin{pmatrix} :\psi^\dagger_A(x)\psi^\dagger_{B'}(y): & :\psi^\dagger_A(x)\psi^{B'}(y): \\ :\psi^A(x)\psi^\dagger_{B'}(y): & :\psi^A(x)\psi^{B'}(y): \end{pmatrix}.$$

If we heuristically view $:B(x)B(y):$ as being obtained by $B(x)B(y) - \mathfrak{H}(x,y)$ and consider the CAR, then this is an antisymmetric matrix. This follows, *e.g.* from $:\psi^\dagger(x)\psi(y): = -:\psi(y)\psi^\dagger(x):$ and it is clear that $:[\psi^\dagger\psi](f):$ can be obtained by smearing $:B(x)B(y):$ with $\mathscr{I}^\oplus_{\Theta\Theta'}(x,y)$. To strengthen this, let us see how Wick's theorem is implemented by $\star_H$. Namely, the usual approach to normal ordering with respect to the Fock representation in a (pure and quasifree) state $\omega$ would give us

$$:[\psi^\dagger\psi](x): :[\psi^\dagger\psi](y): =$$
$$:[\psi^\dagger\psi](x)\,[\psi^\dagger\psi](y): + :\psi^\dagger(x)\psi(y): \omega^-(x,y) + :\psi(x)\psi^\dagger(y): \omega^+(y,x) + \omega^+(y,x)\omega^-(x,y).$$

In contrast, for $\tilde{f}_1$ and $\tilde{f}_2$ defined as in (III.96), we find

$$\tilde{f}_1 \star_H \tilde{f}_2 = \left[\tilde{f}_1 \cdot_a \tilde{f}_2\right] \oplus \text{`one contraction with } \mathfrak{H}\text{'} \oplus \text{`two contractions with } \mathfrak{H}\text{'}.$$

If we now realise that the definition of $\mathfrak{H}(x,y)$ in theorem III.3.2.4 can be rephrased in block matrix form as

$$\mathfrak{H}^{\Theta\Theta'}(x,y) = \begin{pmatrix} 0 & [D_x^* H^+]_{B'}^{\ \ A}(y,x) \\ -[D_y^* H^-]_A^{\ \ B'}(x,y) & 0 \end{pmatrix},$$

we easily see the analogy between $\star_H$ and Wick's theorem.

Regarding states on $\mathcal{W}(DM)$, we can proceed in close analogy to the scalar case. We first realise that the only sensible interpretation of the evaluation of elements like $\widetilde{f} \in \mathcal{C}_{0,\text{ext}}(DM)$ in a Hadamard state $\omega$ is provided by integrating them with the regularised (and, hence, smooth) two-point function $:\omega_2: \doteq \omega_2 - \mathfrak{H}$. We generalise this by defining the regularised $n$-point functions of a *quasifree state* as

$$:\omega_n(x_1,\cdots,x_n): \doteq \begin{cases} 1 & \text{for } n = 0, \\ 0 & \text{for odd } n \text{ and} \\ \sum_{\pi_n \in S_n'} (-1)^{\text{sign}[\pi_n]} \prod_{i=1}^{n/2} (\omega_2 - h) \left( x_{\pi_n(2i-1)}, x_{\pi_n(2i)} \right) & \text{for even } n. \end{cases},$$

where $S_n'$ denotes the set of ordered permutations introduced in definition III.3.1.5. Given an arbitrary $f = \oplus_n \frac{1}{n!} f^{(n)} \in \mathcal{W}(DM)$ and a quasifree Hadamard state $\omega$ on $\mathcal{A}(DM)$, we then define the expectation value of $f$ in $\omega$ as

$$\omega(f) \doteq \sum_n \left\langle :\omega_n:, \frac{1}{n!} f^{(n)} \right\rangle. \tag{III.97}$$

This defines a complex-valued, normalised, linear functional on $\mathcal{W}(DM)$ which extends the action of the quasifree Hadamard state $\omega$ on $\mathcal{A}(DM)$ to $\mathcal{W}(DM)$. Recalling that $\mathcal{C}_0(DM)$ is dense in $\mathcal{C}_{0,\text{ext}}(DM)$ with respect to the Hörmander pseudo topology, we find that positivity of $\omega$ on $\mathcal{W}(DM)$ follows from positivity on $\mathcal{A}(DM)$ by the continuity of $\omega$. In fact, the full proof of theorem III.1 in [HoRu01] can be directly extended to the case of CAR fields, by essentially only inserting a few signs at the correct places.

To close the discussion of the extended algebra of Dirac quantum fields, we remark that objects like $:\psi^\dagger \gamma^\mu \psi:$ and $:\psi^\dagger \nabla_\mu \psi:$ can be easily included in $\mathcal{W}(DM)$ by appropriate generalisations of the spaces $\mathcal{E}_V'(D^{\oplus} M^n)$, as multiplication by smooth sections and derivatives do not increase the wave front set of a distribution (for the same reason, objects like $:\psi^\dagger D\psi:$ are already included in $\mathcal{W}(DM)$).

### III.3.4 Locality and General Covariance

By our brief review of locally covariant quantum field theory in subsection III.1.4, it should be clear how to generalise such concept to the case of vector-valued fields like the Dirac field.

Namely, as already pointed out in the seminal work [BFV03], one achieves this by exchanging the category of globally hyperbolic spacetimes $\mathfrak{Man}$ in definition III.1.4.1 with a more general one including the bundle structure of the vector-valued fields under consideration. In the case of Dirac fields, this means replacing $\mathfrak{Man}$ by the category of spin spacetimes $\mathfrak{SMan}$, see definition I.4.2. Regarding the notion of a locally covariant quantum field as reviewed in definition III.1.4.2, one has to replace the test function category $\mathfrak{Test}$ by a more general one, *e.g.* one incorporating $\mathscr{D}(D^{\oplus}M)$ as a test section space.

In fact, these constructions have been performed to some extent in [Ver01], and in more detail in [San08, San09b] (see also the foundational work in [Dim82]). Particularly, in [San08, San09b] it is found that the double Dirac field $B(f)$, and, hence, also the single ones $\psi^{\dagger}(f_1)$ and $\psi(f_2)$ are locally covariant quantum fields and that the algebra $\mathscr{A}(DM)$ can be re-casted as a locally covariant quantum field theory. These results follow essentially by the fact that only locally covariant concepts have been used in the construction of $\mathscr{A}(DM)$, namely, the bundle structure of Dirac fields, the Dirac equation, and the causal propagator $S$. We recall that we have made two choices in the construction of these objects. Namely, if $M$ is not simply connected, multiple spin structures exist on $M$ [Ger68, Ger70] and we have to choose one to start with. Moreover, we have chosen one specific representation of the Dirac algebra $Cl(3, 1)$ and, hence, a representation of the spin group $Spin_0(3, 1)$ in order to define $DM$ and the associated structures. However, these choices do not interfere with locality and general covariance, as spin structures depend only on the topological properties of a spacetime, and are not affected by the morphisms in $\mathfrak{SMan}$, which are diffeomorphisms in particular. Regarding the representation of $Cl(3, 1)$, one can simply agree to use the same for all spin spacetimes. Nevertheless, the quantum theory of a Dirac field on curved spacetimes is inherently non-unique, and in [San08, San09b] this non-uniqueness has been analysed in great detail. It turns out that such non-uniqueness vanishes if one restricts the relevant algebras to contain only observable elements, although it seems that not all restrictions we have proposed in our definition III.3.1.3 of 'observable' are necessary. In fact it seems to be sufficient to restrict oneself to the subalgebra of $\mathscr{A}(DM)$ generated by an even number of fields $B(f)$. We refer the reader to [San08, San09b] for the details of the mentioned result.

Finally, let us consider how the enlarged algebra $\mathscr{W}(DM)$ fits in the locally covariant framework. We have not repeated the analysis in [HoWa01, HoWa05] of the renormalisation freedom of elements in $\mathscr{W}(M)$ for the case of $\mathscr{W}(DM)$, but we expect that such analysis would yield the following result: for general elements of $\mathscr{W}(DM)$, one would presumably find the renormalisation freedom discovered in [HoWa01, HoWa05] plus additional degrees of freedom in terms of spinor-tensors of appropriate dimension and rank built out of $\gamma$-matrices and the spin curvature tensor $\mathfrak{R}_{\alpha\beta}$. Whereas, if one considers observable elements of $\mathscr{W}(DM)$, then one would most likely find that their renormalisation freedom is exactly of the same kind as found in [HoWa01, HoWa05]. We shall discuss this in more detail in the next chapter for a particular observable Wick polynomial, namely, the stress-energy tensor.

## III.4 The Free Dirac Field in Null Big Bang Spacetimes

In this section, we will for the first time extend the general holographic ideas introduced and applied in [DMP06, DMP09a, DMP09b, DMP09c, Mor06, Mor08] to the case of Dirac fields. Thereby, we will be able to construct states similar to the ones found in section III.2, namely, asymptotic conformal ground and equilibrium states of Hadamard type.

### III.4.1 The Bulk and Boundary Algebras

The first step in the construction of preferred states on the Borchers-Uhlmann algebra $\mathscr{A}(DM_B)$ of the Dirac field on an NBB spacetime $(M_B, g_B)$ by holographic methods is to find a relation between $\mathscr{A}(DM_B)$ and a suitable algebra on the boundary of an NBB spacetimes. Our treatment in section III.2 already suggests that this is most conveniently formulated on the level of solutions of the Dirac equation, rather then on the level of test sections. Let us therefore first explain how the Borchers-Uhlmann algebra $\mathscr{A}(DM_B)$ introduced in definition III.3.1.1 can be equivalently stated in terms of the space of solutions $\mathfrak{S}(M_B)$ of the Dirac equation on $(M_B, g_B)$. To this end, let us recall that in definition III.3.1.1 we have taken a suitable quotient to encode the Dirac equation in $\mathscr{A}(DM_B)$. Particularly, $\mathscr{A}(DM_B)$ is generated by equivalence classes $[f]$ of test sections $f \in \mathscr{D}(DM)$, where two representatives in $[f]$ differ by a solution of the Dirac equation. Hence, an equivalence class $[f]$ is in one-to-one correspondence with an element $u \in \mathfrak{S}(M_B)$ of the form $u = S_B f$, where $S_B$ denotes the causal propagator of the Dirac equation on $(M_B, g_B)$. Moreover, recalling the relation between the inner product $\iota_B$ on $\mathfrak{S}(M_B)$ to the causal propagator $S_B$ found in lemma II.3.2.3 and denoting by $\mathfrak{S}^\dagger(M_B) \subset \mathscr{E}(D^*M)$ the space of solutions of the adjoint Dirac equation, one can formulate $\mathscr{A}(DM_B)$ equivalently in the following way: one considers the algebra generated by tensor products of $\mathfrak{S}(M_B) \oplus \mathfrak{S}^\dagger(M_B)$ and endows it with the locally convex topology induced by the one on $\mathscr{E}(DM) \oplus \mathscr{E}(D^*M)$. Moreover, to encode the CAR, one defines a suitable 'double inner product' on $\mathfrak{S}(M_B) \oplus \mathfrak{S}^\dagger(M_B)$ out of $\iota_B$ and considers an appropriate quotient space which enforces CAR specified by this double inner product. With these considerations in mind, the following definition seems natural.

**Definition III.4.1.1** *Let* $(\mathsf{S}(\mathfrak{J}^-), \mathfrak{i})$ *denote the boundary inner product space, cf. subsection II.4.2 and let* $\mathsf{S}^\dagger(\mathfrak{J}^-)$ *denote the adjoint space*

$$\mathsf{S}^\dagger(\mathfrak{J}^-) \doteq \{u^\dagger \mid u \in \mathsf{S}(\mathfrak{J}^-)\}.$$

*Moreover, let*

$$\mathsf{S}^\oplus(\mathfrak{J}^-) \doteq \mathsf{S}(\mathfrak{J}^-) \oplus \mathsf{S}^\dagger(\mathfrak{J}^-),$$

*let us define an inner product* $\mathfrak{i}^\oplus$ *on* $\mathsf{S}^\oplus(\mathfrak{J}^-)$ *by setting*

$$\mathfrak{i}^\oplus(u_1 + u_2^\dagger, \; v_1 + v_2^\dagger) \doteq \mathfrak{i}(v_2, u_1) + \mathfrak{i}(u_2, v_1),$$

*and let us define a conjugation* $\mathfrak{k}$ *on* $\mathsf{S}^\oplus(\mathfrak{J}^-)$ *via*

$$\mathfrak{k}(u_1 \oplus u_2^\dagger) \doteq u_2 \oplus u_1^\dagger.$$

The **boundary Borchers-Uhlmann algebra** $\mathscr{A}(D\mathfrak{I}^-)$ of the free Dirac field is defined as

$$\mathscr{A}(D\mathfrak{I}^-) \doteq \mathscr{A}_0(D\mathfrak{I}^-)/\mathscr{I}\,,$$

where $\mathscr{A}_0(D\mathfrak{I}^-)$ is the direct sum

$$\mathscr{A}_0(D\mathfrak{I}^-) \doteq \bigoplus_{n=0}^{\infty} \mathsf{S}^{\oplus}(\mathfrak{I}^-)^n$$

$(\mathsf{S}^{\oplus}(\mathfrak{I}^-)^0 \doteq \mathbb{C})$. Elements of $\mathscr{A}_0(D\mathfrak{I}^-)$ are required to be finite linear combinations of tensor powers of elements in $\mathsf{S}^{\oplus}(\mathfrak{I}^-)$ and $\mathscr{A}_0(D\mathfrak{I}^-)$ is equipped with a product defined by the linear extension of the tensor product of $[\mathscr{C}^{\infty}(\mathfrak{I}^-, \mathbb{C}^4) \oplus \mathscr{C}^{\infty}(\mathfrak{I}^-, \mathbb{C}^{*4})]^n$, a *-operation defined by the antilinear extension of $[u^*](x_1, \cdots, x_n) = [\mathfrak{k}^{\otimes n} u](x_n, \cdots, x_1)$, and the topology defined by saying that a sequence $\{u_k\}_k = \{\oplus_n u_k^{(n)}\}_k$ in $\mathscr{A}_0(D\mathfrak{I}^-)$ converges to $u = \oplus_n u^{(n)}$ if $u_k^{(n)}$ converges to $u^{(n)}$ for all $n$ in the locally convex topology of $[\mathscr{C}^{\infty}(\mathfrak{I}^-, \mathbb{C}^4) \oplus \mathscr{C}^{\infty}(\mathfrak{I}^-, \mathbb{C}^{*4})]^n$ and there exists an $N$ such that $u_k^{(n)} = 0$ for all $n > N$ and all $k$. Moreover, $\mathscr{I}$ is the closed *-ideal generated by elements of the form

$$\left( \left[u_1 \oplus u_2^{\dagger}\right] \otimes \left[v_1 \oplus v_2^{\dagger}\right] + \left[v_1 \oplus v_2^{\dagger}\right] \otimes \left[u_1 \oplus u_2^{\dagger}\right] \right) \oplus -\mathrm{i}(u_1 \oplus u_2^{\dagger}, v_1 \oplus v_2^{\dagger}),$$

and $\mathscr{A}(D\mathfrak{I}^-)$ is thought to be equipped with the product, *-operation, and topology descending from $\mathscr{A}_0(D\mathfrak{I}^-)$.

We stress that, as in the scalar case, the just defined algebra is not necessarily closed with respect to its topology. However, the topology is still sufficient to define states on $\mathscr{A}(D\mathfrak{I}^-)$ as continuous linear functionals, which is what we are interested in. Moreover, the following result is not a surprise, as $\mathscr{A}(D\mathfrak{I}^-)$ has been perfectly tailored exactly for this purpose. Namely, employing theorem II.4.2.1, we can directly prove:

**Proposition III.4.1.2** Let $\mathfrak{G} : \mathfrak{S}(M_B) \to \mathsf{S}(\mathfrak{I}^-)$ denote the bulk-to-boundary map introduced in theorem II.4.2.1 and let $[f] \in \mathscr{D}(DM)/(D\mathscr{D}(DM))$ and $[g^{\dagger}] \in \mathscr{D}(D^*M)/(D^*\mathscr{D}(D^*M))$ denote the equivalence classes of $f \in \mathscr{D}(DM)$ and $g^{\dagger} \in \mathscr{D}(D^*M)$ respectively. The map $i_{\mathfrak{G}} : \mathscr{A}(DM_B) \to \mathscr{A}(D\mathfrak{I}^-)$ defined by the tensorialisation of

$$i_{\mathfrak{G}}([f] \oplus [g^{\dagger}]) \doteq \mathfrak{G} S_B f \oplus (\mathfrak{G} S_B g)^{\dagger}$$

is a continuous, unit-preserving, injective *-homomorphism.

Finally, we introduce a natural action of the supertranslations on $\mathfrak{I}^-$ also on the newly obtained algebra $\mathscr{A}(D\mathfrak{I}^-)$. As the boundary inner product defined in subsection II.4.2 is given by the integral with respect to the supertranslation-invariant measure $dv d\mathbb{S}^2$ induced by the Bondi metric $h$, the following result is easily seen to hold exactly as in the scalar case.

**Lemma** III.4.1.3 *Let $\zeta : \mathbb{S}^2 \to \mathbb{R}$ be a smooth function on the two-sphere $\mathbb{S}^2$, let $u$ be an element of $\mathsf{S}(\mathfrak{J}^-)$, and let $\theta$, $\varphi$ denote a coordinate system on $\mathbb{S}^2$. The action*

$$\alpha_{\zeta_t} : \mathsf{S}(\mathfrak{J}^-) \to \mathsf{S}(\mathfrak{J}^-), \quad u(v, \theta, \varphi) \mapsto u\left(v - \zeta(\theta, \varphi)t, \theta, \varphi\right)$$

*induces a $*$-automorphism on $\mathcal{A}(D\mathfrak{J}^-)$. In the case $\zeta \equiv 1$, we shall denote it by $\alpha_t : \mathcal{A}(D\mathfrak{J}^-) \to \mathcal{A}(D\mathfrak{J}^-)$.*

## III.4.2 Preferred Asymptotic Ground States and Thermal States of Hadamard Type

We shall now construct a preferred state on $\mathcal{A}(D\mathfrak{J}^-)$ which is a ground state with respect to $v$-translations. It seems clear what we have to do to obtain such state. We take the Minkowski vacuum of Dirac fields as an inspiration, define a suitable splitting of $\mathsf{S}(\mathfrak{J}^-)$ in positive and negative frequency subspaces, and write down a state with the correct spectral properties.

**Proposition** III.4.2.1 *Let $u$ be an arbitrary element of $\mathsf{S}(\mathfrak{J}^-)$. We define the Fourier-Plancherel transform $\hat{u}$ of $u$ on the level of components as*

$$\hat{u}(k, \theta, \varphi) \doteq \frac{1}{\sqrt{2\pi}} \int_{\mathbb{R}} dv \, u(v, \theta, \varphi) \, e^{ikv}.$$

*Based on this, we define two distributions $w^\pm$ on $\mathsf{S}(\mathfrak{J}^-) \otimes \mathsf{S}^\dagger(\mathfrak{J}^-)$. Namely, let $u_1$, $u_2 \in \mathsf{S}(\mathfrak{J}^-)$ be arbitrary. We set*

$$w^\pm(u_1, u_2^\dagger) \doteq \int_{\mathbb{R} \times \mathbb{S}^2} dk \, d\mathbb{S}^2 \, \Theta(\pm k) \, \hat{u}_2^*(k, \theta, \varphi) \, \hat{u}_1(k, \theta, \varphi).$$

*Here, $*$ denotes the adjoint with respect to the canonical inner product on $\mathbb{C}^4$. The bidistribution $\mathfrak{w}_2^{\mathfrak{J}}$ on $\mathsf{S}^{\oplus}(\mathfrak{J}^-)^2$ defined as*

$$\mathfrak{w}_2^{\mathfrak{J}}(u_1 \oplus u_2^\dagger, \, v_1 \oplus v_2^\dagger) \doteq w^-(u_1, v_2^\dagger) + w^+(v_1, u_2^\dagger),$$

*for arbitrary $u_1 \oplus u_2^\dagger$, $v_1 \oplus v_2^\dagger \in \mathsf{S}^{\oplus}(\mathfrak{J}^-)$ induces an $\alpha_t$-invariant, quasifree state $\mathfrak{w}^{\mathfrak{J}}$ on $\mathcal{A}(D\mathfrak{J}^-)$.*

*Proof.* Continuity of $\mathfrak{w}_2^{\mathfrak{J}}$ follows from the continuity of the inner product on the square-integrable, $\mathbb{C}^4$-valued functions on $\mathbb{R} \times \mathbb{S}^2$. To show positivity (*cf.* (III.76)), we compute

$$\mathfrak{w}_2^{\mathfrak{J}}\left(u_1 \oplus u_2^\dagger, \, (u_1 \oplus u_2^\dagger)^*\right) = \mathfrak{w}_2^{\mathfrak{J}}\left(u_1 \oplus u_2^\dagger, \, u_2 \oplus u_1^\dagger\right) = w^-(u_1, u_1^\dagger) + w^+(u_2, u_2^\dagger) \geq 0.$$

Moreover, to see the CAR property, we recall the definition of the boundary inner product $\mathfrak{i}$ and compute

$$\mathfrak{i}(u_1, u_2) = \int_{\mathfrak{J}^-} dv \, d\mathbb{S}^2 \, u_1^*(v, \theta, \varphi) \, u_2(v, \theta, \varphi) = \int_{\mathbb{R} \times \mathbb{S}^2} dk \, d\mathbb{S}^2 \, \hat{u}_1^*(k, \theta, \varphi) \, \hat{u}_2(k, \theta, \varphi),$$

where the latter equality follows from the unitarity of the Fourier-Plancherel transform on square-integrable functions. Hence, we find

$$w^+(u_2, u_1^\dagger) + w^-(u_2, u_1^\dagger) = \mathrm{i}(u_1, u_2),$$

which is equivalent to the CAR property (here written in unsmeared form for simplicity)

$$\mathfrak{w}_2^\mathfrak{I}(x, y) + \mathfrak{w}_2^\mathfrak{I}(y, x) = \mathrm{i}^\oplus(x, y).$$

Finally, $\alpha_t$-invariance follows trivially from $\widehat{\alpha_t u} = e^{ikt}\hat{u}$ for all $u \in \mathsf{S}(\mathfrak{I}^-)$. $\qquad\square$

We now define a state $\mathfrak{w}^B$ on $\mathscr{A}(DM_B)$ as a pull-back via the bulk-to-boundary *-homomorphism $i_\mathfrak{G}$, viz.

$$\mathfrak{w}^B \doteq \mathfrak{w}^\mathfrak{I} \circ i_\mathfrak{G}.$$

This state is quasifree per definition. Moreover, let $\tilde{f} = f_1 \oplus f_2$, $\tilde{g} = g_1 \oplus g_2 \in \mathscr{D}(D^\oplus M_B)$ be arbitrary. We define the analogues of $\omega^\pm$ in the state $\mathfrak{w}^B$ by setting

$$\mathfrak{w}^B(\tilde{f}, \tilde{g}) \doteq \mathfrak{w}^-(f_1, g_2) + \mathfrak{w}^+(g_1, f_2). \tag{III.98}$$

From this, we obtain

$$\mathfrak{w}^\pm(f, g) = w^\pm\left(\mathfrak{G}S_B f, \left[\mathfrak{G}S_B g^\dagger\right]^\dagger\right),$$

for arbitrary $f \in \mathscr{D}(DM)$, $g \in \mathscr{D}(D^*M)$. We would now like to show that the found state is charge-conjugation invariant, see III.3.1. To this end, we have to assure that

$$\mathfrak{w}^\pm(g^{\dagger c}, f^{\dagger c}) = \mathfrak{w}^\mp(f, g)$$

holds. Let us recall the realisation of the conjugation maps in definition II.3.1.2, the relation of $S_B$ to them (see the end of subsection II.3.1), and the fact that $\mathfrak{G}$ is a (real) conformal rescaling and a smooth extension. Moreover, let us set $u \doteq \mathfrak{G}S_B f$, $v \doteq \mathfrak{G}S_B g^\dagger$. With these preparations, we compute

$$\mathfrak{w}^\pm(g^{\dagger c}, f^{\dagger c}) = w^\pm\left(\mathfrak{G}S_B g^{\dagger c}, \left[\mathfrak{G}S_B f^{\dagger c\dagger}\right]^\dagger\right) = w^\pm\left(v^c, u^{\dagger c}\right)$$

$$= \int\limits_{\mathbb{R}\times\mathbb{S}^2} dk\, d\mathbb{S}^2\, \Theta(\pm k)\, \widehat{u^{\dagger c\dagger}}^* \, \widehat{v^c} = \int\limits_{\mathbb{R}\times\mathbb{S}^2} dk\, d\mathbb{S}^2\, \Theta(\pm k)\, \widehat{\bar{u}}^* \, \widehat{\bar{v}} = \int\limits_{\mathbb{R}\times\mathbb{S}^2} dk\, d\mathbb{S}^2\, \Theta(\mp k)\, \hat{v}^* \, \hat{u}$$

$$= w^\mp\left(u, v^\dagger\right) = \mathfrak{w}^\mp(f, g).$$

As promised in the title of this subsection, we shall now prove that the found state is Hadamard.

**Theorem III.4.2.2**  *The following assertions hold for the charge-conjugation invariant, quasifree state $\mathfrak{w}^B$ on $\mathscr{A}(DM_B)$.*

a) *The two point functions $\mathfrak{w}^{\pm}(f, g)$ defined in (III.98) are distributions in $\mathscr{D}'(D^*M_B \boxtimes DM_B)$ of the form*

$$\mathfrak{w}^{\pm}(f, g) = \lim_{\varepsilon \downarrow 0} \frac{1}{2\pi i} \int\limits_{\mathbb{R}^2 \times \mathbb{S}^2} dv \, dv' \, d\mathbb{S}^2 \, \frac{\left[\mathfrak{G} S_B g^{\dagger}\right]^{*}(v', \theta, \varphi) \, \left[\mathfrak{G} S_B f\right](v, \theta, \varphi)}{v - v' + i\varepsilon} \, .$$

b) $\mathfrak{w}^B$ *fulfils the Hadamard condition (see definition III.3.2.1) and is therefore a Hadamard state.*

*Proof.*    a) The proof proceeds in analogy to the one of proposition III.2.2.4, which in turn is taken from [Mor08]. We recall

$$\mathfrak{w}^{\pm}(f, g) = w^{\pm}\left(\mathfrak{G} S_B f, \left[\mathfrak{G} S_B g^{\dagger}\right]^{\dagger}\right) = \int\limits_{\mathbb{R} \times \mathbb{S}^2} dk \, d\mathbb{S}^2 \, \Theta(\pm k) \, \widehat{\mathfrak{G} S_B g^{\dagger}}^{*}(k, \theta, \varphi) \, \widehat{\mathfrak{G} S_B f}(k, \theta, \varphi)$$

By theorem II.4.2.1, $\mathfrak{G} S_B g^{\dagger}$ and $\mathfrak{G} S_B f$ are square-integrable on $\mathfrak{J}^-$ with respect to $dv \, d\mathbb{S}^2$, we can therefore employ the unitarity of the Fourier-Plancherel transform to compute

$$\mathfrak{w}^{\pm}(f, g) = \int\limits_{\mathfrak{J}^-} dv' \, d\mathbb{S}^2 \, \left[\mathfrak{G} S_B g^{\dagger}\right]^{*}(v', \theta, \varphi) \, \mathscr{F}^{-1}\left[\Theta(\pm k) \widehat{\mathfrak{G} S_B f}\right](v', \theta, \varphi),$$

where $\mathscr{F}^{-1}$ denotes the inverse Fourier-Plancherel transform. To be able to use the convolution theorem on the second factor, we make use of the continuity of $\mathscr{F}^{-1}$ and insert a regularising factor $e^{\mp k\varepsilon}$. Recalling the Fourier-Plancherel transform of the Heaviside step function (see for instance [ReSi75]), we find

$$\mathscr{F}^{-1}\left[\Theta(\pm k) \widehat{\mathfrak{G} S_B f}\right](v', \theta, \varphi) = \lim_{\varepsilon \downarrow 0} \mathscr{F}^{-1}\left[e^{\mp \varepsilon k} \Theta(\pm k) \widehat{\mathfrak{G} S_B f}\right](v', \theta, \varphi)$$

$$= \lim_{\varepsilon \downarrow 0} \frac{1}{2\pi i} \int\limits_{\mathbb{R}} dv \, \frac{\left[\mathfrak{G} S_B f\right](v, \theta, \varphi)}{v - v' \pm i\varepsilon} \, .$$

On account of the decay properties of the smooth and square-integrable functions $\mathfrak{G} S_B g^{\dagger}$ and $\mathfrak{G} S_B f$ found in theorem II.4.2.1, we can insert this in the last identity for $\mathfrak{w}^{\pm}(f, g)$ to find that the resulting integrand is integrable in the joint measure $dv \, dv' \, d\mathbb{S}^2$ and, hence, $\mathfrak{w}^{\pm}(f, g)$ have the asserted form. The same decay properties in combination with the continuity of $S_B$ imply the continuity of $\mathfrak{w}^{\pm}(f, g)$.

b) By lemma III.3.2.2, it is sufficient to prove

$$WF(\mathfrak{w}^{\pm}) = \left\{(x, y, k_x, -k_y) \in T^*M^2 \setminus \{0\} \mid (x, k_x) \sim (y, k_y), \, k_x \triangleright 0\right\}.$$

The proof of this statement can be obtained by repeating the proof of theorem III.2.2.5 with minor modifications, where we recall that the latter proof has essentially been a repetition of the Hadamard proof in [Mor08]. We therefore choose the same notation as in the proof of theorem III.2.2.5 and only sketch the main argument and the modifications necessary to show the assertion considered here. To this end, we recall the explicit form of the bulk-to-boundary map $\mathfrak{G}$ found in theorem II.4.2.1. Namely,

$$\mathfrak{G}S_B f = \frac{\Omega_{\mathbb{M}}^{\frac{3}{2}}}{\Omega_B^{\frac{3}{2}}} S_E \frac{\Omega_{\mathbb{M}}^{\frac{5}{2}}}{a^{\frac{5}{2}}} f = \frac{1}{\sqrt{1+v^2}} S_E \frac{\Omega_{\mathbb{M}}^{\frac{5}{2}}}{a^{\frac{5}{2}}} f \doteq \widetilde{S} f \,, \tag{III.99}$$

where $S_E$ is the causal propagator of the extended Dirac operator $D_E$ on the lower, slit half $(M_E^-, g_E)$ of the Einstein static universe $(M_E, g_E)$ and we refer the reader to section I.3.2 for the definitions of the various conformal factors. What is important here is that the appearing conformal factors are smooth in the regions where they are evaluated, and, hence, $WF(\widetilde{S}) = WF(S_E)$, which in turn equals the by now well-known symmetric wave front set of the causal propagator of the Klein-Gordon operator, hence,

$$WF(\widetilde{S}) = \left\{ (x, y, k_x, -k_y) \in T^*(M_E^-)^2 \setminus \{0\} \mid (x, k_x) \sim (y, k_y) \right\} \,.$$

From this one finds that $\widetilde{S}(x, y)$ can be restricted to $\mathfrak{J}^-$ as its wave front set is conormal to $\mathfrak{J}^-$. We therefore recognise $\mathfrak{w}^\pm(x, y)$ as being the results of composing

$$\mathfrak{T}^\pm \doteq \lim_{\varepsilon \downarrow 0} \frac{1}{v - v' \pm i\varepsilon} \otimes \delta(\vec{\theta}, \vec{\theta}')_{\mathbb{S}^2 \times \mathbb{S}^2}$$

on $\mathfrak{J}^- \times \mathfrak{J}^-$ with two copies of $\widetilde{S}$. Here, $\delta(\vec{\theta}, \vec{\theta}')_{\mathbb{S}^2 \times \mathbb{S}^2}$ denotes the $\delta$-distribution on $\mathbb{S}^2$ and $\mathfrak{T}^-$ has the same wave front set as the distribution $T$ in the proof of theorem III.2.2.5, because $(x - i\varepsilon)^{-1}$ and $(x - i\varepsilon)^{-2}$ have the same wave front set (see for instance [Str09]), whereas $\mathfrak{T}^+$ has the 'flipped' wave front set, i.e.

$$WF(\mathfrak{T}^+(v, v', \vec{\theta}, \vec{\theta}')) = WF(\mathfrak{T}^-(v', v, \vec{\theta}, \vec{\theta}')).$$

We would now like to use the theorems 8.2.13. and 8.2.14. in [Hör90] to compute the wave front set of $\mathfrak{w}^\pm$ (note that this is possible despite of the vector-valued nature of the involved distributions, because we can understand their composition as sums of compositions of scalar distributions). To this end, one introduces an arbitrary diamond-shaped set $N$ in $M_B$ and constructs the related cut-off function $\chi \in \mathscr{D}(M_E)$. This allows us to split $\widetilde{S}$ into

$$\widetilde{S} = \chi \widetilde{S} + (1 - \chi)\widetilde{S} \doteq \mathbb{S} + \mathscr{S} \,,$$

where $\mathscr{S}$ turns out to be smooth and $\mathbb{S}$ is a distribution of compact support on $\mathfrak{J}^- \times N$. Building on this, we can split $\mathfrak{w}^-(f, g)$ into four parts

$$\mathfrak{w}^\pm(f, g) = \mathfrak{w}^\pm_{\mathbb{S}\mathbb{S}}(f, g) + \mathfrak{w}^\pm_{\mathscr{S}\mathbb{S}}(f, g) + \mathfrak{w}^\pm_{\mathbb{S}\mathscr{S}}(f, g) + \mathfrak{w}^\pm_{\mathscr{S}\mathscr{S}}(f, g),$$

whose wave front sets we analyse separately. Per construction, $\mathfrak{w}^{\pm}_{\mathbb{S}\mathbb{S}}(x, y)$ is a composition of two copies of the compactly supported distribution $\mathbb{S}$ and the distributions $\mathfrak{T}^{\pm}$. [Hör90, thm. 8.2.13.] therefore entails that

$$WF(\mathfrak{w}^{\pm}_{\mathbb{S}\mathbb{S}}) \subset \left\{ (x, y, k_x, -k_y) \in T^*M^2 \setminus \{0\} \mid (x, k_x) \sim (y, k_y),\ k_x \vartriangleleft_{\vartriangleright} 0 \right\}.$$

Moreover, via [Hör90, thm. 8.2.14.] one finds that $\mathfrak{w}^{\pm}_{\mathscr{S}\mathbb{S}}$ and $\mathfrak{w}^{\pm}_{\mathbb{S}\mathscr{S}}$ have empty wave front sets, whereas $\mathfrak{w}^{\pm}_{\mathscr{S}\mathscr{S}}$ can directly be computed to have a smooth kernel by the fall-off properties of elements in $\mathsf{S}(\mathfrak{J}^-)$ (which are visible in (III.99)). Altogether, one finds

$$WF(\mathfrak{w}^{\pm}) \subset \left\{ (x, y, k_x, -k_y) \in T^*M^2 \setminus \{0\} \mid (x, k_x) \sim (y, k_y),\ k_x \vartriangleleft_{\vartriangleright} 0 \right\}.$$

Equality of these sets now follows from the CAR property

$$\mathfrak{w}^+(f, g) + \mathfrak{w}^-(f, g) = i S_B(g, f)$$

and the knowledge of $WF(S_B)$ by the argument already given in the proof of lemma III.3.2.2. We therefore find that $\mathfrak{w}^B$ is a Hadamard state on $\big(\mathscr{A}(M_B)$ restricted to the spacetime $(N, g_B)$. To extend this property to the full spacetime $(M_B, g_B)$, we realise that, by the very definition of $N$ as a diamond-shaped set, we can collect several such $N$ to form a *causal normal neighbourhood* $\mathscr{N}(\Sigma)$ of an arbitrary Cauchy surface $\Sigma$ of $(M_B, g_B)$. Here, given a Cauchy surface $\Sigma$ of $(M_B, g_B)$, $\mathscr{N}(\Sigma)$ is called a causal normal neighbourhood of $\Sigma$ if $\Sigma$ is a Cauchy surface for $(\mathscr{N}(\Sigma), g_B)$ and for every pair of points $x, y \in \mathscr{N}(\Sigma)$ with $y \in J^+(x, M_B)$, there is a geodesically convex neighbourhood in $M_B$ which contains $x$ and $y$ [KaWa91]. Hence, we find that $\mathfrak{w}^B$ is a Hadamard state in a causal normal neighbourhood of a Cauchy surface of $M_B$, and by a propagation of singularities argument in combination with a propagation of Hadamard form argument as given in the proof of [SaVe01, thm 5.8.], we obtain the wanted Hadamard property of $\omega^B$ on the full NBB spacetime.

$\square$

After dealing with a ground state on $\mathscr{A}(D\mathfrak{J}^-)$, we now provide a construction of KMS states on $\mathscr{A}(D\mathfrak{J}^-)$. Unsurprisingly, this is achieved by inserting a *Fermi-Dirac distribution* by hand in the spectral decomposition of $\mathfrak{w}^3$, *i.e.* by 'smearing out the Fermi step at $k = 0$'.

**Proposition III.4.2.3** *Let* $u_1, u_2 \in \mathsf{S}(\mathfrak{J}^-)$ *be arbitrary. We define two distributions* $w^{\pm}_{\beta}$ *on* $\mathsf{S}(\mathfrak{J}^-) \otimes \mathsf{S}^{\dagger}(\mathfrak{J}^-)$ *by setting*

$$w^{\pm}_{\beta}(u_1, u_2^{\dagger}) \doteq \int\limits_{\mathbb{R} \times \mathbb{S}^2} dk\, d\mathbb{S}^2\ \frac{\hat{u}_2^*(k, \theta, \varphi)\, \hat{u}_1(k, \theta, \varphi)}{e^{\mp \beta k} + 1}.$$

*The bidistribution* $\mathfrak{w}^3_{\beta, 2}$ *on* $\mathsf{S}^{\oplus}(\mathfrak{J}^-)^2$ *defined as*

$$\mathfrak{w}^3_{\beta, 2}(u_1 \oplus u_2^{\dagger},\ v_1 \oplus v_2^{\dagger}) \doteq w^-_{\beta}(u_1, v_2^{\dagger}) + w^+_{\beta}(v_1, u_2^{\dagger}),$$

*for arbitrary $u_1 \oplus u_2^\dagger$, $v_1 \oplus v_2^\dagger \in \mathsf{S}^\oplus(\mathfrak{I}^-)$ induces an $\alpha_t$-invariant, quasifree state $\mathfrak{w}_\beta^3$ on $\mathscr{A}(D\mathfrak{I}^-)$. Moreover, $\mathfrak{w}_\beta^3$ is a KMS state at inverse temperature $\beta$.*

*Proof.* Note that the above expressions are well-defined, being integrals of an integrable term multiplied by a smooth function which is bounded by 1. Continuity, positivity, and $\alpha_t$-invariance of $\mathfrak{w}_\beta^3$ follow by the same arguments as the ones employed in the analysis of $\mathfrak{w}^3$. Moreover, the CAR property follows by a simple computation from

$$\frac{1}{e^{\beta k} + 1} + \frac{1}{e^{-\beta k} + 1} = 1.$$

To see that the induced state fulfils the KMS condition, let $\tilde{u} \doteq u_1 \oplus u_2^\dagger$, $\tilde{v} \doteq v_1 \oplus v_2^\dagger$ be arbitrary elements of $\mathsf{S}^\oplus(\mathfrak{I}^-)$. We define

$$F(t) \doteq \mathfrak{w}_{\beta,2}^3\left(\tilde{v}, \alpha_t(\tilde{u})\right) = w_\beta^-\left(v_1, \alpha_t(u_2^\dagger)\right) + w_\beta^+\left(\alpha_t(u_1), v_2^\dagger\right)$$
$$= \int\limits_{\mathbb{R} \times \mathbb{S}^2} dk\, d\mathbb{S}^2\, \frac{\hat{u}_2^*(k,\theta,\varphi)\, \hat{v}_1(k,\theta,\varphi)\, e^{-ikt}}{e^{\beta k} + 1} + \frac{\hat{v}_2^*(k,\theta,\varphi)\, \hat{u}_1(k,\theta,\varphi)\, e^{ikt}}{e^{-\beta k} + 1},$$

$$G(t) \doteq \mathfrak{w}_{\beta,2}^3\left(\alpha_t(\tilde{u}), \tilde{v}\right) = w_\beta^-\left(\alpha_t(u_1), v_2^\dagger\right) + w_\beta^+\left(v_1, \alpha_t(u_2^\dagger)\right)$$
$$= \int\limits_{\mathbb{R} \times \mathbb{S}^2} dk\, d\mathbb{S}^2\, \frac{\hat{v}_2^*(k,\theta,\varphi)\, \hat{u}_1(k,\theta,\varphi)\, e^{ikt}}{e^{\beta k} + 1} + \frac{\hat{u}_2^*(k,\theta,\varphi)\, \hat{v}_1(k,\theta,\varphi)\, e^{-ikt}}{e^{-\beta k} + 1}.$$

The above integrands are absolutely integrable and uniformly bounded in $t$, $F(t)$ and $G(t)$ are therefore bounded and continuous. In fact, one may view $F(t)$ and $G(t)$ as inverse Fourier-Plancherel transforms of square-integrable functions and, hence, directly compute their Fourier-Plancherel transforms. In analogy to the proof of proposition III.2.2.6, one finds

$$\widehat{F}(E) = e^{\beta E}\widehat{G}(E)$$

which is equivalent to the KMS condition by lemma III.1.1.4 and the fact that $\mathscr{A}(D\mathfrak{I}^-)$ is a Borchers-Uhlmann algebra. $\qquad\square$

We pull-back $\mathfrak{w}_\beta^3$ by $i_\mathfrak{G}$ to obtain a state

$$\mathfrak{w}_\beta^B \doteq \mathfrak{w}_\beta^3 \circ i_\mathfrak{B}$$

on $\mathscr{A}(DM_B)$. $\mathfrak{w}_\beta^B$ is manifestly quasifree, and one can show that it is charge-conjugation invariant by a similar computation as the one which showed this property in the case of $\mathfrak{w}^B$. Let us note that the spectral decomposition $\mathfrak{w}_\beta^3$ approaches the one of $\mathfrak{w}^3$ for large $k$. Hence, we can expect that $\mathfrak{w}_\beta^B$ shares the Hadamard property enjoyed by $\mathfrak{w}^B$. Let us prove it.

**Theorem** III.4.2.4 *The charge-conjugation invariant, quasifree states $\mathfrak{w}_\beta^B$ on $\mathscr{A}(DM_B)$ fulfil the Hadamard condition and are therefore Hadamard states.*

*Proof.* The proof is essentially a repetition of our proof of theorem III.2.2.7, we therefore only sketch the main steps. We consider the setup and notation of our proof of theorem III.4.2.2. Let therefore $N$ and $\widetilde{S} = \mathbb{S} + \mathscr{S}$ be defined as in the mentioned proof and let $\tilde{f} = f_1 \oplus f_2$, $\tilde{g} = g_1 \oplus g_2$ be arbitrary test sections in $\mathscr{D}(D^\oplus M_B)$ with support in $N$. We introduce the two-point functions on the level of single fields as

$$\mathfrak{w}_\beta^B(\tilde{f}, \tilde{g}) \doteq \mathfrak{w}_\beta^-(f_1, g_2) + \mathfrak{w}_\beta^+(g_1, f_2). \tag{III.100}$$

By lemma III.3.2.2, it is sufficient to prove

$$WF(\mathfrak{w}_\beta^\pm) = \left\{ (x, y, k_x, -k_y) \in T^*M^2 \setminus \{0\} \mid (x, k_x) \sim (y, k_y),\ k_x \gtrless 0 \right\},$$

and, hence, it suffices to prove that

$$d_\beta^\pm \doteq \mathfrak{w}_\beta^\pm - \mathfrak{w}^\pm$$

are distributions specified by a smooth kernel. Let therefore $f \in \mathscr{D}(DM_B)$, and $g \in \mathscr{D}(D^*M_B)$ be arbitrary test sections with support in $N$. A computation yields

$$d_\beta^\pm(f, g) = \int_{\mathbb{R} \times \mathbb{S}^2} dk\, d\mathbb{S}^2 \left( \frac{1}{e^{\mp\beta k} + 1} - \Theta(\pm k) \right) \widehat{\widetilde{S}g^\dagger}^*(k, \theta, \varphi)\, \widehat{\widetilde{S}f}(k, \theta, \varphi)$$

$$= \frac{1}{\sqrt{2\pi}} \int_{\mathbb{R} \times \mathbb{S}^2} dv\, d\mathbb{S}^2 \left[\widetilde{S}g^\dagger\right](v, \theta, \varphi) \left( \mathscr{F}^{-1}\left[ \frac{1}{e^{\mp\beta k} + 1} - \Theta(\pm k) \right] * \left[\widetilde{S}f\right] \right)(v, \theta, \varphi)$$

$$= \int_{\mathbb{R}^2 \times \mathbb{S}^2} dv\, dv'\, d\mathbb{S}^2 \left[\widetilde{S}g^\dagger\right](v, \theta, \varphi)\, R^\pm(v - v')\left[\widetilde{S}f\right](v', \theta, \varphi),$$

where we have set

$$R^\pm(v) \doteq \frac{1}{\sqrt{2\pi}}\, \mathscr{F}^{-1}\left[ \frac{1}{e^{\mp\beta k} + 1} - \Theta(\pm k) \right](v).$$

$R^\pm(v)$ are the inverse Fourier-Plancherel transforms of square-integrable, bounded, and exponentially decaying functions. Hence, $R^\pm(v)$ are square-integrable and smooth and the above computation has been sensible in particular. Based on this, one now employs $\widetilde{S} = \mathbb{S} + \mathscr{S}$ to decompose the distributions $d_\beta^\pm$ into four pieces, which are all found to be individuated by smooth kernels. This follows from the known wave front set and decay properties of $\widetilde{S}$ and $R^\pm$ by suitable applications of [Hör90, thm. 8.2.14.]. Altogether, one finds an upper bound on the wave front set of $\mathfrak{w}_\beta^\pm$ restricted to $N \times N$. The finalising arguments in the proof of theorem III.4.2.2 now yield the wanted conclusion. $\square$

As for the boundary-induced states of the Klein-Gordon field, we would like to have a more explicit interpretation of the newly found Hadamard states in terms of modes. Therefore, we shall now exploit our results found in subsection II.4.1 to interpret $\mathfrak{w}^B$ and $\mathfrak{w}^B_\beta$ as asymptotic conformal vacuum and equilibrium states respectively. For the convenience of the reader, we repeat our findings from subsection II.4.1. Namely, we have shown that an arbitrary solution $u \in \mathfrak{S}(M_B)$ of the Dirac equation with compactly supported initial data of the form $u = S_B f$ with $f \in \mathscr{D}(DM_B)$ can be decomposed as

$$[S_B f](\tau, \vec{x}) = i \int_{\mathbb{R}^3} d\vec{k} \sum_{l=1}^4 \psi^\dagger_{\vec{k},l}(f)\, \psi_{\vec{k},l}(\tau, \vec{x}),$$

where $\psi^\dagger_{\vec{k},l}(f) \doteq \langle \psi^\dagger_{\vec{k},l}, f \rangle$. The Dirac modes $\psi_{\vec{k},l}$ have the form

$$\psi_{\vec{k},l}(\tau, \vec{x}) \doteq \frac{\hat{\mathfrak{D}}^* u_{k,l}(\tau) e^{i\vec{k}\vec{x}}}{(2\pi a(\tau))^{\frac{3}{2}}},$$

where

$$\hat{\mathfrak{D}}^* \doteq \begin{pmatrix} (\partial_\tau - ia(\tau)m)\mathbb{I}_2 & -i\vec{\sigma}\vec{k} \\ -i\vec{\sigma}\vec{k} & (\partial_\tau + ia(\tau)m)\mathbb{I}_2 \end{pmatrix},$$

and

$$u_{k,1} \doteq \begin{pmatrix} \mathfrak{T}_k \\ 0 \\ 0 \\ 0 \end{pmatrix}, \quad u_{k,2} \doteq \begin{pmatrix} 0 \\ \mathfrak{T}_k \\ 0 \\ 0 \end{pmatrix}, \quad u_{k,3} \doteq \begin{pmatrix} 0 \\ 0 \\ \overline{\mathfrak{T}_k} \\ 0 \end{pmatrix}, \quad u_{k,4} \doteq \begin{pmatrix} 0 \\ 0 \\ 0 \\ \overline{\mathfrak{T}_k} \end{pmatrix}.$$

We have shown that the above results make sense once particular regularity properties of the modes $\mathfrak{T}_k(\tau)$ are assured. This turned out to be the case for the asymptotic positive frequency modes found in lemma II.4.1.5. They are of the form

$$\lim_{\tau \to -\infty} e^{ik\tau}\mathfrak{T}_k(\tau) = \frac{1}{\sqrt{2k}}, \qquad \lim_{\tau \to -\infty} e^{ik\tau}\partial_\tau \mathfrak{T}_k(\tau) = -\frac{i}{\sqrt{2}}.$$

We remind the reader that we have only been able to assure sensible $k$-regularity properties of $\mathfrak{T}_k(\tau)$ for sufficiently large (but finite) $-\tau$, the above mode expansion is therefore *a priori* only rigorous in this region of an NBB spacetime. However, the first part of the following result, which analyses the behaviour of the mode decomposition under the bulk-to-boundary map, in combination with the abstractly known Hadamard property of the states $\mathfrak{w}^B$ and $\mathfrak{w}^B_\beta$ will imply that the hereafter obtained mode decompositions of these states hold on the full NBB spacetime.

**Proposition III.4.2.5**   Let $f \in \mathscr{D}(DM_B), g \in \mathscr{D}(D^*M_B)$ be such that their support lies in the region of $M_B$ where the perturbative series defining the modes $\mathfrak{T}_k(\tau)$ are known to converge by lemma II.4.1.5 and let $\psi_{\vec{k},l}$ be the modes resulting from this very lemma.

a) *The mapping of $S_B f$ to $S(\mathfrak{J}^-)$ is of the form*

$$[\mathfrak{G}S_B f](v,\theta,\varphi) = -\frac{iK(-\vec{e}_k)}{\sqrt{2\pi}} \int\limits_{\mathbb{R}} dk\, k \left[ \Theta(k) \begin{pmatrix} \psi^\dagger_{-\vec{k},1}(f) \\ \psi^\dagger_{-\vec{k},2}(f) \\ 0 \\ 0 \end{pmatrix} - \Theta(-k) \begin{pmatrix} 0 \\ 0 \\ \psi^\dagger_{\vec{k},3}(f) \\ \psi^\dagger_{\vec{k},4}(f) \end{pmatrix} \right] e^{-ikv},$$

*where $\vec{e}_k = \partial_k \vec{k}$, $K(-\vec{e}_k)$ is the matrix defined in theorem II.4.2.1 and $\vec{k}$ is thought to be given in polar coordinates $(k,\theta,\varphi)$.*

b) *The single-spinor two-point functions of the preferred Hadamard states $\mathfrak{w}^B$ and $\mathfrak{w}^B_\beta$ can be decomposed as*

$$\mathfrak{w}^\pm(f,g) = \int\limits_{\mathbb{R}^3} d\vec{k} \sum_{l=1}^4 c_l^\pm\, \psi^\dagger_{\vec{k},l}(f)\, \psi_{\vec{k},l}(g),$$

$$\mathfrak{w}^\pm_\beta(f,g) = \int\limits_{\mathbb{R}^3} d\vec{k} \sum_{l=1}^4 \frac{c_l^\pm\, \psi^\dagger_{\vec{k},l}(f)\, \psi_{\vec{k},l}(g)}{e^{-\beta k}+1} + \frac{c_l^\mp\, \psi^\dagger_{\vec{k},l}(f)\, \psi_{\vec{k},l}(g)}{e^{\beta k}+1},$$

*where $c_l^+ \doteq (1,1,0,0)$ and $c_l^- \doteq (0,0,1,1)$.*

*Proof.* Again, we start by showing how b) emerges from a). Namely, inserting the Fourier-Plancherel transform of the result found in a) into

$$\mathfrak{w}^\pm(f,g) = w^\pm\left(\mathfrak{G}S_B f, [\mathfrak{G}S_B g^\dagger]^\dagger\right) = \int\limits_{\mathbb{R}\times\mathbb{S}^2} dk\, d\mathbb{S}^2\, \Theta(\pm k)\, \widehat{\mathfrak{G}S_B g^\dagger}^*(k,\theta,\varphi)\, \widehat{\mathfrak{G}S_B f}(k,\theta,\varphi)$$

trivially leads to the asserted form of $\mathfrak{w}^\pm(f,g)$. The one of $\mathfrak{w}^\pm_\beta(f,g)$ follows analogously.

As a) can be proved with the arguments already used in the proof of proposition III.2.2.8, we only mention the essential steps. To this avail, we recall the form of $\mathfrak{G}$ as devised in theorem II.4.2.1 d) and compute

$$
\begin{aligned}
\left[\mathfrak{G}S_B f\right](v,0,\varphi) &= \lim_{u\to-\infty} \frac{i\sqrt{2}}{\sqrt[4]{1+u^2}} \left(\frac{a}{\Omega_B}\right)^{\frac{3}{2}} \int_{\mathbb{R}^3} d\vec{k} \sum_{l=1}^{4} \psi^\dagger_{\vec{k},l}(f)\, \psi_{\vec{k},l}(\tau,\vec{x}) \\
&= \lim_{u\to-\infty} \frac{-i\,u}{2^{\frac{5}{2}}\pi^{\frac{5}{2}}} \int_{\mathbb{R}^3} d\vec{k} \sum_{l=1}^{4} \psi^\dagger_{\vec{k},l}(f)\, \mathfrak{D}^* u_{k,l}\, e^{i\vec{k}\vec{x}} \\
&= \lim_{u\to-\infty} \frac{-u}{2^{\frac{5}{2}}\pi^{\frac{3}{2}}} \int_0^\infty \int_0^{2\pi} \int_{-1}^1 dk\, d\varphi'\, dc\, k^2 \left[ K(\vec{e}_k) \begin{pmatrix} \psi^\dagger_{\vec{k},1}(f) \\ \psi^\dagger_{\vec{k},2}(f) \\ 0 \\ 0 \end{pmatrix} e^{-i\frac{k}{2}[u(c+1)-v(c-1)]} \right. \\
&\qquad\qquad \left. -K(-\vec{e}_k) \begin{pmatrix} 0 \\ 0 \\ \psi^\dagger_{\vec{k},3}(f) \\ \psi^\dagger_{\vec{k},4}(f) \end{pmatrix} e^{-i\frac{k}{2}[u(c-1)-v(c+1)]} \right],
\end{aligned}
$$

where $c$ denotes the cosine of the angle between $\vec{k}$ and $\vec{x}$, and the computation of $\left[\mathfrak{G}S_B f\right](v,0,\varphi)$ at $\theta=0$ shall be sufficient to obtain a result for arbitrary $\theta$. Namely, as in the proof of proposition III.2.8, we can perform a partial integration with respect to $c$ to cancel the $u$ in front of the integral. This procedure yields two boundary terms at $c=\pm1$ and a new integral with respect to $c$. Due to the rapid decrease of $\psi^\dagger_{\vec{k},l}(f)$ for large $\vec{k}$, two of these three terms always vanish in the limit $u\to-\infty$ on account of the Riemann-Lebesgue lemma. Moreover, at $c=\pm1$, the integrand is degenerate in $\varphi'$ and the associated integration yields a factor of $2\pi$. This also implies that the terms which survive in the limit are independent of $\varphi$, but we can restore the $\varphi$ dependency and at the same time obtain the asserted result for arbitrary $\theta$ by noticing that the integral we have computed has been non-vanishing only if $\vec{k}$ and $\vec{x}$ had been parallel ($l=3,4$) or anti-parallel ($l=1,2$). $\qquad\qquad\square$

Without giving an explicit proof, let us remark that the above result implies that both $\mathfrak{w}^B$ and $\mathfrak{w}^B_\beta$ are homogeneous and isotropic states and that $\mathfrak{w}^B$ is pure in addition.

On the basis of these results, one can interpret $\mathfrak{w}^B$ and $\mathfrak{w}^B_\beta$ as follows. In the massless case, the Dirac field is conformally invariant and, hence, the asymptotic mode expressions for $\mathfrak{T}_k$ are valid on the full NBB spacetime (in fact, on any flat FLRW spacetime). This entails that the states we are considering fulfil

$$
\mathfrak{w}^B(x,y) = \frac{\mathfrak{w}^{\mathbb{M}}(x,y)}{a^{\frac{3}{2}}(\tau_x)a^{\frac{3}{2}}(\tau_y)}, \qquad \mathfrak{w}^B_\beta(x,y) = \frac{\mathfrak{w}^{\mathbb{M}}_\beta(x,y)}{a^{\frac{3}{2}}(\tau_x)a^{\frac{3}{2}}(\tau_y)},
$$

where $\mathfrak{w}^{\mathbb{M}}$ and $\mathfrak{w}^{\mathbb{M}}_{\beta}$ denote respectively the vacuum state and the KMS state of the massless Minkowskian Dirac field on Minkowski spacetime. In the case $m = 0$, our boundary construction therefore naturally yields the conformal vacuum and the conformal temperature states of the Dirac field. One may wonder if there is a natural 'thermometer' like $\omega^{B}_{\beta}(:\phi^2(x):)$ also in the case of massless Dirac fields. Indeed, by dimensional arguments one would assume that $\mathfrak{w}^{B}_{\beta}(:\psi^{\dagger}\psi(x):)$ is proportional to $T^3$. However, it turns out that defining the mentioned Wick monomial by subtraction of the conformal vacuum yields $\mathfrak{w}^{B}_{\beta}(:\psi^{\dagger}\psi(x):) = 0$. Namely, one has

$$\mathfrak{w}^{B}_{\beta}(:\psi^{\dagger}\psi(x):) \doteq Tr\left[\mathfrak{w}^{-}_{\beta}(x,y) - \mathfrak{w}^{-}(x,y)\right]$$

$$= Tr \int_{\mathbb{R}^3} d\vec{k} \left\{ \left(\frac{1}{e^{-\beta k}+1} - 1\right) \sum_{l \in \{3,4\}} \psi_{\vec{k},l}(\tau,\vec{x})\, \psi^{\dagger}_{\vec{k},l}(\tau,\vec{x}) + \left(\frac{1}{e^{\beta k}+1}\right) \sum_{l \in \{1,2\}} \psi_{\vec{k},l}(\tau,\vec{x})\, \psi^{\dagger}_{\vec{k},l}(\tau,\vec{x}) \right\}$$

$$= \frac{1}{(2\pi a)^3} \int_{\mathbb{R}^3} d\vec{k} \left\{ \left(\frac{1}{e^{-\beta k}+1} - 1\right) 2\left(k^2|\mathfrak{T}_k|^2 - |\partial_{\tau}\mathfrak{T}_k|^2\right) + \left(\frac{1}{e^{\beta k}+1}\right) 2\left(|\partial_{\tau}\mathfrak{T}_k|^2 - k^2|\mathfrak{T}_k|^2\right) \right\} = 0.$$

In the case $m \neq 0$, the found states $\mathfrak{w}^{B}$ and $\mathfrak{w}^{B}_{\beta}$ are asymptotic conformal vacuum and equilibrium states in the sense that they are becoming conformally related to the massless Minkowski vacua in the limit to the Big Bang. Particularly, $\mathfrak{w}^{B}_{\beta}(:\psi^{\dagger}\psi(x):)$ is non-vanishing for finite times, but is vanishing towards the Big Bang.

To close, we remark that one can presumably repeat the work of [Pin10] to find that Bogoliubov transformations that give modes differing from the asymptotic positive frequency ones $\mathfrak{T}_k$ by pieces which vanish for large $k$ also yield Hadamard states. However, we shall not discuss Bogoliubov transforms in the Dirac case and possible extensions of the results of [Pin10] here.

IV

# The Backreaction of Quantized Fields in Curved Spacetimes

We have seen how curved spacetimes necessitate new concepts in the context of quantum field theory and, hence, how the background curvature influences quantum fields propagating on it. In this chapter, we will analyse how quantum fields shape the curved spacetime they are living on and we will find that can even generate the full spacetime curvature responsible for the recent history of our cosmos.

## IV.1 The Semiclassical Einstein Equation and Wald's Axioms

We shall start treating the backreaction of quantum fields on the background spacetime by introducing the central equation describing this phenomenon, namely, the *semiclassical Einstein equation*. It reads

$$G_{\mu\nu}(x) = 8\pi G \,\omega(:T_{\mu\nu}(x):), \qquad (IV.101)$$

where the left hand side is given by the standard Einstein tensor $G_{\mu\nu} = R_{\mu\nu} - \frac{1}{2}Rg_{\mu\nu}$, $G$ denotes Newton's gravitational constant, and we have replaced the stress-energy tensor of classical matter by the expectation value of a suitable Wick polynomial $:T_{\mu\nu}(x):$ representing the quantum stress-energy tensor evaluated in a state $\omega$. Considerable work has been invested in analysing how such equation can be derived via a suitable semiclassical limit from some potential quantum theory of gravity. We refer the reader to [FlWa96, sec II.B] for a review of several possibilities and only briefly mention that a possibility to derive (IV.101) is constituted by starting from the *Einstein-Hilbert action*

$$S_{EH} \doteq \frac{1}{16\pi G} \int\limits_M d_g x \, R + S_{\text{matter}} = \frac{1}{16\pi G} \int\limits_M dx \sqrt{|\det g|} \, R + S_{\text{matter}}, \qquad (IV.102)$$

formally expanding a quantum metric and a quantum matter field around a classical (background) vacuum solution of Einstein's equation, and computing the equation of motion for the expected metric while keeping only 'tree-level' ($\hbar^0$) contributions of the quantum metric and 'loop-level' ($\hbar^1$) contributions of the quantum matter field. In this work, we shall not contemplate on whether and in which situation the above mentioned 'partial one-loop approximation' is sensible, but we shall take the following pragmatic point of view: (IV.101) seems to be the simplest possibility to couple the background curvature to the stress-energy of a quantum field in a non-trivial way. We shall therefore consider (IV.101) as it stands and only discuss for which quantum states and Wick polynomial definitions of $:T_{\mu\nu}(x):$ it is a self-consistent one.

To this end, let us first realise that in (IV.101) one equates a 'sharp' classical quantity on the left hand side with a 'probabilistic' quantum quantity on the right hand side. The semiclassical Einstein equation can therefore only be sensible if the fluctuations of the stress-energy tensor $:T_{\mu\nu}(x):$ in the considered state $\omega$ are small. In this respect, we already know that we should consider $\omega$ to be a Hadamard state and $:T_{\mu\nu}(x):$ to be an element regularised by means of a Hadamard bidistribution. Namely, our analysis in the last chapter tells us that this setup at least assures *finite fluctuations* of $:T_{\mu\nu}(x):$ as the pointwise products appearing in the computation of such fluctuations are well-defined distributions once their Hadamard property is assumed. In fact, this observation has been the main motivation to consider Hadamard states in the first place [Wal77, Wal95]. However, it seems one can *a priori* not obtain more than these qualitative observations, and that quantitative statements on the actual size of the fluctuations can only be made *a posteriori* once a solution of (IV.101) is found.

Having agreed to consider only Hadamard states and Wick polynomials constructed by the procedures outlined in the last chapter, two questions remain. Which Hadamard state and which Wick polynomial should one choose to compute the right hand side of (IV.101)? Let us recall that the second question is highly non-trivial, as normal ordering turns out to be ambiguous in curved spacetimes, see [HoWa01, HoWa05] and the last chapter. We have already pointed out at several occasions that one must define the Wick polynomial $:T_{\mu\nu}(x):$ in a local, and, hence, state-independent way. The reason for this is the simple observation that one would like to solve (IV.101) without knowing the spacetime which results from this procedure beforehand, but a state solves the equation of motion and, hence, already 'knows' the full spacetime, thus being a highly non-local object. On account of the above considerations, we can therefore answer the question for the correct Wick polynomial representing $:T_{\mu\nu}(x):$ without having to agree on a specific Hadamard state $\omega$.

To this avail, let us consider the stress-energy tensors of classical matter fields. Given a classical action $S_{\text{matter}}$, the related (Hilbert) stress-energy tensor can be computed as [Wal84, FoRö04]

$$T_{\mu\nu} \doteq \frac{2}{\sqrt{|\det g|}} \frac{\delta S_{\text{matter}}}{\delta g^{\mu\nu}}. \qquad (\text{IV.103})$$

To consider the classical stress-energy tensors of a Klein-Gordon and a Dirac field, we therefore need to introduce their classical actions. They read [FoRö04]

$$S_{\text{KG}} \doteq \int_M d_g x \, L_{\text{KG}} \doteq \int_M dx \sqrt{|\det g|} \left[ \frac{1}{2} \phi_{;\mu} \phi^{;\mu} + \frac{1}{2} (\xi R + m^2) \phi^2 \right], \qquad (\text{IV.104})$$

$$S_{\text{Di}} \doteq \int_M d_g x \, L_{\text{Di}} \doteq \int_M dx \sqrt{|\det g|} \left[ \frac{1}{2} \psi^\dagger D \psi + \frac{1}{2} (D^* \psi^\dagger) \psi \right]. \qquad (\text{IV.105})$$

Computing the related stress-energy tensors according to (IV.103), one finds in the scalar case [Wal84, FoRö04]

$$T_{\mu\nu}^{\text{KG}} = \phi_{;\mu} \phi_{;\nu} + \xi \left( g_{\mu\nu} \Box - \nabla_\mu \nabla_\nu + R_{\mu\nu} \right) \phi^2 - g_{\mu\nu} L_{\text{KG}} \qquad (\text{IV.106})$$

$$= (1 - 2\xi) \phi_{;\mu} \phi_{;\nu} - 2\xi \phi_{;\mu\nu} \phi + \xi G_{\mu\nu} \phi^2 + g_{\mu\nu} \left\{ 2\xi (\Box \phi) \phi + \left( 2\xi - \frac{1}{2} \right) \phi_{;\rho} \phi^{;\rho} - \frac{1}{2} m^2 \phi^2 \right\},$$

whereas, for Dirac fields, the result is (recall our sign convention $(-,+,+,+)$) [FoRö04]

$$T_{\mu\nu}^{\text{Di}} = \frac{1}{2} \left( \psi^\dagger_{;(\mu} \gamma_{\nu)} \psi - \psi^\dagger \gamma_{(\mu} \psi_{;\nu)} \right) - g_{\mu\nu} L_{\text{Di}} \qquad (\text{IV.107})$$

$$= \frac{1}{2} \left( \psi^\dagger_{;(\mu} \gamma_{\nu)} \psi - \psi^\dagger \gamma_{(\mu} \psi_{;\nu)} \right) - \frac{1}{2} g_{\mu\nu} \left( \psi^\dagger D \psi + (D^* \psi^\dagger) \psi \right).$$

Here, $(\cdot, \cdot)$ denotes idempotent symmetrisation of indices and we remark that the computation of $T_{\mu\nu}^{\text{Di}}$ is a quite involved task as one has to take into account the metric dependence of Lorentz frames $e_a$ [FoRö04]. We also point out that one can in principle omit the multiples of the Lagrangean $L_{\text{Di}}$ in the above expression, as this quantity manifestly vanishes on shell, *i.e.* if one assumes that the Dirac fields satisfy the relevant equations of motion $D\psi = 0$, $D^*\psi^\dagger = 0$. However, we have chosen to explicitly write it down as it will be important in the following discussion. A straightforward computation shows that the classical stress energy tensors are covariantly conserved if one imposes the Klein-Gordon equation and the Dirac equations respectively, *i.e.*

$$\nabla^\mu T^{\text{KG}}_{\mu\nu} = 0, \qquad \nabla^\mu T^{\text{Di}}_{\mu\nu} = 0.$$

Moreover, a computation of their trace employing the relevant equations of motion yields

$$g^{\mu\nu} T^{\text{KG}}_{\mu\nu} = (6\xi - 1)\left(\phi\Box\phi + \phi_{;\mu}\phi^{\;\mu}_{;}\right) - m^2\phi^2, \qquad g^{\mu\nu} T^{\text{Di}}_{\mu\nu} = -m\psi^\dagger\psi.$$

Particularly, we see that in the conformally invariant situation, that is, $m = 0$ for Dirac fields and $m = 0$, $\xi = \frac{1}{6}$ for Klein-Gordon fields, the classical stress-energy tensor has vanishing trace in the on-shell case. In fact, one can show that this is a general result, namely, the trace of a classical stress-energy tensor is vanishing on-shell if and only if the respective field is conformally invariant [FoRö04, thm. 5.1].

With the classical stress-energy tensors at hand, its seems natural to take them as a guide for choosing the correct Wick polynomial $:T_{\mu\nu}(x):$ for the semiclassical Einstein equation. To this end, our treatment of normal ordering entails that a Wick polynomial has to be understood as a distribution $u$ with a particular wave front set, whereas its evaluation in a state $\omega$ means to integrate the relevant regularised $n$-point function $:\omega_n:$ with $u$. In the case at hand, we would for instance regard $:(\phi_{;\mu}\phi_{;\nu})(f):$ with $f \in \mathscr{D}(M)$ as the distribution

$$f(x)\nabla_\mu g_\nu^{\;\nu'}\nabla_{\nu'}\delta(x, y),$$

where we recall that $g_\nu^{\;\nu'}$ is the parallel transport (*cf.* subsection III.1.2) along the unique geodesic connection $x$ and $y$ (we implicitly assume that $x$ and $y$ are sufficiently close), and we have deliberately applied the $\nu$ derivative to the $y$-slot to account for the fact that it acts on the second field in the monomial. We interpret $:(\phi_{;\mu}\phi_{;\nu})(f):$ as being regularised with a Hadamard distribution $h$; this entails that the evaluation of $:(\phi_\mu\phi_\nu)(f):$ in a Hadamard state $\omega$ reads

$$\omega\left(:(\phi_{;\mu}\phi_{;\nu})(f):\right) = \left\langle \omega_2(x, y) - \frac{1}{8\pi^2}h(x, y), f(x)\nabla_\mu g_\nu^{\;\nu'}\nabla_{\nu'}\delta(x, y) \right\rangle$$

$$= \left\langle \frac{1}{8\pi^2}\left[\nabla_\mu g_\nu^{\;\nu'}\nabla_{\nu'}w(x, y)\right], f(x) \right\rangle,$$

where $w$ is the smooth state-dependent part of $\omega_2$ and the cornered brackets denote the coinciding point limit, see subsection III.1.2. Let us note that $w(x, y)$ is smooth, the above expression therefore does not have to be integrated with the test function $f$ to give a meaningful result. We are thus free to interpret $\omega(:(\phi_\mu\phi_\nu)(x):)$ as the smooth function

$$\omega\left(:(\phi_{;\mu}\phi_{;\nu})(x):\right) \doteq \frac{1}{8\pi^2}\left[\nabla_\mu g_\nu^{\;\nu'}\nabla_{\nu'}w(x, y)\right].$$

Proceeding like this, we can obtain a canonical expression for $\omega(:T_{\mu\nu}(x):)$ by replacing all field monomials in the classical stress-energy tensor with suitable derivatives of $w$ such that, in the scalar case, we can write

$$\omega\left(:T^{\text{KG,can}}_{\mu\nu}(x):\right) \doteq \frac{1}{8\pi^2}\left[D^{\text{KG,can}}_{\mu\nu}\, w(x,y)\right],$$

where

$$D^{\text{KG,can}}_{\mu\nu} \doteq (1-2\xi)g_{\nu}^{\ \nu'}\nabla_\mu\nabla_{\nu'} - 2\xi\nabla_\mu\nabla_\nu + \xi\, G_{\mu\nu} + g_{\mu\nu}\left\{2\xi\,\Box_x + \left(2\xi - \frac{1}{2}\right)g_{\rho}^{\ \rho'}\nabla^\rho\nabla_{\rho'} - \frac{1}{2}m^2\right\}$$

and the index $^{\text{can}}$ indicates that this is the canonical choice of Wick polynomial $:T^{\text{KG}}_{\mu\nu}(x):$ and related bidifferential operator $D^{\text{KG}}_{\mu\nu}$. What we have just described is the so-called *point-splitting* regularisation scheme [BiDa82, Wal95]. Namely, one takes a quantity which is initially divergent (the expectation value of the non-normal ordered field monomials), writes it as the formal limit of a bitensor ($D^{\text{KG}}_{\mu\nu}$ applied to $\omega_2(x,y)$), subtracts the singular part ($D^{\text{KG}}_{\mu\nu}$ applied to $h(x,y)$) and then finally takes the coinciding point limit of the resulting (sufficiently regular) quantity. Note that one often writes the point-splitting differential operator in such a way that it acts symmetrically on the $x$ and $y$ slots of the regularised two-point function. However, as one is interested in the coinciding point limit, this is not strictly necessary, and we have not done it here.

In the case of Dirac fields, the point-splitting regularisation would proceed analogously. Recalling the Hadamard form in the case of Dirac fields discussed in subsection III.3.2, and defining normal ordering by subtraction of the Hadamard singularity, the canonical choice of a Diracian quantum stress-energy tensor evaluated in a Hadamard state would be

$$\omega\left(:T^{\text{Di,can}}_{\mu\nu}(x):\right) \doteq \frac{1}{8\pi^2}\,Tr\left[D^{\text{Di,can}}_{\mu\nu}\, W(x,y)\right],$$

where

$$D^{\text{Di,can}}_{\mu\nu} \doteq -\widetilde{D}^{\text{Di,can}}_{\mu\nu}D^*_y \doteq \left\{-\frac{1}{2}\gamma_{(\mu}\left(\nabla_{\nu)} - g_{\nu)}^{\ \nu'}\nabla_{\nu'}\right) - \frac{1}{2}g_{\mu\nu}\left(D_y + D^*_x\right)\right\}D^*_y.$$

Note that the trace over spinor indices appears because $W$ and its derivatives are bispinors, and that the minus sign arises because such sign appears in the relation between $\omega^-$ and $H^-$.

As our presentation may have already suggested, it turns out that the expectation valued of the just introduced canonical quantum stress-energy tensors are *not* good candidates for the right hand side of the semiclassical Einstein equation. To understand this, we must of course specify criteria for admissible right hand sides of (IV.101); such axioms have been formulated by Wald in his seminal paper [Wal77]. We remark that from the rather modern point of view we have reviewed and presented in our thesis, it is quite natural and unavoidable that normal ordering in curved spacetimes is ambiguous. However, at the time [Wal77] appeared, workers in the field had computed the regularised stress energy tensor by different methods like adiabatic subtraction, dimensional regularisation, and $\zeta$-function regularisation (see [BiDa82] and also [Mor98]) and differing results had been found. The axioms we shall now review had helped to clarify the case and to understand that in principle *all* employed regularisation schemes were correct in physical terms, and that the apparent differences between them could be understood on clear conceptual grounds. We shall provide Wald's axioms in their modern form and also suitable for both the Klein-Gordon and the Dirac case.

**Definition** **IV.1.0.6**    *We say that* $\omega(:T_{\mu\nu}(x):)$ *fulfils* ***Wald's axioms*** *if the follwing five conditions are fulfilled.*

1. *Given two (not necessarily Hadamard) states $\omega$ and $\omega'$ such that the difference of their two-point functions $\omega_2 - \omega_2'$ is smooth,* $\omega(:T_{\mu\nu}(x):) - \omega'(:T_{\mu\nu}(x):)$ *is equal to*

$$\left[ D_{\mu\nu}^{KG,can} \left( \omega_2(x,y) - \omega_2'(x,y) \right) \right] \qquad \textit{(Klein-Gordon case)},$$

$$Tr \left[ \widetilde{\mathcal{D}}_{\mu\nu}^{Di,can} \left( \omega^-(x,y) - \omega'^-(x,y) \right) \right] \qquad \textit{(Dirac case)}.$$

2. *In Minkowski spacetime $\mathbb{M}$, and in the relevant Minkowski vacuum state $\omega_{\mathbb{M}}$*

$$\omega_{\mathbb{M}} \left( :T_{\mu\nu}(x): \right) = 0.$$

3. *Covariant conservation holds, i.e.*

$$\nabla^\mu \omega \left( :T_{\mu\nu}(x): \right) = 0.$$

4. *$\omega(:T_{\mu\nu}(x):)$ is locally covariant in the following sense: let*

$$\chi : (M,g) \hookrightarrow (M',g') \qquad \textit{(Klein-Gordon case)},$$

$$\chi : (M,g,SM,\rho) \hookrightarrow (M',g',SM',\rho') \qquad \textit{(Dirac case)},$$

*be defined as in subsection I.4 and let $\alpha_\chi$ denote the associated continuous, unit-preserving, injective $*$-morphisms between the relevant enlarged algebras $\mathscr{W}(M,g)$ and $\mathscr{W}(M',g')$. If two states $\omega$ on $\mathscr{W}(M,g)$ and $\omega'$ on $\mathscr{W}(M',g')$ respectively are related via $\omega = \omega' \circ \alpha_\chi$, then*

$$\omega' \left( :T_{\mu'\nu'}(x'): \right) = \chi_* \omega \left( :T_{\mu\nu}(x): \right),$$

*where $\chi_*$ denotes the push-forward of $\chi$ in the sense of covariant tensors.*

5. *$\omega(:T_{\mu\nu}(x):)$ does not contain derivatives of the metric of order higher than 2.*

Let us briefly comment on the relevance of the above listed axioms.

1. In a given Fock-representation of the quantum field, the non-diagonal matrix elements of the formal unrenormalised stress-energy tensor operator in the 'number-basis' are already finite, because their calculation only involves finitely many terms on account of the orthogonality of states with different particle numbers [Wal77, Wal78a]. To regularise the formal stress-energy tensor operator, it is therefore only necessary to subtract an infinite part proportional to the identity operator, thus leaving the non-diagonal matrix elements unchanged[17] – axiom 1 amounts to require such a 'minimal' regularisation. This axiom is also related to so-called *relative Cauchy evolution* of a locally covariant field [BFV03, San08]; since the functional derivative of the relative Cauchy evolution involves the commutator with the stress-energy tensor operator, one could reformulate this axiom on the operator level requiring that any regularisation prescription yields the same relative Cauchy evolution. If we consider Hadamard states, the requirement is equivalent to demanding that the bidifferential operator used in the point-splitting procedure is given by the canonical one of the respective field theory plus a term which does not influence the state dependence of $\omega(:T_{\mu\nu}(x):)$.

2. The motivation behind this axiom is to assure that any regularisation prescription for $\omega(:T_{\mu\nu}(x):)$ should be an extension of the formalism in Minkowski spacetime. However, with our current understanding of curved spacetime quantum field theory, its general framework should rather serve as a guide to Minkowskian quantum field theory than vice versa. Moreover, there are at least two other reasons why one should discard this axiom. On the one hand, this axiom could be interpreted as forcing any 'cosmological constant' present in $\omega(:T_{\mu\nu}(x):)$ to be zero. However, as remarked in [Ful89], this is not necessarily a sensible requirement in cosmological applications. On the other hand, we will see that the only way to assure the validity of this axiom is to fix certain constants in terms of inverse powers of the mass $m$. But this is not a procedure which behaves well in the limit $m \to 0$.

3. This axiom may be obvious, yet it is crucial and the main obstruction in providing a regularisation procedure for $\omega(:T_{\mu\nu}(x):)$ or, in our terms, the correct Wick polynomial representing $:T_{\mu\nu}(x):$. Namely, the left hand side of the semiclassical Einstein equation is constituted by the Einstein tensor $G_{\mu\nu}$, which is covariantly conserved. Consistency of the equation therefore requires the right hand side to be conserved as well.

4. As already remarked at several occasions, a backreaction equation is only sensible if the source term depends only in a local manner on the curvature of the spacetime. Of course a state is a non-local object, and one can not avoid 'some' non-locality in the expectation value $\omega(:T_{\mu\nu}(x):)$. However, one should at least try to obtain the 'minimal' non-locality by constructing $:T_{\mu\nu}(x):$ in a locally covariant way. In fact, this axiom seems to have been an inspiration towards the formulation of locally covariant quantum field theory, as described in the seminal paper [BFV03]. Moreover, this axiom is automatically fulfilled in our case as we define the regularisation by subtracting only the local, geometric part of any Hadamard state.

---

[17]To see the direct relation to the statement of the axiom, consider two normalised Hilbert space vectors $|A\rangle$ and $|B\rangle$. Then $|C^{\pm}\rangle \doteq |A\rangle \pm |B\rangle$ are orthogonal, and $\langle A|T_{\mu\nu}|A\rangle - \langle B|T_{\mu\nu}|B\rangle = \frac{1}{2}(\langle C^+|T_{\mu\nu}|C^-\rangle + \langle C^-|T_{\mu\nu}|C^+\rangle)$. Conversely, every pair of orthogonal vectors can be written in terms of two normalised vectors.

5. Wald originally proposed this axiom in a rather technical and more strict way [Wal77, Wal78a], essentially requiring that $\omega(:T_{\mu\nu}(x):)$ does not depend on derivatives of the metric of order higher than the first. On the one hand, this is motivated by the wish to have a sensible initial value formulation for the (semiclassical) Einstein equations with source, and, on the other hand, one would like to recover the usual Einstein equations in the classical limit $\hbar \to 0$ (recall that $\omega(:T_{\mu\nu}(x):)$ has contributions of order $\hbar^1$), see the enlightening discussion in Wald's original papers [Wal77, Wal78a]). Wald himself had realised, however, that the strict version of this axiom could not even be satisfied in the classical theory and has thus proposed the weaker one stated here. Unfortunately, further examinations have revealed that even this weaker version does not seem possible to fulfil in massless theories without introducing an artificial length scale into the theory; therefore, the axiom has been discarded. We still believe, however, that it could be fulfilled, though only under special circumstances, *e.g.* in cosmological spacetimes. We shall comment on this issue at a later stage.

Using these axioms, Wald could prove that a uniqueness result for $\omega(:T_{\mu\nu}(x):)$ can be obtained. The first and fourth axioms already imply that the results from two different sensible regularisation schemes can only differ by a local curvature tensor. The second and third axiom then imply that this local curvature tensor is conserved and vanishes if the spacetime is locally flat. Requiring that this term has the correct dimension of $m^4$, the possible tensors are presumably [FlTi98] only the ones obtained by varying a Lagrangean of the form

$$m^4 \left( F_1 \left( \frac{R}{m^2} \right) + F_2 \left( \frac{R_{\mu\nu} R^{\mu\nu}}{m^4} \right) + F_3 \left( \frac{R_{\mu\nu\rho\tau} R^{\mu\nu\rho\tau}}{m^4} \right) \right)$$

with respect to the metric, with some dimensionless functions $F_i(x)$. By requiring suitable analyticity properties with respect to the curvature tensors and $m$ it has been shown in [HoWa05] that the only possibilities are $m^4 g_{\mu\nu}$ ($F_1 = 1$) [18], $m^2 G_{\mu\nu}$ ($F_1 = x$) and the three local curvature tensors $I_{\mu\nu}$ ($F_1 = x^2$), $J_{\mu\nu}$ ($F_2 = x$), $K_{\mu\nu}$ ($F_3 = x$) given as

$$I_{\mu\nu} \doteq \frac{1}{\sqrt{|\det g|}} \frac{\delta}{\delta g^{\mu\nu}} \int_M d_g x\, R^2 = g_{\mu\nu}\left(\frac{1}{2}R^2 + 2\Box R\right) - 2R_{;\mu\nu} - 2R R_{\mu\nu}, \tag{IV.108}$$

$$J_{\mu\nu} \doteq \frac{1}{\sqrt{|\det g|}} \frac{\delta}{\delta g^{\mu\nu}} \int_M d_g x\, R_{\alpha\beta}R^{\alpha\beta} = \frac{1}{2}g_{\mu\nu}(R_{\mu\nu}R^{\mu\nu} + \Box R) - R_{;\mu\nu} + \Box R_{\mu\nu} - 2R_{\alpha\beta}R^{\alpha\ \beta}_{\ \mu\ \nu},$$

$$K_{\mu\nu} \doteq \frac{1}{\sqrt{|\det g|}} \frac{\delta}{\delta g^{\mu\nu}} \int_M d_g x\, R_{\alpha\beta\gamma\delta}R^{\alpha\beta\gamma\delta}$$

$$= -\frac{1}{2}g_{\mu\nu}R_{\alpha\beta\gamma\delta}R^{\alpha\beta\gamma\delta} + 2R_{\alpha\beta\gamma\mu}R^{\alpha\beta\gamma}_{\ \ \ \nu} + 4R_{\alpha\beta}R^{\alpha\ \beta}_{\ \mu\ \nu} - 4R_{\alpha\mu}R^{\alpha}_{\ \nu} - 4\Box R_{\mu\nu} + 2R_{;\mu\nu}.$$

_______________

[18]Note that a term proportional to the metric is not allowed if one seeks to fulfil the second axiom. Furthermore, we stress once more that the results in [HoWa05] regarding the restriction of the possible regularisation freedom by demanding analytic dependence on curvature and mass have only been obtained for scalar fields. However, since the stress-energy tensor for Dirac fields is an observable and thus still a 'scalar' field, the results in [HoWa05] can be presumably extended to this case.

In fact, we will show in the next subsections that changing the scale $\lambda$ in the regularising Hadamard bidistributions amounts to changing $\omega(:T_{\mu\nu}(x):)$ exactly by a tensor of this form and, furthermore, the attempt to regularise Einstein-Hilbert quantum gravity at one loop order automatically yields a renormalisation freedom in form of such a tensor as well [tHoVe74]. Having in mind how the semiclassical Einstein equation may be derived, these two arguments are of course related by means of internal consistency.[19] Moreover, using the *Gauss-Bonnet-Chern theorem* in four dimensions, which states that

$$\int_M d_g x R_{\mu\nu\rho\tau}R^{\mu\nu\rho\tau} - 4R_{\mu\nu}R^{\mu\nu} + R^2$$

is a topological invariant and, therefore, has a vanishing functional derivative with respect to the metric [Alt95, tHoVe74], one can restrict the freedom even further by removing $K_{\mu\nu}$ from the list of allowed local curvature tensors as $K_{\mu\nu} = I_{\mu\nu} - 4J_{\mu\nu}$. Finally, the above tensors all have a trace proportional to $\Box R$ and thus the linear combination $I_{\mu\nu} - 3J_{\mu\nu}$ is traceless.

Taking Wald's axioms as a guide, we shall show in the next subsections how a meaningful expectation value of the quantum stress-energy tensors of Klein-Gordon and Dirac fields can be provided. Let us briefly anticipate where the main obstruction lies. Namely, to assure locality of the stress-energy tensor, we have subtracted only the local divergence encoded in the Hadamard form. However, this subtraction term does not fulfil the relevant equations of motion, which have been essential in deriving both the conservation of the classical stress-energy tensor and its vanishing trace in the conformally invariant case. We will see that it is possible to 'restore' the conservation by a judicious choice of Wick polynomial for $:T_{\mu\nu}(x):$. However, one can not assure vanishing trace in the conformally invariant case at the same time. Hence, a well-known quantum anomaly arises, namely, the *trace anomaly*.

---

[19]In fact, at least in the case of scalar fields, the combination of the local curvature tensors appearing as the finite renormalisation freedom in [tHoVe74] is, up to a term which seems to be an artifact of the dimensional regularisation employed in that paper, the same that one gets via changing the scale in the regularising Hadamard bidistribution.

## IV.2 The Quantum Stress-Energy Tensor of a Scalar Field

After some dispute about computational mistakes (see the discussion in [Wal78a]) it had soon been realised that the canonical choice of differential operator mentioned in the previous subsection does not assure conservation of the stress-energy tensor. The obvious solution to this problem has been to compute the expectation value of the canonical stress-energy tensor, calculate its covariant divergence, and then define a new stress-energy tensor expectation value by just subtracting this *conservation anomaly*. It was found that then the trace anomaly arises inevitably, and that every quantum stress-energy tensor (expectation value) which fulfils Wald's axioms necessarily displays this phenomenon.

At a more recent date, Moretti [Mor03] proposed a version of the point-splitting procedure which is conceptually more clear and does not require to bluntly subtract the conservation anomaly. Let us describe the main idea in the case of a Klein-Gordon field. As already anticipated, the conservation anomaly of the canonical stress-energy tensor arises because we have regularised it by means of a singular bidistribution which does not fulfil the equation of motion. As a result, expressions like $\phi P \phi$ which classically vanish on-shell, are on the quantum side represented by Wick polynomials $:\phi P \phi:$ whose expectation value does not vanish, even though the state satisfies the equation of motion. However, as proposed in [Mor03], one can try to turn this disadvantage into an advantage. Namely, one modifies the classical stress-energy by adding a term like $c\, g_{\mu\nu} \phi P \phi$, with an initially undetermined constant $c$. One then constructs the related bidifferential operator $D^c_{\mu\nu}$, computes the covariant divergence of the stress-tensor expectation value, and then hopes that it vanishes for a particular $c$. It turns out that this procedure works very well, and one therefore has a regularised stress-energy tensor in the quantum case which reduces to the canonical one in the classical, on-shell limit.

In [Mor03], the procedure we have just described has been applied to the case of a Klein-Gordon field in any spacetime dimension. We shall review it here in the case of four spacetime dimensions, and then proceed to apply it to the Dirac field in the next subsection. To this avail, we define the stress-energy tensor expectation value in a Hadamard state as

$$\omega\left(:T^{\text{KG}}_{\mu\nu}(x):\right) \doteq \frac{1}{8\pi^2}\left[D^{\text{KG},c}_{\mu\nu} w(x,y)\right] \tag{IV.109}$$

where

$$D^{\text{KG},c}_{\mu\nu} \doteq D^{\text{KG,can}}_{\mu\nu} + c\, g_{\mu\nu} P_x \tag{IV.110}$$
$$= (1-2\xi) g^{\nu'}_\nu \nabla_\mu \nabla_{\nu'} - 2\xi \nabla_\mu \nabla_\nu + \xi\, G_{\mu\nu}$$
$$+ g_{\mu\nu}\left\{2\xi \Box_x + \left(2\xi - \frac{1}{2}\right) g^{\rho'}_\rho \nabla^\rho \nabla_{\rho'} - \frac{1}{2}m^2\right\} + c\, g_{\mu\nu} P_x.$$

The following result can now be shown [Mor03].

***Theorem* IV.2.1** Let $\omega(:T^{KG}_{\mu\nu}(x):)$ be defined as in (IV.109) with $c = -\frac{1}{3}$ and $\omega$ being a Hadamard state on $\mathscr{A}(M)$.

a) $\omega(:T^{KG}_{\mu\nu}(x):)$ is covariantly conserved, i.e.

$$\nabla^\mu \omega\left(:T^{KG}_{\mu\nu}(x):\right) = 0.$$

b) The trace of $\omega(:T^{KG}_{\mu\nu}(x):)$ equals

$$g^{\mu\nu}\omega\left(:T^{KG}_{\mu\nu}(x):\right) = \frac{1}{4\pi^2}\left[v_1\right] - \frac{1}{8\pi^2}\left(3\left(\frac{1}{6}-\xi\right)\square + m^2\right)[w]$$

$$= \frac{1}{2880\pi^2}\left(\frac{5}{2}(6\xi-1)R^2 + 6(1-5\xi)\square R + C_{\alpha\beta\gamma\delta}C^{\alpha\beta\gamma\delta} + R_{\alpha\beta}R^{\alpha\beta} - \frac{R^2}{3}\right)$$

$$+ \frac{1}{4\pi^2}\left(\frac{m^4}{8} + \frac{(6\xi-1)m^2R}{24}\right) - \frac{1}{8\pi^2}\left(3\left(\frac{1}{6}-\xi\right)\square + m^2\right)[w],$$

which, for $m=0$ and $\xi = \frac{1}{6}$, constitutes the **trace anomaly** of the quantum stress-energy tensor proper to the Klein-Gordon field.

c) The conservation and trace anomaly are independent of the chosen scale $\lambda$ in the Hadamard parametrix $h$. Namely, a change

$$\lambda \mapsto \lambda'$$

results in

$$\omega(:T^{KG}_{\mu\nu}(x):) \mapsto \omega(:T^{KG}_{\mu\nu}(x):)' = \omega(:T^{KG}_{\mu\nu}(x):) + \delta T_{\mu\nu},$$

where

$$\delta T_{\mu\nu} \doteq \frac{2\log\lambda/\lambda'}{8\pi^2}\left[D^{KG,c}_{\mu\nu} v\right] = \frac{2\log\lambda/\lambda'}{8\pi^2}\left[D^{KG,can}_{\mu\nu} v\right]$$

$$= \frac{2\log\lambda/\lambda'}{8\pi^2}\left(\frac{m^2(6\xi-1)G_{\mu\nu}}{12} - \frac{m^4}{8}g_{\mu\nu} + \frac{1}{360}(I_{\mu\nu} - 3J_{\mu\nu}) - \frac{(6\xi-1)^2}{144}I_{\mu\nu}\right)$$

is a conserved tensor which has vanishing trace for $m=0$ and $\xi = \frac{1}{6}$.

*Proof.* a) Leaving $c$ undetermined and employing Synge's rule (*cf.* lemma III.1.2.4), we compute

$$8\pi^2\nabla^\mu\omega\left(:T^{KG}_{\mu\nu}(x):\right) = \nabla^\mu\left[D^{KG,c}_{\mu\nu} w\right] = \left[(\nabla^\mu + g^{\mu}_{\mu'}\nabla^{\mu'})D^{KG,c}_{\mu\nu} w\right]$$

$$\left[\left(g^{\nu'}_{\nu}\nabla_{\nu'}P_x + c(g^{\nu'}_{\nu}\nabla_{\nu'}P_x + \nabla_\nu P_x)\right)w\right].$$

Let us now recall that $P_x(h+w) = 0$ and, hence, $P_x w = -P_x h$. Inserting this and the identities found in lemma III.1.2.7, we obtain

$$8\pi^2\nabla^\mu\omega\left(:T^{KG}_{\mu\nu}(x):\right) = -\left[\left(g^{\nu'}_{\nu}\nabla_{\nu'}P_x + c(g^{\nu'}_{\nu}\nabla_{\nu'}P_x + \nabla_\nu P_x)\right)h\right]$$

$$= -(2+6c)\left[v_1\right]_{;\nu}.$$

This proves the conservation for $c = \frac{1}{3}$.

b) It is instructive to leave $c$ undetermined also in this case. Employing Synge's rule and the results of lemma III.1.2.7, we find

$$8\pi^2 g^{\mu\nu}\omega\left(:T^{\mathrm{KG}}_{\mu\nu}(x):\right) = 8\pi^2 g^{\mu\nu}\left[D^{\mathrm{KG},c}_{\mu\nu}\,w\right]$$

$$= (1+4c)[P_x w] + \left(3\left(\frac{1}{6}-\xi\right)\square + m^2\right)[w]$$

$$= -(1+4c)[P_x h] + \left(3\left(\frac{1}{6}-\xi\right)\square + m^2\right)[w]$$

$$= -(1+4c)6[v_1] + \left(3\left(\frac{1}{6}-\xi\right)\square + m^2\right)[w].$$

Inserting $c = -\frac{1}{3}$ yields the wanted result.

c) The proof ensues without explicitly computing $\delta T_{\mu\nu}$ in terms of the stated conserved tensors from the following observation. Namely, a change of scale as considered transforms $w$ by a adding a term $2\log\lambda/\lambda'v$. Hence, our computations in a) and b) entail

$$\frac{8\pi^2}{2\log\lambda/\lambda'}\nabla^\mu\delta T_{\mu\nu} = \left[\left(g^{\nu'}_\nu\nabla_{\nu'}P_x + c(g^{\nu'}_\nu\nabla_{\nu'}P_x + \nabla_\nu P_x)\right)v\right],$$

$$\frac{8\pi^2}{2\log\lambda/\lambda'}g^{\mu\nu}\delta T_{\mu\nu} = (1+4c)[P_x v] + \left(3\left(\frac{1}{6}-\xi\right)\square + m^2\right)[v].$$

The former term vanishes because $P_x v = 0$ as discussed in subsection III.1.2, and the same holds for the latter term if we insert $\xi = \frac{1}{6}$ and $m = 0$. $\qquad\square$

Note that the proof of b) clearly shows that there is a possibility to assure vanishing trace in the conformally invariant case, but this possibility is not compatible with conservation. A related observation is that a trace-anomaly is already present if one considers the canonical stress-energy tensor. Finally, we have stated the result of c) in explicit terms to show how a change of scale in the Hadamard parametrix (almost) exhausts the renormalisation freedom. Moreover, note that the term we have added to the canonical stress energy tensor is state independent and, hence, assure the validity of Wald's first axiom. This follows once more from the observation that $P_x w = -P_x h$.

Regarding the validity of the axiom asking for a vanishing expectation value of the stress-energy tensor in Minkowski spacetime, let us remark that in the massless case this is automatically assured, as then $v$ (and $w$) are identically vanishing (recall the massless vacuum two-point function displayed in (III.49)). If $m \neq 0$, then one can assure a vanishing expectation value by fixing the constant $\lambda$ in the Hadamard parametrix in terms of the inverse mass, see [Mor03] for a detailed calculation. However, as already remarked, such fixing of $\lambda$ would behave badly in the limit to vanishing mass and we do not feel that this axiom should have to be fulfilled in view of cosmological applications.

An observation related to the previous paragraph is the following [Wal78a, Wal95]. If we do not fix the scale $\lambda$ in the Hadamard parametrix, then the explicit form of $\delta T_{\mu\nu}$ provided in the statement of theorem IV.2.1 indicates that we do not fix the 'amount' of the conserved local tensors $I_{\mu\nu}$ and $J_{\mu\nu}$ 'present in $\omega(: T^{KG}_{\mu\nu}(x):)$'. Hence, we do not have any control on these specific higher-than-second derivative terms and Wald's fifth axiom can not be assured. At the same time, we see that we can very well assure the validity of this axiom if we only consider the trace of the stress-energy tensor and if $[w]$ does not explicitly depend on higher derivative terms. This follows from the fact that, in the case $\xi = \frac{1}{6}$, the trace of $\delta T_{\mu\nu}$ does not contain higher derivative terms.

We would also like to point out that the above explicit form of the trace anomaly has also been known before Hadamard point-splitting had been developed. Particularly, the same result had been obtained by means of the so-called *DeWitt-Schwinger expansion* in [Chr76]. This regularising prescription is *a priori* not rigorously defined on general spacetimes and the Hadamard point-splitting computation in [Wal78a] had therefore been the first rigorous derivation of the trace anomaly of the stress-energy tensor proper to the quantized Klein-Gordon field. However, we shall discuss in more detail in the following sections that the status of the DeWitt-Schwinger expansion is not so bad after all.

Finally, let us remark that a treatment of the regularised stress-energy tensor of a scalar field encompassing also the interacting case has been given in [HoWa05]. In this work, the authors have analysed the regularisation freedom of locally covariant Wick polynomials including derivatives by demanding that specific 'Leibniz rules' should be fulfilled. They have shown that the regularisation freedom of locally covariant Wick polynomials is large enough to allow for such Leibniz rules to hold, and this entails in particular that one can have a conserved stress-energy tensor.

## IV.3   The Quantum Stress-Energy Tensor of a Dirac Field

We shall now apply the tidy point-splitting procedure introduced in [Mor03] to the case of Dirac fields. We point out that our results constitute the first computation of the stress-energy tensor of Dirac fields in the rigorous framework of Hadamard states. However, the trace anomaly we find is in principle already known from (apparently non-rigorous) computations in the DeWitt-Schwinger approach [Chr78].

In order to obtain a conserved stress-energy tensor expectation value, we have to explore the classical freedom of the stress-energy tensor also in the Dirac case. Let us observe that the Lagrangean $L_{Di}$ (see (IV.105)) vanishes on shell – this is the reason why we have explicitly written it down when providing the form of the classical stress-energy tensor. A canonical possibility to modify the classical stress-energy tensor and, hence, the related canonical bidifferential operator thus seems to be constituted by adding multiples of $L_{Di}$. This leads us to consider the following definition.

$$\omega\left(:T^{\mathrm{Di}}_{\mu\nu}(x):\right) \doteq \frac{1}{8\pi^2}\,Tr\left[D^{\mathrm{Di},c}_{\mu\nu}\,W(x,y)\right] \tag{IV.111}$$

where

$$D^{\mathrm{Di},c}_{\mu\nu} \doteq D^{\mathrm{Di,can}}_{\mu\nu} - \frac{\tilde{c}}{2}g_{\mu\nu}\left(D^*_x + D_y\right)D^*_y \tag{IV.112}$$

$$\doteq \left\{-\frac{1}{2}\gamma_{(\mu}\left(\nabla_{\nu)} - g^{\nu}_{\nu)}\nabla_{\nu'}\right) - \frac{c}{2}g_{\mu\nu}\left(D_y + D^*_x\right)\right\}D^*_y.$$

Based on this, we can prove the following result.

**Theorem IV.3.1** *Let $\omega(:T^{Di}_{\mu\nu}(x):)$ be defined as in (IV.111) with $c = -\frac{1}{6}$ and $\omega$ being a Hadamard state on $\mathscr{A}(DM)$.*

a) *$\omega(:T^{Di}_{\mu\nu}(x):)$ is covariantly conserved, i.e.*

$$\nabla^\mu\omega\left(:T^{Di}_{\mu\nu}(x):\right) = 0.$$

b) *The trace of $\omega(:T^{Di}_{\mu\nu}(x):)$ equals*

$$g^{\mu\nu}\omega\left(:T^{Di}_{\mu\nu}(x):\right) = -\frac{1}{4\pi^2}\left[V_1\right] + \frac{1}{8\pi^2}m\,Tr\left[D^*_y\,W\right]$$

$$= \frac{1}{2880\pi^2}\left(\frac{7}{2}C_{\mu\nu\rho\tau}C^{\mu\nu\rho\tau} + 11\left(R_{\mu\nu}R^{\mu\nu} - \frac{1}{3}R^2\right) - 6\square R\right)$$

$$- \frac{1}{\pi^2}\left(\frac{m^4}{8} + \frac{m^2R}{48}\right) + \frac{1}{8\pi^2}m\,Tr\left[D^*_y\,W\right],$$

*which, for $m = 0$, constitutes the **trace anomaly** of the quantum stress-energy tensor proper to the Dirac field.*

c) *The conservation and trace anomaly are independent of the chosen scale $\lambda$ in the Hadamard parametrix $H^-$. Namely, a change*

$$\lambda \mapsto \lambda'$$

*results in*

$$\omega(:T^{Di}_{\mu\nu}(x):) \mapsto \omega(:T^{Di}_{\mu\nu}(x):)' = \omega(:T^{Di}_{\mu\nu}(x):) + \delta\,T_{\mu\nu},$$

*where*

$$\delta\,T_{\mu\nu} \doteq \frac{2\log\lambda/\lambda'}{8\pi^2}Tr\left[D^{\mathrm{Di},c}_{\mu\nu}\,V\right] = \frac{2\log\lambda/\lambda'}{8\pi^2}Tr\left[D^{\mathrm{Di,can}}_{\mu\nu}\,V\right]$$

$$= \frac{2\log\lambda/\lambda'}{8\pi^2}\left(\frac{m^4}{2}g_{\mu\nu} - \frac{m^2}{6}G_{\mu\nu} + \frac{1}{60}(I_{\mu\nu} - 3J_{\mu\nu})\right)$$

*is a conserved tensor which has vanishing trace for $m = 0$.*

*Proof.* a) Again, we leave $c$ undetermined to start with. Applying Synge's rule and taking into account that $\{\Re_{\mu\nu}, \gamma^\mu\} = 0$ and $[g_{\nu;\mu}^{\nu'}] = 0$ (*cf.* lemma I.2.2.9 and subsection III.1.2), we get

$$8\pi^2 \nabla^\mu \omega \left(:T_{\mu\nu}^{\mathrm{Di}}(x):\right) = \nabla^\mu Tr\left[D_{\mu\nu}^{\mathrm{Di},c} \, W\right] = Tr\left[(\nabla^\mu + g_{\mu'}^{\mu}\nabla^{\mu'})D_{\mu\nu}^{\mathrm{Di},c} \, W\right]$$

$$= Tr\left[\left\{\frac{1}{4}\left(g_\nu^{\nu'}\nabla_{\nu'} - \nabla_\nu\right)\left(D_x^* D_y^* + \mathfrak{P}_y\right) + \frac{1}{4}\gamma_\nu D_y^*(\mathfrak{P}_y - \mathfrak{P}_x)\right.\right.$$

$$\left.\left. - \frac{c}{2}\left(g_\nu^{\nu'}\nabla_{\nu'} + \nabla_\nu\right)\left(D_x^* D_y^* - \mathfrak{P}_y\right)\right\} \, W\right].$$

Remembering that $-D_y^*(H^- + W)$ is the local two-point distribution of a state, it follows that $H^- + W$ is subject to the distributional differential equations $D_x^* D_y^*(H^- + W) = 0 = \mathfrak{P}_y(H^- + W)$. Thus, we can safely replace $W$ in the above equation by $-H^-$, since every appearing term involves one of the two aforementioned differential operators. Such a procedure yields

$$8\pi^2 \nabla^\mu \omega \left(:T_{\mu\nu}^{\mathrm{Di}}(x):\right) = Tr\left[\left\{\frac{1}{4}\left(\nabla_\nu - g_\nu^{\nu'}\nabla_{\nu'}\right)\left(D_x^* D_y^* + \mathfrak{P}_y\right) + \frac{1}{4}\gamma_\nu\left(\gamma^{\mu'}\nabla_{\mu'} + m\right)(\mathfrak{P}_x - \mathfrak{P}_y)\right.\right.$$

$$\left.\left. - \frac{c}{2}\left(g_\nu^{\nu'}\nabla_{\nu'} + \nabla_\nu\right)\left(\mathfrak{P}_y - D_x^* D_y^*\right)\right\} \, H^-\right].$$

Now we can insert the various coincidence point limits of the differentiated Hadamard bidistribution $H^-$ computed in theorem III.3.2.7 to obtain

$$8\pi^2 \nabla^\mu \omega \left(:T_{\mu\nu}^{\mathrm{Di}}(x):\right) = -(1 + 6c) \, Tr\left[V_1\right]_{;\nu}.$$

This proves conservation for $c = -\frac{1}{6}$.

b) Leaving $c$ undetermined for one last time, we use both the insights on the parallel transport of gamma matrices from subsection III.3.2 and the arguments already employed in the computation of the conservation to get

$$8\pi^2 g^{\mu\nu} \omega \left(:T_{\mu\nu}^{\mathrm{Di}}(x):\right) = g^{\mu\nu} Tr\left[D_{\mu\nu}^{\mathrm{Di},c} \, W\right]$$

$$= Tr\left[\left\{-\left(2c + \frac{1}{2}\right)\left(D_x^* D_y^* - \mathfrak{P}_y\right) + m D_y^*\right\} \, W\right]$$

$$= Tr\left[\left(2c + \frac{1}{2}\right)\left(D_x^* D_y^* - \mathfrak{P}_y\right) H^- + m D_y^* W\right]$$

$$= -6(4c + 1) \, Tr\left[V_1\right] + m \, Tr\left[D_y^* \, W\right].$$

c) A change of scale as considered transforms $W$ by a adding a term $2\log\lambda/\lambda' V$. Hence, our computations in a) and b) entail

$$\frac{8\pi^2}{2\log\lambda/\lambda'}\nabla^\mu\delta T_{\mu\nu} = Tr\left[\left\{\frac{1}{4}\left(g_\nu^{\nu'}\nabla_{\nu'}-\nabla_\nu\right)\left(D_x^*D_y^*+\mathfrak{P}_y\right)+\frac{1}{4}\gamma_\nu D_y^*(\mathfrak{P}_y-\mathfrak{P}_x)\right.\right.$$
$$\left.\left.-\frac{c}{2}\left(g_\nu^{\nu'}\nabla_{\nu'}+\nabla_\nu\right)\left(D_x^*D_y^*-\mathfrak{P}_y\right)\right\} V\right]$$

$$\frac{8\pi^2}{2\log\lambda/\lambda'}g^{\mu\nu}\delta T_{\mu\nu} = Tr\left[\left\{-\left(2c+\frac{1}{2}\right)\left(D_x^*D_y^*-\mathfrak{P}_y\right)+mD_y^*\right\} V\right]$$

The former term vanishes because $V$ is in the kernel of all occurring differential operators as discussed in subsection III.3.2, and the same holds for the latter term if we insert $m = 0$. □

Let us comment on how our definition of $\omega(: T_{\mu\nu}^{\mathrm{Di}}(x):)$ fits into the framework of Wald's axioms. The first axiom requiring state-independence of the procedure is again fulfilled because the additional differential operators we have added allow to replace $W$ by $H^-$ in the relevant terms, thus making the modification of the canonical stress-energy tensor state-independent. In fact, this will be the case quite generally for all spins if one follows the ideas of [Mor03], as one will always consider modifications given by differential operators whose kernel will contain the relevant two-point function of a Hadamard state. In case one insists, the vanishing in Minkowski spacetime can be assured by a similar calculation as performed in the scalar case in [Mor03]. In the massless case, $V \equiv W \equiv 0$ and there is nothing to be done. In the massive case, one fixes the scale in the Hadamard parametrix $\lambda$ in terms of the inverse mass and obtains the wanted result by a straightforward, but cumbersome calculation, see [DHP09] for the details. As in the Klein-Gordon case, one is in the unsatisfactory situation that the lack of a fixed $\lambda$ prevents the fifth axiom to be fulfilled. Namely, as the proof of theorem IV.3.1 c) shows, the conserved local tensors $I_{\mu\nu}$ and $J_{\mu\nu}$ appear inevitably if one changes this scale.

To close, we point out that the structure of many terms related to the stress-energy tensor of Dirac fields seems to be simpler than in the Klein-Gordon case. The reason for this is simply that the Klein-Gordon field is essentially the only field which has a freedom in its coupling to (scalar) curvature. If one chooses conformal coupling $\xi = \frac{1}{6}$, then the structure of the stress-energy tensors of both the Klein-Gordon and the Dirac field is essentially the same. This is one reason why we would like to advertise the choice of conformal coupling whenever one treats Klein-Gordon fields.

## IV.4 The Relation between DeWitt-Schwinger and Hadamard Point-Splitting

We have already remarked at several occasions that the results on the trace anomaly of the quantum stress-energy tensor computed in the scalar case by [Mor03] and in the Dirac case by us have in principle already been known based on point-splitting computations involving the DeWitt-Schwinger expansion. In fact, the trace anomaly for Klein-Gordon, Dirac, and Maxwell fields in general curved spacetimes has been computed in [Chr78] (confirming results of computations in more specific cases). These results have been generalised to field theories of higher spin by providing generalised 'index theorems' in [ChDu79], yet again based on the mentioned DeWitt-Schwinger expansion. This expansion is problematic on general curved spacetimes, because it makes use of results proper to *heat kernel theory* (see for instance [Wa79, Mor99] and references therein). These results, however, are initially only valid on Riemannian spacetimes, and thus can apparently at best be extended by a *Wick rotation* to Lorentzian spacetimes which are analytic and have a Riemannian ('Euclidean') section. The identities used in deriving the DeWitt-Schwinger expansion on general Lorentzian spacetimes are therefore purely formal, moreover, they show infrared divergences in the massless case. Although the workers in the field have been aware of this fact (see for instance the comments at the beginning of section 6.6. in [BiDa82]), the results have been trusted and have found their way into standard monographs like [BiDa82] (*cf.* table 1. on page 179 in that reference). It seems the common sense was that the results of the procedure can be trusted, even though this is not the case for the involved steps. This belief has been strengthened even more after the observation that the Hadamard coefficients appearing in the rigorous treatment of Hadamard point-splitting are essentially the same as the ones appearing in the DeWitt-Schwinger expansion, see for instance [Mor00, Mor99, DeFo06, DeFo08]. In fact, the point of view to date seems to be that the Hadamard approach is rather well-suited for rigorous treatments, while the DeWitt-Schwinger approach is more convenient for actual calculations [DeFo06, DeFo08]. We do not claim to provide a full understanding of the relation between the two approaches, but we would like to show that – at least regarding their application – the approaches are essentially *equivalent*. Particularly, the DeWitt-Schwinger calculations can be put on rigorous grounds without much effort. Our findings are not surprising, but, to our knowledge, they do not seem to have been reported in the explicit form we give here to date. We discuss the example of the Klein-Gordon field and show how the DeWitt-Schwinger calculations yield a stress-tensor which fulfils Wald's conservation, state-independence, and local covariance axioms.

Let us start our treatment by giving a brief and formal account on the derivation of the DeWitt-Schwinger renormalisation of the stress-energy tensor based on the presentation in [Wa79] (see also [Ful89]). The starting point is the *effective action* $S_{\text{eff}}$ defined via a path integral as

$$e^{-S_{\text{eff}}} = \int [d\phi] e^{-\frac{1}{2}\langle \phi, P\phi \rangle},$$

where $P$ is the Klein-Gordon operator, $[d\phi]$ is supposed to denote some (formal) measure on the space of field configurations, and the argument of the exponential in the integrand is the classical action $S_{\text{KG}}$ up to boundary terms. We have written the effective action for the Riemannian case, whereas, in the Lorentzian case, one would put an imaginary unit in front of both actions in the above formula. The relevance of the effective action in our case is the fact that one can formally define the expectation value of the regularised stress-energy tensor in analogy to the classical case as

$$\omega(:T_{\mu\nu}:) \doteq \frac{2}{\sqrt{|\det g|}} \frac{\delta S_{\text{eff}}}{\delta g^{\mu\nu}}.$$

The natural question which arises is where the state dependence of the right hand side is hidden and one can interpret the situation in such a way that the measure $[d\phi]$ is state-dependent [Mor03]. To obtain $\omega(:T_{\mu\nu}:)$ in this picture, one therefore needs to regularise the effective action and then compute the mentioned functional derivative with respect to the metric. Note that, if the regularisation is such that the effective action is diffeomorphism-invariant, than the resulting functional derivative will automatically lead to a conserved $\omega(:T_{\mu\nu}:)$.

Let us proceed to understand why the effective action is divergent and how to regularise it. Ignoring the infinite dimension of the space of field configurations, and evaluating the above integral as a Gaussian integral in a finite dimensional vector space, one finds (up to a constant)

$$e^{-S_{\text{eff}}} = (\det P)^{-\frac{1}{2}} = e^{-\frac{1}{2}Tr\log P},$$

and, hence,

$$S_{\text{eff}} = \frac{1}{2}\int_M d_g x\,[\log P](x,x) \doteq \int_M d_g x\,L_{\text{eff}}(x),$$

where the indentity (valid in finite dimensional cases) $\log \det P = Tr \log P$ has been used, the trace is evaluated by integrating the kernel of the operator $\log P$ at coinciding points, and $L_{\text{eff}}$ is interpreted as the *effective Lagrangean*. The above integral is certainly diverging, as one can already infer from the identity

$$\log P = \lim_{\varepsilon\downarrow 0}\left\{ -\int_\varepsilon^\infty ds\,\frac{e^{-sP}}{s} + (\gamma - \log\varepsilon)\mathbb{I}\right\},$$

where $\gamma$ denotes the Euler-Mascheroni constant and $\mathbb{I}$ is the identity operator. We discard the divergent term proportional to $\mathbb{I}$ (note that $Tr\,\mathbb{I} = \infty$), and define

$$L_{\text{eff}}(x) \doteq -\int_0^\infty ds\,\frac{\left[e^{-sP}\right](x,x)}{s}. \tag{IV.113}$$

The integral kernel $\left[e^{-sP}\right](s,x,y)$ of the appearing exponential of $P$ is called the *heat kernel* because it satisfies the *heat equation*

$$(\partial_s + P_x)\left[e^{-sP}\right](s,x,y) = 0.$$

In the Riemannian case, $P = -\nabla_\mu\nabla^\mu + m^2$ is a positive and symmetric operator on $\mathcal{D}(M,\mathbb{R})$ (and this is also the case if we include a coupling to scalar curvature, provided $\xi$ is positive and $R$ is somehow bounded from below). One can therefore consider a self-adjoint extension of $P$ and define the heat kernel by functional analytic methods, see [Wa79, ReSi75]. In this way, one obtains an expansion for the integral kernel of $e^{-sP}$ on a geodesically convex set given as

[Wa79, Mor99]

$$\left[e^{-sP}\right](s,x,y) = \frac{1}{(4\pi s)^2}\, e^{-\frac{\sigma(x,y)}{2s}} \sum_{n=0}^{\infty} a_n(x,y)s^n \;+\; \text{'smooth biscalar'} \qquad \text{(IV.114)}$$

$$= \frac{1}{(4\pi s)^2}\, e^{-\frac{\sigma(x,y)}{2s}-m^2 s} \sum_{n=0}^{\infty} \alpha_n(x,y)s^n \;+\; \text{'smooth biscalar'},$$

where $a_n(x,y)$ are smooth biscalars which fulfil the recursive differential equations [Ful89, Mor99]

$$a_0 = u\,, \qquad Pa_n - \sigma_{;\mu}a_{n+1}^\mu - \left(\frac{1}{2}\Box\sigma + n - 1\right)a_{n+1} = 0 \qquad \text{(IV.115)}$$

with $u$ denoting the square root of the VanVleck-Morette determinant, and $\alpha_n$ being related to $a_n$ via

$$\alpha_n = \sum_{j=0}^{n} \frac{m^{2j}}{j!}a_{n-j}\,. \qquad \text{(IV.116)}$$

The reason why we have displayed the above expansion in two versions is the following: the second version in terms of $\alpha_n$ is more convenient in the regularisation procedure, because the appearing $m^2$ will be necessary to avoid potential infrared singularities in the integral with respect to $s$ present in $L_{\text{eff}}$. In contrast, the version given in terms of $a_n$ is important to show the relation between the Hadamard coefficients $v_n$ and the $a_n$. Namely, a short computation and comparison with the scalar Hadamard recursion relations discussed in subsection III.1.2 reveals the well-known identity [Mor99, Mor00]

$$a_{n+1} = (-)^{n+1}\,2^{n+1}\,n!\,v_n\,. \qquad \text{(IV.117)}$$

One can show that, in the Riemannian case, the above expansion of the heat kernel is asymptotic for small $s$ and $x = y$, see [Wa79, Mor99], which means that the smooth remainder term vanishes in a controlled way in the limit $s \to 0$. In Mathematics and on Riemannian manifolds, this expansion is called *Minakshisundaram-Plejel expansion* – and the coefficients $a_n$ thus *Minakshisundaram-Plejel coefficients* – while in Physics and on Lorentzian manifolds, the above expansion (with $s$ replaced by $is$) is termed *DeWitt-Schwinger expansion* and $a_n$ are called *DeWitt-Schwinger coefficients* [Ful89]. Note that, in the case of Lorentzian manifolds, there is no general way to define the 'heat kernel' $e^{-sP}$ if $P$ is the Klein-Gordon operator, because $-P$, being hyperbolic and not elliptic, is not a positive operator. Hence, the DeWitt-Schwinger expansion is also not known to have any asymptotic properties whatsoever and one can presumably only take it as a 'clone' of its Riemannian cousin. We also remark that the infinite sum in (IV.114) is not necessarily converging, but this is not relevant for our purposes.

Let us continue discussing the renormalisation of $\omega(:T_{\mu\nu}:)$ via renormalising $L_{\text{eff}}$. Inserting (IV.114) in the definition of $L_{\text{eff}}$ (IV.113), one finds that the contributions due to $n = 0$, $n = 1$, and $n = 2$ lead to divergent integrals with respect to $s$ if $x = y$ (or if $x$ and $y$ are lightlike related in the Lorentzian case). Particularly, we observe UV-divergences at the lower integration limit, while infrared divergences at the upper limit are not present on account of $e^{-m^2 s}$. One therefore defines the *renormalised effective Lagrangean* as

$$L_{\text{eff, ren}}(x) \doteq -\int_0^\infty ds \, \frac{1}{s} \frac{1}{(4\pi s)^2} \, e^{-m^2 s} \sum_{n=3}^\infty \alpha_n(x, x) s^n, \qquad \text{(IV.118)}$$

and $\omega(:T_{\mu\nu}:)$ by the functional derivative of the associated *renormalised effective action*

$$S_{\text{eff,ren}} \doteq \int_M d_g x \, L_{\text{eff, ren}}(x).$$

In the Lorentzian case, one would again define these quantities by replacing $s$ with $is$. We point out two important things. First, we know that the Hadamard coefficients $v_n$ and, hence, $\alpha_n$ are covariant biscalars. Therefore, the renormalised effective Lagrangean is a scalar, and, consequently $\omega(:T_{\mu\nu}:)$ is automatically conserved by its definition as a functional derivative with respect to the metric of a diffeomorphism-invariant quantity [Wal84, app. E]. Secondly, it is clear that $L_{\text{eff, ren}}(x)$ has *no state dependence whatsoever*. This holds because we know that the coefficients $\alpha_n$ are completely specified by local curvature terms and $m$, $\xi$. In fact, we have 'lost' the state dependence by disregarding the smooth remainder term in the expansion of the heat kernel. Defining $\omega(:T_{\mu\nu}:)$ via $L_{\text{eff, ren}}(x)$ therefore completely disregards the state dependence of $\omega(:T_{\mu\nu}:)$. Apart from the appearance of the Hadamard coefficients, there does not seem to be a close relation to our definition of $\omega(:T_{\mu\nu}:)$ in terms of applying a suitable bidifferential operator $D_{\mu\nu}$ to the regularised two-point function and then taking the coinciding point limit. However, one can reformulate the above renormalisation of the effective action in the following way [Chr76, Wa79]. One formally pulls the functional derivative with respect to the metric in the definition of $\omega(:T_{\mu\nu}:)$ under the integral with respect to $s$ and then finds via additional formal steps

$$\omega(:T_{\mu\nu}:) \doteq \frac{2}{\sqrt{|\det g|}} \frac{\delta S_{\text{eff}}}{\delta g^{\mu\nu}} = \left[ \frac{2}{\sqrt{|\det g|}} \frac{\delta \sqrt{|\det g|} P(x, y)}{\delta g^{\mu\nu}} \int_0^\infty ds \, e^{-sP} \right]$$

$$= \left[ \frac{2}{\sqrt{|\det g|}} \frac{\delta S}{\delta g^{\mu\nu} \delta \phi(x) \delta \phi(y)} \int_0^\infty ds \, e^{-sP} \right] = \left[ \frac{\delta T_{\mu\nu}}{\delta \phi(x) \delta \phi(y)} \int_0^\infty ds \, e^{-sP} \right]$$

$$= \left[ D_{\mu\nu}^{\text{can}}(x, y) \left[ P^{-1} \right](x, y) \right],$$

where the outer square brackets denote the coinciding point limit. In the above formal derivation, it has been used that the integral kernel $P(x, y) = \delta(x, y) P_x$ of $P$ is obtained as the second functional derivative of the classical action $S$ with respect to the field $\phi$ and that the canonical differential operator $D_{\mu\nu}^{\text{can}}$ we have considered in section IV.1 is nothing but the second functional derivative of the classical stress-energy tensor with respect to the field $\phi$. In the context of renormalisation of the effective action in Lorentzian spacetimes, one usually considers $P^{-1}$ to be the *Feynman propagator* $\Delta_F$ (note that $P^{-1}$ is not unique). Hence, the divergences of $\omega(:T_{\mu\nu}:)$ computed as above are interpreted to stem from the divergences of the Feynman propagator at coinciding points. To renormalise $\omega(:T_{\mu\nu}:)$ in the Lorentzian case, one therefore inserts the DeWitt-Schwinger expansion (IV.114) of the heat kernel in the integral expression for $P^{-1}$ in terms of $e^{-sP}$ to obtain

$$\Delta_F(x,y) \doteq \lim_{\varepsilon \downarrow 0} \int_0^\infty ds \, \frac{1}{(4\pi s)^2} \, e^{-\frac{\sigma(x,y)+i\varepsilon}{2s} - m^2 s} \sum_{n=0}^\infty \alpha_n(x,y) s^n$$

$$= \lim_{\varepsilon \downarrow 0} \frac{1}{8\pi^2} \sum_{n=0}^\infty \left( \frac{\sigma + i\varepsilon}{2m^2} \right)^{\frac{n-1}{2}} K_{n-1}\left( \sqrt{2m^2(\sigma + i\varepsilon)} \right) \alpha_n(x,y),$$

where the $\varepsilon$-prescription suitable for the Feyman propagator has been inserted and an integral identity for the *modified Hankel function* $K_n$ has been used. Expanding this in powers of $\sigma$, inserting $\alpha_0 = u$, and removing the $\varepsilon$-prescription from the regular terms, we find

$$\Delta_F(x,y) = \frac{1}{8\pi^2} \left\{ \frac{u}{\sigma + i\varepsilon} + \log\left( \frac{(\sigma + i\varepsilon)m^2 e^{2\gamma}}{2} \right) \left( \frac{m^2 u}{2} - \frac{\alpha_1}{2} + \frac{m^4 u\sigma}{8} + \frac{\alpha_2 \sigma}{4} - \frac{m^2 \alpha_1 \sigma}{4} \right) \right.$$

$$\left. - \frac{m^2 u}{2} - \frac{5m^4 u\sigma}{16} + \frac{\alpha_1 \sigma}{2} - \frac{\alpha_2 \sigma}{4} + \frac{\alpha_2}{2m^2} + O\left(\sigma^2 \log(\sigma + i\varepsilon)\right) \right\},$$

and, inserting the relation between $\alpha_n$ and $v_n$ as given in (IV.117) and (IV.116), we see explicitly that, barring the different $\varepsilon$-prescription, $\Delta_F(x,y)$ displays the Hadamard singularity structure. Again we point out that the 'correct' Feynman propagator is always *state-dependent*, while the above expression is manifestly *state-independent*, being essentially only a local curvature expression. Once more, this stems from the fact that one has disregarded the smooth (non-local) remainder in the expansion of the heat kernel. Note however, that, while this smooth remainder term is essentially well-understood in the Riemannian case, this does not hold in the Lorentzian case, as there the whole DeWitt-Schwinger expansion is not rigorous. Hence, in contract to the Hadamard expansion of the two-point function of a Hadamard state, there does not seem to be a possibility to introduce the state-dependence in a rigorous way in the DeWitt-Schwinger renormalisation as we have presented it up to now. Moreover, it is already visible from the few terms we have provided that, in term of the Hadamard series, the smooth term $w$ of the above distribution contains inverse powers of the mass and, hence, diverges in the massless limit. Of course this is particularly also the case for the logarithmic terms, which displays the infrared singularity of the integrals with respect to $s$.

By the well known distributional identities (see for instance [ReSi75, Ful89])

$$\lim_{\varepsilon \downarrow 0} \frac{1}{x + i\varepsilon} = \mathscr{P}\frac{1}{x} + i\pi\delta(x), \qquad \lim_{\varepsilon \downarrow 0} \log(x + i\varepsilon) = \log|x| + \pi i\Theta(-x),$$

where $\mathscr{P}$ denotes the principal value, one finds that $[D_{\mu\nu}^{can}(x,y)[P^{-1}](x,y)]$ is a complex number, and one therefore has to consider its real-part as the 'correct' definition of $\omega(:T_{\mu\nu}:)$. In terms of the Hadamard series we have discussed in subsection III.1.2, this corresponds to consider the symmetric part

$$h^s(x,y) \doteq \frac{1}{2}(h(x,y) + h(y,x))$$

in the definition of the point-splitting prescription. Note that this encodes the same information as the full $h$ in the coinciding point limit.

With the setup we have just described, the regularisation of $\omega\left(:T_{\mu\nu}:\right)$ goes as follows [Chr76]. Applying $D_{\mu\nu}^{can}$ to the real part of $\Delta_F$ given in terms of the DeWitt-Schwinger expansion, one finds that, as in the regularisation of the effective action, divergences come from the terms of order $n = 0$, $n = 1$, and $n = 2$. Hence, one identifies the divergent part of the stress-energy tensor as $D_{\mu\nu}^{can}$ applied to the first three terms in the expansion. However, these terms of course also contain smooth contributions, and in fact one has to take care to not subtract 'too much', otherwise one could spoil the covariant conservation of $\omega(:T_{\mu\nu}:)$ which was automatic in the renormalisation of the effective action. Namely, although in the latter renormalisation one has subtracted the first three terms of the DeWitt-Schwinger expansion as well, some of the subtractions have been zero on account of the vanishing of $\sigma$ in the coinciding point limit. As we now derive the series, we could accidentally introduce these vanishing terms since second derivatives of $\sigma$ do not vanish in the coinciding point limit present in $L_{\text{eff}}(x)$. However, as observed in [Chr76], this can be avoided if one defines the divergent part of $\omega(:T_{\mu\nu}:)$ by applying $D_{\mu\nu}^{can}$ to the real part of $\Delta_F(x,y)$ and then discarding all terms which are proportional to inverse powers of the mass. Proceeding like this, Christensen has computed in [Chr76, Chr78] the divergent part of the quantum stress-energy tensor and has obtained the trace anomaly by computing the negative trace of the divergent part. This follows because the full expression of the stress-energy tensor must have vanishing trace in the conformally invariant case, despite its divergence, it its completely given in terms of the real part of $\Delta_F$, which in turn is a bisolution of the Klein-Gordon equation by its very construction in terms of the DeWitt-Schwinger series. Note that Christensen has introduced the massless limit by replacing the $m^2$ in the logarithmic divergence by an arbitrary scale $\lambda$. This may seem rather ad-hoc, but from our point of view this is a reasonable procedure if we remember that we have defined the Hadamard series with an arbitrary length scale in the logarithm right from the start. Finally, since the smooth terms proportional to inverse powers of the mass had been discarded to assure conservation, the massless limit could be performed in a meaningful way.

Having reviewed the DeWitt-Schwinger point-splitting renormalisation of the stress-energy tensor, let us recapitulate its seeming disadvantages.

a) It has been defined via an expansion of the heat kernel which is not well-defined in general curved Lorentzian spacetimes.

b) It does not take into account the state dependence of $\omega\left(:T_{\mu\nu}:\right)$.

c) It employs a Hadamard series whose smooth part $w$ diverges in the massless limit.

We shall now give a regularisation prescription which closely mimics the one of Christensen, but disposes of the above three problems.

**Theorem** IV.4.1  *Let $\omega_2$ be the two point function of a Hadamard state on $\mathscr{A}(M)$, let*

$$\omega_2^s(x,y) \doteq \frac{1}{2}(\omega_2(x,y) + \omega_2(y,x)),$$

*and let us define for $x$ and $y$ in a geodesically convex neighbourhood*

$$h_{DS}(x,y) \doteq \frac{1}{8\pi^2}\left\{\mathscr{P}\frac{u}{\sigma} + \log\left|\frac{\sigma\, m^2 e^{2\gamma}}{2}\right|\left(\frac{m^2 u}{2} - \frac{\alpha_1}{2} + \frac{m^4 u\sigma}{8} + \frac{\alpha_2\sigma}{4} - \frac{m^2\alpha_1\sigma}{4}\right)\right.$$

$$\left. -\frac{m^2 u}{2} - \frac{5m^4 u\sigma}{16} + \frac{\alpha_1\sigma}{2} - \frac{\alpha_2\sigma}{4} + \frac{\alpha_2}{2m^2}\right\},$$

$$\doteq h_0(x,y) + h_m(x,y),$$

$$h_m(x,y) \doteq \frac{1}{8\pi^2}\frac{\alpha_2}{2m^2}, \quad h_0(x,y) \doteq h_{DS}(x,y) - h_m(x,y).$$

*Moreover, let us split the canonical bidifferential operator*

$$D_{\mu\nu}^{KG,can} = (1-2\xi)g_\nu^{\nu'}\nabla_\mu\nabla_{\nu'} - 2\xi\nabla_\mu\nabla_\nu + \xi G_{\mu\nu} + g_{\mu\nu}\left\{2\xi\Box_x + \left(2\xi - \frac{1}{2}\right)g_\rho^{\rho'}\nabla^\rho\nabla_{\rho'} - \frac{1}{2}m^2\right\}$$

*as*

$$D_{\mu\nu}^{KG,can} \doteq D_{\mu\nu}^0 + D_{\mu\nu}^m, \quad D_{\mu\nu}^m \doteq -\frac{1}{2}m^2 g_{\mu\nu}, \quad D_{\mu\nu}^0 \doteq D_{\mu\nu}^{KG,can} - D_{\mu\nu}^0.$$

*The stress-energy tensor regularisation prescription defined as*

$$\omega\left(:T_{\mu\nu}^{DS}(x):\right) \doteq \left[D_{\mu\nu}^{KG,can}\left(\omega_2^s - h_{DS}\right) + D_{\mu\nu}^0 h_m\right]$$

*fulfils the Wald axioms of conservation, state-independence, and local covariance. Particularly, it displays the trace anomaly*

$$g^{\mu\nu}\omega\left(:T_{\mu\nu}^{DS}(x):\right)\Big|_{m^2=0} = \frac{1}{4\pi^2}[v_1].$$

*Proof.* First of all, let us remark that the regularisation prescription is well-defined, as the relation between $\alpha_n$ and $v_n$ given in (IV.117) and (IV.116) implies that $\omega_2^s - h_{DS}$ is of class $C^3$ (the worst terms in $\omega_2^s - h_{DS}$ are of the form $\sigma^2\log\sigma$). Additionally, the prescription fulfils the requirement of local covariance, since it only involves subtraction of objects given in terms of the DeWitt-Schwinger/Hadamard coefficients. Moreover, state-independence follows manifestly from the definition, as the modification of the canonical prescription is given in terms of $D^0_{\mu\nu} h_m$, which is a state-independent term.

To prove covariant conservation, we recall that, in the proof of theorem IV.2.1, it has been implicitly computed that for any smooth biscalar $B(x,y)$ the following relation holds

$$\nabla^\mu \left[ D^{\mathrm{KG,can}}_{\mu\nu} B \right] = \left[ \nabla_{\nu'} P_x B \right].$$

Applying this to our current case, we find

$$\nabla^\mu \omega \left( :T^{\mathrm{DS}}_{\mu\nu}(x): \right) = \nabla^\mu \left[ D^{\mathrm{KG,can}}_{\mu\nu} (\omega_2 - h_{\mathrm{DS}}) + D^0_{\mu\nu} h_m \right]$$

$$= \left[ \nabla_{\nu'} P_x \left( \omega_2^s - h_{\mathrm{DS}} \right) + \nabla_{\nu'} P^0_x h_m \right] = \left[ \nabla_{\nu'} \left( P_x \omega_2^s - P_x h_{\mathrm{DS}} + P^0_x h_m \right) \right],$$

where we have defined

$$P^0 \doteq -\Box + \xi R.$$

A straightforward computation employing the Hadamard/DeWitt-Schwinger recursion relations yields

$$P h_{\mathrm{DS}} = P^0 h_m.$$

From this and the fact that $\omega_2^s$ naturally solves the Klein-Gordon equation, conservation follows.

By Wald's results [Wal78a], the above findings already imply that $\omega \left( :T^{\mathrm{DS}}_{\mu\nu}(x): \right)$ displays the 'correct' trace anomaly. However, it is instructive to compute it explicitly. To this end, we obtain with steps similar to the ones already taken and using the implicit computational results obtained in the proof of theorem IV.2.1

$$g^{\mu\nu} \omega \left( :T^{\mathrm{DS}}_{\mu\nu}(x): \right) = g^{\mu\nu} \left[ D^{\mathrm{KG,can}}_{\mu\nu} (\omega_2 - h_{\mathrm{DS}}) + D^0_{\mu\nu} h_m \right]$$

$$= \left[ (P_x - m^2) \left( \omega_2^s - h_{\mathrm{DS}} \right) + P^0_x h_m \right] = -m^2 \left[ \omega_2^s - h_{\mathrm{DS}} \right] = -m^2 \left[ \omega_2^s - h_0 \right] + \frac{1}{16\pi^2} \left[ \alpha_2 \right]$$

$$= -m^2 \left[ \omega_2^s - h_0 \right] + \frac{1}{8\pi^2} \left[ 2v_1 - m^2 v_0 + \frac{m^4 u}{4} \right].$$

<div align="right">□</div>

A few comments on the result are in order. First, on practical grounds, the above result is really equivalent to the computation of Christensen in [Chr76, Chr78] because the terms of the DeWitt-Schwinger expansion we have omitted are all proportional to $\sigma^2$ and, hence, vanish upon application of the occurring differential operators in the coinciding point limit. In this sense, we have been able to put his results on firm grounds. Secondly, the modification term $D^0_{\mu\nu} h_m$ does not 'simply cure the conservation anomaly'. In this sense, the regularisation prescription just analysed differs from the one introduced in [Wal78a] and improved in [Mor03] in that it assures conservation in a different way. Namely, conservation does not follow by adding a term by hand or by modifying the classical stress energy tensor. In contrast, it follows by the explicit structure of the smooth term $w$ in the DeWitt-Schwinger two-point function in combination with discarding specific terms proportional to inverse powers of the mass. Recall that the latter procedure has been motivated by analysing carefully the subtractions in the (non-rigorous) regularisation of the effective action. Altogether, we believe that the above result, even though it holds only in the case of scalar fields, tells us that the trace anomaly results obtained by DeWitt-Schwinger calculations can be trusted, or at least most likely completely reproduced in rigorous terms. Moreover, let us remark that the above prescription as we have presented it does not assure that $\omega(: T_{\mu\nu}^{\mathrm{DS}}(x):)$ is smooth, as $\omega_2^s - h_{\mathrm{DS}}$ is *a priori* only known to be twice-differentiable. However, this only affects the state-dependent term in $\omega(: T_{\mu\nu}^{\mathrm{DS}}(x):)$ and not the trace anomaly.

To close, we would like to suggest two other possible ways to put DeWitt-Schwinger computations in general curved Lorentzian spacetimes on rigorous grounds. On the one hand, it is maybe possible to use a deformation argument like the one employed in the proof of proposition III.3.2.5 to deform a sufficiently small region of the Lorentzian spacetime $(M, g)$ under consideration in such a way that one relates it to a portion of an analytical Lorentzian spacetime $(M', g')$ which allows for a Wick rotation. One could compute quantities like $\omega(: T_{\mu\nu}(x):)$ in the analytical spacetime, and thus by locality and general covariance immediately obtain the state-independent part (and hence, the trace anomaly) of $\omega(: T_{\mu\nu}(x):)$ on the full spacetime $(M, g)$. Note that one would have to make sure to perform this steps by means of an analytical spacetime $(M', g')$ which has a non-vanishing Weyl tensor, to be able to grasp the full structure of $\omega(: T_{\mu\nu}(x):)$. A related possibility is to use the local analytic approximation and local Wick rotation techniques developed in [Mor99, Mor00]. This would make it possible to work in the analytic framework without employing a deformation argument.

## IV.5 Cosmological Solutions of the Semiclassical Einstein Equation

With the understanding of $\omega(: T_{\mu\nu}(x):)$ developed in the previous sections, we can try to look for solutions of the semiclassical Einstein equation (IV.101). One can expect that these can be most easily obtained in the case of FLRW spacetimes, as there only one dynamical degree of freedom of the background curvature – the scale factor $a(t)$ – exists. At the same time, this simple situation is already very interesting, as it can offer the possibility to model the evolution of our universe by the effects of quantum matter-energy. Hence, we shall develop in this section the (well-known) analysis of solutions of (IV.101) in the case of flat FLRW spacetimes. In the next section, we will compare the behaviour of these solution with supernova Ia measurements. Although solutions of the semiclassical Einstein equations had already been matched to experiments in past work, see for instance [PaRa99, PaRa01] and references therein, it is safe to claim that the following considerations are the first which take *all* contributions of $\omega(: T_{\mu\nu}(x):)$, namely, the anomalous part, the state-dependent part, and the regularisation freedom into account. In view of the findings of this and the following section, we personally feel that the term 'backreaction' is somehow misleading, as it suggests that the effects of quantum matter-energy are secondary. However, we shall find that one can indeed model the full recent cosmic history by purely quantum effects.

### IV.5.1   The Energy Density of Quantum Fields

In order to analyse the solutions of the semiclassical Einstein equation in the flat FLRW case, let us assume that we consider the expectation value of the stress energy tensor in a homogeneous and isotropic Hadamard state like the ones constructed in the subsections III.2.2 and III.4.2. In this case, the symmetry of the state implies that $\omega(:T_{\mu\nu}(x):)$ has the form of a stress-energy tensor of a perfect fluid, $i.e.$

$$\omega(:T^\mu_{\ \nu}:) = \begin{pmatrix} -\varrho_Q & \\ & p_Q \mathbb{I}_3 \end{pmatrix}$$

characterised by a quantum energy density $\varrho_Q$ and a quantum pressure $p_Q$. We have not computed these quantities up to now explicitly, but let us recall that on the one hand we know very well the trace of $\omega(:T_{\mu\nu}(x):)$ on general spacetimes and that on the other hand we know that $\omega(:T_{\mu\nu}(x):)$ is covariantly conserved. Out of these data, we can directly compute $\varrho_Q$ and $p_Q$. Namely, let us recall that covariant conservation in the flat FLRW case entails

$$\dot{\varrho}_Q + 3H(\varrho_Q + p_Q) = 0,$$

where

$$H \doteq \frac{\dot{a}}{a}$$

is the *Hubble function*. On the other hand, we know that

$$T_Q \doteq g^{\mu\nu}\omega(:T_{\mu\nu}:) = -\varrho_Q + 3p_Q.$$

Combining these equations, we find

$$\frac{\dot{\varrho}_Q}{H} + 4\varrho_Q = -T_Q. \tag{IV.119}$$

This is an ordinary differential equation, we can therefore determine $\varrho_Q$ out of $T_Q$ once an initial condition is given. Let us recall that the stress-energy tensor of classical radiation has vanishing trace because $\varrho_{\mathrm{rad}} = \frac{1}{3}p_{\mathrm{rad}}$, see subsection I.3.1. Hence, any 'radiation component' of the quantum stress-energy tensor would be invisible in (IV.119). This entails than an initial condition can be given by specifying a radiation component of the quantum stress-energy tensor. In more mathematical terms, once a particular solution $\varrho_Q^0$ of (IV.119) is known, the general solution must be of the form

$$\varrho_Q = \varrho_Q^0 + \frac{c}{a^4},$$

where $c$ is some constant because the term $ca^{-4}$, as already discussed in subsection I.3.1, is the general solution of the homogeneous version of (IV.119). Let us stress that $c$ is not a free parameter, on the contrary, it is fixed to a certain value, once the state $\omega$ is specified. Given the quantum energy density $\varrho_Q$, we can solve the semiclassical Einstein equation by solving the first Friedmann equation

$$3H^2 = 8\pi G \varrho_Q$$

as this equation in combination with the covariant conservation is equivalent to the full (semiclassical) Einstein equation. Let us now consider the quantum trace $T_Q$. Our computations of $T_Q$ in the Klein-Gordon and Dirac case in the previous subsections have revealed

$$
\begin{aligned}
T_Q^{\text{KG}} = {} & \frac{1}{2880\pi^2}\left(\frac{5}{2}(6\xi-1)R^2 + 6(1-5\xi)\Box R + C_{\alpha\beta\gamma\delta}C^{\alpha\beta\gamma\delta} + R_{\alpha\beta}R^{\alpha\beta} - \frac{R^2}{3}\right) \\
& + \frac{1}{4\pi^2}\left(\frac{m^4}{8} + \frac{(6\xi-1)m^2R}{24}\right) - \left(3\left(\frac{1}{6}-\xi\right)\Box + m^2\right)\omega\left(:\phi^2:\right), \quad\quad \text{(IV.120)} \\
T_Q^{\text{Di}} = {} & \frac{1}{2880\pi^2}\left(\frac{7}{2}C_{\mu\nu\rho\tau}C^{\mu\nu\rho\tau} + 11\left(R_{\mu\nu}R^{\mu\nu} - \frac{1}{3}R^2\right) - 6\Box R\right) \\
& - \frac{1}{\pi^2}\left(\frac{m^4}{8} + \frac{m^2R}{48}\right) - m\omega\left(:\psi^\dagger\psi:\right),
\end{aligned}
$$

where we have replaced the state dependent terms by expectation values of Wick squares for simplicity, a procedure which is legitimate in view of the regularisation freedom of both the Wick squares and the stress-energy tensor. Namely, let us recall that the finite renormalisation freedom of $:T_{\mu\nu}:$ as discussed in section IV.1 consists of adding arbitrary multiples of the conserved curvature tensors $m^4 g_{\mu\nu}$, $m^2 G_{\mu\nu}$, $I_{\mu\nu}$, and $J_{\mu\nu}$. On the level of traces, this entails that $T_Q$ is determined up to multiples of $m^4$, $m^2R$, and $\Box R$. We have already discussed that $:\phi^2:$ is determined up to multiples of $m^2$ and $R$. Admittedly, we have not discussed the renormalisation freedom of Fermionic Wick polynomials in great detail, but as remarked at several occasions, it can be expected that the an analysis of this subject in the spirit of [HoWa01] would yield results similar to the scalar case for observable quantities. Particularly, one would certainly find that $:\psi^\dagger\psi:$ is determined up to $m^3$ and $mR$ terms. Altogether, we find that the renormalisation freedom of the full $T_Q$ and its state dependent contributions can be treated in a unified way.

Having discussed the renormalisation freedom, let us simplify notation by omitting all terms in $T_Q$ subject to renormalisation freedom in the following. We shall collectively abbreviate them by $T_{RF}$ as long as the discussion of the remaining terms lasts. Hence, from now on,

$$T_{RF} \doteq \alpha m^4 + \beta m^2 R + \gamma \Box R$$

with immaterial constants $\alpha$, $\beta$, and $\gamma$. With this in mind, let us consider (IV.120) in the case of peculiar interest to us, namely, for a flat FLRW spacetime, and, on the scalar field side, with conformal coupling to scalar curvature. Namely, if we consider that the Weyl tensor, being conformally invariant, vanishes in all spacetimes which are conformally related to Minkowski spacetime, if we insert $\xi = 1/6$, and if we express the remaining curvature quantities in term of $H$, (IV.120) becomes

$$T_Q^{\mathrm{KG}} = -\frac{1}{240\pi^2}\left(H^4 + \dot{H}H^2\right) - m^2\omega\left(:\phi^2:\right) + T_{RF},\qquad\text{(IV.121)}$$

$$T_Q^{\mathrm{Di}} = -\frac{11}{240\pi^2}\left(H^4 + \dot{H}H^2\right) - m\,\omega\left(:\psi^\dagger\psi:\right) + T_{RF}.$$

In this simple form, the universal structure of $T_Q$ for free quantum fields of any spin becomes visible. One the one hand, the trace anomaly differs among fields of various spins only by dimensionless coefficients, see [BiDa82, p.179] for a table of coefficients for spins higher than $\frac{1}{2}$ obtained by DeWitt-Schwinger computations. On the other hand, we have the universal renormalisation freedom. Finally, the state-dependent terms are contributing only for massive fields (with the due exception of a non-conformally coupled scalar field). In fact, these very terms are the tricky point in the business of solving the semiclassical Einstein equation, and we shall have to devote a full subsection to their computation.

### IV.5.2 Computation of the State-Dependent Contributions

To be able to compute $\omega(:\phi^2:)$ and $\omega(:\psi^\dagger\psi:)$ explicitly, we need to express them in terms of regularised mode integrals. As shown in [DFP08, Pin10], in case of the conformally coupled scalar field and for pure states, the result is

$$\omega\left(:\phi^2:\right) = \lim_{\vec{y}\to\vec{x}}\frac{1}{a^2(2\pi)^3}\int_{\mathbb{R}^3} d\vec{k}\left(|T_k|^2 - \frac{1}{2\sqrt{k^2 + a^2 m^2}}\right)e^{i\vec{k}(\vec{x}-\vec{y})} + \frac{T_{RF}}{m^2},\qquad\text{(IV.122)}$$

where $T_k$ are the normalised modes of the Klein-Gordon field discussed in subsection II.2.1. While the $T_k$-term is a trivial consequence of the mode expansion of states discussed in subsection III.2.2, the form of the subtracted term can be obtained as follows [Pin10].

Initially and up to the renormalisation freedom,

$$\omega(:\phi^2(x):) \doteq \lim_{y\to x}(\omega_2(x,y) - h(x,y)),\qquad\text{(IV.123)}$$

see the discussion at the end of subsection III.1.3. In the above expression, $x$ and $y$ are thought to be sufficiently close to each other for the local Hadamard parametrix to be well-defined, *cf.* definition III.1.2.5. However, as $\omega_2 - h$ is a smooth function, we are free to choose $y$ and the path along which we approach $x$. Particularly, we can choose $y = (\tau, \vec{y})$ if $x = (\tau, \vec{x})$. The advantage of this is that the half square geodesic distance between such $x$ and $y$ has the simple form

$$\sigma(x,y) = \frac{1}{2}a^2(\tau)(\vec{x} - \vec{y})^2 \doteq a^2(\tau)\sigma_{\mathbb{M}}(x,y),$$

where $\sigma_{\mathbb{M}}(x,y)$ denotes the half square geodesic distance in Minkowski spacetime. To arrive at (IV.122), we need to be able to write $h(t,\vec{x},t,\vec{y})$ in terms of a Fourier integral. To this avail, we note a simple but valueable fact: we have a good knowledge of the two-point function, and, hence, the local Hadamard parametrix of Minkowski spacetime and the corresponding Fourier integral. Namely, for a massive scalar field in the vacuum state one has (see for instance [Mor03])

$$
\begin{aligned}
\omega_2^{\mathbb{M}}(x,y) &= \frac{1}{8\pi^2}\left(\frac{1}{\sigma_{\mathbb{M}}(x,y)} + \frac{m^2}{2}\log\left(\frac{\sigma_{\mathbb{M}}(x,y)}{\lambda^2}\right)\right) - \frac{m^2}{16\pi^2} + \frac{m^2\log(m\lambda)}{8\pi^2} + \mathrm{O}(\sigma_{\mathbb{M}}) \\
&= \frac{1}{8\pi^2}h_{\mathbb{M}}(x,y) - \frac{m^2}{16\pi^2} + \frac{m^2\log(m\lambda)}{8\pi^2} \\
&= \frac{1}{(2\pi)^3}\int_{\mathbb{R}^3} d\vec{k}\,\frac{1}{2\sqrt{k^2+m^2}}\,e^{-i\sqrt{k^2+m^2}(\tau_x-\tau_y)}\,e^{i\vec{k}(\vec{x}-\vec{y})},
\end{aligned}
\tag{IV.124}
$$

where we have omitted the necessary $\varepsilon$-prescription. Moreover, $\lambda$ is the usual scale present in the Hadamard parametrix and the $\mathrm{O}(\sigma_{\mathbb{M}})$ terms are irrelevant since they vanish in the the limit $x \to y$. On the other hand, we know that a massive scalar field in a flat FLRW spacetime corresponds to the conformal transformation of a scalar field in Minkowski spacetime with the time-dependent mass $a(\tau)m$, see subsection II.2.1. As conformal transformations do not affect the wave front set of a distribution and transform causal propagators into causal propagators [Pin09], it follows that the Hadamard form $h(x,y)$ corresponding to the Klein-Gordon operator

$$
P = -\Box + \frac{R}{6} + m^2
$$

in a flat FLRW spacetime and the conformally rescaled Hadamard form $a^{-1}(\tau_x)h_{\mathbb{M},am}(x,y)a^{-1}(\tau_y)$ corresponding to the Klein-Gordon operator

$$
P_{\mathbb{M}} = -\Box_{\mathbb{M}} + a^2 m^2
$$

in Minkowski spacetime must be equal up to a smooth piece! Regarding the Minkowskian quantities, replacing $m$ with $am$ in (IV.124) yields

$$
\begin{aligned}
\frac{1}{8\pi^2}h_{\mathbb{M},am}(x,y) &= \frac{1}{8\pi^2}\left(\frac{1}{\sigma_{\mathbb{M}}(x,y)} + \frac{a^2 m^2}{2}\log\left(\frac{\sigma_{\mathbb{M}}(x,y)}{\lambda^2}\right)\right) + \mathrm{O}(\sigma_{\mathbb{M}}) \\
&= \frac{1}{(2\pi)^3}\int_{\mathbb{R}^3} d\vec{k}\,\frac{1}{2\sqrt{k^2+a^2m^2}}\,e^{i\sqrt{k^2+a^2m^2}(\tau_x-\tau_y)}\,e^{i\vec{k}(\vec{x}-\vec{y})} + \frac{a^2 m^2}{16\pi^2} - \frac{a^2 m^2\log(\frac{am}{\lambda})}{8\pi^2}.
\end{aligned}
\tag{IV.125}
$$

Moreover, by our knowledge of the Hadamard coefficients $u$ and $v$ appearing in $h$, a cumbersome computation indeed reveals [Pin10]

$$
\lim_{\vec{y}\to\vec{x}}\left(h(\tau,\vec{x},\tau,\vec{y}) - \frac{h_{\mathbb{M},am}(\tau,\vec{x},\tau,\vec{y})}{a^2(\tau)}\right) = m^2\log a + \frac{T_{RF}}{m^2},
\tag{IV.126}
$$

Combining (IV.123) with (IV.125) and (IV.126) trivially leads to (IV.122). However, we stress once more that the exact values of $\alpha$ and $\beta$ are immaterial on account of the renormalisation freedom. The above considerations imply that an evaluation of the Wick square in the asymptotic conformal equilibrium states $\omega_\beta^B$ introduced in subsection III.2.2 corresponds to the Fourier integral

$$\omega_\beta^B(:\phi^2:) = \frac{1}{a^2(2\pi)^3} \int_{\mathbb{R}^3} d\vec{k} \left( \frac{e^{\beta k}+1}{e^{\beta k}-1} |T_k|^2 - \frac{1}{2\sqrt{k^2+a^2m^2}} \right) + \frac{T_{RF}}{m^2}. \tag{IV.127}$$

Following the above ideas, we would like to obtain a similar result in the Dirac case. To this avail, we first recall that, up to the renormalisation freedom,

$$\omega(:[\psi^\dagger\psi](x):) \doteq Tr \lim_{y\to x} \left( \omega^-(x,y) + D_y^* H(x,y) \right), \tag{IV.128}$$

see subsection III.3.3. Considering the simplified situation $y = (\tau, \vec{y})$, $x = (\tau, \vec{x})$ once more, proposition III.4.2.5 entails that the relevant Fourier integral for $\omega^-$ in the case of the asymptotic conformal equilibrium state introduced in subsection III.4.2 is

$$Tr \, \mathfrak{w}_\beta^-(\tau, \vec{x}, \tau, \vec{y}) = \frac{1}{(2\pi a(\tau))^3} \int_{\mathbb{R}^3} d\vec{k} \, \frac{e^{\beta k}-1}{e^{\beta k}+1} 2 \left( 2k^2|\mathfrak{T}_k(\tau)|^2 - 1 \right) e^{i\vec{k}(\vec{x}-\vec{y})}. \tag{IV.129}$$

Here, $\mathfrak{T}_k$ are the normalised Dirac modes introduced in subsection II.4.1 and we omit the necessary $\varepsilon$-regularisation prescription for simplicity. To obtain a Fourier expression for the Hadamard subtraction term $-D_y^* H$, we again exploit the conformal relation between flat FLRW spacetimes and Minkowski spacetime. In the latter case and for a Dirac field with mass $m$, the vacuum two-point function $\omega_{\mathbb{M}}^-$ is given by $-(\slashed{\partial}+m)\omega_2^{\mathbb{M}}$, see for instance [Sch95]. Upon taking the trace over spinor indices, the $\slashed{\partial}$ term vanishes. Hence, employing (IV.124), we find

$$\begin{aligned} Tr \, \omega_{\mathbb{M}}^-(x,y) &= -4m \left\{ \frac{1}{8\pi^2} \left( \frac{1}{\sigma_{\mathbb{M}}(x,y)} + \frac{m^2}{2} \log\left( \frac{\sigma_{\mathbb{M}}(x,y)}{\lambda^2} \right) \right) - \frac{m^2}{16\pi^2} + \frac{m^2\log(m\lambda)}{8\pi^2} + O(\sigma_{\mathbb{M}}) \right\} \\ &= -\frac{1}{8\pi^2} Tr \, (\slashed{\partial}+m) h_{\mathbb{M}}(x,y) + \frac{m^3}{4\pi^2} - \frac{m^3\log(m\lambda)}{2\pi^2} \\ &= -\frac{4m}{(2\pi)^3} \int_{\mathbb{R}^3} d\vec{k} \, \frac{1}{\sqrt{k^2+m^2}} e^{i\sqrt{k^2+m^2}(\tau_x-\tau_y)} e^{i\vec{k}(\vec{x}-\vec{y})}, \end{aligned} \tag{IV.130}$$

With the scalar case in mind, we would now like to replace all occurring $m$ in the above equation with $am$ and interpret the resulting quantities as corresponding to a Dirac field with mass $m$ in a flat FLRW spacetime. However, such a replacement is not sufficient in the Dirac case as the 'squared Dirac operator' $\mathfrak{P} = D^*D$ is *not* conformally related to the Minkowskian Klein-Gordon operator with mass $am$, see the discussion in subsection II.4.1. As a result, $h_{\mathbb{M}}$ is not conformally related to $H$. A quick way to see the failure of the mentioned conformal relation and to fix this problem is to notice that the two Hadamard parametrices $h$ and $H$, and, hence, $h_{\mathbb{M}}$ and $H$, differ considerably in their coefficient of the logarithmic singularity. The reason for this is the fact that, although the coinciding point limits of the relevant scalar and Dirac Hadamard coefficients $[v]$ and $[V]$ respectively share the same structure, namely

$$[v] = [V] = \frac{1}{2}\left(m^2 + \left(\xi - \frac{1}{6}\right)R\right),$$

the one of the conformally coupled scalar field does not contain a scalar curvature term because $\xi = \frac{1}{6}$ in this case. In contrast, the coupling to the scalar curvature present in $\mathfrak{P} = D^*D$ is $\xi = \frac{1}{4}$. Recalling that the mass term *inside* the Fourier integral in (IV.130) is responsible for the coefficient of the logarithmic singularity, a natural *ansatz* for the Fourier expression of the Hadamard subtraction term $-D_y^* H$ is the following 'mass replacement': one takes the Fourier integral in (IV.130), replaces the mass term $m$ *outside* of the integral with $am$, and the mass term $m$ *inside* the integral with

$$a\widetilde{m} \doteq a\sqrt{m^2 + \frac{R}{12}}.$$

Hence, one arrives at

$$-\frac{1}{8\pi^2}Tr\left(\slashed{\partial} + am\right)h_{\mathbb{M},a\widetilde{m}}(x,y) =$$

$$= -4am\left\{\frac{1}{8\pi^2}\left(\frac{1}{\sigma_{\mathbb{M}}(x,y)} + \frac{a^2\widetilde{m}^2}{2}\log\left(\frac{\sigma_{\mathbb{M}}(x,y)}{\lambda^2}\right)\right) + O(\sigma_{\mathbb{M}})\right\} \qquad \text{(IV.131)}$$

$$= -\frac{4am}{(2\pi)^3}\int_{\mathbb{R}^3} d\vec{k}\, \frac{1}{2\sqrt{k^2+a^2\widetilde{m}^2}}\, e^{i\sqrt{k^2+a^2\widetilde{m}^2}(\tau_x-\tau_y)}\, e^{i\vec{k}(\vec{x}-\vec{y})} - \frac{a^3m\widetilde{m}^2}{4\pi^2} + \frac{a^3m\widetilde{m}^2\log(\frac{a\widetilde{m}}{\lambda})}{2\pi^2}.$$

The above discussion implies that the conformal rescaling of $Tr\left(\slashed{\partial} + am\right)h_{\mathbb{M},a\widetilde{m}}$ has the potential to differ from $Tr\, D_y^* H$ only by a smooth term. In fact, a computation similar to the one in the scalar case shows that this is the case and that

$$\lim_{\vec{y}\to\vec{x}}\left(Tr\, D_y^* H(\tau,\vec{x},\tau,\vec{y}) - \frac{Tr\left(\slashed{\partial} + a(\tau)m\right)h_{\mathbb{M},a\widetilde{m}}}{a^3(\tau)}\right) = 4m\widetilde{m}^2\log a + \frac{T_{RF}}{m}. \qquad \text{(IV.132)}$$

It is time to reap the rewards of our rather cumbersome considerations and to state the resulting Fourier integral expression for the expectation value of $:\psi^\dagger\psi:$. Namely, combining (IV.128) with (IV.131) and (IV.132) yields

$$\mathfrak{w}_\beta^B(:\psi^\dagger\psi:) = \frac{1}{(2\pi a)^3}\int_{\mathbb{R}^3} d\vec{k}\, \frac{e^{\beta k}-1}{e^{\beta k}+1}\, 2\left(2k^2|\mathfrak{T}_k|^2 - 1\right) + \frac{2am}{\sqrt{k^2+a^2\left(m^2+\frac{R}{12}\right)}} \qquad \text{(IV.133)}$$

$$- \frac{m\left(m^2+\frac{R}{12}\right)\log\left(1+\frac{R}{12m^2}\right)}{4\pi^2} + \frac{T_{RF}}{m}.$$

The last and most difficult input we need for the computation of the expectation values of $:\phi^2:$ and $:\psi^\dagger\psi:$ are explicit expressions for the modes $T_k$ and $\mathfrak{T}_k$. Let us recall what we know about these modes. One the one hand, we know that are subject to ordinary differential equations in the conformal time $\tau$, *viz.*

$$(\partial_\tau^2 + k^2 + a^2 m^2) T_k = 0, \qquad (\partial_\tau^2 + k^2 + a^2 m^2 + i a' m) \mathfrak{T}_k = 0,$$

on the other hand they fulfil suitable normalisation conditions, namely,

$$T_k(\tau) \partial_\tau \overline{T_k(\tau)} - \overline{T_k(\tau)} \partial_\tau T_k(\tau) \equiv i, \qquad |(\partial_\tau + iam)\mathfrak{T}_k|^2 + k^2 |\mathfrak{T}_k|^2 \equiv 1.$$

Moreover, we have specified the asymptotic thermal equilibrium states $\omega_\beta^B$ and $\mathfrak{w}_\beta^B$ by the requirements that the modes are asymptotic, massless, positive frequency modes, *i.e.*

$$\lim_{\tau \to -\infty} e^{ik\tau} T_k = \frac{1}{\sqrt{2k}}, \qquad \lim_{\tau \to -\infty} e^{ik\tau} \partial_\tau T_k = -i\sqrt{\frac{k}{2}},$$

$$\lim_{\tau \to -\infty} e^{ik\tau} \mathfrak{T}_k = \frac{1}{\sqrt{2k}}, \qquad \lim_{\tau \to -\infty} e^{ik\tau} \partial_\tau \mathfrak{T}_k = -\frac{i}{\sqrt{2}}.$$

Nevertheless, these data do not allow us to provide a closed expression for $T_k$ and $\mathfrak{T}_k$. Hence, to be able to obtain any solution of the semiclassical Einstein equation, we have to employ some approximations. In the following, we are interested in matching the solutions of the semiclassical Einstein equation to experimental data in order to model the recent cosmic history. Hence, it is legitimate to choose approximations which are good exactly in this regime. Let us state these in the following.

- Observations imply that the present Hubble rate $H$ is non-vanishing, but that its (cosmological) time derivative $\dot{H}$ and higher order time derivatives are small. Hence, we shall discard derivatives of $H$ in $\varrho_Q$ and thus compute $T_Q$ only up to terms linear in $\dot{H}$.

- The measured Hubble rate is of the order of $10^{-33}$eV [Amm10]. This is many orders larger than any reasonable non-zero mass value one would impose for a quantum field. Hence, we shall discard any $\frac{H}{m}$ terms in $\varrho_Q$. This amounts to omitting $O(H^5)$ and $O(H^3\dot{H})$ terms in $T_Q$.

- Even with the above-mentioned approximations at hand, one finds that, due to the $e^{\beta k}$ factors, the Fourier integrals in the state-dependent terms are impossible to compute analytically. Hence, we shall assume that $\beta m = \frac{m}{T} \ll 1$ and only compute the lowest order $\beta$-dependent terms. On the physical side, this approximation is motivated by the fact that the only temperature scale that appears in the recent cosmological history of the universe, namely, the CMB temperature, if of the order of 1K which corresponds to $10^{-4}$eV.

Note that the approximations regarding the curvature terms are close in spirit to the one employed in [DFP08], although the authors there have not computed the state-dependent contributions to $\varrho_Q$ of order $H^4$. As we would like to compute these terms as well, we choose a different procedure to approximate the mode functions since the approach of [DFP08] would not give an unambiguous answer regarding the $H^4$ terms. To this end, we first note that, for our purposes, we do not need $T_k$ and $\mathfrak{T}_k$, but only their absolute squares. Hence, we employ the differential equations and normalisation conditions fulfilled by the modes themselves to obtain the following differential equations and initial conditions for their absolute squares

$$\partial_\tau^3|T_k|^2 + 4(k^2 + a^2 m^2)\partial_\tau|T_k|^2 + 2a^2 m^2 H|T_k|^2 = 0,$$

$$\lim_{\tau\to-\infty}|T_k|^2 = \frac{1}{2k}, \qquad \lim_{\tau\to-\infty}\partial_\tau|T_k|^2 = \lim_{\tau\to-\infty}\partial_\tau^2|T_k|^2 = 0,$$

$$\partial_\tau^3|\mathfrak{T}_k|^2 + aH\partial_\tau^2|\mathfrak{T}_k|^2 + 4(k^2 + a^2 m^2)\partial_\tau|\mathfrak{T}_k|^2 + 2a^2 m^2 H|\mathfrak{T}_k|^2 - 2aH = 0,$$

$$\lim_{\tau\to-\infty}|\mathfrak{T}_k|^2 = \frac{1}{2k^2}, \qquad \lim_{\tau\to-\infty}\partial_\tau|\mathfrak{T}_k|^2 = \lim_{\tau\to-\infty}\partial_\tau^2|\mathfrak{T}_k|^2 = 0.$$

However, these differential equations have the same flaw as their progenitors, namely, they can not be solved analytically. To find solutions which are approximate in the sense discussed above, we make the *ansatz* that both $|T_k|^2$ and $|\mathfrak{T}_k|^2$ are analytic in $H$ and its time derivatives, *i.e.*

$$|T_k|^2 \doteq g_0(a,k) + \sum_{n=0}^{\infty}\sum_{i_0=0}^{\infty}\cdots\sum_{i_n=0}^{\infty}\left(1 - \delta_{0,\sum_n i_n}\right)g_{i_0,\cdots,i_n}(a,k)\prod_{j=0}^{n}\left(\partial_t^j H\right)^{i_j},$$

$$|\mathfrak{T}_k|^2 \doteq G_0(a,k) + \sum_{n=0}^{\infty}\sum_{i_0=0}^{\infty}\cdots\sum_{i_n=0}^{\infty}\left(1 - \delta_{0,\sum_n i_n}\right)G_{i_0,\cdots,i_n}(a,k)\prod_{j=0}^{n}\left(\partial_t^j H\right)^{i_j}.$$

We stress that, *a priori*, our *ansatz* is solely motivated by the approximation we would like to employ and we do not know if $|T_k|^2$ and $|\mathfrak{T}_k|^2$ really fulfil these analyticity properties. Particularly, it is not clear if this *ansatz* is compatible with the initial conditions for $|T_k|^2$ and $|\mathfrak{T}_k|^2$ in the limit $\tau \to -\infty$. However, we will find that it will be possible to obtain unique solutions for the coefficients once we impose the initial conditions. Finally, one may recognise similarities of our approach with the often employed *adiabatic regularisation scheme*, see for instance [BiDa82]. However, whereas in the adiabatic regularisation approach an *ansatz* related to ours is used to compute the divergent *subtraction* terms in the definition of $\omega(:\phi^2:)$ and other Wick polynomials, we use the low curvature expansion to compute the to-be-regularised original terms which we consider to be made finite by a Hadamard subtraction. Even so, in case of $\omega(:\phi^2:)$, the expression we find for the Hadamard subtraction in Fourier space happens to coincide with the one employed in the adiabatic regularisation scheme, *cf.* [BiDa82].

Recalling which powers and derivatives of $H$ we consider to be relevant in our approximation, it is sufficient to compute the expansion coefficients in the above two power series up to $g_{4,0}$, $g_{2,1}$ and $G_{4,0}$, $G_{2,1}$ respectively. These coefficients can be computed by means of a recursive procedure. Namely, by the very nature of our series expansions, inserting them into the relevant differential equations and collecting terms containing equal powers of $H$ and its derivatives, it turns out that the series coefficients satisfy a first order, inhomogenous ordinary differential equation, *viz*

$$\frac{k^2 + a^2 m^2}{a^2 m^2 H} \dot{g}_{i_0, \cdots, i_n} + g_{i_0, \cdots, i_n} = s_{i_0, \cdots, i_n},$$

$$\frac{k^2 + a^2 m^2}{a^2 m^2 H} \dot{G}_{i_0, \cdots, i_n} + G_{i_0, \cdots, i_n} = S_{i_0, \cdots, i_n},$$

where $s_{i_0, \cdots, i_n}$ and $S_{i_0, \cdots, i_n}$ are source terms which are determined by coefficients of lower order. Hence, all coefficients are uniquely determined up to a solution of the corresponding and universal homogeneous differential equation, which is of the form

$$\frac{C}{\sqrt{k^2 + a^2 m^2}}.$$

Here, $C$ is a dimensionful constant with *zero or negative* mass dimension; the exact value of the mass dimension depends on the coefficient order, it becomes more negative with increasing order. Let us consider the scalar case first. One can compute that the source term $s_0$ is vanishing, hence, $g_0$ is given by the unique solution of the homogeneous equation which fulfils the initial conditions for $|T_k|^2$ towards $t \to 0$, namely,

$$g_0 = \frac{1}{2\sqrt{k^2 + a^2 m^2}}.$$

Considering $g_1$, it turns out that its source term $s_1$ is vanishing as well. The only solution of the homogeneous equation which is compatible with the initial conditions in this case is $g_1 \equiv 0$, because the results of lemma I.3.2.4 imply that $H$ is diverging towards the Big Bang. In fact, the same lemma implies that *all* derivatives of $H$ are diverging in the limit $\tau \to -\infty$, hence, for all higher coefficient orders, the only solution to the equation compatible with the initial conditions is the trivial one. As already anticipated, we can thus obtain unique solutions for all coefficients. However, it is by no means clear that the unique solutions we find are really compatible with the initial conditions. Moreover, the necessary computations are quite involved and barely manageable without the usage of computer algebra systems. Notwithstanding, the result we find for $|T_k|^2$ up to the relevant order is

$$|T_k|^2 = \frac{1}{2\sqrt{k^2 + a^2 m^2}} + H^2 \left( \frac{3 a^4 m^2}{8(k^2 + a^2 m^2)^{\frac{5}{2}}} - \frac{5 a^6 m^4}{16(k^2 + a^2 m^2)^{\frac{7}{2}}} \right)$$

$$+ H^4 \left( \frac{1155 a^{12} m^8}{256(k^2 + a^2 m^2)^{\frac{13}{2}}} - \frac{693 a^{10} m^6}{64(k^2 + a^2 m^2)^{\frac{11}{2}}} + \frac{525 a^8 m^4}{64(k^2 + a^2 m^2)^{\frac{9}{2}}} - \frac{15 a^6 m^2}{8(k^2 + a^2 m^2)^{\frac{7}{2}}} \right)$$

$$+ \dot{H} \frac{a^4 m^2}{8(k^2 + a^2 m^2)^{\frac{5}{2}}} + \dot{H} H^2 \left( \frac{203 a^8 m^4}{32(k^2 + a^2 m^2)^{\frac{9}{2}}} - \frac{231 a^{10} m^6}{64(k^2 + a^2 m^2)^{\frac{11}{2}}} - \frac{43 a^6 m^2}{16(k^2 + a^2 m^2)^{\frac{7}{2}}} \right)$$

and one can check that all terms depending on $H$ and $\dot{H}$ vanish towards the Big Bang because they are multiplied by sufficiently many powers of $a$. Hence, apart from the fact that it is still far from clear if our series *ansatz* for $|T_k|^2$ is converging, we find that it is compatible with the initial conditions towards $\tau \to -\infty$ and that all coefficients can be determined uniquely.

Let us know turn to the Dirac case. Here, the source term $S_0$ corresponding to $G_0$ is non-vanishing, but equals $\frac{1}{2k^2}$. Moreover, already at the lowest order, the trivial solution of the homogeneous equation is the only one compatible with the initial conditions. At higher orders, the considerations we have made in the scalar case again entail that all coefficients have unique solutions, once the initial conditions are imposed. Finally, the result we find for $|\mathfrak{T}_k|^2$ up to the relevant order, *viz*

$$|\mathfrak{T}_k|^2 = \frac{1}{2\sqrt{k^2 + a^2 m^2}\left(\sqrt{k^2 + a^2 m^2} + am\right)} + H^2 \frac{4k^2 a^3 m^3 - a^5 m^5}{16m^2(k^2 + a^2 m^2)^{\frac{7}{2}}}$$

$$+ H^4 \frac{-192 k^6 a^5 m^5 + 712 k^4 a^7 m^7 - 247 k^2 a^9 m^9 + 4a^{11} m^{11}}{256 m^4 (k^2 + a^2 m^2)^{\frac{13}{2}}}$$

$$+ \dot{H} \frac{a^3 m^3}{8m^2 (k^2 + a^2 m^2)^{\frac{5}{2}}} + \dot{H} H^2 \frac{-92 k^4 a^5 m^5 + 131 k^2 a^7 m^7 - 8a^9 m^9}{64 m^4 (k^2 + a^2 m^2)^{\frac{11}{2}}}$$

indeed fulfils the initial conditions, as the coefficients of the $H$-dependent terms contain sufficiently many powers of $a$.

By now we have collected all ingredients to compute the state dependent terms present in $T_Q$. To take the low-temperature approximation into account, we split the integrals present in (IV.122) and (IV.133) into a $\beta$-dependent and a $\beta$-independent part. In the former, we rewrite the integral in terms of the dimensionless variable $y = k(am)^{-1}$. As the $e^{\beta k} = e^{am\beta y}$ factors appear in the denominator of the relevant integrands, we find that these integrands are essentially only non-vanishing for $y < (am\beta)^{-1} \ll 1$. Hence, we can obtain an expansion in $T/m$ by expanding the $\beta$-dependent integrands in $y$. Proceeding in this way, and computing the resulting integrals, we find that the state dependent terms in the approximation employed read

$$-m^2 \omega_\beta^B(:\phi^2:) = \frac{H^4 + 7\dot{H}H^2}{240\pi^2} - \frac{mT^3\zeta(3)}{a^3\pi^2} + T_{RF},$$

$$-m\,\mathfrak{w}_\beta^B(:\psi^\dagger\psi:) = \frac{-50\dot{H}m^2 + 11H^4 + 47\dot{H}H^2}{240\pi^2} - \frac{3mT^3\zeta(3)}{a^3\pi^2} + T_{RF},$$

where $\zeta$ denotes the $\zeta$-function. Moreover, to obtain the mentioned result for $\mathfrak{w}_\beta^B(:\psi^\dagger\psi:)$, we have expanded the logarithmic term present in (IV.133) up to the relevant orders of $H$ and $\dot{H}$.

### IV.5.3 Classification of the Cosmological Solutions

Inserting our results for the state-dependent terms evaluated in the asymptotic equilibrium states in (IV.121), we find

$$T_Q^{\mathrm{KG}} = \frac{\dot{H}H^2}{40\pi^2} - \frac{mT^3\zeta(3)}{a^3\pi^2} + T_{RF},$$

$$T_Q^{\mathrm{Di}} = \frac{-50\dot{H}m^2 + 36\dot{H}H^2}{240\pi^2} - \frac{3mT^3\zeta(3)}{a^3\pi^2} + T_{RF}.$$

An important observation is that the state-dependent terms have exactly cancelled the $H^4$ contributions of the trace anomaly, both in the scalar and in the spinor case. To see the relevance of this point, let us recall the approximation we have employed. Namely, we have decided to discard terms containing derivatives of $H$ in the quantum energy density $\varrho_Q$ which arises out of $T_Q$ by means of $\dot{\varrho}_Q/H + 4\varrho_Q = -T_Q$. To obtain the the contributions to $\varrho_Q$ stemming from the $\dot{H}$ terms in the above expressions in a closed form, we proceed in the following way. If we add suitable higher derivatives of $H$ to $T_Q$, we can integrate the result in a closed form to find that the $\dot{H}$ terms in $T_Q$ in combination with the added terms generate $\dot{H}$ terms in $\varrho_Q$. Hence, all $\dot{H}$ terms in $T_Q$ contribute $\dot{H}$ terms to $\varrho_Q$ up to higher derivative terms, but these are immaterial in our context anyway. Consequently, in our approximation,

$$T_Q^{\mathrm{KG}} = -\frac{mT^3\zeta(3)}{a^3\pi^2} + T_{RF},$$

$$T_Q^{\mathrm{Di}} = -\frac{3mT^3\zeta(3)}{a^3\pi^2} + T_{RF},$$

and we see that, up to potential contributions from $T_{RF}$ which we anticipate to be irrelevant for our current discussion, the traces only contain terms which look like classical matter! However, in previous works, *e.g.* [Wal78b, DFP08, Kok09], it was found that the anomalous $H^4$ term which is not present in classical matter-energy contributions significantly alters the solution structure of the Einstein equations in the flat FLRW case. Hence, we find that the contribution from massive quantum fields in our asymptotic equilibrium states cancels the only anomalous term in $T_Q$ which remains in our approximation. It seems that a similar result is at the heart of the works [PaRa99, PaRa01], though the authors there compute the state dependent contributions by means of partially resumming the DeWitt-Schwinger series, which is manifestly state-independent, see the discussion in section IV.4. In contrast, it seems that in [And85, And86], where essentially the same quantum state we consider (though at zero temperature) is used to evaluate the state dependent terms of a massive scalar field, the conclusion is drawn that the state-dependent terms *double* the anomalous $H^4$ term rather than *cancelling* it.

Motivated by the findings in the preceding paragraph and by the pragmatic fact that we do not know the non-anomalous contribution of $T_Q$ for vector fields yet, we choose the following field content for solving the semiclassical Einstein equations: we consider $N_0$ massive and conformally coupled scalar fields, $N_{1/2}$ massive Dirac fields, and $N_1$ massless vector fields. All fields are considered to be free fields, and for notational simplicity we assume that all scalars and Dirac fields have the same mass $m$. Note that the latter simplification is not essential for the interpretation of our solutions, as masses only rescale field multiplicities in the terms we consider. Moreover, we assume that the quantum state of the scalars and Dirac fields is the relevant asymptotic equilibrium state for a common temperature $T \ll m$. Considering the vector field results listed in [BiDa82, p. 179] and writing out the $T_{RF}$ terms, we thus arrive at the following total stress energy trace

$$T_Q = -\frac{31N_1}{120\pi^2}\left(H^4 + \dot{H}H^2\right) - \frac{N_0 m T^3 \zeta(3)}{a^3 \pi^2} - \frac{3N_{1/2} m T^3 \zeta(3)}{a^3 \pi^2}$$
$$+ \alpha m^4 + \beta m^2 \left(\dot{H} + 2H^2\right) + \gamma \left(\dddot{H} + 6\ddot{H}H + 4\dot{H}^2 + 12\dot{H}H^2\right).$$

As already anticipated, the terms which are subject to the finite renormalisation freedom of $T_Q$ do not contribute $H^4$ terms to $T_Q$. Let us consider these terms in more detail. The term containing the high derivatives of $H$ is unsurprisingly the $\Box R$ contribution to the renormalisation freedom. One can straightforwardly check that its contribution to the quantum energy density $\varrho_Q$ is

$$\varrho_{Q,\Box R} = \frac{1}{2}\left(2\ddot{H}H - \dot{H}^2 + 6\dot{H}H^2\right).$$

In [Wal78b, DFP08], this term has been discarded in order to fulfil Wald's fifth axiom, which requires that higher-than-second derivatives of the metric are not present in $\omega(:T_{\mu\nu}:)$. Equivalently, one could discard it in order to assure the existence of stable solutions of the semiclassical Einstein equation. However, although these equations certainly have runaway solutions once $\Box R$ terms are included, they still maintain at least one family of stable ones, see for instance [And83, Kok09]. Indeed, general procedures to consistently select the 'stable subspace' of solutions to higher-derivative equations containing the $\Box R$ exist, cf. [PaSi93, FlWa96]. In any case, we will discard these terms on the basis of the approximation we would like to employ (see also the 'quasi-de Sitter' approximation in [Kok09]).

The $\dot{H} + 2H^2$ term coming from the $m^2 R$ renormalisation freedom contributes a $H^2$ term to the quantum energy density, while the trace anomaly term leads to a $H^4$ contribution. Altogether, we find that the total energy density induced by our quantum field content is

$$\varrho_Q = \frac{31N_1}{30\pi^2}H^4 + \frac{N_0 m T^3 \zeta(3)}{a^3 \pi^2} + \frac{3N_{1/2} m T^3 \zeta(3)}{a^3 \pi^2} + \alpha m^4 + \frac{\beta}{2}H^2 + \frac{c}{a^4},$$

where $c$ is the integration constant which arises upon solving (IV.119) for $\varrho_Q$. Physically, it determines the magnitude of an effective radiation component present in the quantum energy density. As such, it is fixed once we have chosen a state for the massless vector fields. However, as we are not able to fix such a state within our analysis, we shall consider $c$ to be a free parameter in the following and we refer the reader to [And85, Pin10] for further discussions regarding this freedom.

To finally solve the semiclassical Einstein equation, we insert the found quantum energy density in the first Friedmann equation $3H^2 = 8\pi G \varrho_Q$ to obtain an algebraic equation of fourth order in $H$, namely,

$$H^4 - 2H_*^2 H^2 + \frac{C_1}{a^4} + \frac{C_2}{a^3} + C_3 = 0, \qquad (\text{IV.134})$$

where

$$H_*^2 \doteq \frac{15\pi^2 \left( \frac{\beta}{2} - \frac{3}{8\pi G} \right)}{31 N_1}, \quad C_1 \doteq \frac{30\pi^2 c}{31 N_1}, \qquad (\text{IV.135})$$

$$C_2 \doteq \frac{30\zeta(3) m T^3 \left( N_0 + 3 N_{1/2} \right)}{31 N_1}, \quad C_3 \doteq \frac{30\pi^2 \alpha m^4}{31 N_1}.$$

The algebraic equation can be solved immediately, thus leading to the following solution of the semiclassical Einstein equation

$$H_\pm^2(a) \doteq H_*^2 \pm \sqrt{H_*^4 - \frac{C_1}{a^4} - \frac{C_2}{a^3} - C_3}. \qquad (\text{IV.136})$$

Note that, in view of the Friedmann equation $3H^2 = 8\pi G \varrho$, these solutions can be regarded as constituting the *effective* quantum energy density (multiplied by $\frac{8\pi G}{3}$) which arises upon inserting the solutions for $H$ in $\varrho_Q$. Assuming that the universe is expanding forever, we observe that the solution we have found has the two asymptotes

$$H_\pm^2 \doteq H_*^2 \pm \sqrt{H_*^4 - C_3}.$$

Although cosmological time is not a good astrophysical observable, one can integrate the above solution to obtain $H(t)$. A generic plot of such solution is shown in figure IV.3. We can see the anticipated asymptotes very well. Moreover, we find that only the upper branch displays a Big Bang behaviour (it goes back until $t = 0$), whereas the lower branch becomes singular at finite time. Note that the singularity in this case is not of Big Bang type because of $a \neq 0$. This is best seen in the explicit form (IV.136) – the non-Big Bang singularities arise when $H_-^2(a)$ fails to be real because the argument of the square root becomes negative. However, it is clear that we can not 'trust' our solutions up to this point, as there $\dot{H}$ diverges and the approximation we have employed breaks down. Indeed, it is well-known that this singularity is not present if one includes higher derivatives terms, see for instance [And83, Kok09].

The phenomenon that we have just discussed in the context of the lower branch can not happen on the upper branch, at least if we require that $\dot{H} < 0$ (in accord with observations of the recent cosmic history) and that $C_2$ is suitably bounded by $C_1$. Namely, in this case, it turns out that the 'radiation component' contained in $\varrho_Q$ must have *negative energy density* ($C_3 < 0$). However, one should not assign any direct physical significance to this, as this is only an intrinsic component of $\varrho_Q$ which has the same functional behaviour as radiation, the full quantum energy density is still positive. It should therefore not be interpreted as a weird form of classical radiation.

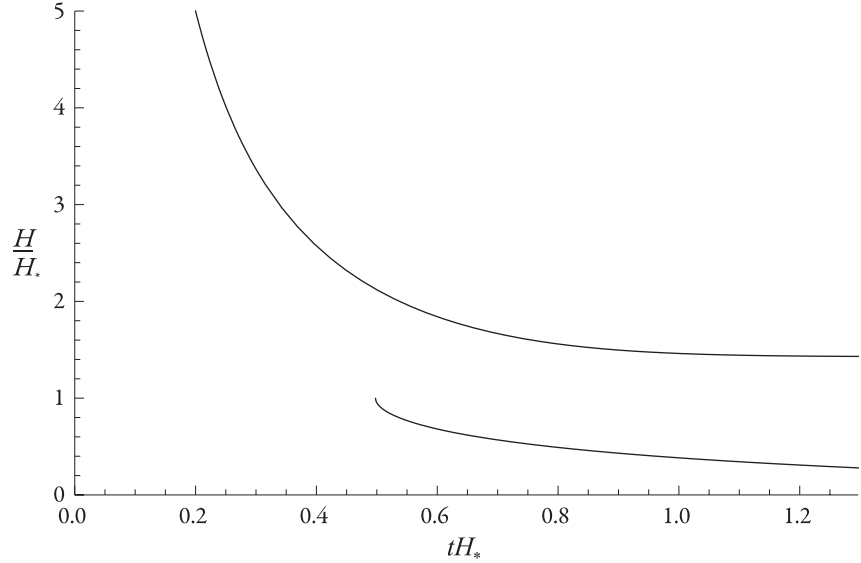

*Figure IV.3:* Generic plot of $H(t)$ obtained by integrating $H_\pm(a)$ as provided in (IV.136). $H$ is given in units of $H_*$, while $t$ is given in units of $H_*^{-1}$. The upper branch corresponds to $H_+(t)$ with $C_3 = C_2 = -C_1 = 0.1H_*^4$, the lower branch corresponds to $H_-(t)$ with $C_1 = C_2 = C_3 = 0.15H_*^4$. As observations indicate that $\dot{H} < 0$ in the recent cosmic history, we consider solutions with $\dot{H} > 0$ to be unphysical and have not plotted them.

The most important observation is that both the lower and the upper branch display the behaviour of an expansion dominated by dynamical dark energy at late times. Moreover, we can finally prove the often anticipated result, namely, that the upper branch is in fact a null Big Bang spacetime and, hence, displays a power-law inflationary behaviour.

**Proposition** IV.5.1    *The solution of the semiclassical Einstein equation given as*

$$H_+^2(a) \doteq H_*^2 + \sqrt{H_*^4 - \frac{C_1}{a^4} - \frac{C_2}{a^3} - C_3}$$

*with $C_3 < 0$ is a null Big Bang spacetime according to definition I.3.1.1. Namely, integrating $H_+(a)$ to obtain $a(t)$ from the differential equation $\dot{a}(t) = a(t)H_+(a(t))$, $a(0) = 0$, one obtains a scale factor $a(t)$ with the following properties.*

*a) There exist constants    $C_0 \in (0, \infty)$,    $t_0 > 0$    such that    $a(t) \leq C_0 t$    $\forall\, t \leq t_0$.*

*b) For all $\varepsilon \in (0, 1)$, there exist constants $C_\varepsilon \in (0, \infty)$, $t_\varepsilon > 0$ fulfilling $C_\varepsilon t^{1+\varepsilon} \leq a(t)$    $\forall\, t \leq t_\varepsilon$*

*c) All derivatives of $a$ with respect to $t$ are bounded at the origin.*

*Proof.* In order to obtain the bounds on $a(t)$, we derive the algebraic solution (IV.134) of $H(a)$ with respect to $t$. Rearranging terms, this leads us to the following differential equation.

$$-\frac{\dot{H}}{H^2} = \frac{H^2 - 2H_*^2 - \frac{C_3 + C_2(4a^3)^{-1}}{H^2}}{H^2 - H_*^2}.$$

We know that $H$ diverges at the origin, hence, we can obtain the following bounds for a suitable $t_0 > 0$, $\varepsilon > 0$, and all $t < t_0$

$$1 \leq \frac{d}{dt}\left(\frac{1}{H}\right) \leq \frac{1}{1+\varepsilon}.$$

Integrating this with the initial condition $H^{-1}(0) = 0$, we obtain

$$t \leq \frac{1}{H} \leq \frac{1}{1+\varepsilon}\, t\,, \qquad \forall\, t < t_0$$

If we now invert these estimates and insert the result into

$$a(t) = a_0\, e^{-\int_t^{t_0} H(t')\,dt'},$$

the assertion is proved with $t_0 = t_\varepsilon$.

To show the boundedness of $t$-derivatives of $a$, we first consider

$$\dot{a} = aH_+ = H_*\sqrt{a^2 + \sqrt{a^4\left(1 + \frac{C_3}{H_*^4}\right) - \frac{C_1}{H_*^4} - \frac{aC_2}{H_*^4}}}$$

and observe that it is manifestly bounded at $a = 0$. Regarding higher derivatives of $a$ we compute

$$\partial_t^{n+1} a = \partial_t^n aH_+ = (aH_+\partial_a)^n aH_+$$

and find that boundedness of all $t$-derivatives of $a$ follows from boundedness of all $a$-derivatives of $aH_+$. However, the latter is manifestly assured, as $aH_+$ is on account of $C_1 < 0$ a smooth function of $a$ in an open neighbourhood of $a = 0$. □

This result has already been observed and employed in [Pin10] and it shows that the constructions we have performed on NBB spacetimes have been sensible. Of course the approximation we have employed breaks down when $H$ and its derivatives becomes large, *i.e.* close to the Big Bang. However, numerical analyses generically yield the possibility to have power-law inflationary solutions also with higher derivative terms included, see [And83, And85]. Moreover, in [Pin10] it has been shown that, taking the full state dependence into account, though still discarding $\Box R$ terms to fulfil Wald's fifth axiom, analytic solutions which constitute null Big Bang spacetimes can be found. For a recent treatment which considers the full state dependence for scalar fields with not necessarily conformal coupling to curvature, see [ElGo10]. Finally, let us point out that the asymptotic equilibrium states are *a priori* not well-defined on the lower branch solutions. However, as already remarked, one can avoid the non-Big Bang singularity on the lower branch by going beyond the low-curvature approximation we have employed [And83, And85], and we suspect that the full analysis of the asymptotically defined states can be extended to this more general case. Moreover, in computing the $T$-dependent contribution of the state, we have used that $T \ll am$ is equivalent to $T \ll m$. This equivalence certainly fails in the limit to the Big Bang. In fact, as discussed in the subsections III.2.2 and III.4.2, we expect that any $T$-dependent contribution in $\omega_\beta^B(:\phi^2:)$ is proportional to $T^2$ close to the Big Bang, while the same contribution in $\mathfrak{w}_\beta^B(:\psi^\dagger\psi:)$ is vanishing in this regime.

Let us comment on the relation of the solutions found in this subsection to previous works on the subject. The first work analysing solutions of the semiclassical Einstein equation in the cosmological scenario has been [Sta80], see also [Vil85]. In fact, the solution found in [Sta80] constitutes the first inflationary model ever suggested [Vil85]. In more detail, in [Sta80], the analysis we have performed here is done in the case of a conformally invariant field without taking the renormalisation freedom or state-dependent terms into account. Hence, only the trace anomaly enters $\varrho_Q$, and the resulting solution found corresponds to the upper branch, though being unstable and 'decaying' into the lower branch on account of the presence of the $\Box R$ term. The same analysis has been performed in [Wal78b], though with the $\Box R$ term removed by means of the regularisation freedom. Both the upper and the lower branch of the solution are discussed, but without taking the regularisation freedom in terms of the field masses times $G_{\mu\nu}$ into account, the upper branch of the solution turns out to give a $H$ which is very large, in fact, at Planck scales. To see this, note that $H_*$ defined as in (IV.135) is essentially the Planck. Hence, in [Wal78b], the lower branch is taken as the sole physically sensible solution, and it is argued that the singularity at finite $a$ stems from the breakdown of the semiclassical approximation. Similarly, in the more recent work [Kok09], the semiclassical Einstein equation provided by considering only the trace anomaly is analysed and all possible prefactors of $\Box R$ in $\varrho_Q$ are taken into account. Generalising the results of [Sta80], it is found that for a negative coefficient of $\Box R$ in $\varrho_Q$, the lower branch is stable but shows underdamped oscillations, while the upper branch is unstable in that case. In contrast, a positive coefficient of $\Box R$ in $\varrho_Q$ yields a stable upper branch and an unstable lower branch. However, as in in [Wal78b], it is argued that the upper branch is always at Planck scale and, hence, unphysical. Confirming the results of [DFP08], [Kok09] find that the case of vanishing prefactor of $\Box R$ in $\varrho_Q$ displays two stable asymptotes. Quite generally one can safely claim that most works which appeared after [Sta80], for instance [And83, And84, Vil85, KoPr08, Kok09, NoOd, NOT05, Sha03], have only considered the cosmological effects of conformally coupled quantum fields. Hence, only the trace anomaly and the indeterminacy of the $\Box R$ term have been taken into account in these works, while all $m$-dependent finite regularisation and all state-dependent terms have been neglected.

Admittedly, there have been attempts to go beyond the trace anomaly and to consider the cosmological effects of quantum fields with non-vanishing mass. Indeed, massive scalar fields have been considered in [And85, And86] and extensive numerical analyses of the solutions have been performed, however, without considering the full renormalisation freedom of $\omega$(: $T_{\mu\nu}$:) and without comparing the solutions with (at that time unavailable) measurements. In [PaRa99, PaRa01] (see also the references therein), the effects of a non-conformally coupled and/or massive scalar field have been analysed and it has been found that an initially 'classical' cosmological expansion is altered in the low-curvature regime by quantum effects which mimic dynamical dark energy and can be matched to experimental data. Particularly, it seems that the authors find quantum effects to be important only at late times and low curvatures. Our impression is that these results do not hold once at least one conformally couple scalar field or any massless higher-spin field is included. Namely, in this case the trace anomaly terms, not being cancelled by the state-dependent terms, will certainly be important at high curvatures. Finally, our result is a generalisation of the results in [DFP08], as we consider arbitrary fields, compute the state dependent terms up to the curvature orders present in the trace anomaly, provide the solution for $C_3 \neq 0$, and consider experimental data to perform a more detailed analysis. Particularly, we share the new point of view taken in [DFP08] that the *upper branch* can be considered physical as well on account of the renormalisation freedom. We remark that the upper branch differs from the lower one by giving an interpretation of a late time cosmological constant in terms of fundamentally different quantum effects, whereas the lower branch can be interpreted as adding small quantum contributions to a 'classical' cosmological constant. We shall analyse these issues in a quantitative manner in the next section.

Finally, we point out the following fact. In contrast to the common folklore assumption, *quantum field theory does not predict a Planck scale cosmological constant*. It should be clear from our presentation that quantum field theory on curved spacetimes *does not predict any value of the cosmological constant*. Rather, the careful analysis of such theory shows that there are renormalisation degrees of freedom (determining the value of an effective cosmological constant), which, with the current lack of a full theory of quantum gravity, can only be fixed by experiment.

## IV.6 Comparison with Supernova Ia Measurements

Before we proceed to match both the lower and the upper branch solutions found in the preceding section to recent experimental data, we shall briefly introduce the relevant astrophysical observables. As of today, it is standard knowledge that our universe is expanding, which manifests itself in a measured non-vanishing *Hubble constant* $H_0 \doteq H(t_{\mathrm{now}})$, see the recent review [FrMa10]. To measure the expansion of the universe, one first considers that the light emitted from a distant object at time $t_{\mathrm{em}}$ at wavelength $\lambda_{\mathrm{em}}$ is redshifted by the expansion of the universe to a lower wavelength $\lambda_{\mathrm{now}}$, where the relative deviation between the two wavelengths is given by [Wal84, Dod03]

$$z \doteq \frac{\lambda_{\mathrm{em}} - \lambda_{\mathrm{now}}}{\lambda_{\mathrm{em}}} = \frac{a(t_{\mathrm{now}})}{a(t_{\mathrm{em}})} - 1.$$

$z$ is called the *redshift*, and setting $a(t_{\text{now}}) \doteq 1$, we have the well-known relation

$$z(t) = \frac{1}{a(t)} - 1 \,.$$

For nearby objects, the usual Doppler relation holds and the redshift is equal to the *recession speed* $v$ of the object (in units where the speed of light is set to 1), *i.e.*

$$z = v = \frac{d(a\,r)}{d\,t} = H_0\, r \,,$$

where $r$ is the comoving distance to the emitting object and $a\,r$ is its *physical distance* [Wal84, chap. 5.3.]. Hence, to determine the current Hubble rate $H_0$, one has to measure redshift and distance of nearby objects. The former is not a problem, one just matches the measured spectra of the considered objects to known template spectra, *e.g.* known emission or absorption line patterns. However, measuring the distance is non-trivial. A possibility is to know the intrinsic brightness of an object, measure its observed brightness, and argue that the difference must be due to the distance of the object. To be able to perform such a procedure, one needs *standard candles*. These are objects which are by some common physical mechanism known to have an empirical relationship between a property which can be measured very well and their intrinsic brightness. An example are *Cepheid variables*, these are stars which show a periodic variability in their brightness, and one knows that there is a correlation between their intrinsic brightness and their period. To determine such correlations, one naturally needs a couple of these objects which are so close nearby that their distance can be measured by some direct means. Indeed, Cepheid variables are among the objects one has used to determine the Hubble constant, see for instance [Dod03, FrMa10] for details.

However, one is not only interested in the current Hubble rate $H_0$, *i.e.* the current expansion *speed* of the universe, but also in its *acceleration* (and ultimately of course, in its complete evolution). To measure the acceleration of the universe and the full behaviour of $a(t)$, one needs to consider distant standard candles and the correct distance–redshift relationship. It turns out that this is given by [Dod03]

$$\mu \doteq m - M = 5 \log\left(\frac{d_L}{10\text{pc}}\right) + K \,, \tag{IV.137}$$

where $m$ is the *observed magnitude*, $M$ is the *absolute (intrinsic) magnitude*, $K$ is a suitable constant, 'pc' denotes the unit Parsec (1pc=$3.08 \times 10^{16}$m), and $d_L$ is the *luminosity distance* defined as

$$d_L(z) \doteq (1+z) \int\limits_0^z d\,z' \, \frac{1}{H(z')} \,.$$

Moreover, the standard candles used in this context are supernovae of type Ia. These are believed to emerge from a twin star system, where one star is a white dwarf, and the other one is a red giant. Without going too much into details, let us mention that the white dwarf slowly accretes matter from the red giant until it reaches the *Chandrasekhar limit* and collapses [HiNi00]. It has been observed that there is a strong correlation between the maximum brightness of the lightcurve resulting from this collapse and its shape, particularly, the width of the curve. Hence, to determine the intrinsic brightness $M$ of a supernova explosion, one fits its measured lightcurve into a template and reads off the brightness of the peak. There are various templates that are used, see for instance [Amm10], and these 'lightcurve-fitters' are 'trained' with data from nearby supernovae whose distance has been determined by other means. We would like to point out that the status of supernovae as standard candles is not undoubted, as the details of the related collapse are not completely understood. Particularly, it is believed that there might be a $z$-dependence of the lightcurves which is not taken into account, or an intrinsic reddening due to the particular surrounding (host galaxy) of the individual supernovae. However, we shall not be concerned with these issues here, and refer the reader to [HiNi00, Amm10] and references therein for details.

Supernova measurements indeed show that our universe is accelerating and, because they constitute the only experimental data which is available in a whole redshift interval rather than only at selected special points, their analysis is a major part of the input to the currently ruling *ΛCDM cosmological concordance model* which entails that the current matter-energy density of the universe is constituted by 70% dark energy/cosmological constant and 30% dust. [Amm10]. We shall now proceed to compare the late time cosmological evolution given by the solutions of the semiclassical Einstein equation with the measured supernova Ia data, under the hypothesis that only the quantum energy density $\varrho_Q$ found in the previous subsection and no classical energy density is present. To this end, we choose the recent *Union2 compilation* provided in [Amm10]. This is a selected compilation of various supernova surveys with a total number of 556 supernovae ranging from $z = 0.01$ up to $z = 1.55$. Note that a data point at $z = 1$ corresponds to a supernova which happened roughly 9 billion years ago.

We start with the upper branch and state its parametrisation in terms of the redshift, *viz.*

$$H_+(z) = \sqrt{H_*^2 + \sqrt{H_*^4 - C_1(1+z)^4 - C_2(1+z)^3 - C_3}}\,.$$

To underline that the dark energy-behaviour of the upper branch deviates significantly from a cosmological constant, we discard the 'bare cosmological constant' contribution in $H^+(z)$ by setting $C_3 = 0$. We determine the best-fit parameters of our model by minimising[20]

$$\chi^2(H_*, C_1, C_2) \doteq \sum_{i \in \{\text{Data}\}} \left( \frac{\mu_{i,\text{obs}}(z_{\text{obs}}) - \mu(z)}{\Delta\mu_{i,\text{obs}}} \right)^2,$$

where $\mu_{i,\text{obs}}$ denote the $\mu$ of specific supernovae listed in the Union2 compilation tables, $\Delta\mu_{i,\text{obs}}$ their errors, $\mu(z)$ the parameter-dependent $\mu$ provided by our solution $H(z)$ via (IV.137). It is possible to remove the constant $K$ present in $\mu$ from the fitting procedure by *marginalisation*, see for instance [Wei10]. This is indeed necessary to deduce $H_0$ from the supernova data. However, we shall not do so, as we are mainly interested in the time variation of $H$ rather than in $H_0$. One usually states the result of such a fit in the form $\chi^2/\text{dof}$, where dof denotes the *degrees of freedom* of a fit, calculated as the number of data points minus the number of fit parameters plus one. To compare our results with the concordance model, we also provide the fit to this scenario, which is given by the solution of the Friedmann equation

$$3H^2 = 8\pi G \left( \varrho_{\text{CC}} + \frac{\varrho_{\text{matter}}}{a^3} \right)$$

and, hence, by

$$H(z) = \sqrt{\frac{8\pi G}{3}} \sqrt{\varrho_{\text{CC}} + \varrho_{\text{matter}}(1+z)^3}.$$

---

[20]We are very grateful to Christian Hambrock for providing us his $\chi^2$-fit script.

The above described fitting procedure yields the best fit parameters $H_* = 0.570 H_0$, $C_1 = -0.273 H_0^4$, $C_2 = 0.270 H_0^4$. Moreover, we find that the upper branch of the solution of the semiclassical Einstein equation determined in the previous subsection is capable of fitting the Union2 set as good as the concordance model ($\chi^2_{\min}/\text{dof} = 0.98$ in both cases), but not better, see figure 5. Nevertheless, we point out that the upper branch solution provides an *explanation* of the origin of dark energy in that it shows that *every free and massless quantum field displays generically the behaviour of a late time cosmological constant*. Therefore, our results are not bound to a specific model, but rather an inevitable consequence of quantum fields propagating on curved spacetimes, provided one considers the semiclassical Einstein equation to be a sensible equation. Supernova data fits are certainly only a first step to determine the actual physical relevance of the upper branch, and further checks are necessary. Particularly, we note that, because of the additional square root induced by the anomalous $H^4$ term, there is no component in the effective energy density $3H_+^2(z)(8\pi G)^{-1}$ which scales like classical matter, *i.e.* $a^{-3}$. However, it is well-known that a sufficiently large effective equation of state parameter $w_{\text{eff}} = p_{\text{total}}/\varrho_{\text{total}} > -\frac{1}{3}$ is necessary to account for structure formation, and it is usually assumed that structure formation happens in the matter-dominated phase (see for instance [Dod03]). Hence, it seems natural to ask if $w_{\text{eff}} > -\frac{1}{3}$ is possible on the upper branch and we can answer this question to the positive, see figure IV.4. Admittedly, there are still large differences between the standard matter-dominated phase in $\Lambda$CDM and the $w_{\text{eff}} > -\frac{1}{3}$ phase in the upper branch solution, and further analyses are necessary to give a definite answer on whether the upper branch can really describe the cosmological reality.

Let us now consider the lower branch, which, given in terms of $z$, reads

$$H_-(z) = \sqrt{H_*^2 - \sqrt{H_*^4 - C_1(1+z)^4 - C_2(1+z)^3 - C_3}}.$$

For a sensible physical interpretation, we assume that the parameters are such that the non-Big Bang singularity, where the feasibility of our solution breaks down, occurs at $z \gg 1$. This amounts to require that $H_*^4$ is much larger than $C_i$, $i = 1, 2, 3$. Hence, we can expand $H_-(z)$ as

$$H_-(z) \doteq \sqrt{K_0 + K_1(1+z)^4 + K_2(1+z)^3 + K_3(1+z)^6 + K_4(1+z)^7 + O\left(\frac{C_i^3}{H_*^{10}}\right)},$$

where $K_i$ are given in terms of the $C_i$ and $H_*$. As already anticipated, it is visible that the lower branch can be phenomenologically interpreted as $\Lambda$CDM plus quantum corrections, where the latter are here constituted by the $K_3$, $K_4$ and the omitted terms. To investigate the magnitude of the quantum corrections in the best fit, we set $K_1 = 0 \Leftrightarrow C_1 = 0 \wedge K_4 = 0$, as radiation is usually thought to have no influence on the recent cosmological expansion in $\Lambda$CDM. The error bars in the data seem to be insufficient to give a definite answer on the magnitude of the quantum corrections. In fact, the parameter space is so degenerate that any $K_3$ ranging from zero to $10^{-3}H_0^4$ gives an equally good fit, whereas the other best fit parameters are $K_0 = 0.73H_0^2$, $K_2 = 0.27H_0^2$. Hence, on the currently available supernova Ia data, the lower branch is indistinguishable from $\Lambda$CDM. However, it is important to point out that the dark energy behaviour of the lower branch is still of dynamical nature and deviates from a pure cosmological constant at large $z$. Moreover, as

$$K_2 = \frac{C_2 C_3 + 2C_2 H_*^4}{4H_*^6} \propto \left(3N_{1/2} + N_0\right)mT^3,$$

we find that a thermal component of the quantum energy density displays the scaling behaviour of classical matter. However, having a $T^3$ rather than a volume dependence, it can *not* be interpreted as a density of massive non-relativistic particles. As already mentioned, field masses $m$ and field multiplicities $N_0$, $N_{1/2}$ are degenerate, hence, one can not infer a specific mass, temperature, or field multiplicity from the best fit. Nevertheless, the mentioned scaling behaviour of the thermal contribution to $\varrho_Q$ opens the tantalising possibility that dark matter, at least on cosmological scales, is not a new, weakly interacting quantum field which is not present in the standard model of particle physics, but maybe only a low-temperature effect of standard model fields. A deeper analysis of such topic is certainly in order, and we hope to return to this issue at a later occasion.

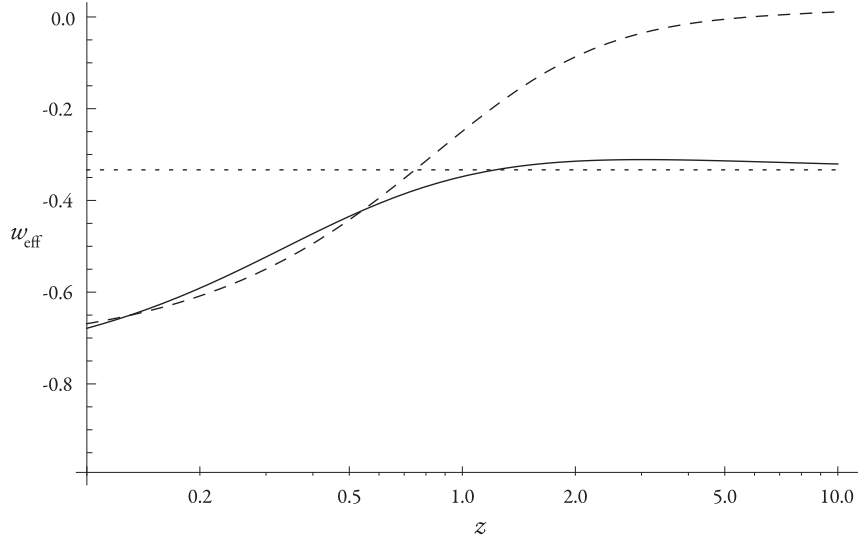

*Figure IV.4:* Plot of the effective equation of state parameter $w_{\text{eff}} \doteq \frac{p_{\text{total}}}{\varrho_{\text{total}}} = -1 - \frac{2}{3}\frac{\dot{H}}{H^2}$ for the best fit upper branch (solid line) and the $\Lambda$CDM model/best fit lower branch (dashed line). The dotted line displays the minimum $w_{\text{eff}}$, $-\frac{1}{3}$, necessary for structure formation.

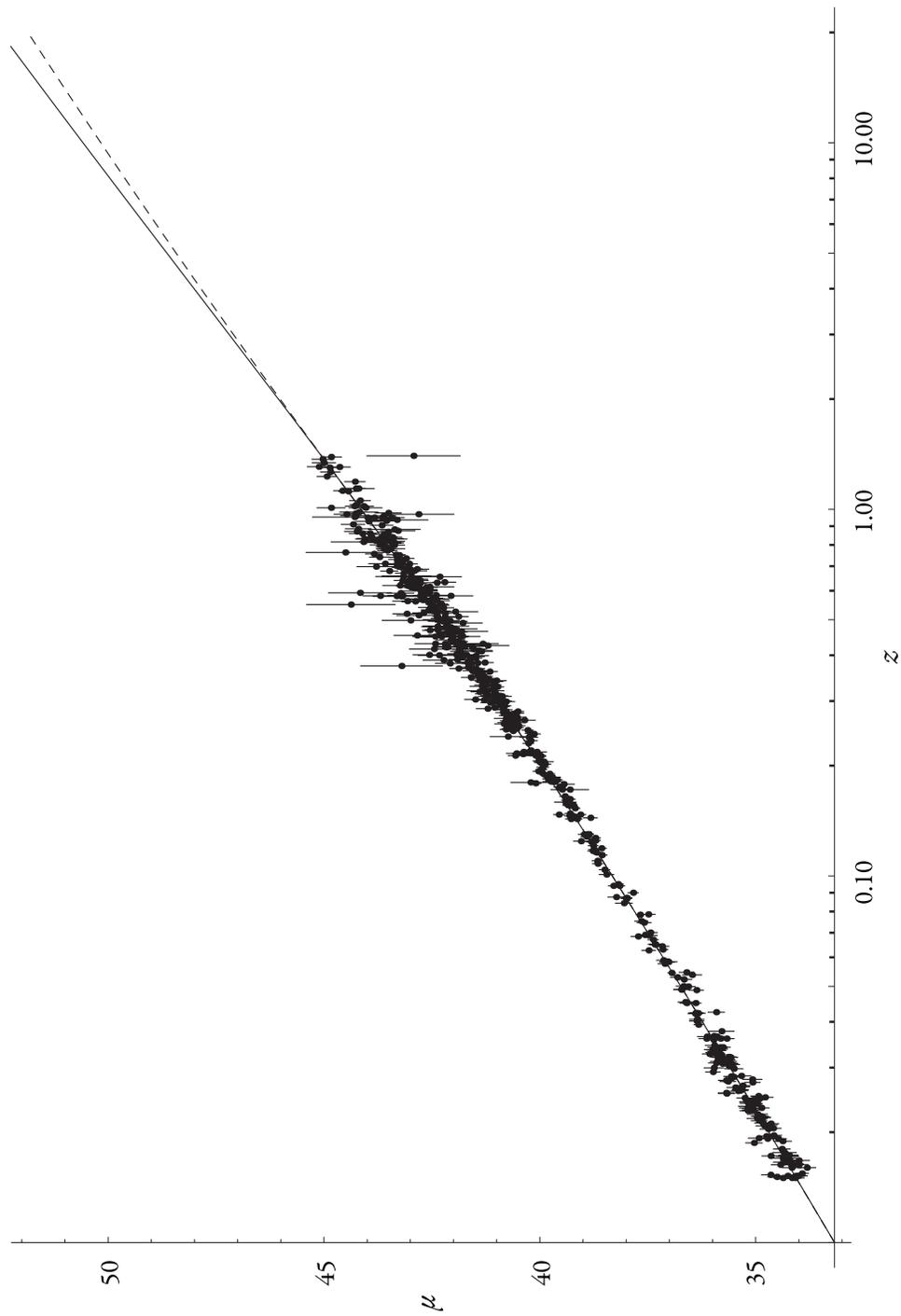

*Figure 5.* Best fits of the Union2 set [Amm10]. Dashed line: best fit $\Lambda$CDM model/best fit lower branch solution. Solid line: best fit upper branch solution.

# Conclusions

We shall now give a brief synopsis of our results. To start with, we have been able to successfully generalise the deformation quantization approach which had been well-developed for scalar fields [BDF09] also to the case of Dirac fields in curved spacetimes. As a result, we have constructed the enlarged algebra of observables of the Dirac fields. The starting point of this construction has been the classical algebra of observables constituted by antisymmetric smooth and compactly supported sections and endowed with an antisymmetric tensor product. This has been promoted to the algebra of quantum observables of the Dirac field by replacing the antisymmetric classical product with a quantum product $\star_S$ encoding the canonical anticommutation relations determined by the causal propagator $S(x, y)$ of the Dirac equation. In a subsequent step, this basic algebra of observables has been enlarged by considering, on the one hand, compactly supported distributions with a specific wavefront set and, on the other hand, a new $\star$-product $\star_H$ obtained by replacing $S$ by a Hadamard parametrix $H(x, y)$. We have argued that this new algebra is well-defined, contains local and covariant Wick polynomials like $:\psi^\dagger\psi:$, and that its product $\star_H$ provides an implementation of Wick's theorem in a way which is independent of any state or its Hilbert space representation.

In a next step, we have considered an important element of the enlarged algebra of observables, namely, the stress-energy tensor $:T_{\mu\nu}:$ of the quantum Dirac field. Following the concepts introduced and applied in [Mor03] for the Klein-Gordon case, we have been able to provide a conceptually clean and rigorous point-splitting regularisation prescription of the expectation value $\omega(:T_{\mu\nu}:)$ of $:T_{\mu\nu}:$ in an arbitrary Hadamard state $\omega$ also in the Dirac case. The obstruction that one had to overcome is the following: as one is interested in obtaining a local and covariant definition of $:T_{\mu\nu}:$, one defines the regularisation of $\omega(:T_{\mu\nu}:)$ by means of subtracting only the Hadamard singularity of the two-point function $\omega_2(x, y)$ of the Hadamard state $\omega$, but not the full $\omega_2$. Hence, the conservation of $\omega(:T_{\mu\nu}:)$, which in a naive approach would be assured on account of $\omega_2$ being a bisolution of the Dirac equation, is not automatically fulfilled. To retain the local covariance of $\omega(:T_{\mu\nu}:)$ but still assure its covariant conservation, we have modified the classical stress energy tensor by adding terms which classically vanish on shell but have non-vanishing locally covariant quantum Wick polynomial counterparts. Thereby, a $\omega(:T_{\mu\nu}:)$ has been obtained which can be understood to yield the original classical stress-energy tensor in the classical, on-shell limit. As is well-known due to results of Wald [Wal78a], a local, covariant, and covariantly conserved definition of $\omega(:T_{\mu\nu}:)$ inevitably leads to a trace anomaly, namely, the trace of $\omega(:T_{\mu\nu}:)$ is not vanishing in the conformally invariant, *i.e.* massless, case. Indeed, we have been able to compute the trace anomaly of $\omega(:T_{\mu\nu}:)$ explicitly. In order to achieve this and to construct a conserved $\omega(:T_{\mu\nu}:)$, detailed and cumbersome calculations of the Hadamard coefficients of Dirac fields had been necessary.

The Dirac trace anomaly we have found is matching previous computations performed by means of the DeWitt-Schwinger point-splitting prescription in [Chr78]. This is somehow not a surprise, because it had been well-known that the Hadamard and DeWitt-Schwinger coefficients are essentially the same. However, it seems that a strict and clear relation between the Hadamard and DeWitt-Schwinger prescriptions has never been given explicitly, and we have managed to fill this gap. Particularly, we have been able to prove that the DeWitt-Schwinger regularisation can be formulated entirely in a well-defined way.

Apart from the above mentioned results, valid on general curved spacetimes, we have also studied a special class of flat Robertson-Walker spacetimes. Such spacetimes are entirely determined by the behaviour of their scale factor $a$, and we have chosen a subclass where $a(\tau)$ expressed in terms of the conformal time $\tau$ is exponentially vanishing towards $\tau \to -\infty$. This has entailed in particular that these spacetimes possess a Big Bang at finite cosmological time $t$, and that the corresponding constant-time hypersurface is lightlike. Hence, we have termed these spacetimes null Big Bang spacetimes (NBB). NBB spacetimes had already been studied in [DFP08, Pin10], particularly, it has been shown that these spacetimes arise naturally as the solutions of the semiclassical Einstein equation

$$G_{\mu\nu} = 8\pi G \omega \left( :T_{\mu\nu}: \right)$$

sourced by the stress-energy tensor of quantized Klein-Gordon fields. Moreover, in [Pin10], it had been shown that NBB spacetimes possess a distinguished Hadamard state for Klein-Gordon fields, which can be interpreted as an asymptotic conformal vacuum state towards $\tau \to -\infty$. We have added to these results by constructing asymptotic conformal ground states for the Klein-Gordon field and asymptotic conformal ground and equilibrium states for the Dirac field, where all these states have been shown to be of Hadamard type. To achieve these results, we have applied and generalised the holographic methods developed in [DMP06, DMP09a, DMP09b, DMP09c, Mor06, Mor08] for the case of Klein-Gordon fields. The underlying idea is the following. One considers spacetimes which possess a suitable, highly symmetric, null hypersurface, in our case constituted by the Big Bang hypersurface $\mathfrak{I}^-$ of NBB spacetimes. One seeks a way to map classical solutions of a field equation in a well-defined way to $\mathfrak{I}^-$ and then uses this classical map to obtain a map on the level of quantum fields. In this way, one constructs a full-fledged quantum field theory on $\mathfrak{I}^-$ and subsequently exploits the high symmetry of $\mathfrak{I}^-$ to identify preferred quantum states, *i.e.* ground states and KMS states in our case. Finally, using the relation between quantum fields on the bulk NBB spacetime and the same objects on $\mathfrak{I}^-$, it is possible to pull-back the preferred ground and equilibrium states on $\mathfrak{I}^-$ to obtain states for the bulk quantum fields, which then can be interpreted as asymptotic conformal ground and equilibrium states.

In the last stages of the thesis, we have combined the somehow different lines of research constituted by the extended algebra and stress-energy tensor of Dirac fields on general curved spacetimes on the one hand, and the analysis of preferred states on NBB spacetimes on the other hand, in that we have considered solutions of the semiclassical Einstein equation on flat Robertson-Walker spacetimes and for quantum fields of arbitrary spin. In contrast to most earlier works on the subject and generalising the results in [DFP08], we have taken into account the cosmological effects of both the often analysed trace anomaly and the rarely considered state-dependent contribution and finite renormalisation freedom induced only by massive fields. As a result, we have found solutions of the semiclassical Einstein equation which are partially of power-law inflationary type, but all display a late time de Sitter behaviour, where all these features are independent of the considered field content and hold for arbitrary spins. Moreover, we have compared the cosmological dynamics provided by these solutions with measurements of supernova Ia data. In this way, we have found that both families of semiclassical solutions explain the data as good as the cosmological concordance model. Moreover, the upper branch solution differs fundamentally from the $\Lambda$CDM concordance model, while the lower branch can be interpreted as the $\Lambda$CDM model plus quantum corrections. Thus, quantum fields on curved spacetimes provide a natural and dynamical explanation of dark energy which is difficult to discard. Furthermore, it is often said that quantum field theory predicts a too large value of the cosmological constant. We find that this is by no means true. In fact, as it is well-known within the community of quantum field theory on curved spacetime, this theory does not predict any value of the cosmological constant. In contrast, it has inherent renormalisation degrees of freedom which have to be fixed by experiment, and, we find that these can indeed be fixed in a way which perfectly matches the cosmological evolution inferred from supernova Ia measurements.

To close, let us comment on some prospects and open questions. First, it is mandatory to extend the analysis we have performed to the more general and realistic case of interacting fields. While we do not expect new results on the qualitative level regarding the recent cosmological history, one might anticipate that interactions among matter fields become important in the early universe. In this respect, it is necessary to find ways to compute the state-dependent terms also at high curvatures, maybe by means of numerical computations in the spirit of [And85, And86]. Finally, we have found that a thermal component in the quantum energy density mimics the behaviour of dark matter on cosmological scales. This phenomenon definitely deserves more attention and a deeper analysis, as it might ultimately lead to the insight that what be believe to be dark matter is only a tiny thermal effect of standard model quantum fields.

# Acknowledgements

An erster Stelle gilt mein Dank dem Betreuer meiner Dissertation, Professor Klaus Fredenhagen. Ich bin ihm zutiefst dankbar, dass er angeregt und es mir ermöglicht hat, auf dem äußerst spannenden Gebiet kosmologischer Anwendungen der Quantenfeldtheorie auf gekrümmten Raumzeiten zu arbeiten. Bedanken möchte ich mich auch für die aufschlußreichen Diskussionen, die zahlreichen wertvollen Hinweise und Hilfestellungen und die Geduld die er in der Anfangszeit mit mir gehabt hat. Nicht zuletzt bin ich ihm dankbar für seine äußerst klaren und undogmatischen Antworten auf fundamentale Fragestellungen und für die zahlreichen Dinge, die ich in meiner Zeit in Hamburg lernen durfte.

I am very grateful to Valter Moretti for agreeing to be the second referee of my thesis despite of his limited time. I deeply thank him for his numerous valuable comments and suggestions and especially for the great hospitality he offered me together with Romeo Brunetti and little Bianca during my stay in October 2009 in Trento.

It is an honour to thank Professor Robert Wald for agreeing to be the third referee of my thesis.

Thanks to my comrades Claudio Dappiaggi and Nicola Pinamonti. Thanks for the great projects we have been working on together, thanks for the many laughs, and thanks a lot for sharing the spirit (but not the spirits).

Ein besonderer Dank gilt meinem Kollegen Kai Keller für die schöne Zeit und nicht zuletzt dafür, dass er am Ende meiner Promotionszeit in rührender Weise alles und jeden mit allen Mitteln von mir ferngehalten und mir damit ein konzentriertes Arbeiten ermöglicht hat. Danke auch an Paola Arias, Bruno Chilian, Thomas Creutzig, Andreas Degner, Elisabeth Duarte-Monteiro, Juliane Grossehelweg, Jan Heisig, Marc Hoge, Muharrem Küskü, Sebastian Jakobs, Benjamin Lang, Pedro Lauridsen-Ribeiro, Falk Lindner, Martin Porrmann, Katarzyna Rejzner, Jan Schlemmer, Lilja Schmidt, Kolja Them, Ole Vollertsen, Jan Weise und Matthias Westrich für eine schöne gemeinsame, lehrreiche und lustige Zeit und besonders für das rührende Mitfiebern am Ende.

Mein herzlicher Dank gilt Jan Möller für das Vorschlagen des gemeinsamen spannenden Projektes, das für mich den persönlichen Höhepunkt meiner Promotionszeit darstellt. Danke auch für die zahlreichen herzlichen, hartnäckigen, fundamentalen und äußerst lehrreichen Diskussionen.

Der größte Dank gilt zweifellos meinen Eltern, meiner Schwester Julia und meiner Freundin Kathrin. Danke für die gemeinsame vergangene und zukünftige Zeit, danke für eure unersetzliche Geduld und Unterstützung. Die vorliegende Arbeit ist euch gewidmet. ♬

⋆ ⋆ ⋆

*Begin at the beginning, and go on till you come to the end: then stop.*

The *King of Hearts* (presumably adressing acknowledgement-only-readers) in: Lewis Carroll, *Alice's Adventures in Wonderland*